\documentclass{cernrep}
\usepackage{texnames}
\usepackage[T1]{fontenc}
\long\def\comment#1{ }
\newcommand{\beq}{\begin{equation}}
\newcommand{\eeq}{\end{equation}}
\newcommand{\nn}{\nonumber\\}
\newcommand{\order}[1]{\mathcal{O}{(#1)}}
\newcommand{\Lam}{\Lambda_{{\rm QCD}}}
\newcommand{\abar}{\bar{\alpha}_s}

\newcommand{\bmk}{\bm{k}}
\newcommand{\bmp}{\bm{p}}
\newcommand{\bmq}{\bm{q}}
\newcommand{\bmx}{\bm{x}}
\newcommand{\bmy}{\bm{y}}
\newcommand{\bmb}{\bm{b}}
\newcommand{\bmr}{\bm{r}}

\begin{document}
\title{QCD in heavy ion collisions\footnote{Based on lectures
presented at the 2011 European School of High--Energy Physics
(ESHEP2011), 7-20 September 2011, Cheile Gradistei, Romania.}}

\author{Edmond Iancu}

\institute{Institut de Physique Th\'{e}orique de Saclay, F-91191
Gif-sur-Yvette, France}

\maketitle 

\begin{abstract}
These lectures provide a modern introduction to selected topics in the
physics of ultrarelativistic heavy ion collisions which shed light on the
fundamental theory of strong interactions, the Quantum Chromodynamics.
The emphasis is on the partonic forms of QCD matter which exist in the
early and intermediate stages of a collision --- the colour glass
condensate, the glasma, and the quark--gluon plasma --- and on the
effective theories that are used for their description. These theories
provide qualitative and even quantitative insight into a wealth of
remarkable phenomena observed in nucleus--nucleus or deuteron--nucleus
collisions at RHIC and/or the LHC, like the suppression of particle
production and of azimuthal correlations at forward rapidities, the
energy and centrality dependence of the multiplicities, the ridge effect,
the limiting fragmentation, the jet quenching, or the dijet asymmetry.

\end{abstract}

\tableofcontents

\bigskip
\section{Introduction} \label{sec:intro}

With the advent of the high--energy colliders RHIC (the Relativistic
Heavy Ion Collider operating at RHIC since 2000) and the LHC (the Large
Hadron Collider which started operating at CERN in 2008), the physics of
relativistic heavy ion collisions has entered a new era: the energies
available for the collisions are high enough --- up to 200~GeV per
interacting nucleon pair at RHIC and potentially up to 5.5~TeV at the LHC
(although so far one has reached `only' 2.76~TeV) ---- to ensure that
{\em new forms of QCD matter}, characterized by high parton densities,
are being explored by the collisions. These new forms of matter refer to
both the wavefunctions of the incoming nuclei, prior to the collision,
which develop high gluon densities leading to {\em colour glass
condensates}, and the partonic matter produced in the intermediate stages
of the collision, which is expected to form a {\em quark--gluon plasma}.
The asymptotic freedom property of QCD implies that these high--density
forms of matter are {\em weakly coupled} (at least in so far as their
bulk properties are concerned) and hence can be studied via controlled
calculations within perturbative QCD. But such studies remain difficult
and pose many challenges to the theorists: precisely because of their
high density, these new forms of matter are the realm of {\em collective,
non--linear phenomena}, whose mathematical description often transcends
the ordinary perturbation theory. Moreover, there are also phenomena
(first revealed by the experiments at RHIC) which seem to elude a
weak--coupling description and call for non--perturbative techniques.

These challenges stimulated new ideas and the development of new
theoretical tools aiming at a fundamental understanding of {\em QCD
matter under extreme conditions}~: high energy, high parton densities,
high temperature. The ongoing experimental programs at RHIC and the LHC
provide a unique and timely opportunity to test such new ideas, constrain
or reject models, and orient the theoretical developments. Over the last
decade, the experimental and theoretical efforts have gone hand in hand,
leading to a continuously improving physical picture, which is by now
well rooted in QCD. The purpose of these lectures is to provide an
introduction to this physical picture, with emphasis on those aspects of
the dynamics for which we are confident to have a reasonably good
(although still far from perfect) understanding from first principles,
i.e. from the Lagrangian of Quantum Chromodynamics. These aspects concern
the {\em partonic stages} of a heavy ion collision, at sufficiently early
times. These are also the stages to which refers most of the experimental
and theoretical progress over the last decade.

\section{Stages of a heavy ion collision: the case
for effective theories} \label{sec:stages}

The theoretically motivated space--time picture of a heavy ion collision
(HIC) is depicted in \Fref{fig:HIC}. This illustrates the various forms
of QCD matter intervening during the successive phases of the collision:

\begin{enumerate}

\item Prior to the collision, and in the center-of-mass frame (which
    at RHIC and the LHC is the same as the laboratory frame), the two
    incoming nuclei look as two Lorentz--contracted `pancakes', with
    a longitudinal extent smaller by a factor $\gamma\sim 100$ (the
    Lorentz boost factor) than the radial extent in the transverse
    plane. As we shall see, these `pancakes' are mostly composed with
    {\em gluons} which carry only tiny fractions $x\ll 1$ of the
    longitudinal momenta of their parent nucleons, but whose density
    is rapidly increasing with $1/x$. By the uncertainty principle,
    the gluons which make up such a high--density system carry {\em
    relatively large transverse momenta}. A typical value for such a
    gluon in a Pb or Au nucleus is $k_\perp\simeq 2$~GeV for
    $x=10^{-4}$. By the `asymptotic freedom' property of QCD, the
    gauge coupling which governs the mutual interactions of these
    gluons is {\em relatively weak}. This gluonic form of matter,
    which is dense and weakly coupled, and dominates the wavefunction
    of any hadron (nucleon or nucleus) at sufficiently high energy,
    is {\em universal} --- its properties are the same form all
    hadrons. It is known as the {\em colour glass condensate} (CGC).

\item At time $\tau=0$, the two nuclei hit with each other and the
    interactions start developing. The `hard' processes, i.e. those
    involving relatively large transferred momenta $Q\gtrsim 10$~GeV,
    are those which occur faster (within a time $\tau\sim 1/Q$, by
    the uncertainty principle\footnote{Throughout these notes, we
    shall generally use the natural system of units $\hbar=c=k_B=1$,
    so in particular there is no explicit factor $\hbar$ in the
    uncertainty principle. Yet, in some cases, we shall restore this
    factor for more clarity.}). These processes are
responsible for the production of `hard
    particles', i.e. particles carrying transverse energies and
    momenta of the order of $Q$. Such particles, like (hadronic)
    jets, direct photons, dilepton pairs, heavy quarks, or vector
    bosons, are generally the most striking ingredients of the final
    state and are often used to characterize the topology of the
    latter --- \eg one speaks about `a dijet event', cf.
    \Fref{fig:events} left, or `a photon--jet' event, cf.
    \Fref{fig:events} right.

\begin{figure}[t]
\begin{center}
\includegraphics[width=.9\textwidth]{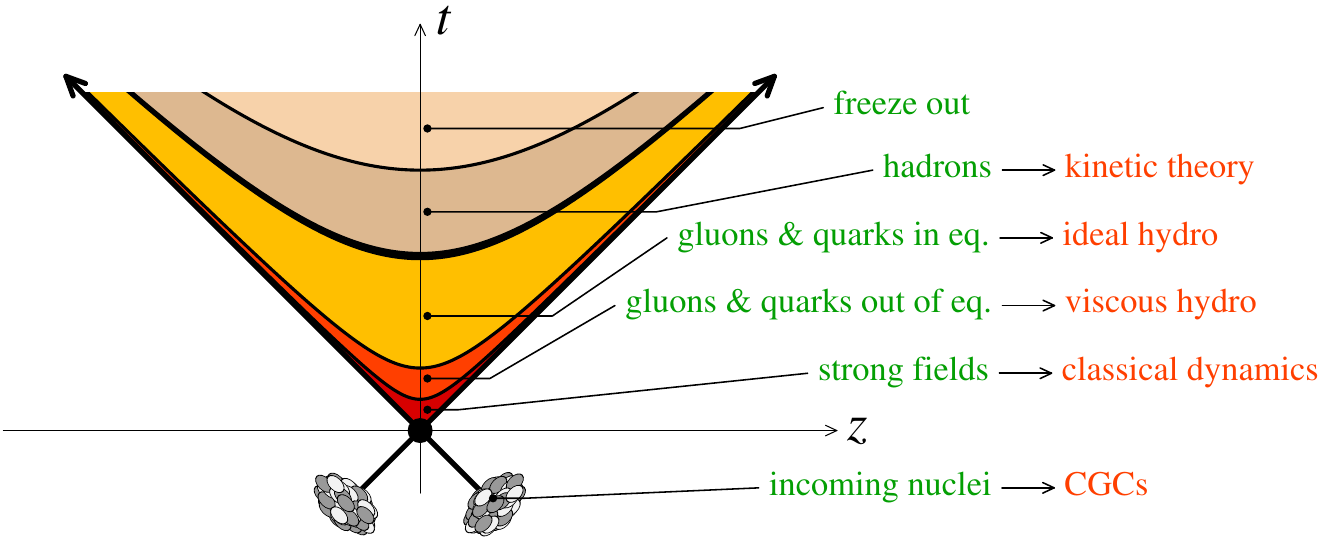}
\caption{Schematic representation of the various stages of a HIC as a function
of time $t$ and the longitudinal coordinate  $z$ (the collision axis).
The `time' variable which is used in the discussion in the text is the
{\em proper time} $\tau\equiv\sqrt{t^2-z^2}$, which has a Lorentz--invariant
meaning and is constant along the hyperbolic curves separating
various stages in this figure.} \label{fig:HIC}
\end{center}
\end{figure}

\item At a time $\tau\sim 0.2$~fm/c, corresponding to a `semi-hard'
    transverse momentum scale $Q\sim 1$~GeV, the bulk of the partonic
    constituents of the colliding nuclei (meaning the gluons
    composing the respective CGCs) are liberated by the collision.
    This is when most of the `multiplicity' in the final state is
    generated; that is, most of the hadrons eventually seen in the
    detectors are produced via the fragmentation and the
    hadronisation of the initial--state gluons liberated at this
    stage. But before ending up in the detectors, these partons
    undergo a complex evolution. Just after being liberated, they
    form a relatively dense medium, whose average density energy in
    Pb+Pb collisions at the LHC is estimated as $\varepsilon \gtrsim
    15$~GeV/fm$^3$; this is about 10 times larger than the density of
    nuclear matter and 3 times larger than in Au+Au collisions at
    RHIC. This non--equilibrium state of partonic matter, which
    besides its high density has also other distinguished features to
    be discussed later, is known as the {\em glasma}.

\begin{figure}[ht]
\begin{center}\centerline{
\includegraphics[width=.5\textwidth]{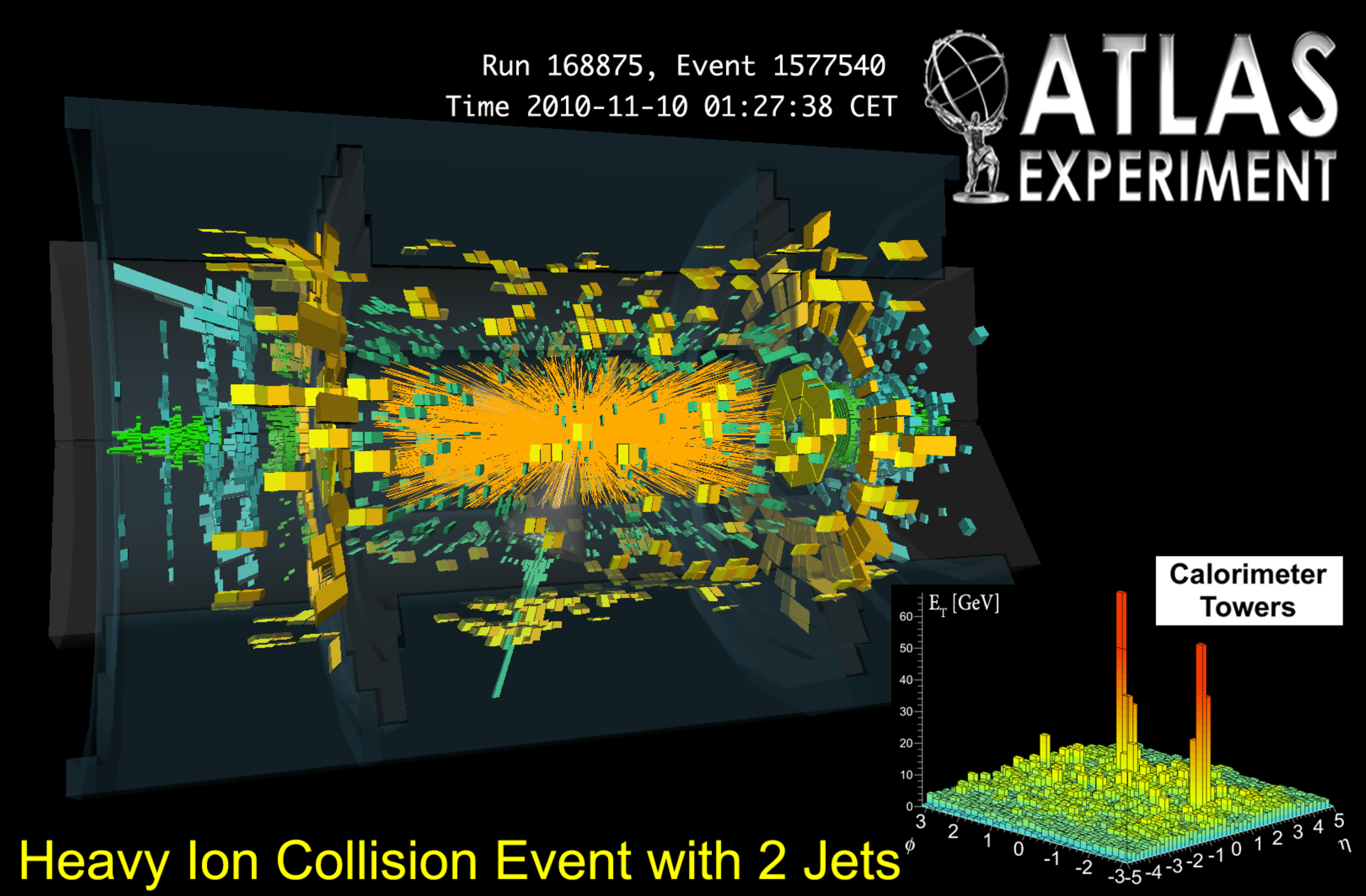}\qquad
\includegraphics[width=.4\textwidth]{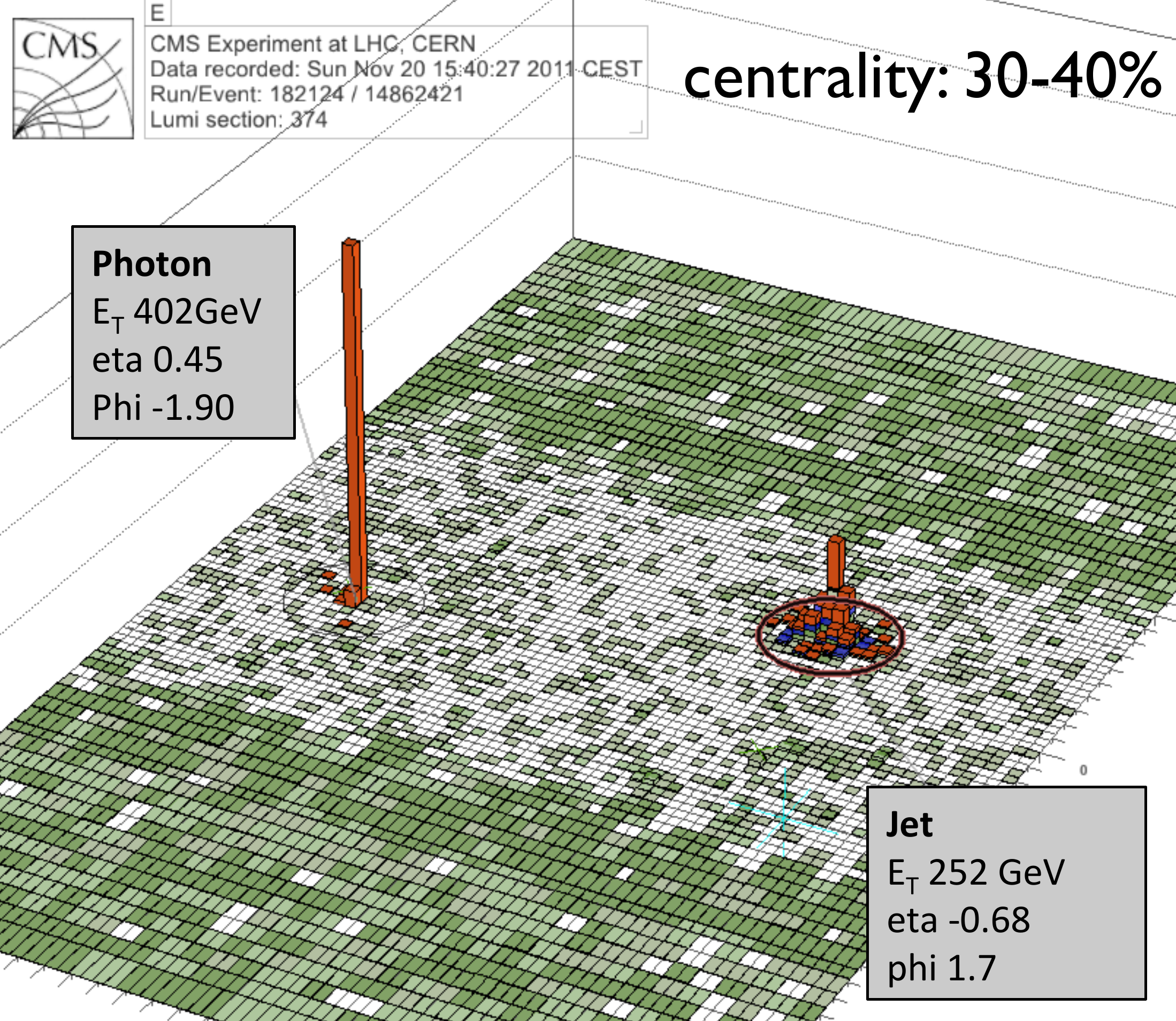}
}
\caption{\sl A couple of di--jets events in Pb+Pb collisions at ATLAS
(left) and CMS (right).}
\label{fig:events}
\end{center}
\end{figure}

\item If the produced partons did not interact with each other, or if
    these interactions were negligible, then they would rapidly
    separate from each other and independently evolve (via
    fragmentation and hadronization) towards the final--state
    hadrons. This is, roughly speaking, the situation in
    proton--proton collisions. But the data for heavy ion collisions
    at both RHIC and the LHC exhibit collective phenomena (like the
    `elliptic flow' to be discussed later) which clearly show that
    the partons liberated by the collision {\em do} actually interact
    with each other, and quite strongly. A striking consequence of
    these interactions is the fact that this partonic matter rapidly
    approaches towards {\em thermal equilibrium}~: the data are
    consistent with a relatively short thermalization time, of order
    $\tau\sim 1$~fm/c. This is striking since it requires rather
    strong interactions among the partons, which can compete with the
    medium expansion: these interactions have to redistribute energy
    and momentum among the partons, in spite of the fact that the
    latter separate quite fast away from each other. Such a rapid
    thermalization seems incompatible with perturbative calculations
    at weak coupling and represents a main argument in favour of a
    new paradigm: the dense partonic matter produced in the
    intermediate stages of a HIC may actually be a {\em strongly
    coupled fluid}.

\item The outcome of this thermalization process is the
    high--temperature phase of QCD known as the {\em quark--gluon
    plasma}. The abundant production and detailed study of this phase
    is the Holy Grail of the heavy ion programs at RHIC and the LHC.
    The existence of this phase is well established via theoretical
    calculations on the lattice, but its experimental production
    within a HIC is at best ephemeral: the partonic matter keeps
    expanding and cooling down (which in particular implies that the
    temperature is space and time dependent, i.e. thermal equilibrium
    is reached only {\em locally}) and it eventually hadronizes ---
    the `coloured' quark and gluons get trapped within colourless
    hadrons. Hadronization occurs when the (local) temperature
    becomes of the order of the critical temperature $T_c$ for
    deconfinement, known from lattice QCD studies as $T_c\simeq
    150\div 180$~MeV. In Pb+Pb collisions at the LHC, this is
    estimated to happen around a time $\tau\sim 10$~fm/c.

\item For larger times $10\lesssim\tau\lesssim 20$~fm/c, this
    hadronic system is still relatively dense, so it preserves local
    thermal equilibrium while expanding. One then speaks of a {\em
    hot hadron gas}, whose temperature and density are however
    decreasing with time.

\item Around a time $\tau\sim 20$~fm/c, the density becomes so low
    that the hadrons stop interacting with each other. That is, the
    collision rate becomes smaller than the expansion rate. This
    transition between a fluid state (where the hadrons undergo many
    collisions) and a system of free particles is referred to as the
    {\em freeze--out}. From that moment on, the hadrons undergo free
    streaming until they reach the detector. One generally expects
    that the momentum distribution of the outgoing particles is
    essentially the same as their thermal distribution within the
    fluid, towards the late stages of the expansion, just before the
    freeze--out. This assumption appears to be confirmed by the data:
    the particle spectra as measured by the detectors can be well
    described as {\em thermal} (Maxwell--Boltzmann) distributions,
    with only few free parameters, like the fluid temperature and
    velocity at the time of freeze--out. This is generally seen as an
    additional argument in favour of thermalization, but one must be
    cautious on that, since the mechanism of hadronisation itself can
    lead to spectra which are apparently thermal. As a matter of
    fact, the freeze--out temperature extracted from the ratios of
    particle abundances at RHIC appears to be the same, $T_f\simeq
    170$~MeV, in both Au+Au and p+p collisions, while of course no
    QGP phase is expected in p+p.

\end{enumerate}

Although extremely schematic, this simple enumeration of the various
stages of a HIC already illustrates the variety and complexity of the
forms of matter traversed by the QCD matter liberated by the collision on
its way to the detectors. In principle, all these forms of matter and
their mutual transformations admit an unambiguous theoretical description
in the framework of Quantum Chromodynamics, which is the fundamental
theory of strong interactions. But although this theory exists with us
for about 40 years, it is still far from having delivered all its
secrets. Indeed, in spite of the apparent simplicity of its Lagrangian,
which looks hardly more complicated than that of the quantum
electrodynamics (the theory of photons and electrons), the QCD dynamics
is considerably richer and more complicated --- which is why it can
accommodate so many phases! What renders the theoretical study of HIC's
so difficult is the extreme complexity of the relevant forms of hadronic
matter, characterized by high (parton or hadron) densities and strong
collective phenomena. For a theorist, the most efficient way to try and
organize this complexity is to build {\em effective theories}.

An `effective theory' should not be confused with a `model': its main
purpose is not to provide a heuristic description of the data using some
physical guidance together with a set of free parameters. Rather, it aims
at a fundamental understanding and its construction is always guided by
the underlying fundamental theory --- here, QCD. Specifically, an
effective theory is a simplified version of the fundamental theory which
includes the `soft' ({\em i.e.} low energy and momentum) degrees of
freedom (d.o.f.) required for the description of the physical phenomena
occurring at a relatively large space--time scale, but ignores the `hard'
d.o.f. with higher energies and momenta. More precisely, the hard modes
cannot be totally ignored --- they interact with the soft modes and thus
affect the properties of the effective theory ---, rather they are
`integrated out' via some coarse--graining (or `renormalization group')
procedure, which can be perturbative or non--perturbative.

If the coupling is weak ($g\ll 1$), the `hard--soft' interactions can be
treated in perturbation theory and then the effective theory emerges as a
controlled approximation to the original theory. This generally amounts
to computing Feynman graphs with hard loop momenta and soft external
legs. By the uncertainty principle, the hard modes are localized on short
space--time distances, so their net effect is to provide {\em
quasi--local vertices}, or `parameters' --- like effective masses and
couplings --- in the effective Lagrangian for the soft modes. But even at
weak coupling, one often has to deal with a large, or even infinite,
number of Feynman graphs {\em at any given order} in $g$, because the
contributions due to individual graphs are enhanced by the large
disparity of scales between the hard and soft d.o.f. and/or by the high
density of medium constituents. This is where the effective theory is
most useful: it allows us to `resum' (modulo some approximations) a large
number of Feynman graphs of the original field theory and replace their
effects by a small number of `parameters' in the effective Lagrangian.

When the coupling is relatively strong, $g\gtrsim 1$, standard
perturbation theory (the expansion in powers of $g$) is bound to fail and
the construction of effective theories becomes more problematic. So long
as the coupling is just {\em moderately} strong, say $g\sim\order{1}$,
there is still hope that some insightful resummations of the perturbation
theory, as based on the proper identification of the relevant d.o.f., may
reasonably work --- we shall later encounter some examples in that sense.
If, in some regime, the coupling happens to be even stronger,
perturbation theory brings no guidance anymore, and there is no
systematic method to construct effective theories. They can merely be
postulated on the basis of general physical considerations, like the {\em
symmetries} of the fundamental theory. In such a case, the effective
masses or coupling constants are generally treated as free parameters, to
be matched against the data or, in some cases, against {\em lattice QCD}
calculations. Effective theories may also emerge for rather deep and
unexpected reasons, as we shall see on the example of the {\em
gauge/string duality} later on.

If in the previous discussion we mentioned both {\em weak} and {\em
strong} coupling scenarios, is because in QCD --- and indeed in any of
the fundamental field theories in Nature --- the coupling `constant' is
{\em not} fixed: it `runs' with the typical momenta exchanged in the
interactions, meaning that it is different when probing the physics on
different space--time scales. What is essential about QCD is the property
of {\em asymptotic freedom} : the fact that the coupling becomes weaker
on shorter distances, or with increasing momentum transfer. Given our
experience with electromagnetism, this property may look
counterintuitive. In QED, the electric charge of the nucleus inside an
atom is well known to be {\em screened} by the surrounding electron
cloud, so that the atom appears electrically neutral from far away.
Similarly, the electric charge of an electron is screened by
electron--positron virtual pairs which pump up from the vacuum, with the
result that the {\em effective} charge  $\alpha_{\rm e.m.}(R)\equiv
e^2(R)/4\pi$ decreases with the distance $R$ from the electron. But in
QCD, there is {\em anti}--screening: the effective colour charge of a
quark or gluon, as measured by its coupling `constant' $\alpha_s(R)\equiv
g^2(R)/4\pi$, increases with $R$ or with decreasing the transferred
momentum $Q$. (Recall that $Q\sim 1/R$ by the uncertainty principle.)
Specifically, one has $\alpha_s(Q^2)\simeq 0.1$ when $Q=100$~GeV (the
typical scale for electroweak physics and also for hadronic jets at the
LHC). This `asymptotic freedom'  is, of course, the ultimate reason
behind the success story of perturbative QCD in relation with `hard'
processes. It also justifies the use of perturbation theory for
integrating out the `hard' d.o.f. in the construction of effective
theories for the `soft' ones.

But there is also the reverse of the medal: with decreasing $Q$ below
100~GeV, the QCD coupling is increasing, albeit slowly, according to
 \beq\label{run}
\alpha_s(Q^2)\,\equiv\, \frac{g^2(Q^2)}{4\pi}\,=\, \frac{4\pi N_c}{(11
N_c-2N_f)
 \ln({Q^2}/{\Lam^2})}\,, \eeq
so that e.g. $\alpha_s(Q^2)\simeq 0.4$ when $Q=2$~GeV. Formally,
\Eref{run} predicts that the coupling diverges when $Q=\Lam$, but this
equation cannot be trusted for $Q\lesssim 1$~GeV, as it has been obtained
in perturbation theory. The fate of the QCD coupling for $Q\sim\Lam$ is
still under debate, but various non--perturbative approaches suggest that
$\alpha_s(Q^2)$ should (roughly) saturate at a value close to one. For
all purposes, this is very strong coupling (e.g. it corresponds to
$g\simeq 3$).

After this digression through the general scope of an effective theory
and the QCD running coupling, let us return to the main stream of our
presentation, namely, the phases of QCD as probed in a HIC. Some key
ideas, that will be succinctly mentioned here and developed in more
detail in the remaining part of these lectures, are as follows:

\texttt{(i)} The different stages of a HIC involve different forms of
hadronic matter with specific active degrees of freedom. Their
theoretical description requires different effective theories.

\texttt{(ii)} During the early stages of the collision --- the colour
glass condensate and the glasma --- the parton density is very high, the
typical transverse momenta are semi--hard (a few GeV), and the QCD
coupling is moderately weak, say $\alpha_s\sim 0.3$. In this case,
perturbation theory is (at least, marginally) valid, but it goes beyond a
straightforward expansion in powers of $\alpha_s$. The construct the
corresponding effective theory, one needs to resum an infinite class of
Feynman graphs which are enhanced by high--energy and high gluon density
effects. This has been done in the recent years, led to a formalism
--- {\em the CGC effective theory} --- which offers a unified description
from first principles for both the nuclear wavefunctions prior to the
collision and the very early stages of the collision. A key ingredient in
this construction is the proper recognition of the relevant d.o.f. : {\em
quasi--classical colour fields}. The concept of {\em field} is indeed
more useful in this high--density environnement than that of {\em
particle}, since the phase--space occupation numbers are large ($\gg 1$),
meaning that the would--be `particles' overlap with each other and thus
form coherent states, which are more properly described as classical
field configurations.

\texttt{(iii)} At later stages, the partonic matter expands, the
phase--space occupation numbers decrease, and the concept of {\em
particle} becomes again meaningful: the classical fields break down into
particles. If these particles are weakly coupled (as one may expect by
continuity with the previous stages), then their subsequent evolution can
be described by {\em kinetic theory}. This is an effective theory which
emerges under the assumption that the mean free path between two
successive collisions is much longer than any other microscopic scale
(like the duration of a collision or the Compton wavelength
$\lambda=1/k_\perp$ of a particle). Over the last years, kinetic theory
has been extensively derived from QCD at weak coupling, but the results
appear to be deceiving: for instance, they cannot explain the rapid
thermalization suggested by the data at RHIC and the LHC. (The
thermalization times predicted by perturbative QCD are much larger,
$\tau\gtrsim 10~$fm/c.) Several alternative solutions have been proposed
so far, but the final outcome is still unclear. One of these proposals is
that the softer modes, which keep large occupation numbers and should be
better described as classical fields, become {\em unstable} due to the
anisotropy in the momentum distribution of the harder particles (which in
turns follows from the disparity between longitudinal and transverse
expansions). But numerical simulations of the coupled system soft
fields--hard particles leads to thermalization times which are still too
large. Another suggestion is that the partonic matter is (moderately)
{\em strongly coupled} --- the QCD coupling could indeed become larger,
because of the system being more dilute. In such a scenario, a candidate
for an effective theory is the {\em AdS/CFT correspondence}, to be
discussed later.

\texttt{(iv)} Assuming (local) thermal equilibrium, and hence the
formation of a {\em quark--gluon plasma} (QGP), the question is whether
this plasma is weakly or strongly coupled. The maximal temperature of
this plasma, as estimated from the average energy density, should be
around $T\sim 500\div600$~MeV; so the respective coupling is moderately
strong: $\alpha_s\sim 0.3\div0.4$ or $g\sim 1.5\div 2$. The thermodynamic
properties (like pressure or energy density) of a QGP at {\em global}
thermal equilibrium within this range of temperatures are by now well
known from numerical calculations on a lattice and can serve as a
baseline of comparison for various effective theories. If the coupling is
weak, one has to use the {\em Hard Thermal Loop effective theory} (HTL),
a version of the kinetic theory which describes the long--range (or
`soft') excitations of the QGP. This effective theory lies at the basis
of a physical picture of the QGP as a gas of {\em weakly--coupled
quasi--particles} --- quarks or gluons with temperature--dependent
effective masses and couplings. Using this picture as a guideline for
reorganizations of the perturbation theory, one has been able to
reproduce the lattice data quite well. Thus, the thermodynamics appears
to be consistent with a weak--coupling picture for the QGP, although this
picture is considerably more complicated than that emerging from naive
perturbation theory (the strict expansion in powers in $g$). Yet, this is
not the end of the problem, as we shall see.

\texttt{(v)} The QGP created in the intermediate stages of a HIC is
certainly not in {\em global} thermal equilibrium, but only in a {\em
local} one: it keeps expanding. Under very general assumptions, the
effective theory describing this flow is {\em hydrodynamics}. The
corresponding equations of motion are simply the conservation laws for
energy, momentum, and other conserved quantities (like the electric
charge or the baryonic number), and as such they are universally valid.
But these equations also involve `parameters', like the {\em
viscosities}, which describe dissipative phenomena occurring during the
flow and which depend upon the specific microscopic dynamics. The values
of these parameters are very different at weak vs. strong coupling. A
meaningful way to characterize the strength of dissipation is via the
dimensionless viscosity--over--entropy-density ratio $\eta/s$. (This
ratio is dimensionless when using natural units; otherwise it has the
dimension of $\hbar$.) Remarkably, the {\em elliptic flow} data at RHIC
and the LHC to be later discussed suggest a very small value for this
ratio, which is inconsistent with the present calculations at weak
coupling, based on kinetic theory. On the other hand, such a small ratio
is naturally emerging at strong coupling, as shown by calculations within
the AdS/CFT correspondence. The smallness of $\eta/s$ represents so far
the strongest argument in favour of a {\em strongly coupled quark--gluon
plasma} (sQGP). This may look contradictory with the previous conclusions
drawn from thermodynamics. But one should remember that the QCD coupling
depends upon the relevant space--time scale and that hydrodynamics refers
to the long--range behaviour of the fluid, as encoded in its softest
modes. By contrast, thermodynamics is rather controlled by the hardest
modes --- those with typical energies and momenta of the order of the
(local) temperature. So, it is not inconceivable that a same system look
effectively weakly coupled for some phenomena and strongly coupled for
some others.

\texttt{(vi)} Another strategy for studying the hadronic matter produced
in a HIC refers to the use of {\em hard probes}. These are particles with
large transverse energies (say, $E_\perp\gtrsim 20$~GeV at the LHC),
which are produced in the very first instants of a collision and then
cross the QCD matter liberated at later stages along their way towards
the detector. Some of these particles, like the (direct) photons and the
dilepton pairs, do not interact with this matter and hence can be used a
baseline for comparaisons. But other particles, like quarks, gluons, and
the jets initiated by them, {\em do} interact, and by measuring the
effects of these interactions --- say, in terms of energy loss, or the
suppression of multi--particle correlations --- one can infer
informations about the properties of the matter they crossed. The RHIC
data have demonstrated that semi--hard partons can lose a substantial
fraction of the transverse energy via interactions in the medium (`jet
quenching'), thus suggesting that these interactions can be quite {\em
strong}. These results have been confirmed at the LHC, which moreover
found that even {\em very} hard jets ($E_\perp\gtrsim 100$~GeV) can be
strongly influenced by the medium, in the sense that they get strongly
defocused : the energy distribution in the polar angle with respect to
the jet axis becomes much wider after having crossed the medium. This is
visible for the photon--hadron di--jet event in the right panel of
\Fref{fig:events}~: the photon and the parton which has initiated the
hadronic jet have been created by a hard scattering, so they must have
been balanced in transverse momentum at the time of their creation. Yet,
the central peak in the hadronic jet, which represents the final `jet'
according to the conventional definition, carries much less energy than
the photon jet. This is interpreted as the result of energy transfer to
large polar angles (outside the conventional `jet' definition) via
in--medium interactions. In order to study such interactions, in
particular high--density effects like {\em multiple scattering} and {\em
coherence}, it is again useful to build effective theories. In that case
too, it is not so clear whether the physics is controlled by {\em mostly
weak coupling}, or a {\em mostly strong} one. (By itself, the jet {\em
is} hard, but its coupling to the relatively soft constituents of the
medium may be still governed by a moderately strong coupling.) In fact,
the unexpectedly strong jet quenching observed at RHIC is sometimes
interpreted as another evidence for strong coupling behaviour. {\em
Moderately} strong coupling turns out to be the most difficult situation
to deal with, so in these lectures we shall rather describe the effective
theories proposed in the limiting situations of weak and, respectively,
strong coupling. Within perturbative QCD, this is known as {\em
medium--induced gluon radiation}. At strong coupling, it again relies on
AdS/CFT.

\section{The Color Glass Condensate}
\label{sec:CGC}

This chapter is devoted to the {\em early stages} of an ultrarelativistic
heavy ion collision (HIC), that is, the wavefunctions of the energetic
nuclei prior to the collision and the partonic matter liberated by the
collision. As already mentioned, these early stages are the realm of
high--density, coherent, forms of QCD matter, characterized by high gluon
occupation numbers. Such forms of matter can be described in terms of
{\em strong, semi--classical, colour fields}. In what follows, we shall
explain this theoretical description, starting with the perhaps more
familiar {\em parton picture} of QCD scattering at high energy.

\subsection{The QCD parton picture}
\label{sec:DIS}

\begin{figure}[t]
\begin{center}\centerline{
\includegraphics[width=.3\textwidth]{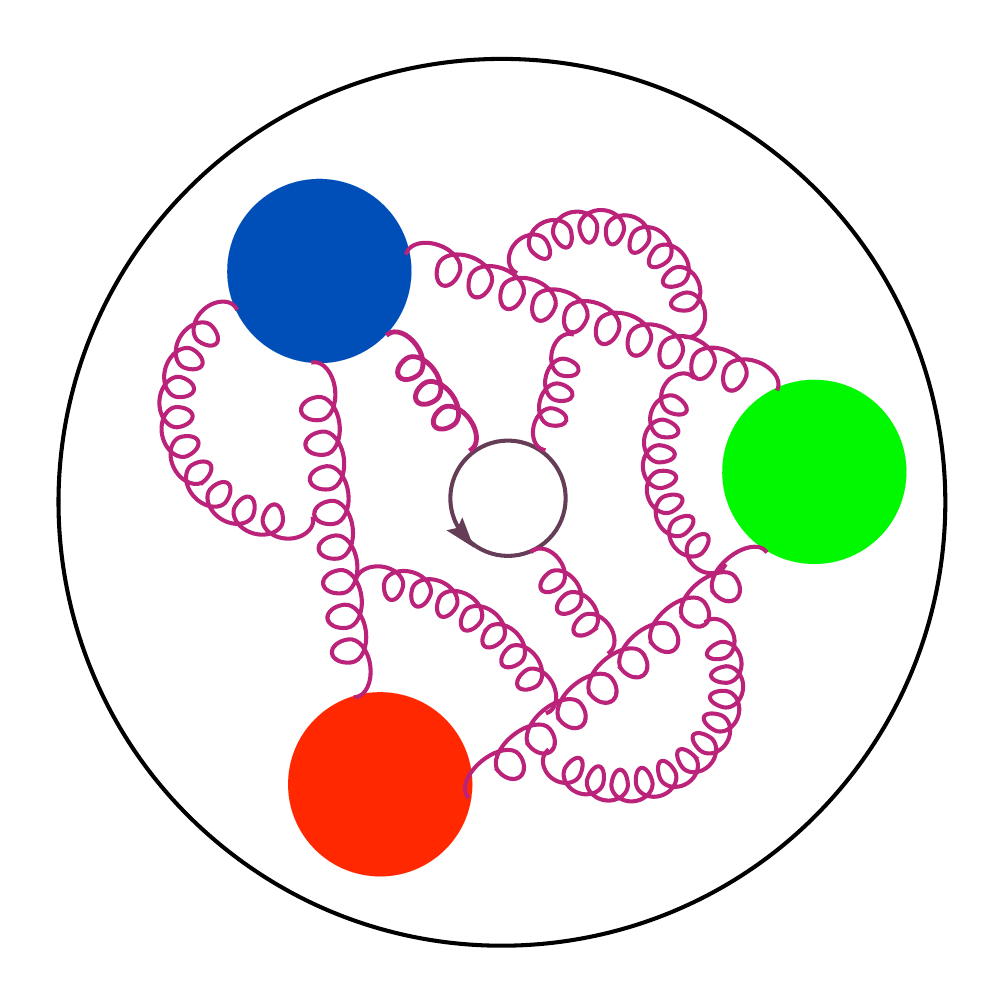}\qquad
 \qquad
\includegraphics[width=.35\textwidth]{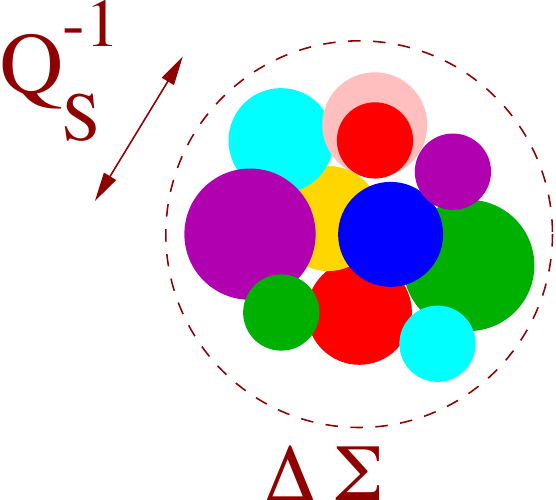}
}
\caption{\sl Left: a cartoon of the proton structure in its rest frame.
Right: cartoon of one saturation disk in the infinite momentum frame (this
will be discussed in \Sref{sec:glasma}).}
\label{fig:hadron}\vspace*{-1.cm}
\end{center}
\end{figure}

The microscopic structure of a hadron depends upon the resolution scales
which are used to probe it, that is, upon the kinematics of the
scattering process. It furthermore depends upon the Lorentz frame in
which the hadron is seen: unlike physical observables, like
cross--sections, which are boost invariant, the physical interpretation
of these observables in terms of partons depends upon the choice of a
frame. This is best appreciated by first looking at a hadron (say, a
proton) in its rest frame (RF), where the proton 4--momentum reads
$P_0^\mu=(M,0,0,0)$. The proton has the quantum numbers of a system of
three quarks --- the `valence quarks'
--- which are bound by confinement in a colour singlet state. But this
binding proceeds via the exchange of gluons, which in turn can generate
additional quark--antiquark pairs (see \Fref{fig:hadron}). All these
partons are `virtual', meaning that they keep appearing and disappearing,
and have typical energies and momenta of order $\Lam$, since this is the
scale where the QCD coupling becomes of $\order{1}$ and thus the binding
is most efficient. Clearly, such fluctuations are non--perturbative.
$\Lam$ is also the typical scale for vacuum fluctuations, like a
quark--antiquark pair pumping up from the vacuum and then being
reabsorbed. By the uncertainty principle, such fluctuations have
lifetimes and sizes of order $1/\Lam$, of the same order as the proton
size itself. Under these conditions, it makes no sense to speak about
`hadronic substructure' : the hadronic fluctuations are ephemeral,
delocalized over the whole proton volume, and cannot be distinguished
from the vacuum fluctuations having the same kinematics and quantum
numbers.

\begin{figure}[ht]
\begin{center}{
\includegraphics[width=1.05\textwidth]{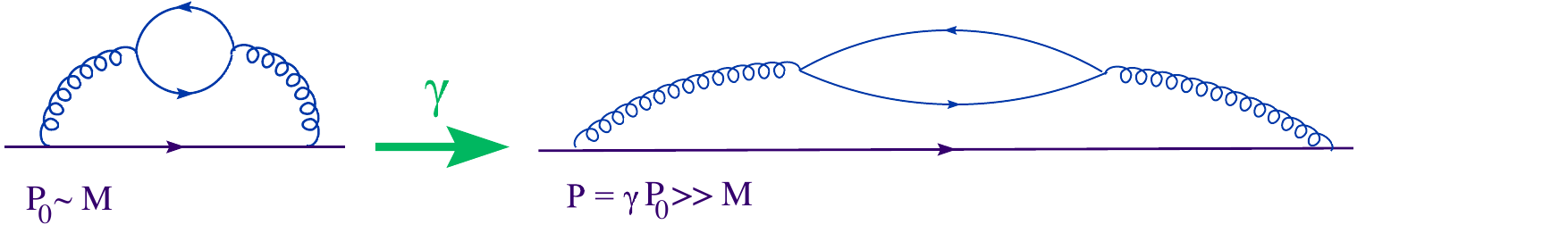}
}
\caption{\sl A hadronic fluctuation in the hadron rest frame (left) and in the
infinite momentum frame (right).}
\label{fig:boost}\vspace*{-.5cm}
\end{center}
\end{figure}

However, the situation changes if one observes the same hadron in a frame
which is boosted by a large Lorentz factor $\gamma\gg 1$ w.r.t. the rest
frame. Then the hadron 4--momentum reads $P^\mu=(E,0,0,P)$ with
$E=\sqrt{P^2+M^2}\simeq P$. (We have chosen the boost along the $z$ axis
and denoted $P_z=P$.) In this boosted frame, conventionally referred to
as the {\em infinite momentum frame} (IMF), the lifetime of the hadronic
fluctuations is enhanced by Lorentz time dilation (see \Fref{fig:boost}),
 \beq\label{lifeIMF}
 \Delta t_{\rm IMF}\,=\,\gamma\Delta t_{\rm RF}\,\sim\,\frac{\gamma}{\Lam}\,,
 \eeq
so these fluctuations are now well separated from the those of the vacuum
(which have a lifetime $\sim1/\Lam$ in any frame, since the vacuum is
boost invariant). The lifetime \eqref{lifeIMF} is much larger than the
duration of a typical collision process (see below); so, for the purpose
of scattering, the hadronic fluctuations can be viewed as {\em free,
independent quanta}. These quanta are the {\em partons} (a term coined by
Feynman). It then becomes possible to {\em factorize} the cross--section
(say, for a hadron--hadron collision) into the product of {\em parton
distribution functions} (one for each hadron partaking in the collision),
which describe the probability to find a parton with a given kinematics
inside the hadronic wavefunction, and {\em partonic cross--sections},
which, as their name indicates, describe the collision between subsets of
partons from the target and the projectile, respectively. If the momentum
transferred in the collision is hard enough, the partonic cross--sections
are computable in perturbation theory. The parton distributions are {\em
a priori} non--perturbative, as they encode the information about the
binding of the partons within the hadron. Yet, there is much that can be
said about them within perturbation theory, as we shall explain. To that
aim, one needs to better appreciate the role played by the {\em
resolution} of a scattering process. In turn, this can be best explained
on the example of a simpler process: the electron--proton {\em deep
inelastic scattering} (DIS).

\begin{figure}[t]
\begin{center}\centerline{
\includegraphics[width=.5\textwidth]{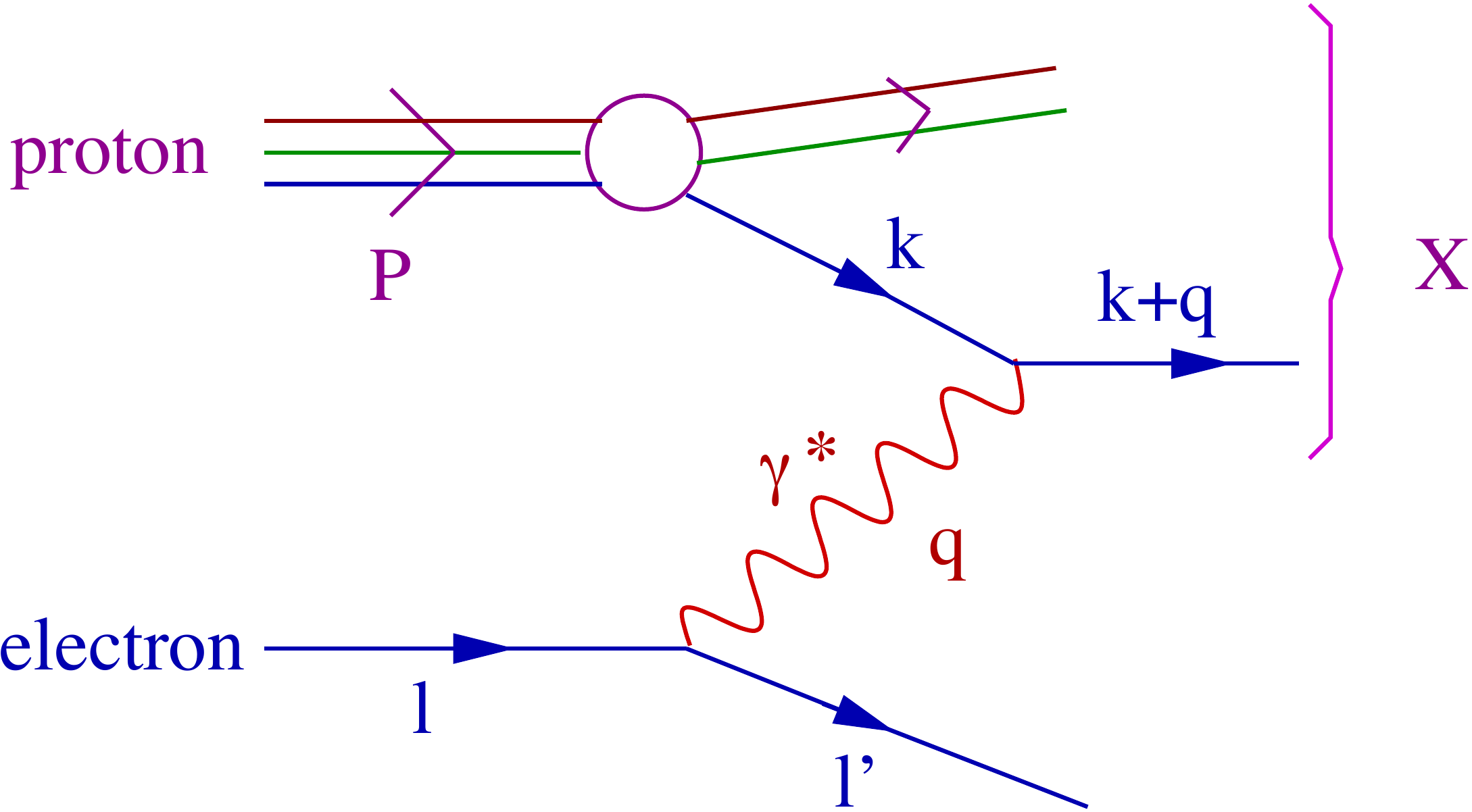}\qquad
\includegraphics[width=.35\textwidth]{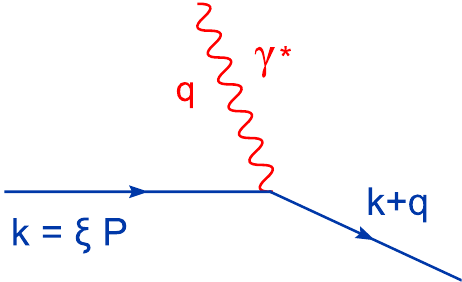}
}
\caption{\sl Left: the DIS process. Right: the absorption of the virtual photon
by a quark.}
\label{fig:DIS}
\end{center}\vspace*{-.9cm}
\end{figure}

The DIS process is illustrated in \Fref{fig:DIS} (left): an electron with
4--momentum $\ell_\mu$ scatters off the proton by exchanging a virtual
photon ($\gamma^*$) with 4--momentum $q_\mu$ and emerges after scattering
with 4--momentum $\ell^{'}_{\mu}=\ell_\mu-q_\mu$. The exchanged photon is
{\em space--like} :
 \beq
 q^2=(\ell-\ell')^2=-2\ell\cdot\ell'=
 -2E_{\ell}E_{\ell'}(1-\cos\theta_{\ell\ell'})\,\equiv\,-Q^2\qquad
 \mbox{with}\quad Q^2 > 0,\eeq
with $E_{\ell}=|\bm{\ell}|$, $E_{\ell'}=|\bm{\ell}'|$, and
$\theta_{\ell\ell'}=\angle(\bm{\ell}, \bm{\ell}')$. The positive quantity
$Q^2$ is referred to as the `virtuality'. The {\em deeply inelastic}
regime corresponds to $Q^2 \gg M^2$, since in that case the proton is
generally broken by the scattering and its remnants emerge as a
collection of other hadrons (denoted by $X$ in \Fref{fig:DIS}). The
(inclusive) DIS cross--section involves the sum over all the possible
proton final states $X$ for a given $\ell'$.

A space--like probe is very useful since it is well localized in space
and time and thus provides a snapshot of the hadron substructure on
controlled, transverse and longitudinal, scales, as fixed by the
kinematics. Specifically, we shall argue that, when the scattering is
analyzed in the proton IMF, the virtual photon measures partons which are
localized in the transverse plane within an area $\Sigma\sim 1/Q^2$ and
which carry a longitudinal momentum $k_z=xP$, where $x$ is the {\em
Bjorken variable} :
 \beq\label{bj}
 x\,\equiv\,\frac{Q^2}{2(P\cdot q)}\,=\,\frac{Q^2}{s+Q^2-M^2}\,,\eeq
where $s\equiv (P+q)^2$ is the invariant energy squared of the
photon+proton system. That is, the two kinematical invariants $Q^2$ and
$x$, which are fixed by the kinematics of the initial state ($\ell,\,P$)
and of the scattered electron ($\ell'$), completely determine the
transverse size ($\sim 1/Q$) and the {\em longitudinal momentum fraction}
($x$) of the parton that was involved in the scattering. This parton is
necessarily a quark (or antiquark), since the photon does not couple
directly to gluons. But the DIS cross--section allows us to indirectly
deduce also the gluon distribution, as we shall see.

As a first step in our argument, consider a quark excitation of the
hadron, viewed in the IMF. This quark is a virtual fluctuation which has
been boosted together with the proton, so its virtuality and its
transverse momentum are both small as compared to its longitudinal
momentum $k_z =\xi P$. (We temporarily denote with $\xi$ the fraction of
the proton longitudinal momentum which is carried by the quark.) So, for
most purposes, one can treat the quark as a nearly on--shell excitation
with 4--momentum as $k^\mu \simeq \xi P^\mu =(\xi P, 0, 0, \xi P)$ and
$k^2\equiv k_\mu k^\mu \simeq 0$. (More precisely, $k^2\simeq k_\perp^2
\sim \Lam^2$.) Such an excitation has a relatively large lifetime, which
can be estimated as in \Eref{lifeIMF} :
 \beq\label{lifetime}
 \Delta t_{\rm fluct}\,=\,\gamma\Delta t_{\rm RF}
 \,\simeq\,\frac{2k_z}{k_\perp^2}\,=\,
 \frac{2\xi P}{k_\perp^2}\,,\eeq
where $\Delta t_{\rm RF}\sim 2/k_\perp$ is the lifetime of the
fluctuation in the hadron rest frame and $\gamma=k_z/k_\perp$ is the
boost factor from the RF to the IMF.

Consider now the absorption of the virtual photon by the quark, cf.
\Fref{fig:DIS} right. The quark is liberated by this collision, meaning
that it is put on shell; so we can write
 \beq\label{xi} (k+q)^2\,=\,0 \, \Longrightarrow \, - Q^2 +
2\xi P\cdot q \,=\,0  \, \Longrightarrow\,
 \xi\,=\,\frac{Q^2}{2(P\cdot q)}\,=\,x\,,\eeq
where we have also used $k^2\approx 0$, as discussed before. We see that
the collision identifies the longitudinal momentum fraction $\xi$ of the
participating quark with the Bjorken--$x$ kinematical variable, as
anticipated. From now on, we shall use the notation $x$ for both
quantities.

To also clarify the {\em transverse resolution} of the virtual photon, we
first need an estimate for the {\em collision time}. This is the typical
duration of the partonic process $q+\gamma^*\to q$ (cf. \Fref{fig:DIS}
right) and is given by the uncertainty principle: $\Delta t_{\rm
coll}\sim 1/\Delta E$, where $\Delta E=q_0+|\bmk+\bmq|-|\bmk|$ is the
energy difference at the photon emission vertex. To estimate $\Delta E$,
it is convenient to choose a space--like photon with zero energy and only
transverse momentum: $q^\mu=(0,\bmq_\perp,0)$. Then
 \beq
 \Delta E=|\bmk+\bmq|-|\bmk| =\sqrt{ (xP)^2+ q_\perp^2}-xP\simeq\,\frac{
 q_\perp^2}{2xP}\quad\Longrightarrow\quad \Delta t_{\rm coll}\simeq
 \frac{2xP}{Q^2}.\eeq
(Note that $Q^2 = q_\perp^2$ for the virtual photon at hand.) In order to
be `found' by the photon, a quark excitation must have a lifetime larger
than this collision time:
 \beq
 \Delta t_{\rm fluct}\,\simeq\,
 \frac{2 x P}{k_\perp^2}\,>\,\Delta t_{\rm coll}\simeq \frac{2xP}{Q^2}
  \, \Longrightarrow\, k_\perp^2\,< \,Q^2\,.\eeq
Hence, the virtual photon can discriminate only those partons having
transverse momenta smaller than its virtuality $Q$. By the uncertainty
principle, such partons are localized within a transverse area $\sim
1/Q^2$, as anticipated after \Eref{bj}.

The previous considerations motivate the following formula for the DIS
cross--section  :
 \beq\label{sigmagamma}
 \sigma_{\gamma^* p}(x,Q^2)\,=\,
 \frac{4\pi^2 \alpha_{\rm em}}{Q^2}\,F_2(x,Q^2)\,,\eeq
where the first factor in the r.h.s. is the elementary cross--section for
the photon absorbtion by a quark (or an antiquark), whereas the second
factor --- the {\em structure function} $F_2(x,Q^2)$ --- is the sum of
the {\em quark and antiquark distribution functions}, weighted by the
respective electric charges squared
 \begin{align}\label{F2}
 F_2(x,Q^2)&\,=\,\sum_{f} e_f^2\,\big[xq_f(x,Q^2) + x\bar
 q_f(x,Q^2)\big]\,,\nn
 q_f(x,Q^2)&\,\equiv\,\frac{\rmd N_{f}}{\rmd x}(Q^2)
 \,=\,\int^{Q}\!
 \rmd^2\bmk_\perp\, \frac{\rmd N_{f}}{\rmd x\rmd^2\bmk_\perp}\,.
 \end{align}
That is, $q_f(x,Q^2)\rmd x$ is the number of quarks of flavor $f$ with
longitudinal momentum fraction between $x$ and $x+\rmd x$ and which
occupy a transverse area $1/Q^2$.

One may naively think that the condition $k_\perp^2< Q^2$ is trivially
satisfied, since the partons confined inside the hadron have transverse
momenta $k_\perp\sim\Lam$, whereas $Q^2\gg \Lam^2$ by the definition of
DIS. If that were the case, the structure function $F_2(x,Q^2)$ would be
independent of $Q^2$ --- a property known as {\em Bjorken scaling}.
However, the DIS data show that Bjorken scaling holds only approximately
and only in a limited range of values for $x$, namely for $x\gtrsim 0.1$.
This can be understood as follows : the typical transverse momenta are
$\sim\Lam$ only for the valence quarks and, more generally, for the
partons with relatively large longitudinal momentum fractions. But
virtual quanta with much larger values for $k_\perp$ can be generated via
radiative processes like {\em bremsstrahlung}. Such quanta have very
short lifetimes, but so long as $k_\perp^2 < Q^2$, they can still
contribute to DIS. Also, they generally have small values of $x$, as they
share all together the longitudinal momenta of their parents partons.
Hence, we expect the {\em parton evolution} via bremsstrahlung to lead to
an increase in the parton distributions at large values of $k_\perp$ and
small values of $x$. This evolution is responsible for the violations of
the Bjorken scaling seen in the data and, more generally, for the {\em
DGLAP evolution} \cite{DGLAP,Ellis:1991qj} of the parton distribution
functions with increasing $Q^2$. It is furthermore responsible for the
rapid growth in the gluon distribution with decreasing $x$ and the
formation of a {\em colour glass condensate} at high energy. This will be
further discussed in \Sref{sec:evol}.

\subsection{Particle production at the LHC: why small $x$ ?}
\label{sec:eta}

Before we turn to a discussion of parton evolution, let us explain here
why we shall be mostly interested in partons with small longitudinal
momentum fractions $x\ll 1$. As it should be clear from \Eref{bj}, small
values of $x$ correspond to the {\em high--energy} regime at $s\gg Q^2$.
The conceptual importance of this regime will be explained later, but for
the time being let us discuss it from the experimental point of view.
Very small values of $x$, as low as $x=10^{-5}\div 10^{-4}$, have been
already reached in the e+p collisions at HERA, but in that context they
were associated (because of the experimental constraints) with rather
small values of the transferred momentum $Q^2$. Namely, the HERA data at
$x\le 10^{-4}$ correspond to values $Q^2 < 1$~GeV$^2$ which are only
marginally under control in perturbation theory. Because of that, the DIS
data at HERA remained inconclusive for a check of our theoretical
understanding of the physics at small $x$.

\begin{figure}[t]
\begin{center}\centerline{
\includegraphics[width=0.6\textwidth]{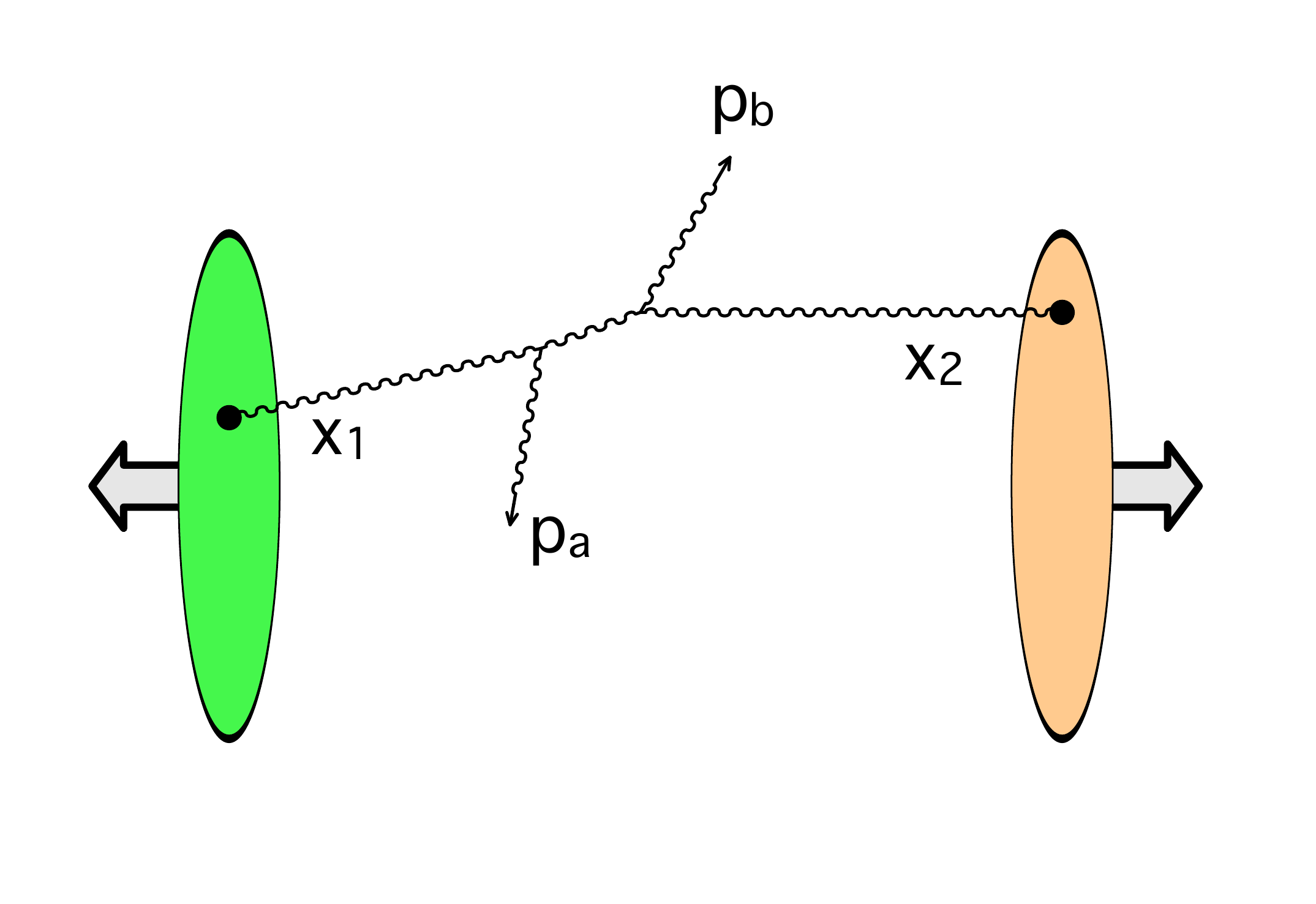}
 \includegraphics[width=0.4\textwidth]{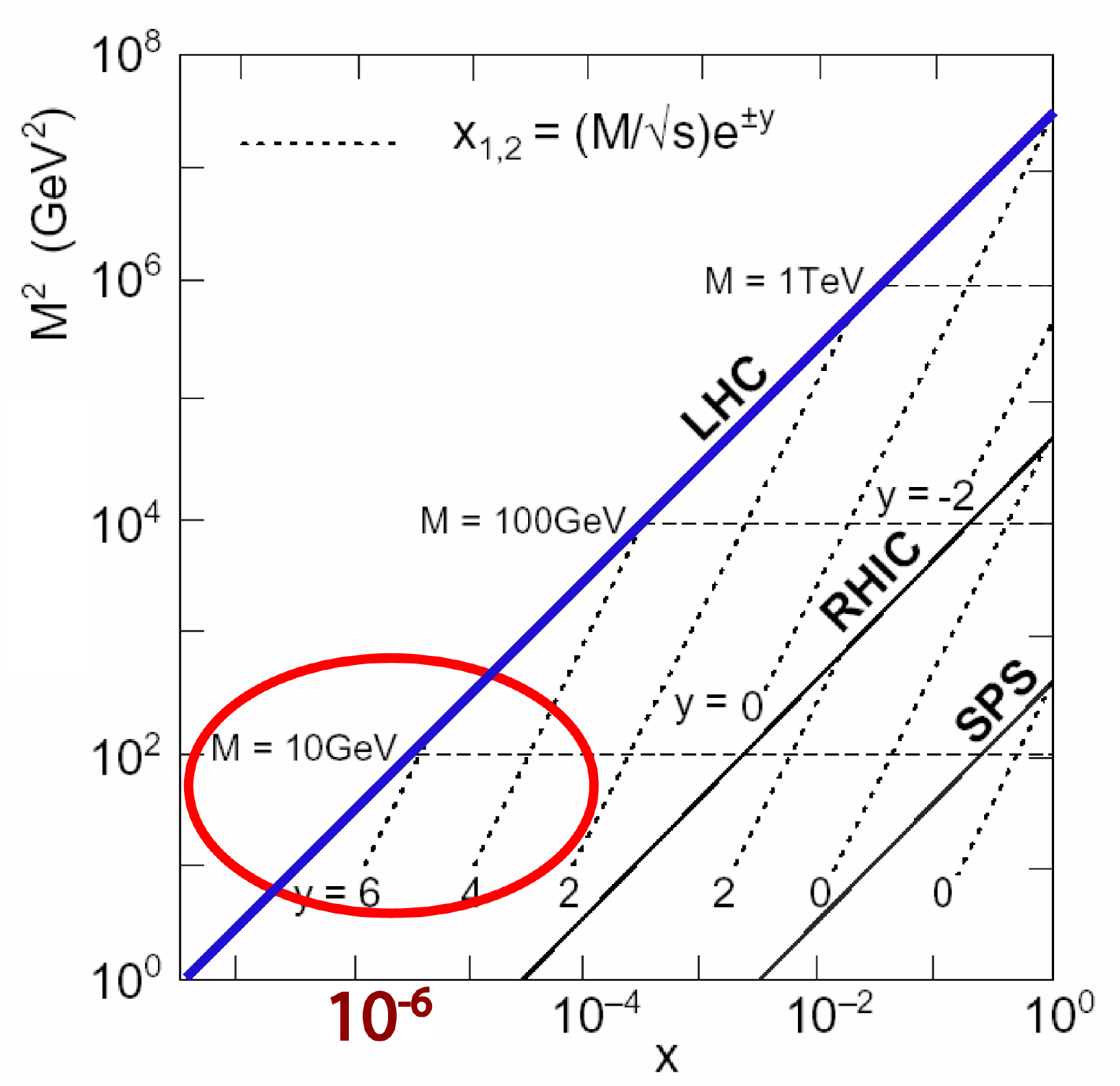}
 }
 \caption{\sl Left: two particle production in hadron--hadron collision.
 Right: the kinematical domain accessible at the LHC, as compared to
 RHIC and the SPS; the small--$x$ region is highlighted.\label{fig:2part}}
\end{center}\vspace*{-.8cm}
\end{figure}

But the situation has changed with the advent of the new hadron--hadron
colliders, RHIC and, especially, the LHC. Given the much higher available
energies, the bulk of the particle production (with semi--hard transverse
momenta) in these experiments is controlled by partons with $x\le
10^{-3}$. Moreover, for special kinematical conditions to be shortly
specified, one can probe values as low as $x\sim 10^{-6}$ with truly hard
momentum transferts, such as $Q^2=10$~GeV$^2$.

To describe the kinematics of particle production, it is useful to
introduce a new kinematical variable, the {\em rapidity} $y$, which is an
alternative for the longitudinal momentum. For an on--shell particle with
4--momentum $p^\mu=(E,\bm{p}_\perp,p_z)$, the rapidity is defined as
 \beq\label{ydef}
 y\,\equiv\,\frac{1}{2}\,\ln\frac{E+p_z}{E-p_z}\,\Longrightarrow\,
 E=m_\perp \cosh y,\quad p_z =m_\perp \sinh y\,,\eeq
where $m_\perp\equiv\sqrt{m^2+p^2_\perp}$ is the `transverse mass' and
$E^2=m^2_\perp+p_z^2$. Note that $y$ is positive for a `right--mover'
($p_z>0$) and negative for a `left--mover' ($p_z<0$). In fact, one has
$v_z=p_z/E=\tanh y$, so $y$ is simply related to the longitudinal boost
factor: $\gamma=\cosh y$. A similar quantity which is perhaps more useful
in the experiments (since easier to measure) is the {\em
pseudo--rapidity}
 \beq\label{etadef}
 \eta\,\equiv\,\frac{1}{2}\,\ln\frac{p+p_z}{p-p_z}\,=\,
 -\ln
\,\tan\frac{\theta}{2}\,,
 \eeq
where $p=|\bmp|$ is the magnitude of the 3--momentum vector. As shown by
the second equality above, $\eta$ is directly related to the polar angle
$\theta$ made by the particle with the longitudinal axis
($\cos\theta={p_z}/{p}$). For massless particles or for ultrarelativistic
ones (whose masses can generally be ignored), the two rapidities coincide
with each other, as manifest by comparing Eqs.~\eqref{ydef} and
\eqref{etadef}.

\begin{figure}[t]
\begin{center}\centerline{
\includegraphics[width=.40\textwidth]{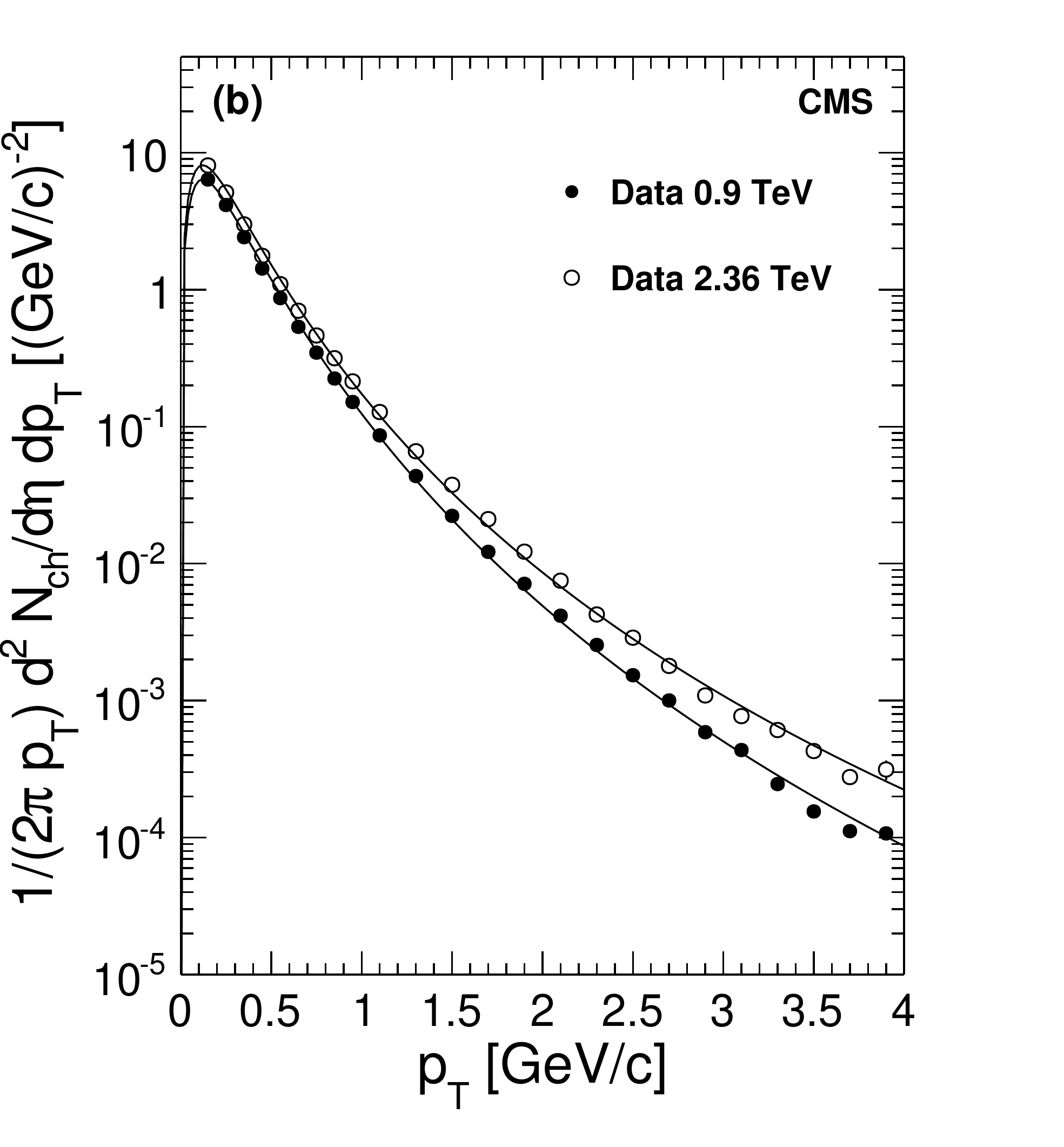}\qquad
 \quad
\includegraphics[width=.44\textwidth]{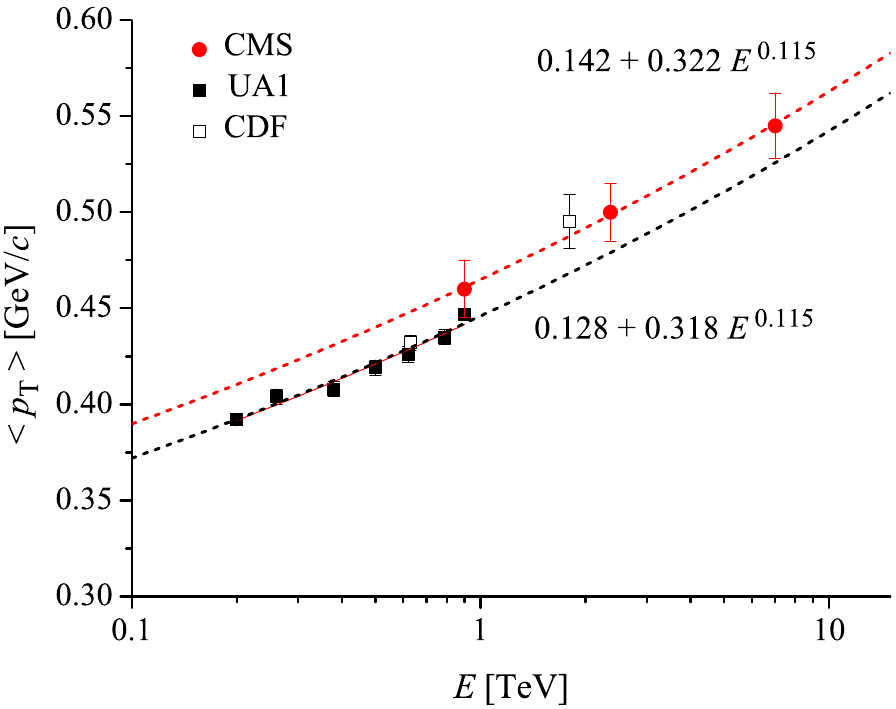}
}
\caption{\sl Transverse momentum dependence of the single inclusive particle
production in hadron--hadron collisions. Left: Pb+Pb collisions at the LHC.
Right: p+p collisions at the LHC and p+$\overline{\mbox{p}}$ collisions
at SPS and CDF.} \label{fig:avept}\vspace*{-.5cm}
\end{center}
\end{figure}

Consider now the process illustrated in \Fref{fig:2part} (left), i.e. the
production of a pair of particles in a partonic subcollision of a
hadron--hadron scattering. In the center--of--mass (COM) frame, the two
partons partaking in the collision have 4--momenta $k_i^\mu=x_iP_i^\mu +
k^\mu_{i\perp}$ where $i=1,2$, $P_1^\mu=(P,0,0,P)$, $P_2^\mu=(P,0,0,-P)$,
and $k^\mu_{i\perp}=(0,\bmk_{i\perp},0)$. Notice that $P=\sqrt{s}/2$. The
two outgoing particles will be characterized by the respective transverse
momenta, $\bmp_{a\perp}$ and $\bmp_{b\perp}$, and rapidities, $y_a$ and
$y_b$. Energy--momentum conservation implies
$\bmp_{a\perp}+\bmp_{b\perp}= \bmk_{1\perp}+\bmk_{2\perp}$ and
 \beq\label{2pkin}
  x_1\,=\,\frac{p_{a\perp}}{\sqrt{s}}\,\rme^{y_a}
+\frac{p_{b\perp}}{\sqrt{s}}\,\rme^{y_b}\,,\qquad x_2
\,=\,\frac{p_{a\perp}}{\sqrt{s}}\,\rme^{-y_a}+
\frac{p_{b\perp}}{\sqrt{s}}\,\rme^{-y_b}\,.
 \eeq
For particle production at RHIC or the LHC, the average transverse
momentum of a hadron in the final state is below 1~GeV ; moreover, 99\%
of the `multiplicity' (i.e. of the total number of produced hadrons) has
$p_\perp\le 2$~GeV (see \Fref{fig:avept}). For $p_{a,b\,\perp}=1$~GeV and
central rapidities $y_{a,b}\simeq 0$, \Eref{2pkin} implies
 \beq
 x_i\simeq 10^{-2} \ \mbox{at RHIC}\ (\sqrt{s}=200\,\mbox{GeV}),\qquad
 \quad x_i\simeq 4\times 10^{-4} \ \mbox{at the LHC}\
 (\sqrt{s}=2.76\,\mbox{TeV}),
 \eeq
where in the case of the LHC we have chosen the maximal COM energy {\em
per nucleon pair} that has been reached so far in Pb+Pb collisions. Thus,
the bulk of the particle production is initiated by partons carrying
small values of $x$, as anticipated. Moreover, one of these values ($x_1$
or $x_2$) can be made much smaller by studying particle production at
either {\em forward}, or {\em backward}, rapidities. The rapidities are
`forward' when both $y_a$ and $y_b$ are positive and relatively large,
that is, the final particles propagate essentially along the same
direction as the original hadron `1'~; then their production probes very
small values of $x_2$ in the wavefunction of the hadron `2' and
comparatively large values of $x_1$. At the LHC, one can probe values as
small as $x_2\sim 10^{-6}$ for $p_\perp\sim 10$~GeV, as indicated in the
r.h.s. of \Fref{fig:2part}.

We finally discuss the cross--section for the production of a pair of
hadrons. When the transverse momenta $p_{a,b\,\perp}$ are large enough,
one can ignore the `intrinsic' transverse momenta $\bmk_{1,2\,\perp}$ of
the colliding partons. Then the transverse momentum conservation
$\bmp_{a\perp}+\bmp_{b\perp}\simeq 0$ implies that the outgoing particles
propagate back--to--back in the transverse plane, i.e. they make an
azimuthal angle $\Delta\phi\simeq\pi$. The associated cross--section
admits the following {\em collinear factorization}, analogous to
\Eref{F2} for DIS
 \begin{eqnarray}\label{coll}
  \frac{\rmd \sigma}{\rmd p_{\perp}^2
\rmd y_1 \rmd y_2}=\sum_{ij}
x_1 f_{i}(x_1,\mu^2)\, x_2 f_{j}(x_2,\mu^2)\,
\frac{\rmd \hat\sigma_{ij}}{\rmd p^2_{\perp}}\,,
 \end{eqnarray}
where $xf_{i}(x,\mu^2)$ are parton distributions for all species of
partons ($i=q,\bar q,g$), $\mu^2$ is the factorization scale, and ${\rmd
\hat\sigma_{ij}}/{\rmd p^2_{\perp}}$ is the cross--section for the
(relatively hard) partonic process $i+j\to a+b$. Leading--order
perturbative QCD yields ${\rmd \hat\sigma}/{\rmd
p^2_{\perp}}\propto\alpha_s^2/p^4_{\perp}$ at high energy. So, if one
tries to compute the total multiplicity by integrating over all values of
$p_\perp^2$ (say, for $y_a\sim y_b\sim 0$), then one faces a quadratic
infrared divergence from the limit $p_\perp^2\to 0$. One may think that
this divergence is cut off at $p_\perp\sim\Lam$, since this is the
typical value expected for the intrinsic momenta $k_{1,2\,\perp}$. But
then one would conclude that the bulk of the particle production, even at
very high energy, is concentrated at very soft transverse momenta, of the
order of the confinement scale $\Lam$. Moreover, the average $p_\perp$
would be independent of the energy (since of $\order{\Lam}$). These
conclusions are however contradicted by the data in \Fref{fig:avept},
which rather show that $\langle p_\perp\rangle\simeq 0.5$~GeV is about 2
to 3 times larger than $\Lam$ at the LHC energies and, remarkably, it
clearly rises with the COM energy $E=\sqrt{s}$. This conflict between the
data and the prediction \eqref{coll} of collinear factorization clearly
shows that the latter cannot be extrapolated down to lower values for
$p_\perp$, say of order 1~GeV. The proper way to describe this {\em
semi--hard} region within (perturbative) QCD will be explained in the
next subsection. The main outcome of that analysis will be to introduce a
new infrared cutoff in the problem, which is dynamically generated ---
via gluon evolution with decreasing $x$ --- and rises as a power of the
energy. This is the {\em saturation momentum}.

\subsection{Gluon evolution at small $x$}
\label{sec:evol}

\begin{figure}[t]
\begin{center}
\includegraphics[width=0.6\textwidth]{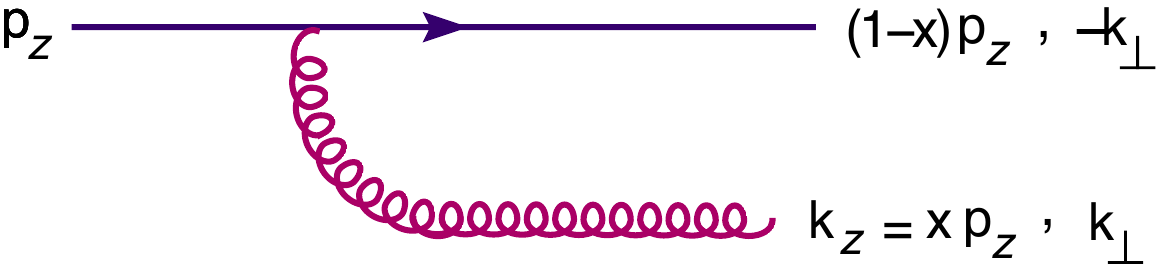}
 \caption{\sl Gluon bremsstrahlung out of a parent quark to lowest order in
pQCD. \label{onegluon}}
\end{center}\vspace*{-.5cm}
\end{figure}

In perturbative QCD, parton evolution proceeds via bremsstrahlung, which
favors the emission of {\em soft} and {\em collinear} gluons, i.e. gluons
which carry only a small longitudinal momentum fraction $x\ll 1$ and a
relatively small transverse momentum $k_\perp$. \Fref{onegluon}
illustrates one elementary step in this evolution: the emission of a
gluon which carries a fraction $x=k_z/p_z$ of the longitudinal momentum
of its parent parton (quark or gluon). For $x\ll 1$ and to lowest order
in $\alpha_s$, the differential probability for this emission (obtained
as the modulus squared of the amplitude represented in \Fref{onegluon})
reads
\begin{eqnarray}\label{brem} \rmd P_{\rm
    Brem}\,\simeq\,C_R\frac{\alpha_s(k_\perp^2)
    }{\pi^2}\,\frac{\rmd^2k_\perp}{k_\perp^2}\,\frac{\rmd
    x}{x}\; ,
\end{eqnarray}
where $C_R$ is the SU$(N_c)$ Casimir in the colour representation of the
emitter: $C_A=N_c$ for a gluon and $C_F=(N_c^2-1)/2N_c$ for a quark.
($N_c$ is the number of colours, which is equal to 3 in real QCD, but it
is often kept as a free parameter in theoretical studies, because many
calculations simplify in the formal limit $N_c\gg 1$. The results
obtained in this limit provide insightful, qualitative and
semi--quantitative informations about real QCD.) \Eref{brem} exhibits the
collinear ($k_\perp\to 0$) and soft ($x\to 0$) singularities mentioned
above, which result in the enhancement of gluon emission at small
$k_\perp$ and/or $x$. If the emitted parton with small $x$ were a quark
instead a gluon, there would be no small $x$ enhancement, only the
collinear one. This asymmetry, due to the spin--1 nature of the gluon,
has the remarkable consequence that the small--$x$ part of the
wavefunction of any hadron is built mostly with gluons.

As manifest on \Eref{brem}, parton branching is suppressed by a power of
$\alpha_s(k_\perp^2)$, which is small when $k_\perp\gg \Lam$. But this
suppression can be compensated by the large phase--space available for
the emission, which equals $\ln(Q^2/\Lam^2)$ for the emission of a parton
(quark or gluon) with transverse momentum $k_\perp\ll Q$ and,
respectively, $\ln(1/x)$ for that of a gluon with longitudinal momentum
fraction $\xi$ within the range $x\ll \xi\ll 1$. Hence, for large
$Q^2\gg\Lam^2$ and/or small $x\ll 1$, such radiative processes are not
suppressed anymore and must be resummed to all orders. Depending upon the
relevant values of $Q^2$ and $x$, one can write down evolution equations
which resum either powers of $\alpha_s\ln(Q^2/\Lam^2)$, or of
$\alpha_s\ln(1/x)$, to all orders. The coefficients in these equations
represent the elementary splitting probability and can be computed as
power series in $\alpha_s$, starting with the leading--order result in
Eq.~(\ref{brem}).

\begin{figure}[t]
\begin{center}\centerline{
\includegraphics[width=.25\textwidth]{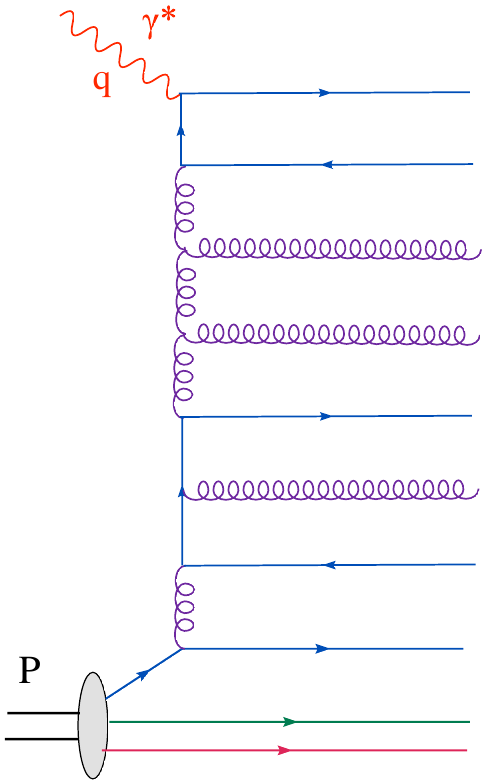}\qquad
 \qquad\qquad
\includegraphics[width=.27\textwidth]{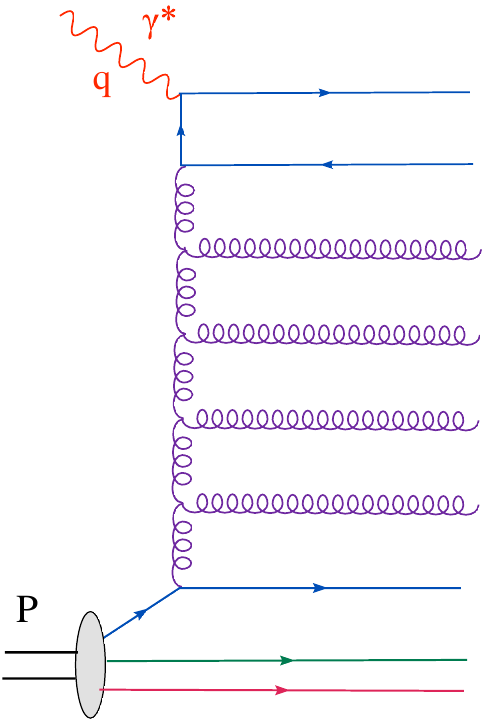}
}
\caption{\sl Parton evolution in perturbative QCD. The parton cascade
 on the right involves only gluons (at intermediate stages)
 and is a part of the BFKL resummation
 at small $x$ (see the discussion around \Eref{eq:unintp}).}
\label{fig:cascade}
\end{center}\vspace*{-.8cm}
\end{figure}

The evolution with increasing $Q^2$ is described by the DGLAP equation
(from Dokshitzer, Gribov, Lipatov, Altarelli and Parisi)
\cite{DGLAP,Ellis:1991qj}. This evolution mixes quarks and gluons (see
Fig.~\ref{fig:cascade}.a), which in particular allows us to reconstruct
the gluon distribution from the experimental results for $F_2$. The
small--$x$ evolution, on the other hand, involves only gluons and
corresponds to resumming ladder diagrams like those in
Fig.~\ref{fig:cascade}.b in which successive gluons are strongly ordered
in $x$ (see below). Both evolutions lead to an increase in the number of
partons at small values of $x$ (and a decrease at large values $x\gtrsim
0.1$), but the physical consequences are very different in the two cases:

\texttt{(i)} When increasing $Q^2$, one emits partons which occupy a
smaller transverse area $\sim 1/Q^2$, as shown in \Fref{HERA-gluon}
(right). The decrease in the area of the individual partons is much
stronger than the corresponding increase in their number. Accordingly,
the {\em occupation number} in the transverse plane {\em decreases} with
increasing $Q^2$, meaning that the partonic system becomes {\em more and
more dilute}. Accordingly, the partons may be viewed as {\em
independent}. This observation lies at the basis of the conventional
parton picture, which applies for sufficiently high $Q^2$ (at a given
value of $x$).

The parton occupation number mentioned above yields the proper measure of
the parton density in the hadron. It can be estimated as [{\em the number
of partons with a given value of $x$}] $\times$ [{\em the area occupied
by one parton}] divided by [{\em the transverse area of the hadron}],
that is (for gluons, for definiteness),
 \beq\label{occup}
 n(x,Q^2)\,\simeq\,\frac{xg(x,Q^2)}{Q^2R^2}\,,\qquad\mbox{with}\quad
  g(x,Q^2)\,\equiv\,\frac{\rmd N_{g}}{\rmd x}(Q^2)
 \,=\,\int^{Q}\!
 \rmd^2\bmk_\perp\, \frac{\rmd N_{g}}{\rmd x\rmd^2\bmk_\perp}
 \,,\eeq
where $R$ is the hadron radius in its rest frame (so its transverse area
is $\sim \pi R^2$ in any frame). The numerator in the above definition of
the occupation number, that is
 \beq\label{xGx}
 xg(x,Q^2)\,\equiv\,x\,\frac{\rmd N_{g}}{\rmd x}\,=\,k_z\,
 \frac{\rmd N_{g}}{\rmd k_z}\,\simeq\,
 \frac{\Delta N_{g}}{\Delta z\,\Delta k_z}\,,
 \eeq
is known as the {\em gluon distribution}. The last estimate above follows
from the uncertainty principle: partons with longitudinal momentum
$k_z=xP$ are delocalized in $z$ over a distance $\Delta z\simeq 1/k_z$.
Hence, the gluon distribution yields the number of gluons {\em per unit
of longitudinal phase--space}, which is indeed the right quantity for
computing the occupation number. Note that gluons with $x\ll 1$ extends
in $z$ over a distance $\Delta z\sim 1/xP$ which is much larger than the
Lorentz contracted width of the hadron, $R/\gamma\sim 1/P$. This shows
that the image of an energetic hadron as a `pancake', that would be
strictly correct if the hadron was a classical object, is in reality a
bit naive: it applies for the valence quarks with $x\sim \order{1}$
(which carry most of the total energy), but not also for the small--$x$
partons (which are the most numerous, as we shall shortly see).

\texttt{(ii)} When decreasing $x$ at a fixed $Q^2$, one emits mostly
gluons which have smaller longitudinal momentum fractions, but which
occupy, roughly, the same transverse area as their parent gluons (see
\Fref{HERA-gluon} right). Then the gluon occupation number, \Eref{occup},
{\em increases}, showing that {\em the gluonic system evolves towards
increasing density.} As we shall see, this evolution is quite fast and
eventually leads to a breakdown of the picture of independent partons.

\begin{figure*}[t]
\begin{center}
\centerline{
\includegraphics[width=0.35\textwidth]{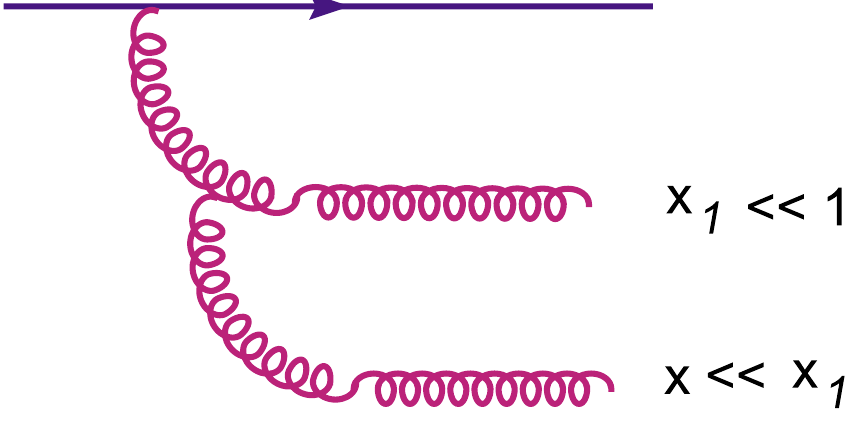}
\includegraphics[width=0.35\textwidth]{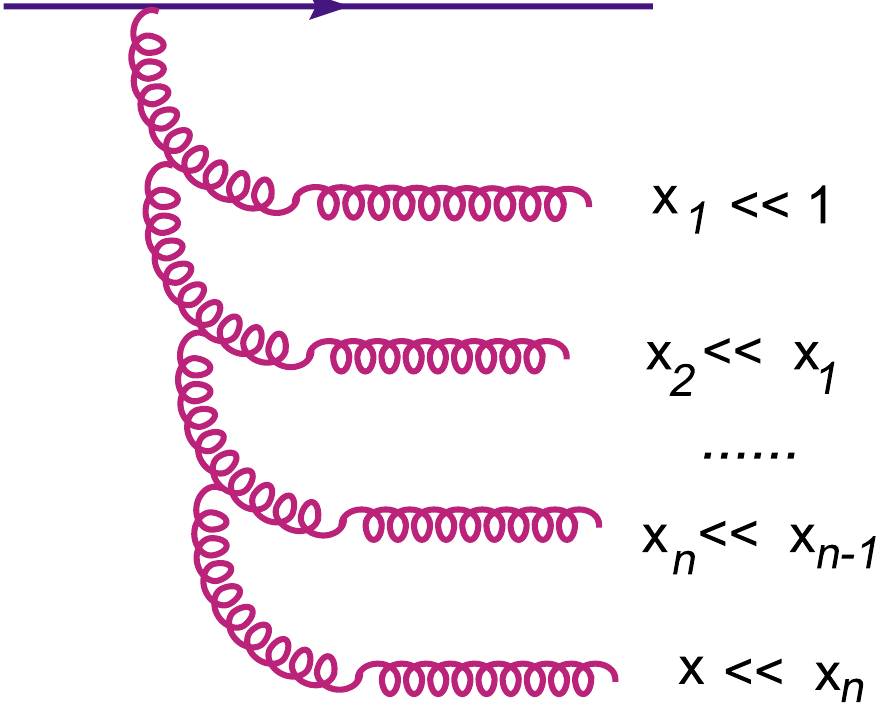}\quad
} \caption{\sl Gluon cascades produced by the high--energy (BFKL)
evolution of the proton wavefunction.\label{fig:BFKL}}
\end{center}\vspace*{-.8cm}
\end{figure*}

\begin{figure*}[t]
\begin{center}\centerline{
 \includegraphics[width=0.45\textwidth]{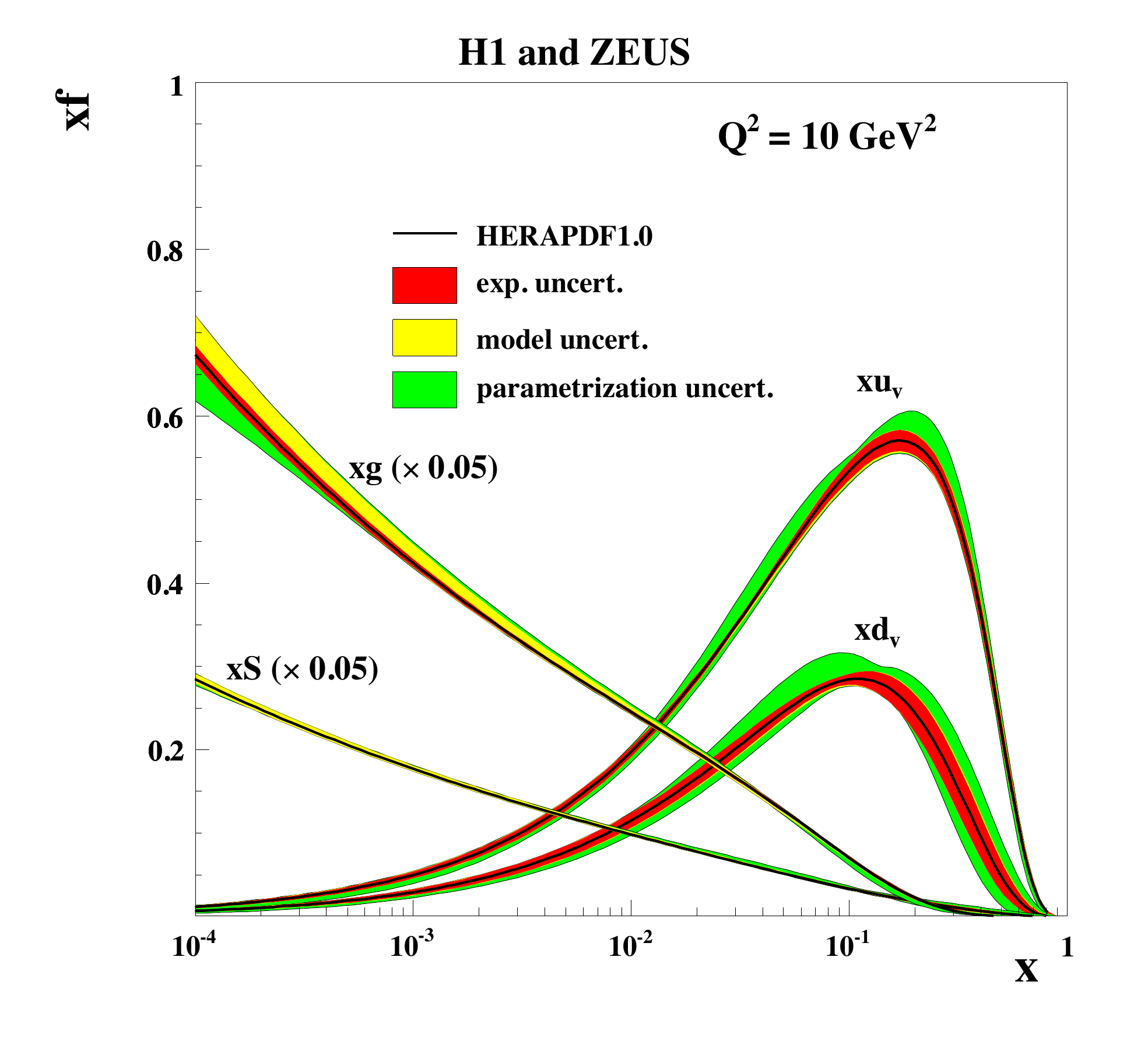}
 \includegraphics[width=0.5\textwidth]{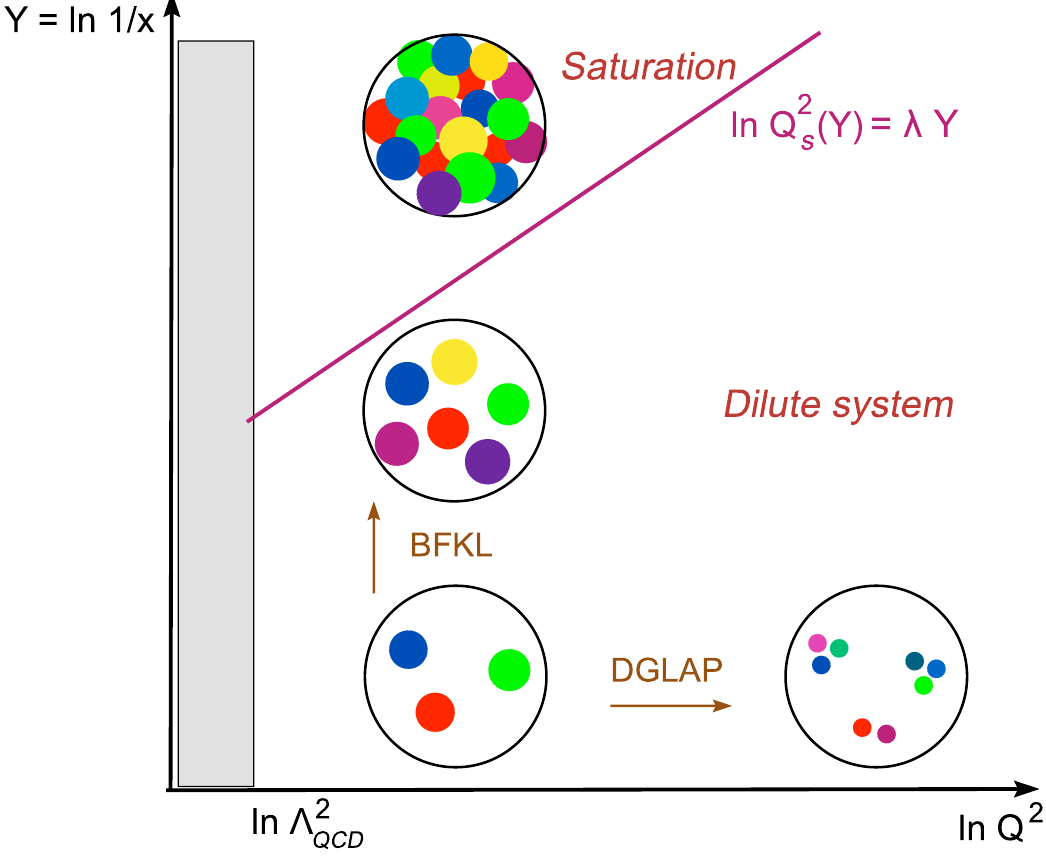}}
\caption{\sl Left: the $1/x$--evolution of the gluon, sea quark, and
valence quark distributions for $Q^2=10$ GeV$^2$, as measured at HERA
(combined H1 and ZEUS analysis \cite{Gwenlan:2009kr}). Note that the
gluon and sea quark distributions have been reduced by a factor of 20 to
fit inside the figure. Right: the `phase--diagram' for parton evolution
in QCD; each coloured blob represents a parton with transverse area
$\Delta x_\perp \sim 1/Q^2$ and longitudinal momentum $k_z=xP$. The
straight line $\ln Q_s^2(x)=\lambda Y$ is the saturation line, cf.
\Eref{QsxA}, which separates the dense and dilute regimes.
 \label{HERA-gluon}}
\end{center}\vspace*{-1.cm}
\end{figure*}

In order to describe the small--$x$ evolution, let us start with the
gluon distribution generated by a single valence quark. This can be
inferred from the bremsstrahlung law in \Eref{brem} (the emission
probability is the same as the number of emitted gluons) and reads
 \begin{eqnarray}\label{xGxp}
x \frac{\rmd N_g}{\rmd x}(Q^2)\,=\,\frac{\alpha_s
 C_F}{\pi}\,\int_{\Lam^2}^{Q^2}\frac{\rmd k_\perp^2}{k_\perp^2}
 \,=\,\frac{\alpha_s C_F}{\pi}\,\ln\left(\frac{Q^2}{\Lam^2}\right)\; ,
\label{eq:Ng-1}
\end{eqnarray}
where we have ignored the running of the coupling --- formally, we are
working to leading order (LO) in pQCD where the coupling can be treated
as fixed --- and the `infrared' cutoff $\Lam$ has been introduced as a
crude way to account for confinement: when confined inside a hadron, a
parton has a minimum virtuality of $\order{\Lam^2}$. In \Eref{xGxp} it is
understood that $x\ll 1$. In turn, the soft gluon emitted by the valence
quark can radiate an even softer gluon, which can radiate again and
again, as illustrated in figure~\ref{fig:BFKL}.  Each emission is
formally suppressed by a power of $\alpha_s$, but when the final value of
$x$ is tiny, the smallness of the coupling constant can be compensated by
the large available phase--space, of order $\ln(1/x)$ per gluon emission.
This evolution leads to an increase in the number of gluons with $x\ll
1$.

For a quantitative estimate, consider the first such correction, that is,
the two--gluon diagram in Fig.~\ref{fig:BFKL} left: the region in
phase--space where the longitudinal momentum fraction $x_1$ of the
intermediate gluon obeys $x\ll x_1\ll 1$ provides a contribution of
relative order
 \beq
 \frac{\alpha_s N_c}{\pi}\,\int_{x}^1\frac{\rmd x_1}{x_1}
 \,=\,\abar\ln\frac{1}{x}\,,\qquad \abar\equiv\frac{\alpha_s N_c}{\pi}\,.
 \eeq
When $\abar\ln(1/x)\sim 1$, this becomes of $\order{1}$, meaning that
this two--gluon diagram contributes on the same footing as the single
gluon emission in Fig.~\ref{onegluon}. A similar conclusion holds for a
diagram involving $n$ intermediate gluons {\em strongly ordered} in $x$,
cf. Fig.~\ref{fig:BFKL} right, which yields a relative contribution of
order
 \begin{eqnarray}\label{ngluons}
 \abar^n\,\int_{x}^1\frac{\rmd x_n}{x_n}
 \int_{x_n}^1\frac{\rmd x_{n-1}}{x_{n-1}}\cdots
 \int_{x_2}^1\frac{\rmd x_1}{x_1}
 \,=\,\frac{1}{n!}\left(\abar\ln\frac{1}{x}\right)^n
\; .
 \end{eqnarray}
When $\abar\ln(1/x)\gtrsim 1$, the correct result for the gluon
distribution at leading order is obtained by summing contributions from
all such ladders. As clear from \Eref{ngluons}, this sum {\em
exponentiates}, modifying the integrand of Eq.~(\ref{eq:Ng-1}) into
 \begin{eqnarray}\label{eq:unintp}
x \frac{\rmd N_g}{\rmd x \rmd k_\perp^2}\,\sim\,\frac{\alpha_s
 C_F}{\pi}\,\frac{1}{k_\perp^2}\,\rme^{\omega\abar Y}\; ,\qquad
 Y\equiv\ln \frac{1}{x}\;,
 \end{eqnarray}
where $\omega$ is a number of order unity which cannot be determined via
such simple arguments. The variable $Y$ is the rapidity difference
between the final gluon and the original valence quark and it is often
simply referred to as `the rapidity'. The quantity in the l.h.s. of
\Eref{eq:unintp} is the number of gluons per unit rapidity and with a
given value $k_\perp$ for the transverse momentum, a.k.a. the {\em
unintegrated gluon distribution}\footnote{The occupation number
\eqref{occup} is more correctly defined as the unintegrated gluon
distribution per unit transverse area: $n(Y,\bmk_\perp)= {\rm d}
N_g/({\rmd Y {\rm d}^2\bmk_\perp{\rm d}^2\bmb_\perp})$ where $\bmb_\perp$
(the `impact parameter') is the transverse position of a gluon with
respect to the center of the hadron.}.

To go beyond this simple power counting argument, one must treat more
accurately the kinematics of the ladder diagrams and include the
associated virtual corrections. The result is the {\em BFKL equation}
(from Balitsky, Fadin, Kuraev, and Lipatov) \cite{BFKL} for the evolution
of the unintegrated gluon distribution with $Y$. The solution of this
equation, which resums perturbative corrections $(\abar Y)^n$ to all
orders, confirms the exponential increase in Eq.~(\ref{eq:unintp}),
albeit with a $k_\perp$--dependent exponent and modifications to the
$k_\perp^{-2}$--spectrum of the emitted gluons.

An important property of the BFKL ladder is its {\em coherence in time} :
the lifetime of a parton being proportional to its value of $x$, $\Delta
t\simeq 2k_z/k_\perp^2\propto x$, cf. \Eref{lifetime}, the `slow' gluons
at the lower end of the cascade have a much shorter lifetime than the
preceding `fast' gluons. Therefore, for the purposes of small--$x$
dynamics, fast gluons with $x'\gg x$ act as {\em frozen colour sources
emitting gluons at the scale $x$}. Because these sources may overlap in
the transverse plane, their colour charges add coherently, giving rise to
a large colour charge density. The {\em average} colour charge density is
zero by gauge symmetry but {\em fluctuations} in the colour charge
density --- as measured in particular by the unintegrated gluon
distribution --- are nonzero and increase rapidly with $1/x$, cf.
\Eref{eq:unintp}.

This growth is indeed seen in the data: e.g., the HERA data for DIS
confirm that the proton wavefunction at $x< 0.01$ is totally dominated by
gluons (see \Fref{HERA-gluon} left). However, on physical grounds, such a
rapid increase in the gluon distribution cannot go on for ever (that is,
down to arbitrarily small values of $x$). Indeed, the BFKL equation is
{\em linear} --- it assumes that the radiated gluons do not interact with
each other, like in the conventional parton picture. While such an
assumption is perfectly legitimate in the context of the
$Q^2$--evolution, which proceeds towards increasing diluteness, it
eventually breaks down in the context of the $Y$--evolution, which leads
to a larger and larger gluon density. As long as the gluon occupation
number \eqref{occup} is small, $n\ll 1$, the system is dilute and the
mutual interactions of the gluons are negligible. When $n\sim\order{1}$,
the gluons start overlapping, but their interactions are still weak,
since suppressed by $\alpha_s\ll 1$. The effect of these interactions
becomes of order one only when $n$ is as large as $n\sim
\order{1/\alpha_s}$. When this happens, non--linear effects (to be
shortly described) become important and stop the further growth of the
gluon distribution. This phenomenon is known as {\em gluon saturation}
\cite{Gribov:1984tu,Mueller:1985wy,McLerran:1993ni}. An important
consequence of it is to introduce a new transverse--momentum scale in the
problem, the {\em saturation momentum} $Q_s(x)$, which is determined by
\Eref{occup} together with the condition that $n\sim {1/\alpha_s}$~:
 \beq\label{Qsat}
 n\big(x,Q^2=Q_s^2(x)\big)\,\sim \,\frac{1}{\alpha_s}\ \Longrightarrow
 \ Q_s^2(x)\,\simeq\,\alpha_s\frac {xg\big(x,Q_s^2(x)\big)}{R^2}\,.\eeq
Except for the factor $\alpha_s$, the r.h.s. of \Eref{Qsat} is recognized
as the density of gluons per unit transverse area, for gluons localized
within an area $\Sigma\sim 1/Q_s^2(x)$ set by the saturation scale.
Gluons with $k_\perp\le Q_s(x)$ are at saturation: the corresponding
occupation numbers are large, $n\sim 1/\alpha_s$, but do not grow anymore
when further decreasing $x$. Gluons with $k_\perp\gg Q_s(x)$ are still in
a dilute regime: the occupation numbers are relatively small $n\ll
1/\alpha_s$, but rapidly increasing with $1/x$ via the BFKL evolution.
The separation between the saturation (or dense, or CGC) regime and the
dilute regime is provided by the {\em saturation line} in
\Fref{HERA-gluon} right, to be further discussed below.

\begin{figure}[t]
\begin{center}
\centerline{
\includegraphics[width=0.35\textwidth]{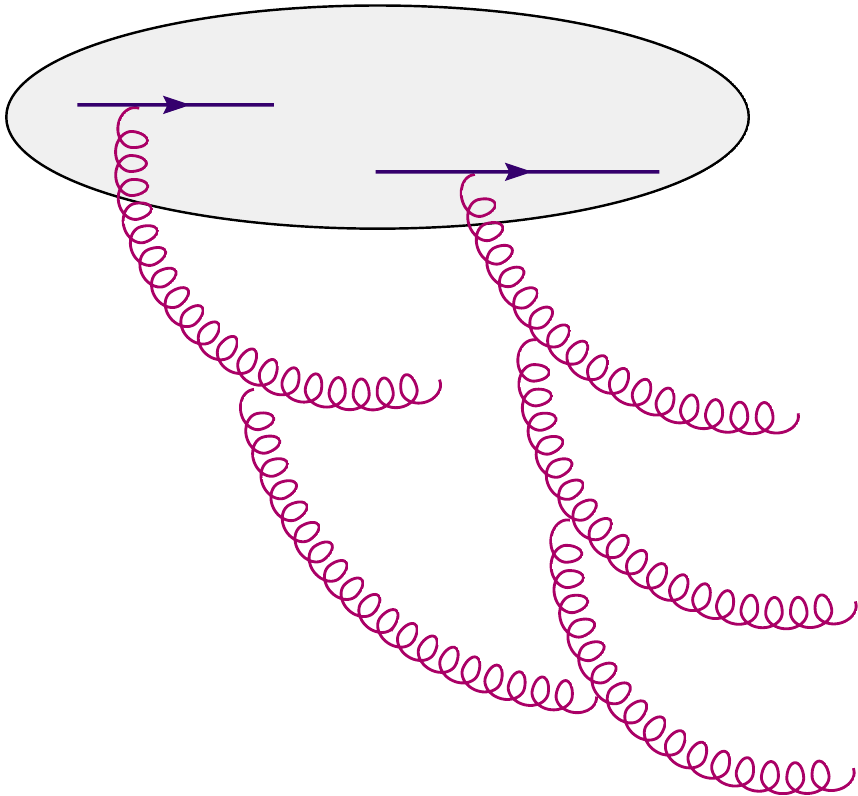}\qquad
\quad \includegraphics[width=0.45\textwidth]{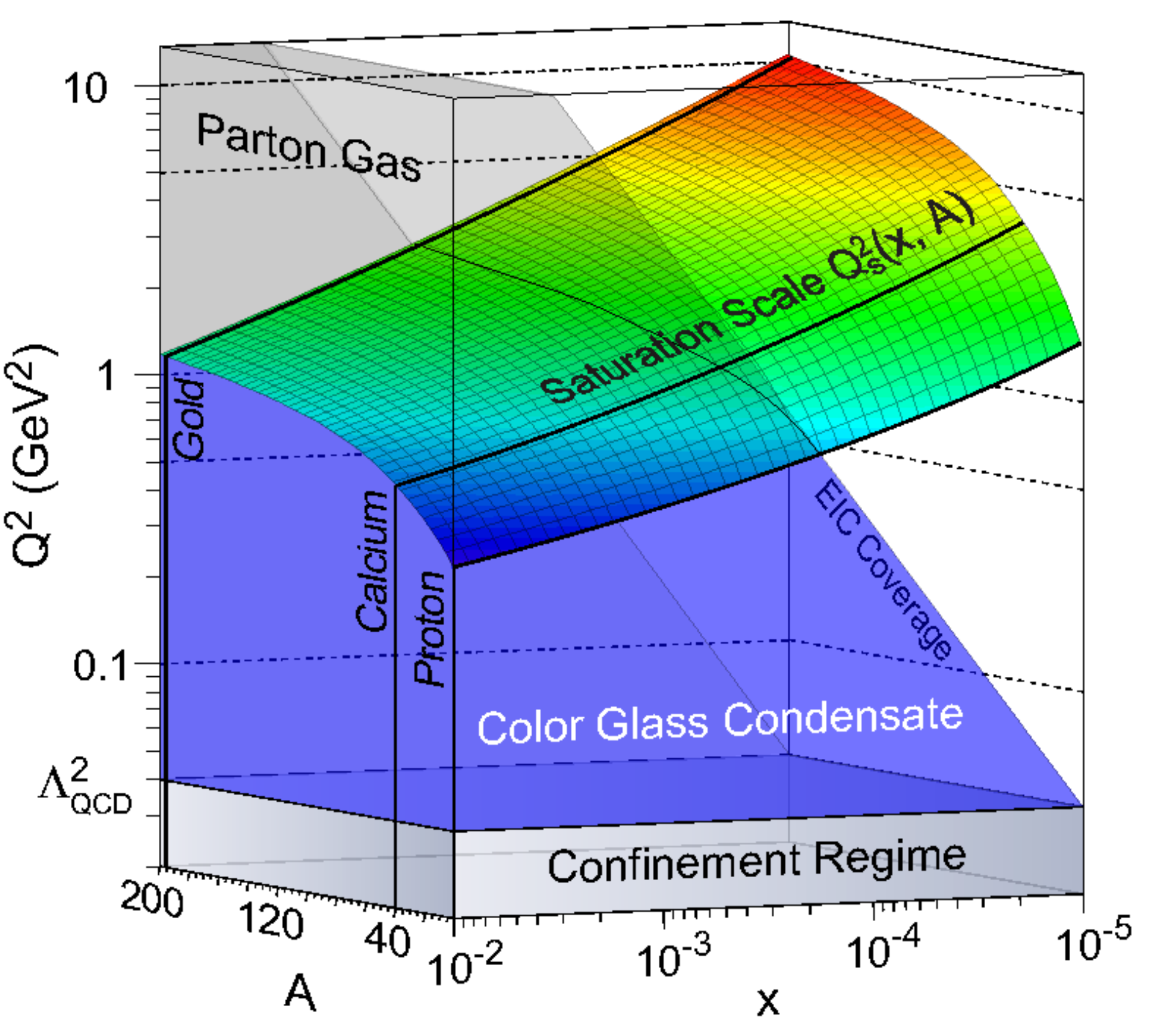} 
} \caption{\sl Left: $gg\to g$ recombination process leading to saturation.
Right: the saturation momentum $Q_s(x,A)$ as a function of the
longitudinal momentum fraction $x$ and of the atomic number $A$.
\label{fig:Qsat}}
\end{center}\vspace*{-.8cm}
\end{figure}

The microscopic interpretation of \Eref{Qsat} can be understood with
reference to \Fref{fig:Qsat} (left) : gluons which have similar values of
$x$ (and hence overlap in the longitudinal direction) and which occupy a
same area $\sim 1/Q^2$ in the transverse plane can {\em recombine} with
each other, with a cross--section $\sigma_{gg\to g}\simeq\alpha_s/Q^2$.
After taking also this effect into account, the change in the gluon
distribution in one step of the small--$x$ evolution (i.e. under a
rapidity increment $Y\to Y+\rmd Y$) can be schematically written as
 \begin{equation}
 \frac{\partial }{\partial Y}\,xg(x,Q^2)\,=\,
 \omega \abar xg(x,Q^2)
 -\, \abar\,\frac{\alpha_s}{Q^2 R^2} \,{\big[xg(x,Q^2)\big]^2}
 \; .
 \label{eq:GLR}
 \end{equation}
The overall factor of $\abar$ in the r.h.s. comes from the differential
probability $\propto\abar \rmd Y$ to emit one additional gluon in this
evolution step, cf. \Eref{brem}. The first term, linear in $xg(x,Q^2)$,
represents the BFKL evolution; by itself, this would lead to the
exponential growth with $Y$ shown in \Eref{eq:unintp}. The second term,
quadratic in $xg(x,Q^2)$, is the rate for recombination. This is formally
suppressed by one factor $\alpha_s$, but it becomes as important as the
first term when $Q^2$ is of the order of the saturation momentum
$Q_s^2(x)$ introduced in \Eref{Qsat}. When that happens, the r.h.s. of
\Eref{eq:GLR} vanishes, and then the gluon distribution stops growing
with $Y$. The above argument, due to Gribov, Levin and Ryskin back in
1983 \cite{Gribov:1984tu}, is a bit oversimplified (and the actual
evolution equation is considerably more complicated than \Eref{eq:GLR};
see the review papers
\cite{Iancu:2002xk,Mueller:2001fv,Iancu:2003xm,JalilianMarian:2005jf,Weigert:2005us,Triantafyllopoulos:2005cn,Gelis:2010nm,Lappi:2010ek}
and the discussion in \Sref{sec:jimwlk} below), but it has the merit to
illustrate in a simple way the physical mechanism at work: the gluon
occupation numbers saturate because the non--linear effects associated
with the high gluon density compensate the bremsstrahlung processes.

Remarkably, \Eref{Qsat} implies that the saturation momentum increases
with $1/x$, since so does the gluon distribution for $k_\perp\gtrsim
Q_s(x)$, cf. \Eref{eq:unintp}. So, for sufficiently small values of $x$
(say, $x\le 10^{-4}$ in the case of a proton), one expects $Q_s^2(x)\gg
\Lam^2$. In that case, the (semi)hard scale $Q_s(x)$ supplants $\Lam$ as
an infrared cutoff for the calculation of physical observables like the
multiplicity (cf. the discussion at the end of \Sref{sec:eta}). This has
the remarkable consequence that, for sufficiently high energy, the bulk
of the particle production can be computed in perturbation theory. But
the proper framework to perform this calculation is not standard pQCD as
based on the collinear factorization, but the CGC effective theory which
includes the non--linear physics of gluon saturation. This will be
discussed in the next subsection.

Gluon occupancy is further amplified if instead of a proton we consider a
large nucleus with atomic number $A\gg 1$.  The corresponding gluon
distribution $xg_A(x,Q^2)$ scales like $A$, since gluons can be radiated
by any of the $3A$ valence quarks of the $A$ nucleons. Since the nuclear
radius scales like $R_A\sim A^{1/3}$, \Eref{occup} implies that the gluon
occupation number scales as $A^{1/3}$. This factor is about 6 for the Au
and Pb nuclei respectively used at RHIC and the LHC. Thus, for a large
nucleus, saturation effects become important at larger values of $x$ than
for a proton. This explains why ultrarelativistic heavy ion collisions
represent a privileged playground for observing and studying the effects
of saturation.

\Fref{fig:Qsat} (right) summarizes our current expectations for the value
and the variation of the saturation momentum. The dependence upon $x$ is
by now known to next-to-leading-order (NLO) accuracy
\cite{Triantafyllopoulos:2002nz} --- that is, by resumming radiative
corrections $\alpha_s\big[\alpha_s\ln (1/x)\big]^n$ to all orders
together with non--linear effects. The result can be roughly expressed as
 \beq\label{QsxA}
 Q^2_s(x,A)\, \simeq \,Q_0^2\,A^{1/3}
 \left(\frac{x_0}{x}\right)^{\lambda}\,,
 \qquad\mbox{with}\quad \lambda=0.2\div 0.3\,,
 \eeq
with the power $\lambda$ known as the {\em saturation exponent}. The
overall scale $Q_0^2$, which has the meaning of the proton saturation
scale at the original value $x_0$, is non--perturbative and cannot be
computed within the CGC effective theory. (The latter governs only the
evolution from $x_0$ down to $x\ll x_0$.) In practice, this is treated as
a free parameter which is fitted from the data. The fits yield $Q_0\simeq
0.5$~GeV for $x_0=10^{-2}$. \Fref{fig:Qsat} shows that for $x=10^{-5}$ (a
typical value for forward particle production at the LHC), $Q_s\simeq
1$~GeV for the proton, while $Q_s\simeq 3$~GeV for the Pb nucleus. This
difference is significant: while 1~GeV is only marginally perturbative,
3~GeV is sufficiently `hard' to allow for controlled perturbative
calculations. This confirms the usefulness of HIC as a laboratory to
study saturation.

\begin{figure}[t]
\begin{center}\centerline{
\includegraphics[width=.4\textwidth]{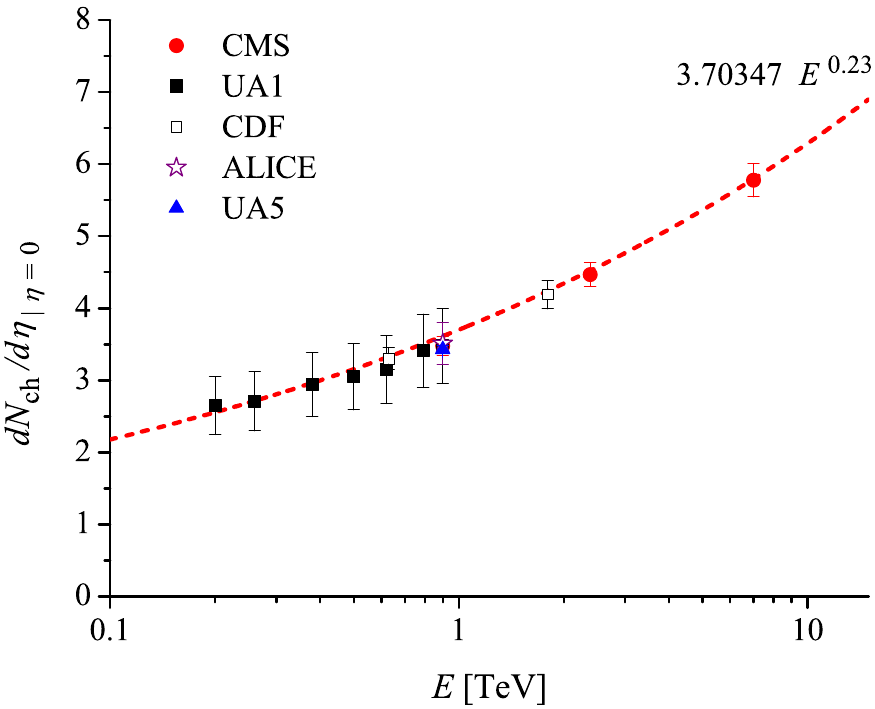}\qquad \quad
 \includegraphics[width=.5\textwidth]{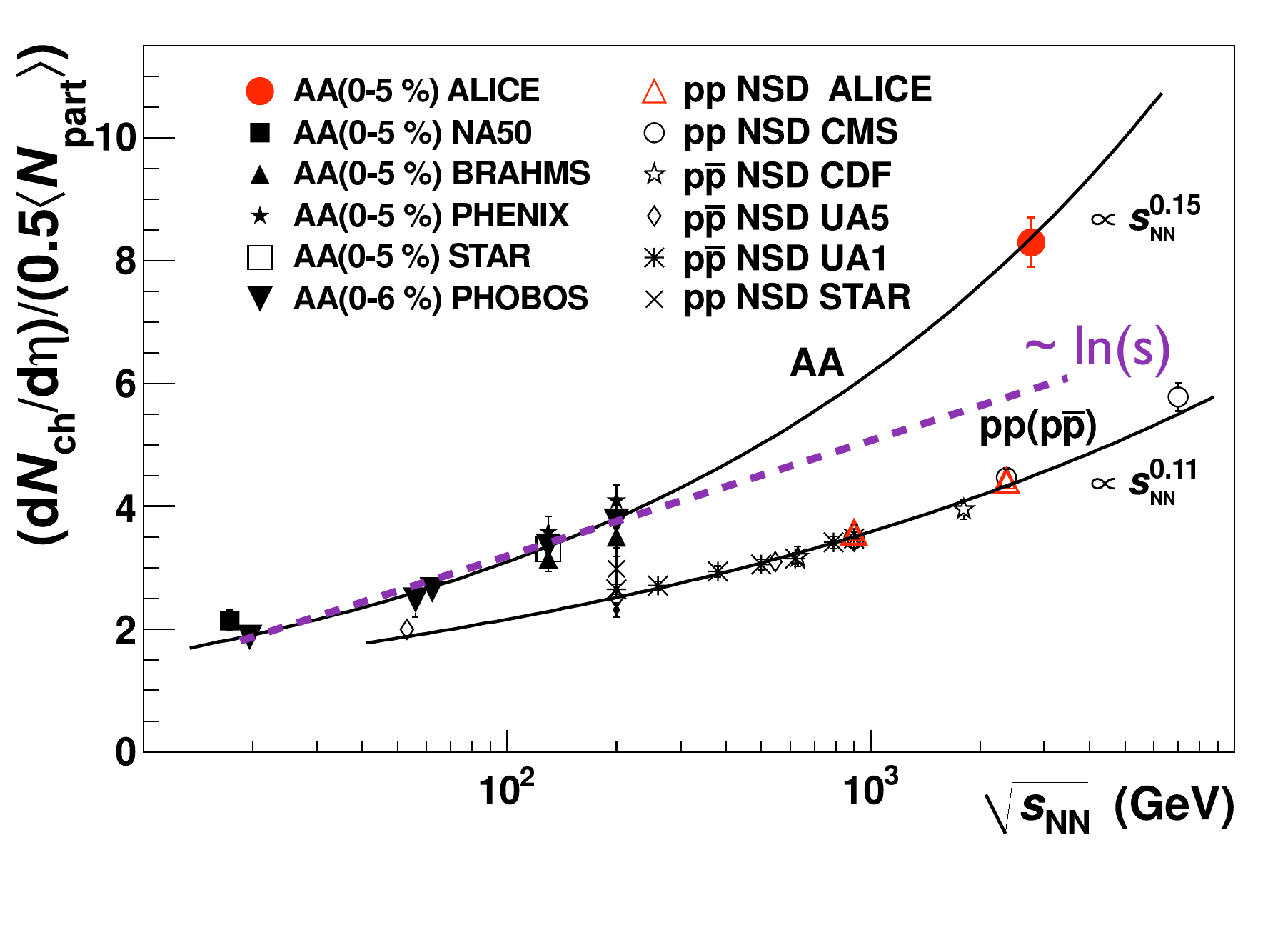}
}
\caption{\sl Charged particles multiplicity at central (pseudo)rapidity $\eta=0$
as a function of the COM energy for p+p and nucleus--nucleus collisions.
The data are consistent with a power--law increase with $\sqrt{s}$ (with
an exponent which is slightly larger for A+A than for p+p), but
appear to exclude a logarithmic law $\propto\ln s$.
} \label{fig:multiplicity}
\end{center}\vspace*{-.8cm}
\end{figure}

Before we conclude this subsection, let us notice some robust predictions
of the saturation physics, which do not require a detailed theory and can
be directly checked against the data. One of them refers to the
energy--dependence of the average transverse momentum of the produced
particles: as shown in \Fref{fig:avept}, this grows like a power of
$E=\sqrt{s}$, with an exponent which is fitted from the data as $0.115$.
This is consistent with expectations based on gluon saturation
\cite{McLerran:2010ex}. Indeed, prior to the collision, the gluon
distribution inside the hadron wavefunction is peaked at $k_\perp\sim
Q_s$ (see \Fref{fig:Phi} below and the related discussion) and these
gluons are then released in the final state. We thus expect the average
$p_\perp$ of the produced hadrons to scale like $Q_s(x)$ evaluated at the
appropriate value of $x$, that is, $x=p_\perp/E$ (cf. \Eref{2pkin}). This
argument implies $\langle p_\perp\rangle\propto E^{\lambda/2}$, which is
indeed consistent with the data in \Fref{fig:avept} (right) together with
the estimate in \eqref{QsxA} for the saturation exponent. Another
prediction of this kind refers to the particle multiplicity in the final
state $\rmd N/\rmd\eta$, say, at central (pseudo)rapidity $\eta=0$. By
the above argument, this is dominated by gluons with $Q^2\simeq Q_s^2(x)$
and hence it is proportional to the respective gluon distribution, that
is, to $Q_s^2(x)$ itself (cf. \Eref{Qsat}) : $\rmd
N/\rmd\eta\,\propto\,Q_s^2(E)\sim E^\lambda$. Once again, this appears to
be consistent with the data for both p+p and A+A collisions, as shown in
\Fref{fig:multiplicity}.

\subsection{The CGC effective theory}
\label{sec:jimwlk}

The partonic form of matter made with the saturated gluons is known as
the {\em colour glass condensate}
\cite{Iancu:2002xk,Mueller:2001fv,Iancu:2003xm,JalilianMarian:2005jf,Weigert:2005us,Triantafyllopoulos:2005cn,Gelis:2010nm,Lappi:2010ek}.
\begin{itemize}

\item This is {\em coloured} since gluons carry the `colour' charge
    of the non--Abelian group SU$(3)$.

\item It is a {\em glass} because of the separation in time scales,
    due to Lorentz time dilation, between the `slow' gluons at small
    $x$ and their `fast' sources at larger $x$. The sources appear as
    `frozen' over the characteristic time scales for the dynamics at
    small $x$, but they can vary over much larger time scales, as set
    by their own, comparatively large, longitudinal momenta. A system
    which behaves as a solid on short time scales and as a fluid on
    much longer ones, is a glass.

\item It is a {\em condensate} because the saturated gluons and their
    sources have high occupation numbers $n(x,k_\perp)\sim
    1/\alpha_s$ and their colour charges add coherently to each
    other, as explained in \Sref{sec:evol} in relation with the BFKL
    ladder. A coherent quantum state with high occupancy can be in a
    first approximation described as a {\em classical field} (here, a
    colour field), which is the most generic example of a condensate.

\end{itemize}

\noindent Because of its high density, the CGC is {\em weakly coupled}
and thus it can be studied within perturbative QCD. This is strictly
correct for sufficiently small values of $x$, such that $Q_s^2(x)\gg
\Lam^2$ and hence $\alpha_s(Q_s^2)\ll 1$, but it remains marginally true
for the phenomenology at RHIC and, especially, the LHC, where the
saturation momentum is semi--hard, cf. \Fref{fig:Qsat} (right). Based on
that, an {\em effective theory} has been explicitly constructed, which
resums an infinite series of Feynman graphs of the ordinary perturbation
theory --- those which are enhanced by either the large logarithm
$\ln(1/x)$, or by the high gluon density. This theory governs the
dynamics of the gluons with a given, small, value of $x$, while the
gluons at larger values $x'\gg x$ have been `integrated out' in
perturbation theory. In order to describe its mathematical structure, it
is useful to recall that the gluon field in QCD is represented by a
non--Abelian vector potential $A^\mu_a(x)$ where the upper index $\mu$
refers to the 4 Minkowski coordinates and the subscript $a$ is a colour
index in the adjoint representation of SU$(N_c)$ and can take $N_c^2-1=8$
values.

The CGC effective theory may be viewed as a non--linear generalization of
the BFKL evolution, but in fact it is much more complex than just a
non--linear evolution equation (say, like that in \Eref{eq:GLR}). The
BFKL equation applies to the unintegrated gluon distribution (or
occupation number), which is a Fourier transform of the 2--point
function\footnote{See \Eref{Phi} for a more precise definition of the
unintegrated gluon distribution in the presence of non--linear effects.}
$\langle A^i_a(x) A^i_a(y)\rangle$ of the colour fields within the hadron
(The average refers to the hadron wavefunction and the upper index $i$
with $i=1,2$ indicates the transverse directions.) This quantity offers
more information than the standard parton distributions like $xg(x,Q^2)$
--- it also describes the distribution of gluons in transverse momentum,
and not only in $x$ ---, but it still does not probe {\em many--body
correlations} in the gluon distribution, as the higher $n$--point
functions with $n\ge 4$ would do. The restriction to the 2--point
function is justified so long as the system is dilute and gluons do not
interact with each other. But this cannot encode the non--linear physics
of saturation, which is sensitive to higher $n$--point functions and
hence to correlations. In fact, to correctly describe gluon saturation,
one needs to control $n$--point functions with {\em arbitrarily high
$n$}. This can be understood as follows: the fact that the occupation
numbers are $n\sim\order{1/\alpha_s}$ at saturation, means that the
colour field strengths are as large as $A^i_a\sim\order{1/g}$, and then
there is no penalty for inserting arbitrary powers of $A^i_a$. Indeed,
any such an insertion is accompanied by a factor of $g$. (Recall that
interactions in QCD enter via the covariant derivative
$D^\mu=\partial^\mu-igA^\mu$.) So, the CGC effective theory is truly {\em
an infinite hierarchy of coupled evolution equations} describing the
simultaneous evolution of all the $n$--point functions. Remarkably
enough, this hierarchy can be summarized into a single, {\em functional},
evolution equation for the {\em CGC weight function} --- a functional
generalization of the `unintegrated' gluon distribution that will be
shortly discussed.

\begin{figure}[t]
\begin{center}
\centerline{
\includegraphics[width=0.65\textwidth]{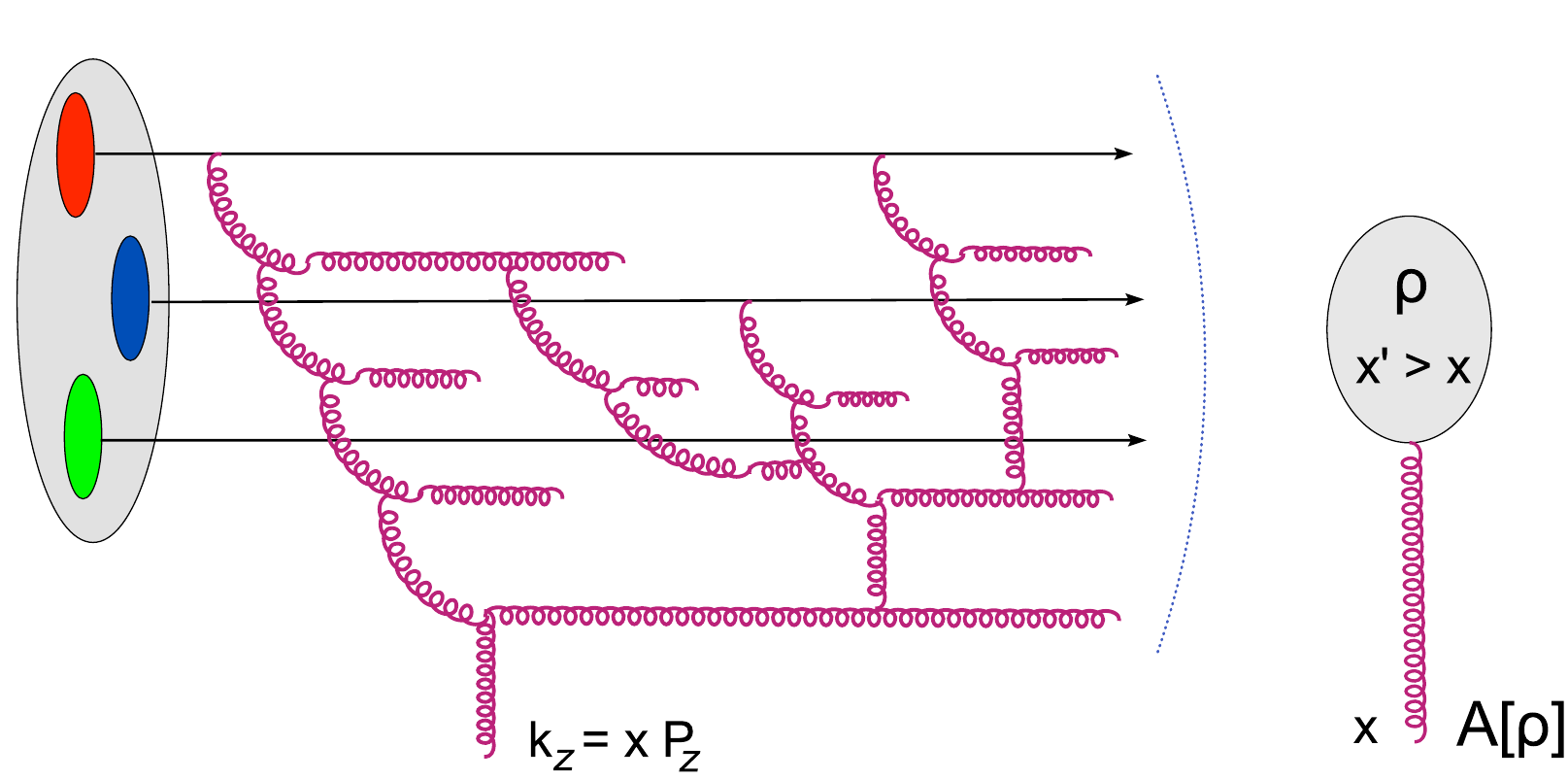}}
 \caption{\sl Schematic representation of the d.o.f. involved in the CGC effective
 theory and of the quantum evolution which is taken into account in this
 theory. A newly emitted gluon with a small longitudinal momentum
 fraction $x\ll 1$ rescatter off the gluon field $A^\mu_a[\rho]$ created in the
 previous steps by the gluons with larger values $x'\gg x$,
 effectively represented by their global colour charge density $\rho_a$.
  \label{fig:CGC}}
\end{center}\vspace*{-.8cm}
\end{figure}

The key ingredient for such an economical description is the proper
choice of the relevant degrees of freedom: as already mentioned, the
small--$x$ gluons with high occupation numbers $n\sim 1/\alpha_s$ can be
treated {\em semi--classically} to leading order in $\alpha_s$ --- that
is, they can be described as {\em classical colour fields} $A^\mu_a(x)$
radiated by {\em colour sources} representing the faster gluons with
$x'\gg x$. This distinction between `classical fields' (= the small--$x$
gluons for which the effective theory is built) and their `sources' (=
the large--$x$ gluons which are integrated out in the construction of the
effective theory) is illustrated in \Fref{fig:CGC}. The effective theory
based on this separation is valid to LO in $\alpha_s$, but to all orders
in $\alpha_s\ln(1/x)$ and in the classical field $A^\mu_a\sim
\order{1/g}$.

The mathematical structure of the CGC theory is rather complex and it
will be only schematically described here. To that aim, it is convenient
to switch to {\em light--cone vector notations}. Namely, for any
4--vector such as $x^\mu$, $p^\mu$, $A^\mu_a$ etc. we shall define its
light--cone (LC) components as
 \beq\label{LC}
 x^+\equiv \frac{1}{\sqrt{2}}\,(x^0 +x^3)\,,\quad
 x^-\equiv \frac{1}{\sqrt{2}}\,(x^0 -x^3)\,,\quad x^\mu_{\rm LC}
 = (x^+,x^-,\bmx_\perp)\,.
 \eeq
In LC notations, the scalar product reads $k\cdot x\equiv k_\mu x^\mu=
k^+x^- + k^-x^+ -\bmk_\perp\cdot\bmx_\perp$.

To see the usefulness of these notations, consider a right--moving
ultrarelativistic hadron, with $P^\mu\simeq (P,0,0,P)$ : this propagates
at nearly the speed of light along the trajectory $x^3=t$. In LC
notations, the 4--momentum $P^\mu_{\rm LC}\simeq (\sqrt{2}P,0,0,0)$ has
only a `plus' component, while the trajectory reads simply $x^-=0$. The
same holds for any of the large--$x$ partons which move
quasi--collinearly with the hadron and serve as sources for the
small--$x$ gluons that we are interested in. In the semi--classical
approximation, these small--$x$ gluons are described as the solution to
the Yang--Mills equations (the non--Abelian generalization of the Maxwell
equations) having these `fast' gluons as sources:
 \beq\label{YM}
 D_\nu^{ab} F_b^{\nu\mu}(x)\,=\,\delta^{\mu +}\rho^a(x^-,\bmx_\perp)\,.
 \eeq
In this equation, the l.h.s. features the covariant derivative
$D_\nu^{ab}=\partial^\nu-gf^{abc}A^\nu_c$ and the field strength tensor
$F_a^{\nu\mu} =\partial^\nu A^\mu_a-\partial^\mu
A^\nu_a-gf^{abc}A^\nu_bA^\mu_c$ associated with the classical colour
field, while the r.h.s. is the {\em colour current} of the `fast' gluons:
$J^\mu_a=\delta^{\mu+}\rho_a$, with $\rho^a(x^-,\bmx_\perp)$ their colour
charge density. The latter is localized in $x^-$ near $x^-=0$ and is
independent of time (hence of $x^+$), because these fast charges are
`frozen' by Lorentz time dilation. But the distribution of these charges
in transverse space is {\em random}, since the fast gluons can be in any
of the quantum configurations produced at the intermediate stages of the
gluon evolution down to $x$. The proper way to describe this randomness
is to give the {\em probability} to find a specific configuration
$\rho_a(x^-,\bmx_\perp)$ of the colour charge density. This probability
is a functional of $\rho_a(x^-,\bmx_\perp)$, known as {\em the CGC weight
function} and denoted as $W_Y[\rho]$, with $Y=\ln(1/x)$. This functional
is {\em gauge--invariant}, which in particular ensures that
$\langle\rho_a(x^-,\bmx_\perp)\rangle=0$, as it should.

To the accuracy of interest, all the observables relevant for the
scattering off the small--$x$ gluons are represented by gauge--invariant
operators built with the classical field $A^\mu_a$. If ${O}[A]$ is such
an operator, then its hadron expectation value is computed by averaging
over all the configurations of $\rho$ with the CGC weight function:
  \beq \langle {O[A]}\rangle_Y \equiv \int
{\mathcal D}\rho  \, W_Y[\rho]\, {O}\big[A[\rho]\big]\,,
 \label{average}
 \eeq
where $A^\mu_a[\rho]$ is the solution to \Eref{YM}.

The expectation value \eqref{average} depends upon the rapidity
$Y=\ln(1/x)$ via the corresponding dependence of the weight function
$W_Y[\rho]$. The latter is obtained by successively integrating the
quantum gluon fluctuations in layers of $x$, down to the value of
interest. One step in this evolution corresponds to the emission of a new
gluon (with a probability $\order{\alpha_s}$ per unit rapidity) out of
the preexisting ones. But unlike in the BFKL evolution, where gluons with
different rapidities do not `see' each other, in the context of the CGC
evolution, the newly emitted gluon is allowed to interact with the strong
colour field radiated by `sources' (gluons and valence quarks) with
higher values of $x$ (see \Fref{fig:CGC}). Accordingly, the change in the
CGC weight function in one evolution step is {\em non--linear} in the
background field $A^\mu_a[\rho]$, and hence in the colour charge density
$\rho_a$. This procedure generates a {\em functional evolution equation}
for $W_Y[\rho]$ with the schematic form (see
\cite{Iancu:2002xk,Iancu:2003xm} for details)
 \beq\label{jimwlk}
 {\partial W_Y[\rho] \over {\partial Y}}\,=\,H_{\rm
 JIMWLK}\left[\rho,{\delta \over {\delta \rho}}\right]\,W_Y[\rho]\,,
 \eeq
where the JIMWLK Hamiltonian (from Jalilian-Marian, Iancu, McLerran,
Weigert, Leonidov, and Kovner \cite{JKLW,CGC}) $H_{\rm JIMWLK}$ is
non--linear in $\rho$ to all orders (thus encoding the rescattering
effects in the emission vertex) but quadratic in the functional
derivatives $\delta/{\delta \rho}$ (corresponding to the fact that there
is only one new gluon emitted in each step in the evolution). In the
dilute regime, where parametrically $g\rho\ll 1$, the non--linear effects
are negligible, the JIMWLK Hamiltonian can be expanded to quadratic order
in $\rho$, and then it describes the BFKL evolution. But for $g\rho\sim
1$, the non--linear effects encoded in $H_{\rm JIMWLK}$ prevent the
emission of new gluons; this is {\em gluon saturation}.

Eqs.~\eqref{YM}--\eqref{jimwlk} are the central equations of the CGC
effective theory. When completed with an initial condition at the
rapidity  $Y_0$ at which one starts the high--energy evolution, they
fully specify the gluon distribution in the hadron wavefunction,
including all its correlations. The initial condition $W_{Y_0}[\rho]$ is
not determined by the effective theory itself, rather one must resort on
some model. For a large nucleus ($A\gg 1$) and for $Y_0=4\div 5$
(corresponding to $x_0\sim 0.01$), a reasonable initial condition is
provided by the McLarren--Venugopalan (MV) model \cite{McLerran:1993ni},
which assumes that the `fast' colour sources are the $N_c\times A$
valence quarks, which radiate independently from each other (since they
are typically confined within different nucleons). The corresponding
weight function is a Gaussian in $\rho_a$.

By taking a derivative w.r.t. $Y$ in \Eref{average} and using
\Eref{jimwlk} for $W_Y$, one can deduce evolution equations for all the
observables of interest. In general, these equations do not form a closed
set; rather, they form an infinite hierarchy (originally derived by
Balitsky \cite{Balitsky:1995ub}) which couples $n$--point functions with
arbitrarily large values of $n$. In practice, this hierarchy can be
truncated via mean field approximations
\cite{Iancu:2002xk,Kovchegov:2008mk,Iancu:2011nj}, leading to closed but
non--linear equations, in particular the Balitsky--Kovchegov equation
\cite{Kovchegov:1999yj}, that can be explicitly solved. It is also
possible to numerically solve the functional JIMWLK equation
\eqref{jimwlk}, by first reformulating this as a stochastic process (a
functional Langevin equation) \cite{Blaizot:2002xy} which can be
simulated on a lattice
\cite{Rummukainen:2003ns,Lappi:2011ju,Dumitru:2011vk}.

In order to describe a scattering cross--section, the CGC effective
theory developed so far must be combined with a factorization scheme.
This will be described in the next subsection.

\subsection{Particle production from the CGC}

\begin{figure}[t]
\begin{center}
\centerline{
\includegraphics[width=0.5\textwidth]{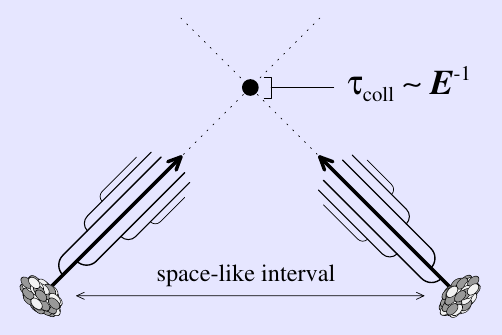}
\qquad\includegraphics[width=0.46\textwidth]{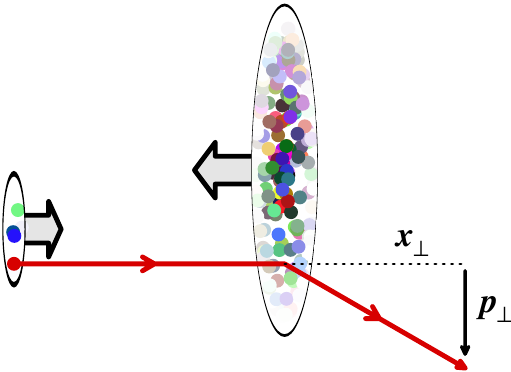}}
 \caption{\sl Left: The space--time picture of a high--energy collisions,
 illustrating the factorization of the cross--section. Right: particle
 production in a proton--nucleus (or `dilute--dense') collision.
  \label{fig:factor}}
\end{center}\vspace*{-.8cm}
\end{figure}

Let us start with some general remarks on factorization in scattering at
high energies: this is a generic consequence of {\em causality}. For a
hadron--hadron collision in the COM frame, the collision time $\Delta
t_{\rm coll}\sim 1/\sqrt{s}$ is much shorter than the lifetime
\eqref{lifetime} of the partons participating in the collisions, which is
proportional to the parton longitudinal momentum $xP\sim\sqrt{s}$. Hence,
these partons have been produced long time before the collision, at a
time where the two incoming hadrons were causally disconnected from each
other (see \Fref{fig:factor} left). Accordingly, the respective parton
distributions have evolved {\em independently} from each other and thus
they are {\em universal} --- i.e. independent of the scattering process
that is used to probe them. This argument is purely kinematic and hence
it remains true in the presence of QCD interactions leading to {\em
parton evolution} or {\em gluon saturation}. However, the precise form of
the factorization formula depends upon the kinematics and the structure
of the process at hand and it is different when probing dense or dilute
parts of the hadron wavefunction.

\texttt{(i)} In the {\em dilute regime}, which corresponds to the
situation where the transverse momenta $p_\perp$ of the produced partons
are significantly larger than the saturation momenta in the two hadrons
as evaluated at the relevant values of $x$, cf. \Eref{2pkin}, the
partonic subprocess involves merely a {\em binary collision} (cf.
\Fref{fig:2part} left) : one parton in one projectile interacts with one
parton in the other projectile, to produce the final state. Then, the
cross--section depends only upon the {\em parton densities} (the 2--point
correlations of the quark and gluon fields) in the incoming hadrons and
the factorization formula takes a rather simple form: the hadronic
cross--section is the convolution of two parton distribution functions
(one for each hadron) times the cross--section for the partonic
subprocess.

Even in this case, one needs to distinguish between two types of
factorizations, depending upon the kinematics of the final state:

\begin{itemize}

\item If the relevant values of $x$ are not that small (say $x\gtrsim
    0.01$), then the parton evolution with decreasing $x$ can be
    neglected and one can use the {\em collinear factorization} : the
    partons are assumed to move collinearly with the incoming hadrons
    (that is, one neglects their `intrinsic' transverse momenta
    $\sim{\Lam}$) and the parton distributions like $xg(x,\mu^2)$
    depend only upon the longitudinal momentum fractions and upon the
    transverse resolution $\mu^2$ (the `factorization scale') of the
    hard, partonic, subprocess. The dependence upon $\mu^2$ reflects
    the DGLAP evolution with increasing virtuality $Q^2$. \Eref{coll}
    provides an example of collinear factorization.

\item At smaller values of $x$, such that $\alpha_s\ln(1/x)\gtrsim
    1$, the small--$x$ evolution becomes important, leading to an
    increase in the density of gluons and in their average transverse
    momentum. Because of that, the gluons cannot be considered as
    `collinear' anymore : their distribution in transverse momenta
    must be explicitly taken into account. That is, one has to use
    the {\em `unintegrated' gluon distribution}, whose evolution with
    $1/x$ is described by the BFKL equation, cf. \Eref{eq:unintp}.
    The corresponding factorization formula, known as {\em
    $k_T$--factorization}, involves a convolution over the transverse
    momenta of the participating gluons. This is in fact a limiting
    case of the CGC factorization to be described shortly --- namely
    its dilute limit, in which the saturation effects can be
    neglected.

\end{itemize}

\texttt{(ii)} The {\em dense regime} corresponds to collisions which
probe saturation effects in at least one of the incoming hadrons. This
happens when the transverse momenta of some of the produced particles are
comparable with the saturation momentum at the relevant values of $x$. In
such a case, the partons from one projectile scatter off a dense gluonic
system (a CGC) in the other projectile, so they typically undergo {\em
multiple scattering}. This is a non--linear effect similar to saturation:
each additional scattering represents a correction of order $\alpha_s n$
to the cross--section, which for $n\sim 1/\alpha_s$ is an effect of order
one. (As usual, $n=n(x,k_\perp)$ denotes the gluon occupation number in
the dense projectile and is of $\order{1/\alpha_s}$ when $k_\perp\lesssim
Q_s(x)$.) So, when a parton scatters off a CGC, the multiple scattering
series must be resummed to all orders. This resummation involves
arbitrarily many insertions of the strong colour field $A^\mu_a$ which
represents the CGC, which implies that the associated cross--section is
sensitive to multi--gluon correlations ($n$--point functions of the field
$A^\mu_a$ with $n\ge 2$). Clearly, the multiple scattering cannot be
encoded in the (collinear of $k_T$) factorization schemes alluded to
above, which involve only the respective 2--point functions
--- the parton distributions.

Fortunately, there is an important simplification which occurs at high
energy and which permits to compute multiple scattering to all orders: an
energetic parton is not significantly deflected by the scattering and
thus can be assumed to preserve a straight line trajectory throughout the
collision. This is known as the {\em eikonal approximation}. To explain
it in a simple, but phenomenologically relevant, setting, consider first
{\em proton--nucleus (p+A) collisions}, cf. \Fref{fig:factor} right. This
is an example of a {\em dense--dilute scattering}, that is, a collision
in which a projectile which is relatively dilute (the `proton') and can
therefore be treated in the collinear or the $k_T$ factorization,
scatters off a dense target (the `nucleus'). Then the partons from the
dilute projectile undergo multiple scattering off the strong colour field
of the dense target. For the kinematical conditions at RHIC or the LHC,
this `dense--dilute' scenario is optimally realized in the case of
particle production at {\em forward rapidities}, that is, in the
fragmentation region of the proton (deuteron at RHIC).

To be specific, consider the production of a light quark with rapidity
$y>0$ and semi--hard transverse momentum $p_\perp$. The production
mechanism is as follows: a quark from the proton, with relatively large
longitudinal momentum fraction $x_1=(p_\perp/\sqrt{s})\rme^{y}$ and
negligible transverse momentum, scatters off the gluons with $x_2\simeq
(p_\perp/\sqrt{s})\rme^{-y}$ and $k_{2\perp}\sim Q_s(x_2)$ from the
nucleus --- which form a dense system because $x_2\ll 1$ and $A\gg 1$ ---
and thus accumulates a final transverse momentum $p_\perp\sim Q_s(x_2)$.
Within the CGC effective theory, the nucleus in a given scattering event
is described as a classical colour field $A^\mu_a$, off which the quark
scatters with the $S$--matrix
 \beq\label{Seik}
 S_{\alpha\beta}=\langle\beta|\,{\rm T}
 \exp\Big\{i\int\rmd^4y {\mathcal{L}}_{\rm int}(y)\Big\}|\alpha\rangle
 \quad\mbox{with}\quad {\mathcal{L}}_{\rm int}(y)=j^\mu_a(y)A_\mu^a(y)
 \quad\mbox{and}\quad  j^\mu_a(y)=g\bar\psi\gamma^\mu t^a\psi\,.\eeq
Here ${\mathcal{L}}_{\rm int}(y)$ is the Lagrangian density for the
interaction between the colour current $j^\mu_a$ of the quark and the
colour field $A_\mu^a$ of the target, and $\alpha$ ($\beta$) represents
the ensemble of the quantum numbers characterizing the state of the quark
prior to (after) the collision. The quark deflection angle reads
$\theta\simeq p_\perp/E_1$ with $E_1=x_1\sqrt{s}/2$ (the quark energy).
For the kinematics of interest we have $p_\perp\ll E_1$, so this angle is
small, $\theta\ll 1$, as anticipated. Hence, one can assume that the
quark keeps a {\em fixed transverse coordinate} $\bmx_\perp$ while
crossing the nucleus. This is the eikonal approximation. In this
approximation, $\alpha=(\bmx_\perp, i)$ and $\beta=(\bmx_\perp, j)$,
where $i$ and $j$ are colour indices in the fundamental representation of
SU$(N_c)$ which indicate the quark colour states before and respectively
after the scattering. Also, assuming the quark to be a left--mover (and
hence the nucleus to be a right mover, as in \Eref{YM}), one can write
$j^\mu_a(y)\simeq \delta^{\mu-}
gt^a\delta(y^+)\delta^{(2)}(\bmy_\perp-\bmx_\perp)$. Then \Eref{Seik}
reduces to
 \beq\label{Wilson}
 \langle \bmx_\perp, j| \,S\,|\bmx_\perp, i\rangle\,\simeq\,
 V_{ij}(\bmx_\perp)
 \quad\mbox{with}\quad V(\bmx_\perp)\,\equiv {\rm T}\exp\Big\{ig\int \rmd x^-
 A^+_a(x^-,\bmx_\perp) t^a\Big\}\,,\eeq
where the `path--ordering' symbol T denotes the ordering of the colour
matrices $A^+_a(x^-, \bmx_\perp)t^a$ in the exponent, from right to left,
in increasing order of their $x^-$ arguments. The integration runs
formally over all the values of $x^-$, but in reality this is restricted
to the longitudinal extent of the nucleus, which is localized near
$x^-=0$ because of Lorentz contraction\footnote{More precisely, the
small--$x$ gluons which participate in the scattering are delocalized
within a distance $\Delta x^-\sim 1/(x_2 P)$ around $x^-=0$, as explained
after \Eref{xGx}.}. The path--ordered exponential $V$ is a colour matrix
in the fundamental representation, also known as a {\em Wilson line}. It
shows that the only effect of the scattering in the high energy limit is
to {\em `rotate' the colour state of the quark} while the latter is
crossing the nucleus. If instead of a quark, one would consider the
scattering of a gluon, the corresponding $S$--matrix would be again a
Wilson line, but in the {\em adjoint} representation ($t^a\to T^a$). When
the target field is weak, $gA^+\ll 1$, one can expand the exponential in
\Eref{Wilson} in powers of $gA^+$, thus generating the multiple
scattering series. But when $gA^+\sim 1$, as is the case for a target
where the gluons are at saturation, such an expansion becomes useless,
since all the terms count on the same order. In such a case, one has to
work with the all--order result, as compactly encoded in the Wilson line.

The above considerations also show that multiple scattering at high
energy is most conveniently treated in the transverse {\em coordinate}
representation: the successive collisions modify the transverse momentum
of the partonic projectile, but do not significantly alter its transverse
coordinate (or `impact parameter'). But the interesting observable is the
cross--section for producing a quark (or gluon) with a given transverse
momentum $p_\perp$ and rapidity $y$. This is obtained by multiplying the
amplitude $V_{ij}(\bmx_\perp)$ with the complex conjugate amplitude
$V_{ji}^\dagger(\bmy_\perp)$ for a quark at a {\em different} impact
parameter $\bmy_\perp$, and then taking the Fourier transform
$\bmx_\perp-\bmy_\perp\to \bmp_\perp$. This yields (for the forward
kinematics of interest here)
 \beq\label{quarkprod}
 \frac{\rmd N_q}{\rmd y\,\rmd^2\bmp_\perp}\,\simeq\,x_1f_q(x_1,p_\perp^2)
 \int \rmd^2\bmr_\perp\,\rme^{-i\bmr_\perp\cdot\bmp_\perp}
 \,\frac{1}{N_c}\left\langle {\rm tr} V(\bmx_\perp)
 V^\dagger(\bmy_\perp)\right \rangle_Y.\eeq
The quark distribution $x_1f_q(x_1,p_\perp^2)$ gives the probability to
find a quark `collinear' with the proton, with longitudinal momentum
fraction $x_1$, on the resolution scale $p_\perp^2$ set by the partonic
scattering. Within the integral, we have defined $\bmr_\perp\equiv
\bmx_\perp-\bmy_\perp$. Furthermore, the colour trace has been generated
by summing over the final colour indices ($\sum_j$) and averaging over
the initial ones ($(1/N_c)\sum_i$), and the brackets denote the average
over the target wavefunction, evolved up to the rapidity $Y=\ln(1/x_2)$.
In the CGC formalism, this target average is computed according to
\Eref{average}, that is, as an average over the colour charge density of
the `fast' sources which are responsible for the target field $A^+_a(x^-,
\bmx_\perp)$ via \Eref{YM}:
 \beq S_Y(\bmx_\perp,\bmy_\perp)\,\equiv\,
 \frac{1}{N_c}\left\langle {\rm tr} V(\bmx_\perp)
 V^\dagger(\bmy_\perp)\right \rangle_Y = \int {\mathcal D}\rho \,
 W_Y[\rho]\,\,\frac{1}{N_c}\,{\rm tr} \big(V(\bmx_\perp)
 V^\dagger(\bmy_\perp)\big)\,.
 \label{aveWilson}
 \eeq
This 2--point function of the Wilson lines is recognized as the
$S$--matrix for the scattering between a {\em colour dipole} (a
quark--antiquark pair in an overall colour singlet state) and the CGC.
The non--linear effects enter this $S$--matrix at two levels:
\texttt{(i)} via {\em multiple scattering} for the quark and the
antiquark, as described by the respective Wilson lines, and \texttt{(ii)}
via {\em gluon saturation} in the target wavefunction, as encoded in the
CGC weight function. In this context, the target saturation momentum
$Q_s(A,Y)$ also plays the role of the {\em unitarization scale} for the
colour dipole: the dipole scattering becomes strong (meaning that
$|S_Y(r_\perp)|\ll 1$) when the dipole size $r_\perp$ is of order $1/Q_s$
or larger. Indeed, so long as the dipole is relatively small, such that
$r_\perp\ll 1/Q_s$, it predominantly scatters off the gluon modes with
$k_\perp\sim 1/r_\perp \gg Q_s$, which are dilute. The corresponding
target field is weak ($gA^+\ll 1$), the Wilson lines are close to one,
and so is the dipole $S$--matrix. Namely, for $r_\perp Q_s\ll 1$ one
finds $1-S_Y(r_\perp)\propto \big(r_\perp^2Q_s^2(Y)\big)^{\gamma_s}$ with
$\gamma_s\simeq 0.63$. On the other hand, a large dipole with
$r_\perp\gtrsim 1/Q_s(Y)$ probes the high--density gluon modes with
$k_\perp \lesssim Q_s(Y)$; the associated colour fields are strong,
$gA^+\sim 1$, so the Wilson lines are rapidly oscillating and their
product averages out to a very small value: $|S_Y(r_\perp)|\ll 1$ when
$r_\perp Q_s\gg 1$.

The dipole $S$--matrix provides a convenient framework to study high
energy evolution and saturation since, in the limit where the number of
colours is large $N_c\gg 1$, it obeys a relatively simple equation for
the evolution with $Y$ --- a non--linear generalization of the BFKL
equation known as the Balitsky--Kovchegov (BK) equation
\cite{Kovchegov:1999yj}. Originally deduced within Mueller's `dipole
picture' \cite{Mueller:1993rr,Mueller:2001fv} (an insightful
reformulation of the BFKL evolution valid at large $N_c$), the BK
equation also emerges from the large--$N_c$ limit of the Balitsky--JIMWLK
hierarchy. Remarkably, the BK equation is presently known to
next--to--leading order (NLO) accuracy
\cite{Balitsky:2006wa,Kovchegov:2006vj,Balitsky:2008zza}. This is
important in view of phenomenological studies of both deep inelastic
scattering --- at high energy, the DIS cross--section \eqref{sigmagamma}
can be related to the dipole $S$--matrix \cite{GolecBiernat:1998js} and
the NLO corrections are essential in order to achieve a good description
of the HERA data at small $x$
\cite{Iancu:2003ge,Kowalski:2006hc,Goncalves:2006yt,Tribedy:2010ab,Albacete:2010sy}
--- and of particle production in dense--dilute scattering, which is the
topics of interest for us here.

Specifically, Eqs.~\eqref{quarkprod}--\eqref{aveWilson} express the {\em
CGC factorization} for inclusive quark production in dense--dilute
scattering
\cite{Kovchegov:1998bi,Kovchegov:2001sc,Blaizot:2004wu,JalilianMarian:2005jf}.
\Eref{quarkprod} is usually written as
 \beq\label{ktfact}
 \frac{\rmd N_q}{\rmd y\,\rmd^2\bmp_\perp}\,\simeq\,
 \frac{\alpha_s}{p_\perp^2}\,
 x_1f_q(x_1,p_\perp^2)\,\Phi_A(Y,\bmp_\perp,\bmb_\perp),
 \eeq
where the first factor $\alpha_s/{p_\perp^2}$ is the cross--section for
the elementary $q+g\to q$ scattering (the would-be partonic subprocess in
the single scattering limit) while
 \beq\label{Phi}
 \Phi_A(Y,\bmp_\perp,\bmb_\perp)\equiv
 \frac{p_\perp^2}{\alpha_s}
 \int \rmd^2\bmr_\perp\,\rme^{-i\bmr_\perp\cdot\bmp_\perp}
 \,\frac{1}{N_c}\left\langle {\rm tr} V(\bmx_\perp)
 V^\dagger(\bmy_\perp)\right \rangle_Y,\eeq
plays the role of a {\em generalized unintegrated gluon distribution}
(here, for the nucleus) at impact parameter $\bmb_\perp
=(\bmx_\perp+\bmy_\perp)/2$. When $p_\perp\gg Q_s(A,Y)$, \Eref{Phi} can
be evaluated in the single scattering approximation, as obtained by
expanding the product of Wilson lines to quadratic order in $A^+$. In
that (dilute) regime, $\Phi_A$ reduces indeed to the usual unintegrated
gluon distribution --- the one which enters the $k_T$--factorization and
obeys the BFKL equation. But for lower momenta $p_\perp\lesssim
Q_s(A,Y)$, the non--linear effects become essential, as already discussed
in relation with the dipole scattering.

\begin{figure}[t]
\begin{center}
\centerline{
\includegraphics[width=0.48\textwidth]{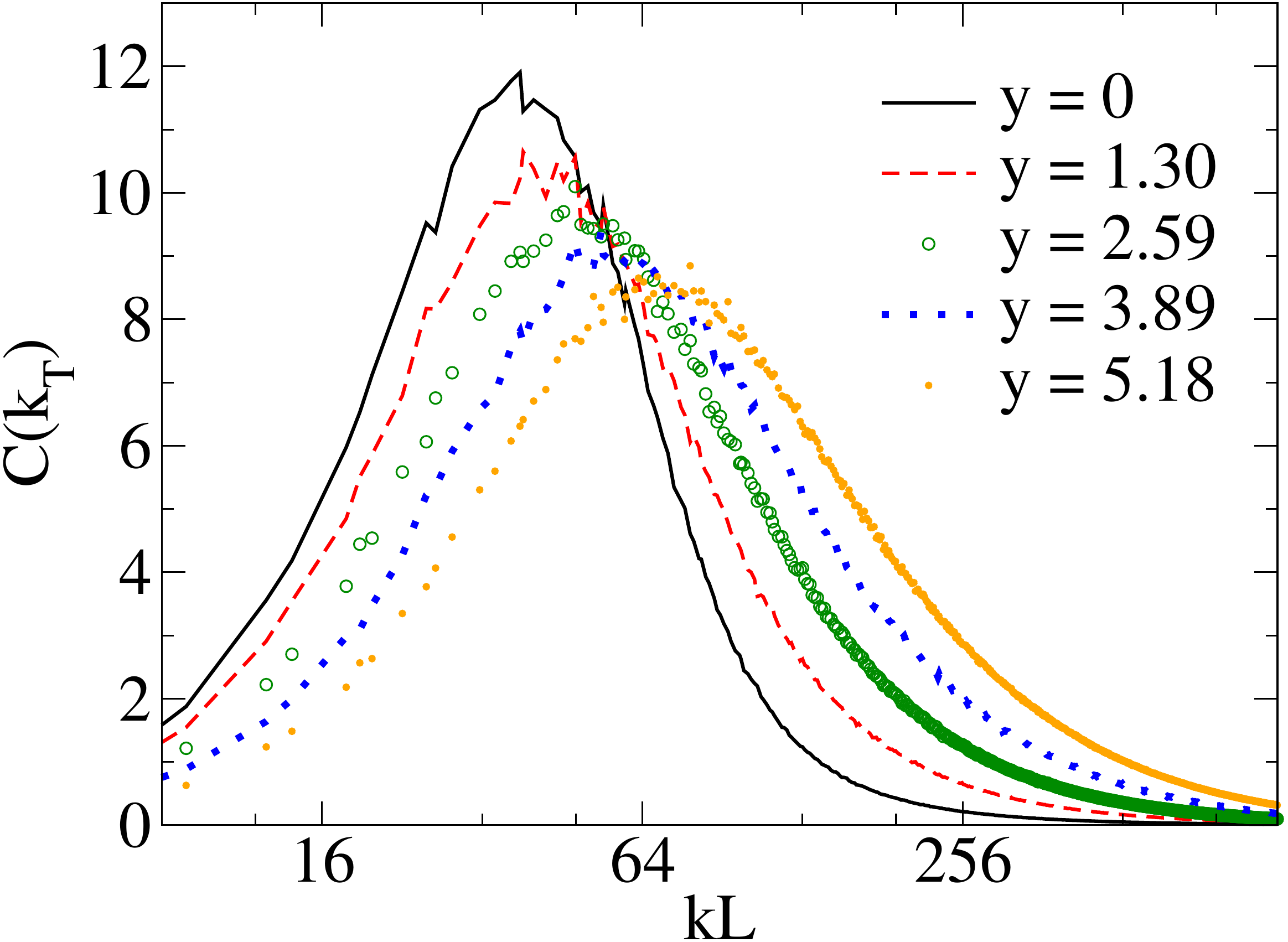}
\qquad\includegraphics[width=0.48\textwidth]{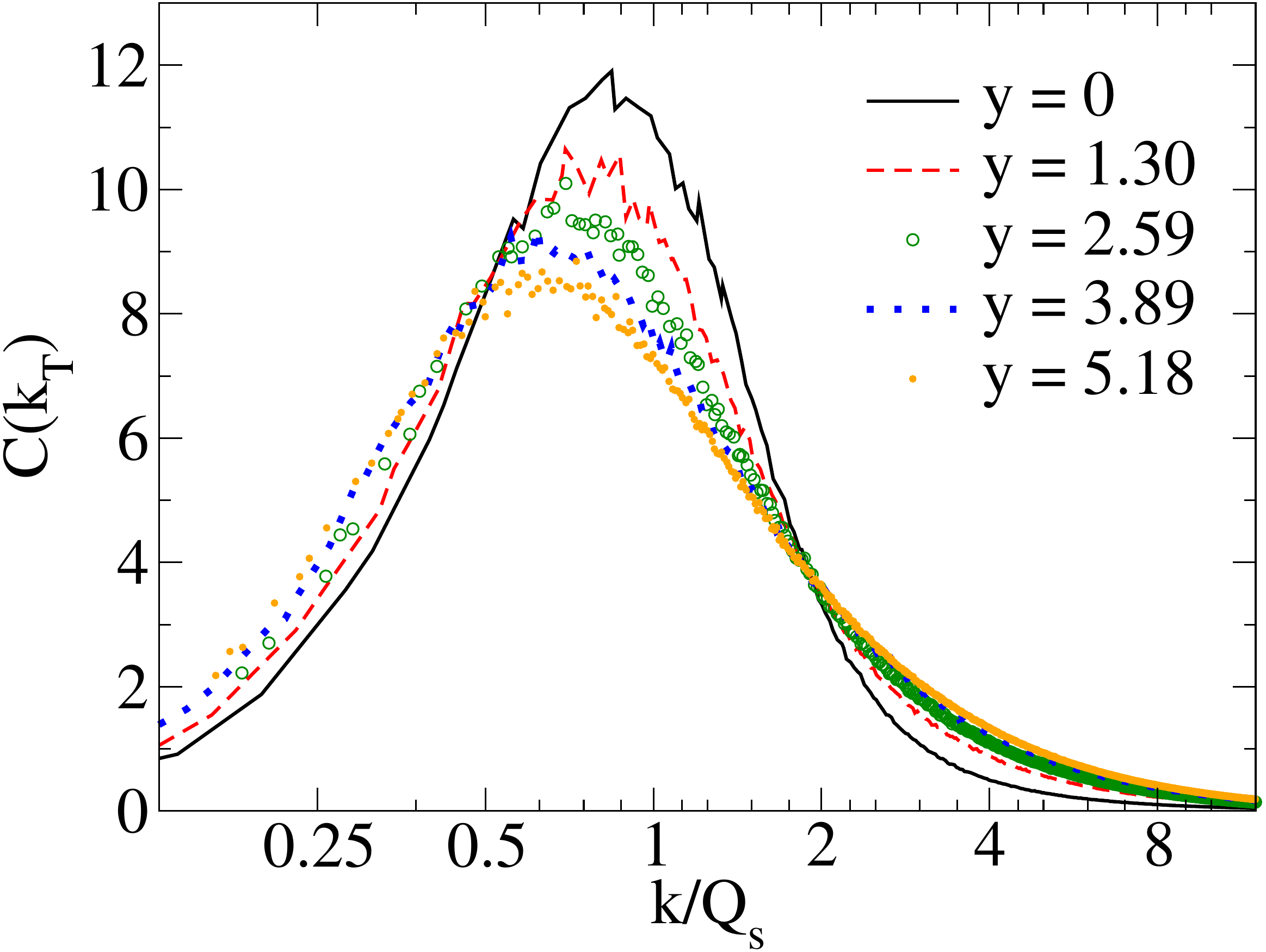}}
 \caption{\sl The results of the JIMWLK evolution for the generalized gluon
distribution in \Eref{Phi}, parameterized as
$\Phi(Y,\bmk_\perp)\equiv (1/{\alpha_s})C
(Y,\bmk_\perp)$. Left: as a function of $k_\perp L$, with $L$ the transverse
size of the system. Right: as a function of the scaling variable
$k_\perp/Q_s(Y)$, to exhibit geometric scaling (the curves
corresponding to different values of $Y$ fall approximately on top of
each other). From Ref.~\cite{Lappi:2011ju}.
  \label{fig:Phi}}
\end{center}\vspace*{-.8cm}
\end{figure}

\Fref{fig:Phi} shows the CGC prediction for the generalized gluon
distribution \eqref{Phi}, as obtained via the numerical resolution of the
JIMWLK equation with initial conditions at $Y=0$ of the
McLerran--Venugopalan type (and with a running coupling)
\cite{Lappi:2011ju}. The two plots exhibit a pronounced peak at a special
value of the transverse momentum $p_\perp$ which increases with $Y$~:
this special value is, of course, the saturation momentum $Q_s(A,Y)$. In
fact, it is quite easy to understand these plots in the light of our
previous discussion of the dipole scattering. The Fourier transform in
\Eref{Phi} is controlled by the competition between the complex phase
$\rme^{-i\bmr_\perp\cdot\bmp_\perp}$ and the dipole $S$--matrix
$S_Y(r_\perp)$. For large momenta $p_\perp\gg Q_s$, the complex
exponential limits the integration to small dipole sizes $r_\perp\lesssim
1/p_\perp\ll 1/Q_s$, for which the scattering is weak:
$1-S_Y(r_\perp)\sim \big(r_\perp^2Q_s^2(Y)\big)^{\gamma_s}$. Then the
Fourier transform yields
 \beq\label{scaling}
 \Phi_A(Y,p_\perp)\,\simeq\,\frac{1}{\alpha_s}\left(
 \frac{Q_s^2(A,Y)}{p_\perp^2}\right)^{\gamma_s}\qquad\mbox{for}\quad
 p_\perp\gg Q_s(A,Y).\eeq
The difference $1-\gamma_s\simeq 0.37$ is an {\em anomalous dimension}
introduced by the high energy evolution. (Without this evolution, one
would have the bremsstrahlung spectrum $\propto 1/p_\perp^2$, cf.
\Eref{eq:unintp}.) For lower momenta $p_\perp < Q_s$, the integral over
$r_\perp$ in \Eref{Phi} is limited by the dipole $S$--matrix to values
$r_\perp\lesssim 1/Q_s < 1/p_\perp$. (Recall that $S_Y(r_\perp)\ll 1$
when $r_\perp\gg 1/Q_s$.) Then
 \beq\label{Philow}
 \Phi_A(Y,p_\perp)\,\simeq\, \frac{p_\perp^2}{\alpha_s}
  \int \rmd^2\bmr_\perp\,\Theta\big(1/Q_s-r_\perp\big)\,\simeq\,
  \frac{1}{\alpha_s} \,\frac{p_\perp^2}{Q_s^2(A,Y)}
  \qquad\mbox{for}\quad
  p_\perp\lesssim Q_s(A,Y)
  \,.\eeq
Eqs.~\eqref{scaling} and \eqref{Philow} explain the pronounced peak in
the gluon distribution at $p_\perp\simeq Q_s(Y)$, as visible in
\Fref{fig:Phi}. This in turn implies that the cross--section for particle
production is dominated by semi--hard gluons with $p_\perp\sim Q_s(Y)$
and hence can be computed in perturbation theory. Another important
consequence of saturation, which is visible too in Eqs.~\eqref{scaling}
and \eqref{Philow} (and is numerically tested in the right panel of
\Fref{fig:Phi}), is {\em geometric scaling}
\cite{Stasto:2000er,Iancu:2002tr,Mueller:2002zm,Munier:2003vc}: the
unintegrated gluon distribution depends upon the two kinematical
variables $p_\perp$ and $Y$ only via the dimensionless ratio
$p_\perp/Q_s(Y)$. This scaling has important consequences for the
phenomenology, to be discussed in the next section.

Note also that, as a result of the non--linear physics,  the generalized
`gluon distribution' \eqref{Phi} is a {\em process--dependent quantity}~:
it depends not only upon the gluon density in the target, but also upon
the nature of the partonic subcollision. For instance, if instead of the
quark production, one would consider the production of a gluon, the
`fundamental' Wilson lines in \Eref{Phi} would be replaced by `adjoint'
ones \cite{Kovchegov:1998bi,Kovchegov:2001sc,Blaizot:2004wu}. Also, if
one considers a more complicated final state --- say, the production of a
pair of partons
--- then the analog of \Eref{Phi} will involve a pair of Wilson lines for
each of the partons partaking in the collision (one such a line in the
direct amplitude and another one in the complex conjugate amplitude).
More examples in that sense can be found in
Refs.~\cite{Kovner:2001vi,Blaizot:2004wv,JalilianMarian:2004da,Baier:2005dv,Marquet:2007vb,Dominguez:2011wm}.
See also \cite{Avsar:2012hj} for a recent, comprehensive, discussion of
the relation between the CGC factorization and the $k_T$--factorization.
These considerations show that, in the presence of non--linear effects,
the notion of `parton distribution' ceases to be useful: the observables
involve higher $n$--point functions of the gluon fields (generally, via
the product of Wilson lines) which moreover couple with each other under
the non--linear, JIMWLK, evolution. The complete information about the
gluon correlations and their evolution with $Y$ (to leading--logarithmic
accuracy at least) is encoded in the CGC weight function.

\begin{figure}[t]
\begin{center}
\centerline{
\includegraphics[width=0.6\textwidth]{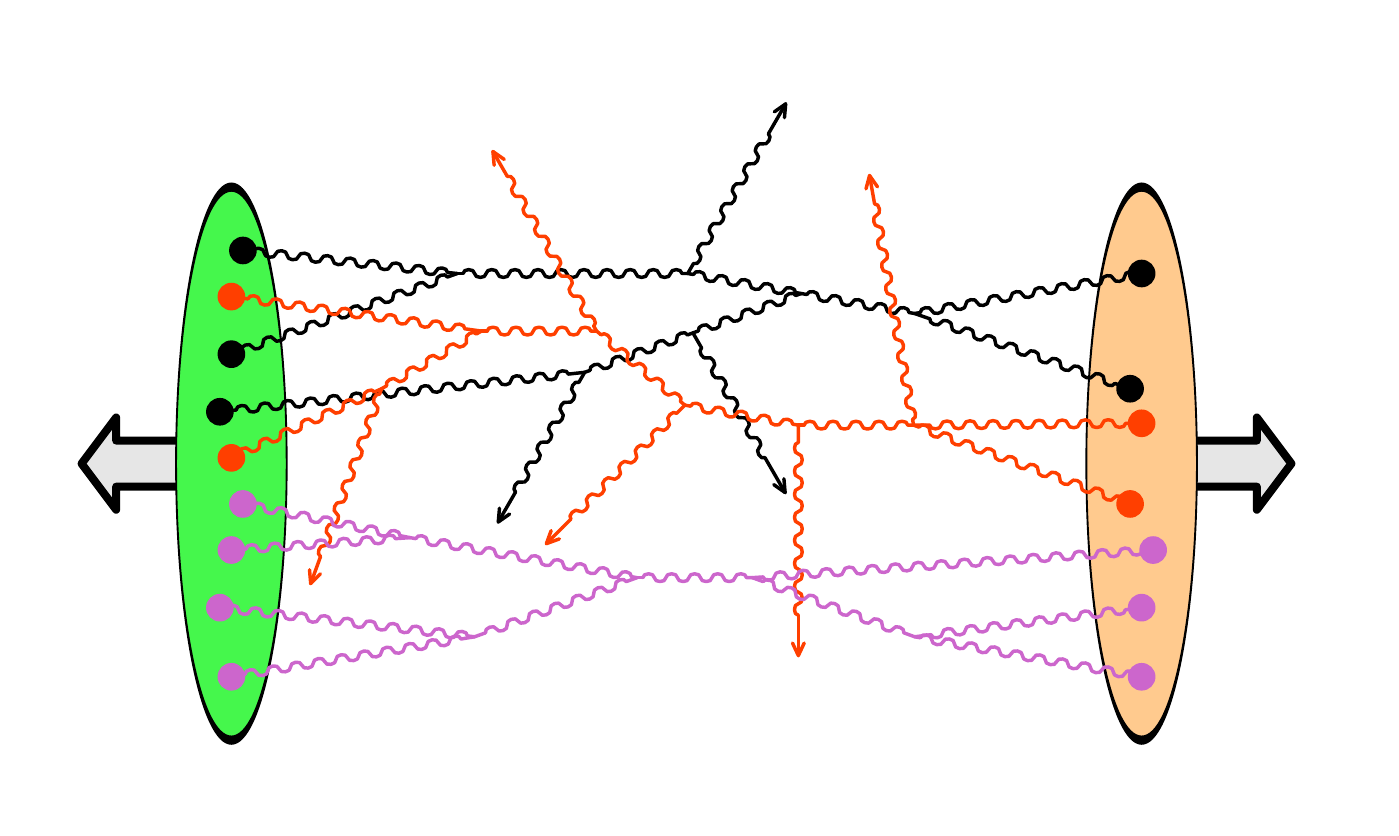}
\includegraphics[width=0.4\textwidth]{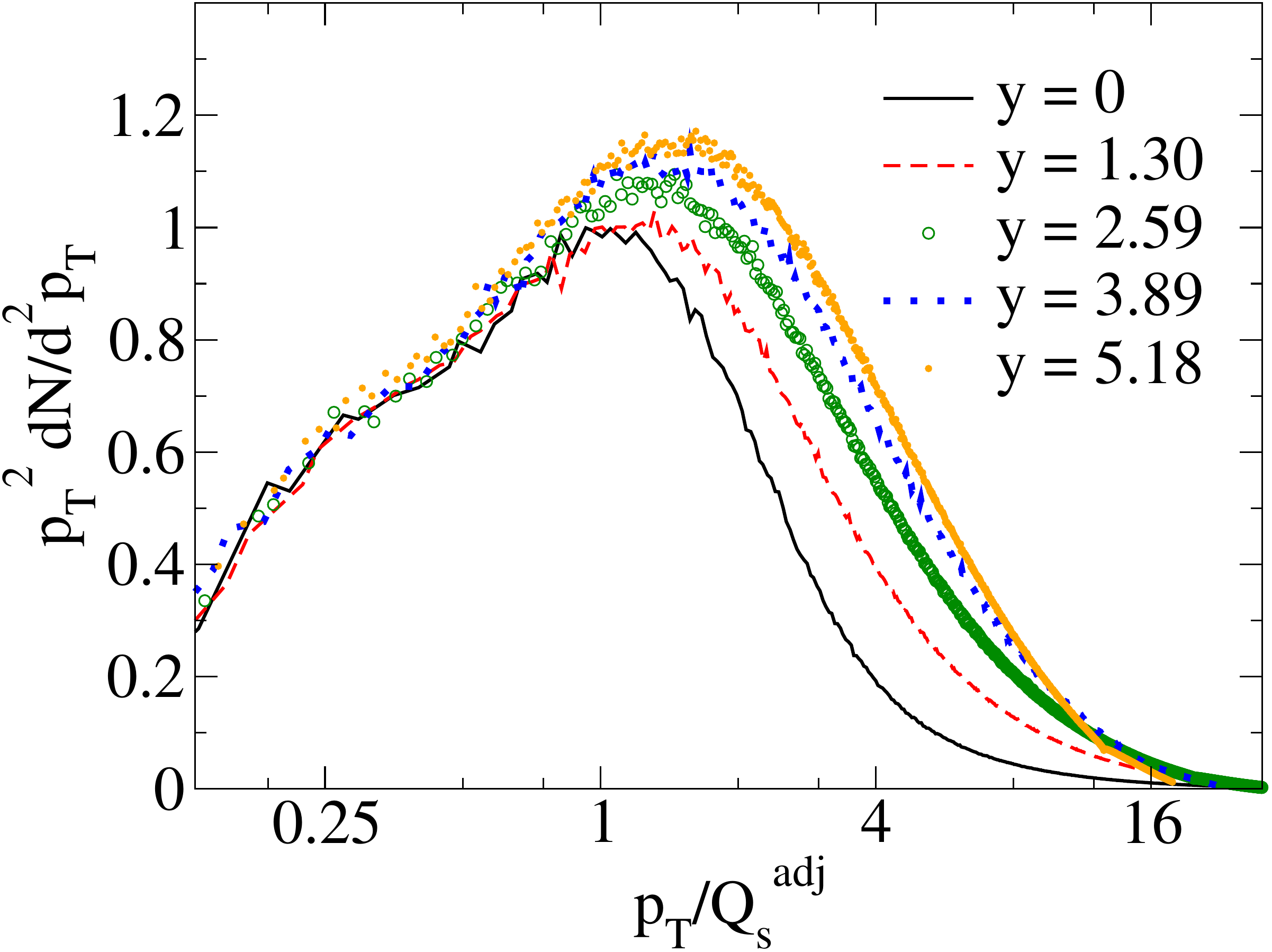}}
 \caption{\sl Left panel:
 a cartoon of the `dense--dense' collision between two
 heavy nuclei, which illustrates the complexity of the process.
 Non--linear effects enter via gluon saturation (in both nuclei)
 and multiple scattering. Right panel: the gluon spectrum generated
 by this collision within the CGC formalism
 (from Ref.~\cite{Lappi:2011ju}).
  \label{fig:AA}}
\end{center}\vspace*{-.8cm}
\end{figure}

We now turn to {\em nucleus--nucleus ($A+A$) collisions}, which for the
typical kinematical conditions at RHIC and the LHC represents an example
of {\em dense--dense scattering}~: the wavefunctions of both nuclei
develop saturation effects which influence the production of particles
with semi--hard transverse momenta. Hence, in order to compute the bulk
of particle production, one must take into account the many--body
correlations associated with gluon saturation in both nuclei, together
with the multiple scattering between these two saturated gluon
distributions. Once again, this complex problem can be addressed within
the CGC formalism and in the eikonal approximation. Not surprisingly, the
treatment of the two nuclear projectiles is now {\em symmetric}: they are
both described as colour glass condensates, with weight functions
$W_{Y_1}[\rho_1]$ and, respectively, $W_{Y_2}[\rho_2]$. And their
collision is described as the scattering between two classical
distributions of colour charges moving against each other.

To be more specific, consider inclusive gluon production in the COM
frame, where the nuclei have rapidities $\pm y_{\rm beam}$. If the
produced gluon has rapidity $y$, it will probe the evolutions of the two
nuclear wavefunctions up to rapidities $Y_1=y_{\rm beam}-y$ and
$Y_2=y_{\rm beam}+y$, respectively. This motivates the following CGC
factorization for the spectrum of the produced gluons \cite{Gelis:2008rw}
 \beq\label{CGCfact}
\left\langle\frac{\rmd N}{\rmd y\,\rmd^2\bmp_\perp}\right\rangle \,=\int
[\mathcal{D}{\rho_1} \mathcal{D}{\rho_2}]\, {W_{y_{{\rm
beam}-y}}[\rho_1]}\,{W_{y_{{\rm beam}+y}}[\rho_2]}\, \frac{\rmd N}{\rmd
 y\,\rmd^2\bmp_\perp}\bigg |_{\rm class},\eeq
where the last factor inside the integrand,  $({\rmd N}/{\rmd
y\rmd^2\bmp_\perp})|_{\rm class}$, represents the spectrum produced in
the scattering between two given configurations of classical colour
sources (the `fast partons') --- one for each nucleus. More precisely, as
discussed in relation with \Eref{YM}, the right--moving nucleus is
described in a given event as a colour current having only a `plus'
component: $J^{\mu,a}_1=\delta^{\mu+}\rho^a_1$, with the charge density
$\rho_1^a$ localized near $x^-=0$ (due to Lorentz contraction) and
independent of $x^+$ (by Lorentz time dilation). Similarly, the
left--moving nucleus is represented by a colour current with only a
`minus' component: $J^{\mu,a}_2=\delta^{\mu-}\rho^a_2$, with $\rho_2^a$
localized near $x^+=0$ and independent of $x^-$. At a classical level,
the `scattering' between these two currents is described by the solution
$A^\mu_a$ to the Yang--Mills equation including both types of sources
(compare to \Eref{YM}) :
 \beq\label{YMAA}
 D_\nu^{ab} F_b^{\nu\mu}(x)\,=\,\delta^{\mu +}\rho^a_1(x^-,\bmx_\perp)
 +\delta^{\mu -}\rho^a_2(x^+,\bmx_\perp)\,.
 \eeq
This equation describes multiple scattering because it is non--linear:
the collision begins at $x^+=x^-=0$ (i.e. $t=z=0$) and for positive
values of $x^+$ and $x^-$, the solution $A^\mu_a$ is non--linear to all
orders in both $\rho_1$ and $\rho_2$. This solution cannot be computed
analytically, but numerical solutions are by now available
\cite{Krasnitz:2001qu,Krasnitz:2002mn,Lappi:2003bi,Lappi:2011ju}. The
cross--section $({\rmd N}/{\rmd y\rmd^2\bmp_\perp})|_{\rm class}$ for
particle production is obtained via the Fourier transform of this
classical solution, that is, by projecting the field $A^\mu_a$ onto modes
with transverse momentum $\bmp_\perp$. Finally, the average over the CGC
weight functions of the two nuclei, cf. \Eref{CGCfact}, is numerically
performed. It is this last procedure which introduces the dependence of
the cross--section upon the rapidity $y$, via the corresponding
dependence of the two weight functions. Note that, in line with the
general philosophy of the CGC formalism, the only quantum effects to be
included in the calculation are those associated with the high--energy
evolution of the projectile wavefunctions prior to scattering, which are
enhanced by the large logarithms $Y_i=\ln(1/x_i)$, with $i=1,\,2$. The
final outcome of this calculation is the gluon spectrum displayed in
\Fref{fig:AA} (right panel) \cite{Lappi:2011ju}. This is very similar to
the `unintegrated' gluon distribution in any of the incoming nuclei
(compare to \Fref{fig:Phi}), in particular, it is peaked at a value of
$p_\perp$ of the order of the saturation momentum and which increases
with $y$. Some further consequences of the solution to \Eref{YMAA} will
be discussed in \Sref{sec:glasma}.

Note finally an important difference between p+A (dilute--dense) and A+A
(dense--dense) collisions: in the former, the particles produced by the
collision do not interact with each other, but merely evolve via
fragmentation and hadronisation towards the final hadrons observed in the
detectors; by contrast, in A+A collisions the partonic medium created in
the early stages of the collisions is very dense, so these partons keep
interacting with each other --- one then speaks about {\em final state}
interactions, as opposed to the {\em initial state} ones, which were
associated with high density effects like saturation in the incoming
wavefunctions. These `final state' interactions redistribute the partons
in energy and momentum, which makes it difficult to verify the
early--stages spectrum, as given by the aforementioned CGC calculations,
against the measured hadron yield. Yet, the CGC formalism has the virtue
to provide the {\em initial conditions} for the subsequent dynamics, at a
time $\tau\simeq 1/Q_s \sim 0.2$~fm/c. So, its predictions can be at
least indirectly tested, via calculations of the final state effects
which include initial conditions of the CGC type. We shall return to such
issues later on.

\subsection{Some experimental signatures of the CGC in HIC}
\label{sec:pA}

Let us consider now some phenomenological applications of the previous
results for `dilute--dense' scattering
\cite{Iancu:2003xm,JalilianMarian:2005jf,Gelis:2010nm,Lappi:2010ek} (and
Refs. therein). We start with the $R_{pA}$ {\em ratio}, defined as the
ratio between particle production in p+A collisions and that in p+p
collisions for the same kinematics; schematically,
 \beq\label{RpA}
 R_{pA}(\eta,p_\perp) \,\equiv\,\frac{1}{N_{coll}}\,
 \frac{
\frac{\rmd N_h}{\rmd^2p_\perp
   \rmd \eta}\Big |_{pA}}
   {\frac{\rmd N_h}{\rmd^2p_\perp \rmd
 \eta}\Big |_{pp}}\,,\eeq
where the subscript $h$ denotes the hadron species and $N_{coll}$ is the
number of binary proton--nucleon collisions in the p+A scattering at a
given impact parameter, as computed under the assumption that the various
nucleons inside the nucleus scatter independently from each other. For
relatively central collisions (i.e. small impact parameters; see
\Fref{fig:geometry}), one has $N_{coll}\simeq A^{1/3}$. The normalization
in \Eref{RpA} is such that $R_{pA}$ would be equal to one if the p+A
collision was a superposition of $A$ incoherent p+p collisions.
Conversely, any deviation in $R_{pA}$ from unity is an indication of
coherence (high--density) effects in the nuclear wavefunction. Such a
deviation is clearly seen in the respective RHIC data at {\em forward
rapidities} ($\eta >0$). More precisely, RHIC performed deuteron--gold
(d+Au) collisions\footnote{For d+Au collisions the number of binary
collisions in \Eref{RpA} should be evaluated as  $N_{coll}\simeq
2A^{1/3}$ since the projectile deuteron involves 2 nucleons, i.e. twice
as much as the proton.} at $\sqrt{s}=200$~GeV per nucleon pair and
measured the ratio $R_{\rm d+Au}$ for semi--hard momenta $p_\perp= 1\div
5$~GeV and for rapidities $\eta=0\div 4$ in the deuteron fragmentation
region \cite{Arsene:2004ux,Adams:2006uz}.

\begin{figure}[t]
\begin{center}\centerline{
\includegraphics[width=.95\textwidth]{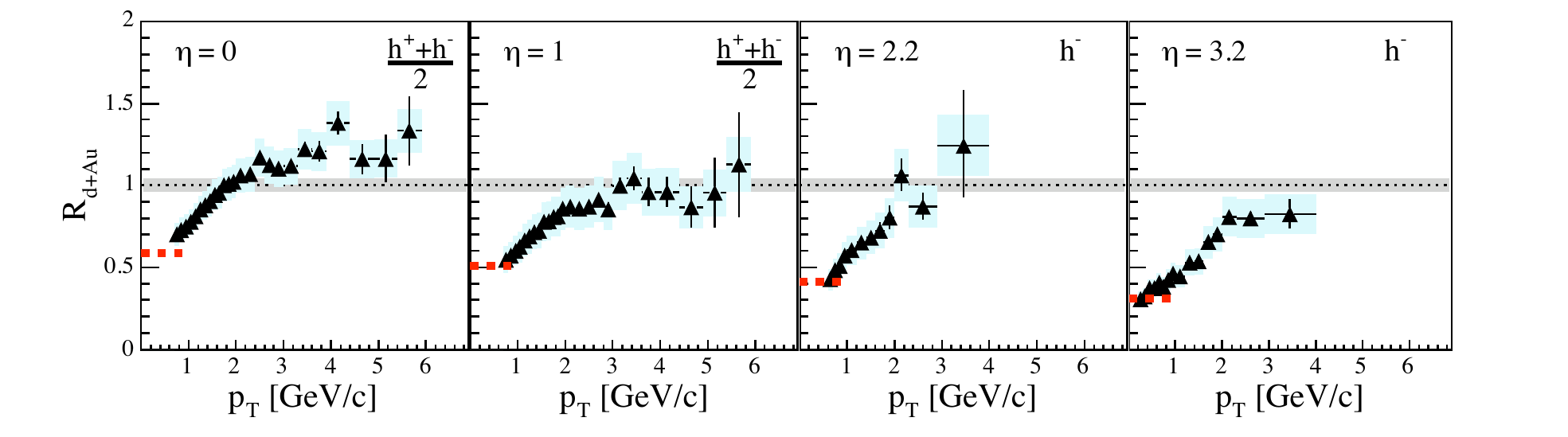}}
\caption{\sl BRAHMS results for d+Au collisions at
 200 GeV/nucleon \protect{\cite{Arsene:2004ux}} : the ratio
$R_{\rm dAu}$ for charged hadrons as a function of $p_\perp$ for
central and forward pseudorapidities.
} \label{fig:dAu}
\end{center}\vspace*{-8mm}
\end{figure}

The results of the corresponding analysis by BRAHMS are shown in
\Fref{fig:dAu}. For $p_\perp\gtrsim 2$~GeV, they show an {\em
enhancement} ($R_{\rm d+Au} > 1$) for $\eta=0$, known as the `Cronin
peak', which however disappears when increasing $\eta$, leading to {\em
suppression} ($R_{\rm d+Au} < 1$) at $\eta >1$. E.g. for $\eta=3.2$, one
finds $R_{\rm d+Au}\simeq 0.6\div 0.8$ for $p_\perp= 2\div 4$~GeV. This
behaviour can be understood in terms of saturation in the nuclear gluon
distribution and its evolution with $Y$
\cite{Kharzeev:2002pc,Albacete:2003iq,Iancu:2004bx,Kharzeev:2004yx}. For
central rapidities $\eta\simeq 0$ and $p_\perp= 2$~GeV, one probes
$x_2\sim 10^{-2}$, which is large enough for the high--energy evolution
to be negligible. Then the gluon density in the nucleus is large juste
because of the many ($3A$) valence quarks acting as sources for gluon
radiation. An incoming parton from the deuteron scatters off this dense
gluonic system and thus acquires an additional transverse momentum of
order $Q_s(A,x_2)\sim 1$~GeV. This yields a shift in the spectrum of the
produced particles towards higher values of $p_\perp$, leading to the
Cronin peak. For forward rapidities, say $\eta=3.2$ and $p_\perp= 2$~GeV,
one has $x_2\sim 10^{-4}$ and then the high--energy evolution with
$Y=\ln(1/x_2)$ is important. In that case, one can use the factorization
\eqref{ktfact} for both d+Au and for the p+p collision which serves as a
benchmark, leading to
 \beq\label{dAu}
 R_{\rm d+Au}\,\simeq\,\frac{1}{A^{1/3}}\,\frac{\Phi_A(Y,p_\perp)}
 {\Phi_p(Y,p_\perp)}\,.\eeq
For this kinematics, the saturation effects are important in the gold
nucleus (since the nuclear saturation scale $Q_s(A,Y)\sim 2$~GeV is
comparable with $p_\perp$) and lead to a slow down of the evolution with
$Y$. On the other hand, saturation is still negligible for the proton, so
the corresponding unintegrated gluon distribution $\Phi_p(Y,p_\perp)$
rises rapidly with $Y$, according to the BFKL evolution. Hence, when
increasing $\eta$ (and thus $Y$), the denominator in \Eref{dAu} rises
much faster than the numerator there, leading to a decrease in the ratio.
This is precisely the trend seen in the data.

In particular, for sufficiently high $Y$, such that both the nucleus and
the proton in the ratio in \Eref{dAu} are at saturation, and for
transverse momenta comparable or slightly larger to the nuclear
saturation momentum ($p_\perp\gtrsim Q_s(A,Y)\simeq A^{1/3} Q_s(p,Y)$),
one can use \Eref{scaling} for the unintegrated gluon distributions in
both targets. This yields a particularly simple result,
 \beq\label{dAuscaling}
 R_{\rm d+Au}\,\simeq\,\frac{1}{A^{1/3}}\,
 \left(\frac{Q_s^2(A,Y)}{Q_s^2(p,Y)}\right)^{\gamma_s}
 \,\simeq\,A^{-\frac{1-\gamma_s}{3}}
 \,,\eeq
which is independent of both the transverse momentum (within a limited
range above the nuclear saturation momentum) and the rapidity. It would
be interesting to test this prediction in p+A collisions at the LHC,
where the coverage in both $Y$ and $p_\perp$ will be larger than at RHIC.

\begin{figure}[t]
\begin{center}\centerline{
 \includegraphics[width=.40\textwidth]{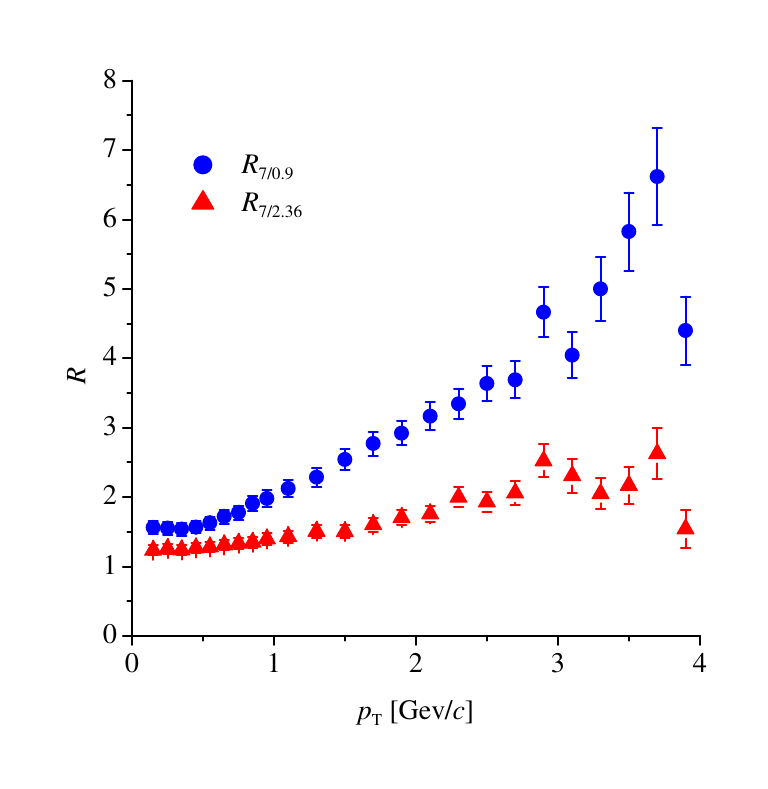}\qquad
 \quad
 \includegraphics[width=.44\textwidth]{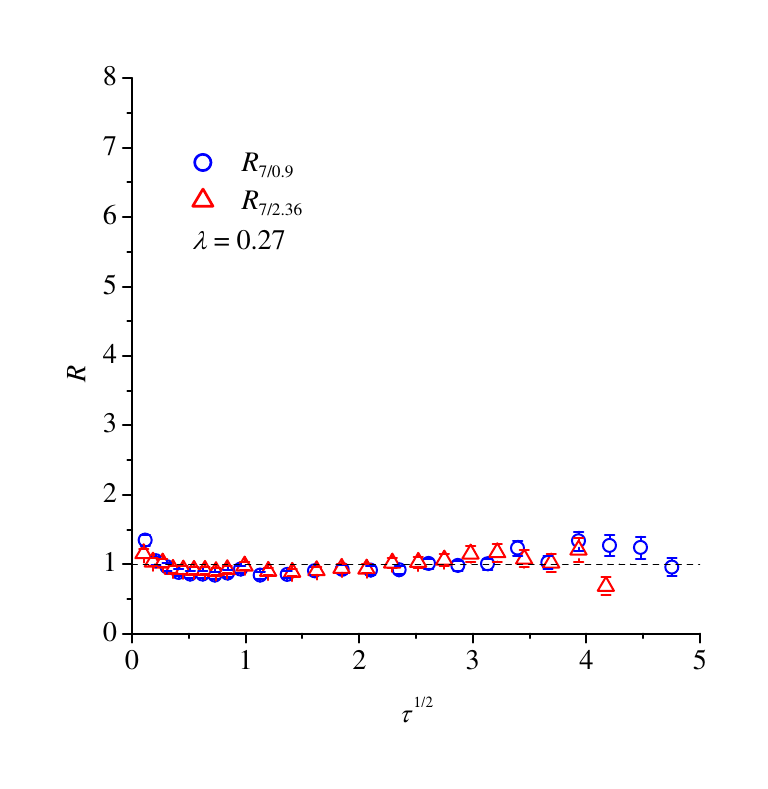}
}
\caption{\sl The ratio $R_{E_1/E_2}$ between particle spectra at energies
$E_1$ and $E_2$, as measured for three energies at the LHC, is plotted as
a function of $p_\perp$ (left panel) and of the scaling variable
$\tau=p_\perp/Q_s(Y)$ (right panel). From Ref.\cite{Praszalowicz:2011rm}.
} \label{fig:ptscaling}
\end{center}\vspace*{-8mm}
\end{figure}

Another important consequence of saturation, which is manifest in
Eqs.~\eqref{scaling}--\eqref{Philow} and has been implicitly used in
deriving \Eref{dAuscaling}, is geometric scaling in the gluon
distribution. Via \Eref{ktfact}, this suggests a similar scaling in the
spectrum of the produced particles. This prediction has been tested
against the LHC data for p+p collisions \cite{Praszalowicz:2011rm}, with
the results shown in \Fref{fig:ptscaling}. There one can see the ratio
 \beq\label{ratio}
 R_{E_1/E_2} (p_\perp,Y)
 \,=\,\frac{\big({\rmd N}/{\rmd^2 p_\perp
 \rmd \eta}\big)\big|_{E_1}}{\big({\rmd N}/{\rmd^2
p_\perp\rmd \eta}\big)\big|_{E_2}}
 \eeq
between the measured spectra for single--inclusive charged hadron
production at two COM energies, $E_1=\sqrt{s_1}$ and $E_2=\sqrt{s_2}$,
and for midrapidities: $|\eta|\le 2.4$. More precisely, one displays two
such ratios, as obtained by combining the LHC data for three different
energies: $\sqrt{s}=0.9,\,2.36,$ and 7~GeV. (Note that these energies are
high enough for the saturation effects to be important in the proton
wavefunction for $\eta\simeq 0$ and semi--hard transverse momenta.) If
the spectrum $({\rmd N}/{\rmd^2 p_\perp \rmd \eta})$ scales as a function
of $\tau\equiv p_\perp/Q_s(Y)$, then the ratio \eqref{ratio} should be
equal to one when plotted as a function of $\tau$. This expectation is
indeed met by the data, as shown in the r.h.s of \Fref{fig:ptscaling}. At
this point, it is worth noting that geometric scaling has been first
observed in the HERA data for DIS \cite{Stasto:2000er} and that the
experimental search for this remarkable behaviour has been inspired by
the theoretical ideas about gluon saturation.

\begin{figure}[t]
\begin{center}\centerline{
\includegraphics[width=.4\textwidth]{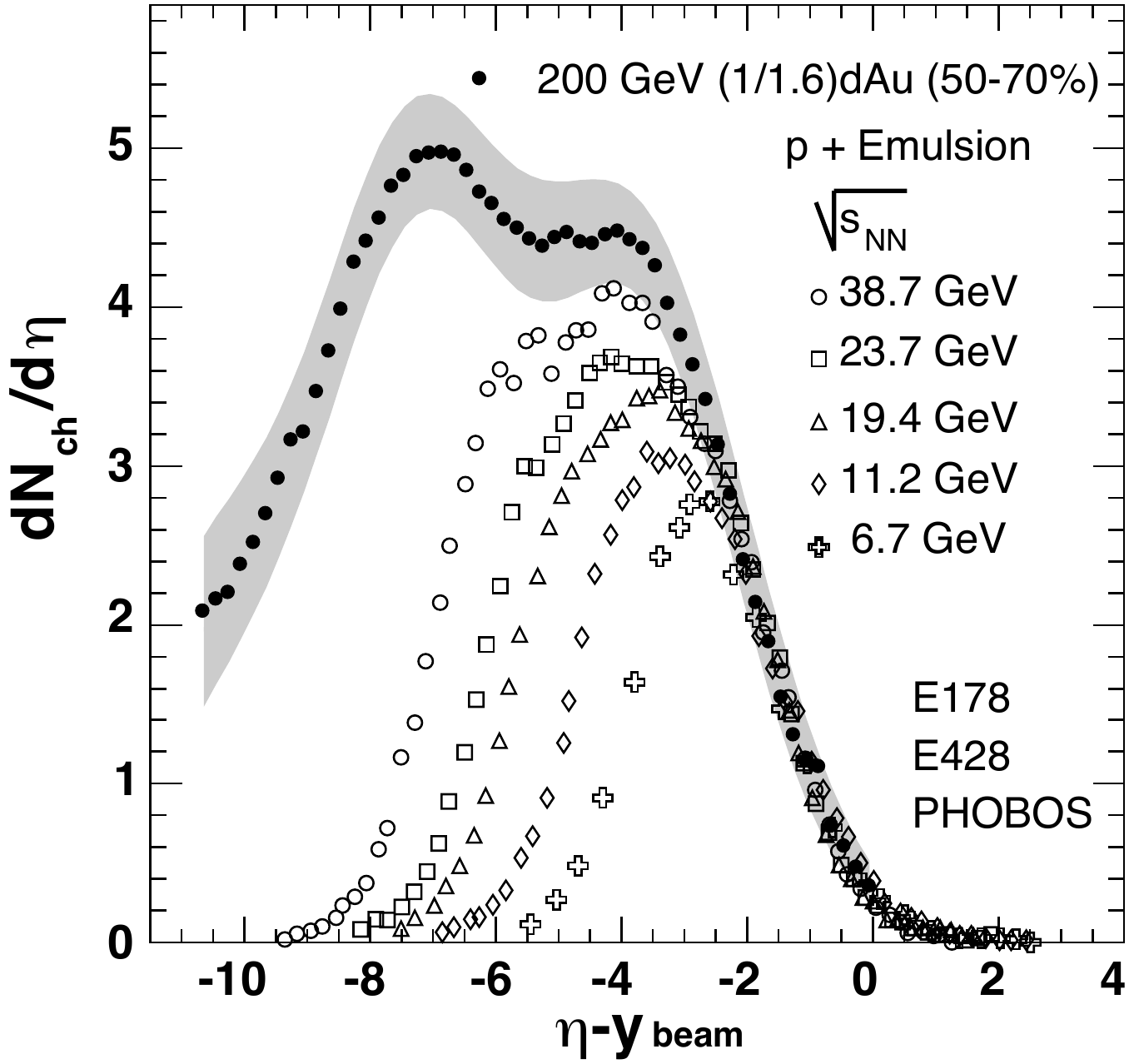}\quad
\includegraphics[width=.55\textwidth]{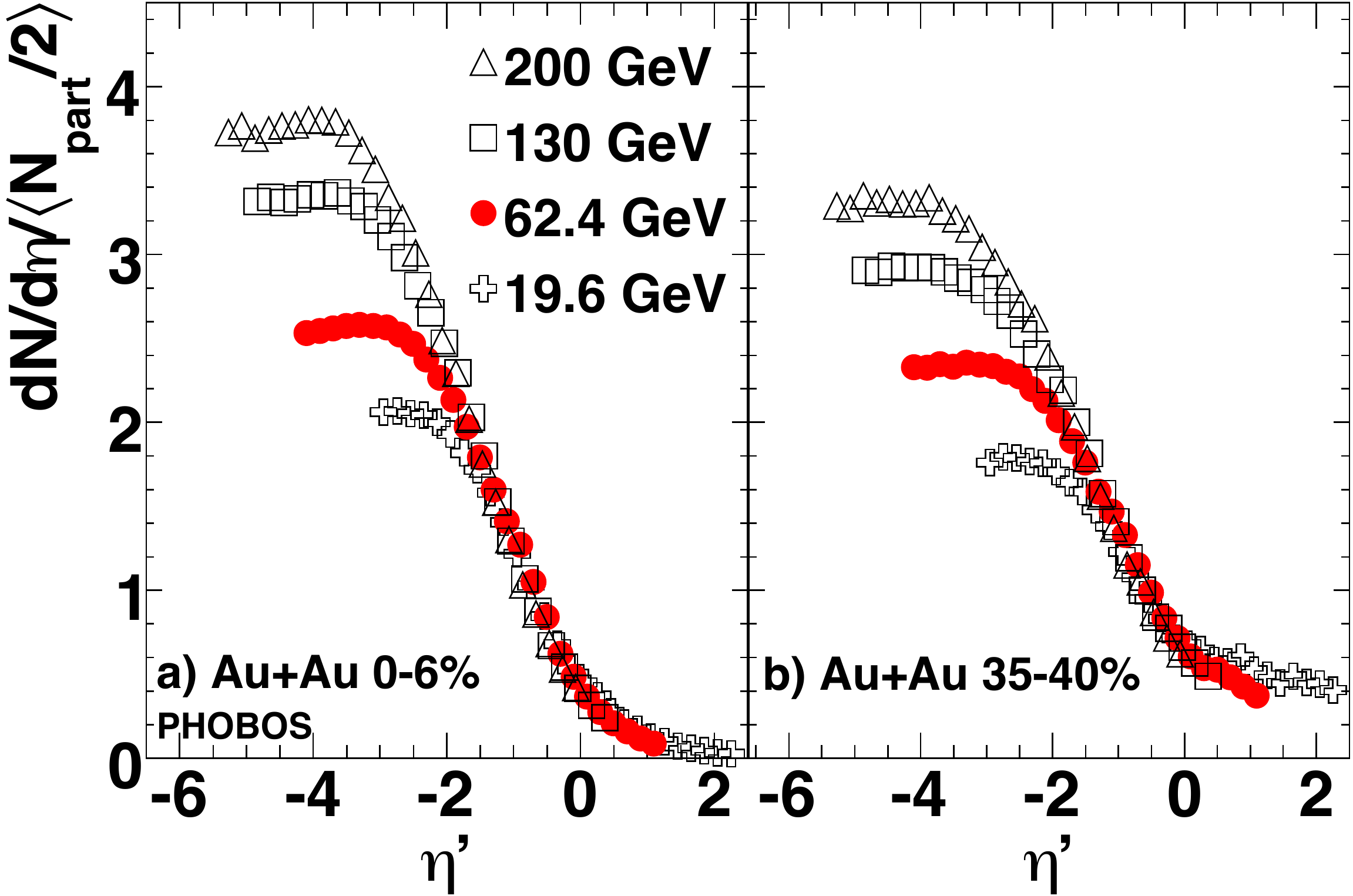}
}
\caption{\sl \label{limfrag} Pseudorapidity density distributions of charged
particles emitted in d+Au collisions (left panel) and Au+Au collisions
(right panel, separately for central and peripheral impact
parameters), at various
energies, as a function of the $\eta'=\eta-y_{\rm beam}$ variable.
In the right panel, the impact parameter is
represented by the centrality bins: 0-6\%
(i.e. the 6\% most central data) and respectively 35-40\%.
From Ref.~\cite{Veres:2005ix}.}
\end{center}\vspace*{-8mm}
\end{figure}

Another remarkable regularity in the data which is naturally explained by
the CGC is the {\em limiting fragmentation} : when the charged particle
rapidity distribution $\rmd N/\rmd \eta$ is plotted as a function of the
variable $\eta'=\eta-y_{\rm beam}$ --- the rapidity difference between
the produced particle and the dilute projectile (say, the deuteron in the
case of d+Au collisions) ---, then the distribution turns out to be
independent of the collision energy over a wide range around $\eta'=0$,
whose extent is increasing with $\sqrt{s}$ (see \Fref{limfrag}). Note
that $\eta'\simeq 0$ corresponds to forward rapidities ($\eta>0$)
according to our previous terminology, to which \Eref{ktfact} applies.
Moreover, for such rapidities, even a nucleus--nucleus collision may be
viewed as `dilute--dense', in the sense that one of the nuclei is probed
at large $x_1$, where it looks dilute.

To understand limiting fragmentation on the basis of \Eref{ktfact},
notice that \texttt{(i)} $\eta-y_{\rm beam}\simeq \ln x_1$ (the rapidity
of the produced particle is roughly equal to that of the fast parton
which initiated the scattering), \texttt{(ii)} the `multiplicity' $\rmd
N/\rmd \eta$ is obtained by integrating the spectrum \eqref{ktfact} over
all values of $\bmp_\perp$, \texttt{(iii)} this integral is dominated by
$p_\perp\sim Q_s(Y)$ (since the gluon distribution is strongly peaked at
$Q_s$, cf. \Fref{fig:Phi}), and \texttt{(iv)} the result of the
integration is very weakly dependent upon $Y=\ln(1/x_2)$, by geometric
scaling (indeed, $\Phi_A(Y,p_\perp)$ is roughly a function of $\tau\equiv
p_\perp/Q_s(Y)$, which is evaluated at $\tau=1$ when performing the
integral). These arguments imply that $\rmd N/\rmd \eta$ depends upon
$x_1\simeq \exp(\eta-y_{\rm beam})$ but is approximately independent of
$x_2$ (and hence of the total collision energy) within the range of
`forward rapidities'. This is precisely the property called limiting
fragmentation, as visible in \Fref{limfrag} for both d+Au and Au+Au
collisions at RHIC (PHOBOS) \cite{Back:2004je,Veres:2005ix}.

As a final application for `dilute--dense' scattering, we shall consider
the production of a pair of hadrons at forward rapidities. The kinematics
has been already explained in relation with \Eref{2pkin}~: when both
$\eta_a$ and $\eta_b$ are positive and large, the produced hadrons
explore small values $x_2\ll 1$ in the target wavefunction and hence they
can experience high density effects --- multiple scattering and gluon
saturation. As before, these effects are important when the transverse
momenta $p_{a\perp}$ and $p_{b\perp}$ of the two hadrons are comparable
to the target saturation momentum. The respective cross--section admits a
factorization similar to \Eref{quarkprod}, where however the target
expectation value now involves the trace of the product of {\em four}
Wilson lines: two for the produced partons in the direct amplitude, and
two for the same partons in the complex conjugate amplitude. Hence, the
generalized `gluon distribution' of \Eref{Phi} gets now replaced by a
4--point function, known as a {\em colour quadrupole}
\cite{JalilianMarian:2004da,Baier:2005dv,Marquet:2007vb,Dominguez:2011wm,Iancu:2011nj}.

A convenient way to study high--density effects in this setup is to
measure {\em di--hadron correlations in the azimuthal angle}, which is
the angle indicating the direction of propagation in the transverse
plane. If the medium effects are negligible, two relatively hard
particles ($p_\perp\gg \Lam$) are produced back--to--back
($\bmp_{a\perp}+\bmp_{b\perp}\simeq 0$). So, if the trigger detects one
of these particles together with its fragmentation products (a `jet'),
then by measuring the particle distribution in the same event one should
find another `jet' at a relative angle $\Delta\Phi\simeq \pi$. On the
other hand, if the target looks dense on the transverse resolution scale
of the produced particles, then the $p_\perp$--distribution gets broaden
via multiple scattering and the peak corresponding to the `away jet' at
$\Delta\Phi\simeq \pi$ gets smeared, or it even disappears.

 \begin{figure}[t]
\centerline{
\includegraphics[width=.44\textwidth]{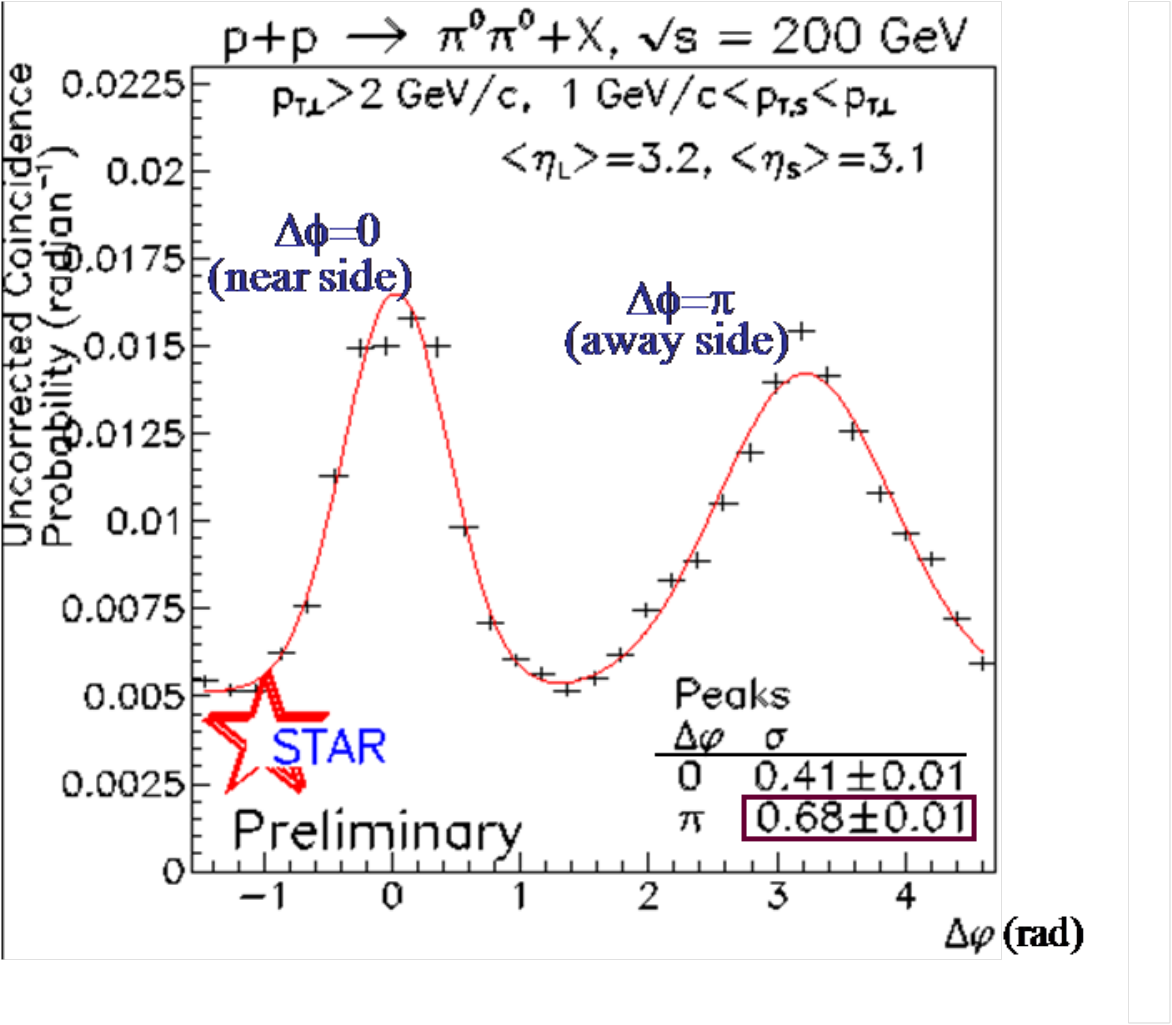}\quad
\includegraphics[width=.45\textwidth]{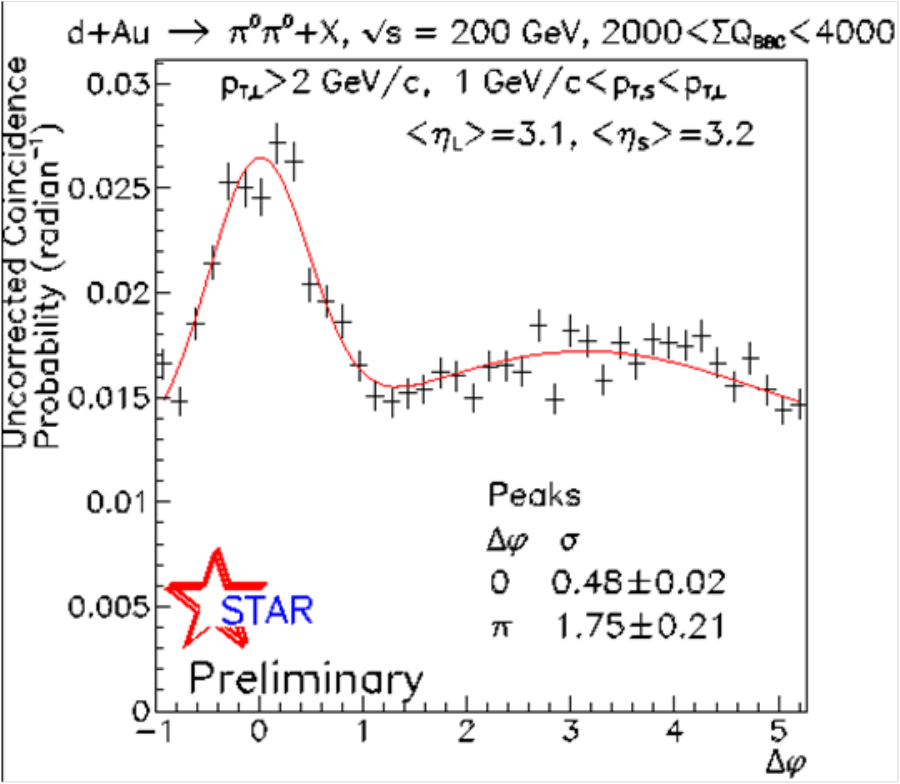}
}  \caption{\sl  Di--hadron azimuthal correlations at forward rapidities
($\eta_a,\,\eta_b\simeq 3$) and semi--hard
transverse momenta as measured in p+p (left panel) and d+Au (right panel)
collisions at RHIC, for $\sqrt{200}$~GeV per nucleon pair.
 \label{fig:dijet}}   \vspace*{-.2cm}\end{figure}

This scenario has been indeed confirmed by the RHIC data as measured by
STAR \cite{Braidot:2010ig,Braidot:2011zj}. \Fref{fig:dijet} shows the
experimental results for di--hadron production at forward rapidities
($\eta_{a,b}\simeq 3$) and semi--hard transverse momenta in both p+p and
d+Au collisions. As already discussed in relation with \Eref{RpA}, for
this kinematics one expects the saturation effects to be negligible for a
proton target, but important for a gold nucleus. And indeed, the data for
azimuthal correlations in \Fref{fig:dijet} show a pronounced `away' peak
in the p+p collisions (the left panel), but a strongly suppressed one in
the d+Au ones (the right panel). Such a suppression was actually
predicted by the CGC effective theory
\cite{Marquet:2007vb,Albacete:2010pg} and its experimental observation at
RHIC is one the most compelling evidences in favour of gluon saturation
available so far.

\begin{figure}[t]
\begin{center}
\centerline{
 \includegraphics[width=0.65\textwidth]{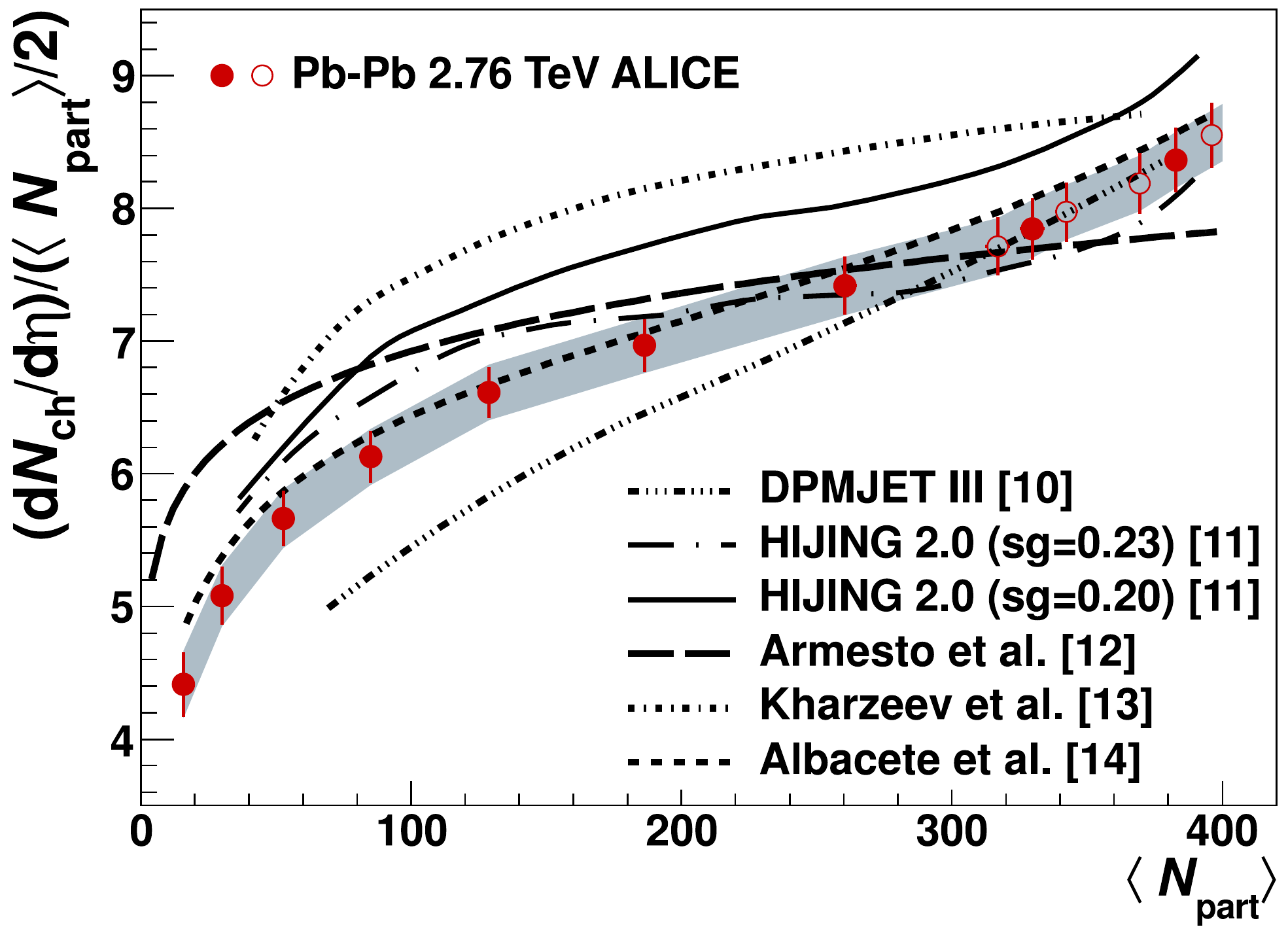}}
 \caption{\sl The LHC data (ALICE \cite{Aamodt:2010cz}) for the midrapidity
 ($\eta=0$) charged particle multiplicity in  Pb+Pb
collisions at $\sqrt{s}=2.76$~TeV/nucleon pair,
normalized by the number of participants
$N_{\rm part}$ and plotted as a function of $N_{\rm part}$ ---
hence of the centrality of the collisions, which increases (meaning that
the impact parameter $b_\perp$ decreases)
with increasing $N_{\rm part}$. The predictions of six theoretical calculations
are also shown. The three lowest ones, marked as `Armesto et al',
`Kharzeev et al', and `Albacete et al', are inspired by the CGC effective
theory and explicitly include gluon saturation. The three other models
include some of the effects of saturation, in the form of an energy--dependent
`infrared' cutoff at low transverse momenta, $p_\perp^{\rm min}
\sim E^\lambda$, which effectively plays the role of $Q_s$.
  \label{fig:multipli}}
\end{center}\vspace*{-.8cm}
\end{figure}

Turning to A+A (or `dense--dense') collisions, we recall that, in that
case, the CGC framework provides the initial conditions for the
subsequent evolution of the liberated partonic matter, but not also the
spectrum of the produced hadrons. Still, predictions can be more for more
inclusive quantities like the total multiplicity, which are expected to
be less affected by `final state' interactions. In particular,
\Fref{fig:multipli} exhibits the centrality dependence of the
multiplicity of the charged particles as measured at the LHC in Pb+Pb
collisions at $\sqrt{s}=2.76$~TeV/nucleon pair, together with the
respective predictions of various theoretical models. The centrality of a
collision refers to the relative impact parameter $b_\perp$ of the two
projectiles in the transverse plane (see \Fref{fig:geometry}). For A+A
collisions, this is often parameterized in terms of the `number of
participants' --- the number of incoming nucleons from the two nuclei in
the region where the nuclei overlap with each other (the `interaction
region'). Clearly, central collisions ($b_\perp\simeq 0$) involve more
participants than the peripheral (large $b_\perp$) ones. Although one
cannot compute $\rmd N_{\rm ch}/\rmd \eta$ fully from first principles
(as this also requires some information about the distribution of
nucleons within the nuclear disk), one can easily estimate the dependence
of this quantity upon $N_{\rm part}$ within the framework of the CGC
effective theory. As discussed in relation with \Fref{fig:AA}, the
partons produced in the early stages of a A+A collision are typically
gluons with transverse momenta $p_\perp\sim Q_s$, which have been
liberated by the collision. So the multiplicity $\rmd N_{\rm ch}/\rmd
\eta$ near $\eta=0$ is proportional to the number of such gluons which
were present in the initial nuclei, within their region of overlapping:
 \beq\label{multiplicityNpart}
 \frac{\rmd N_{\rm ch}}{\rmd \eta}\Big |_{\eta=0}
 \,\propto\,S\,\frac{xg_A(x,Q_s^2)}
 {\pi R_A^2}\,\propto\,\frac{1}{\alpha_s(Q_s^2)}\,S\,Q_s^2(A,E)
 \,\sim\,N_{\rm part}\,E^\lambda\ln\frac{Q_s^2(A,E)}{\Lam^2}\,.\eeq
Here, $S$ is the transverse area of the interaction region and we have
used \Eref{Qsat} for the (nuclear) saturation momentum together with the
fact that $S\,Q_s^2(A,E)$ is proportional to $N_{\rm part}$ and it grows
with the COM energy $E=\sqrt{s}$ like $E^\lambda$ (cf. \Eref{QsxA} and
the discussion of \Fref{fig:multiplicity}). Hence, the ratio $(\rmd
N_{\rm ch}/\rmd \eta)/N_{\rm part}$ is expected to be only weakly
dependent upon $N_{\rm part}$, via the corresponding dependence of the
running coupling: $1/{\alpha_s(Q_s^2)}\sim \ln ({Q_s^2}/{\Lam^2})\sim \ln
N_{\rm part}$. This is in good agreement with the data, as shown in
\Fref{fig:multipli}. (For the most refined calculation to date, whose
results are indicated in \Fref{fig:multipli} by the curve denoted as
`Albacete et al', see Ref.~\cite{ALbacete:2010ad}.)

\begin{figure}[t]
\begin{center}
\centerline{
\includegraphics[width=0.95\textwidth]{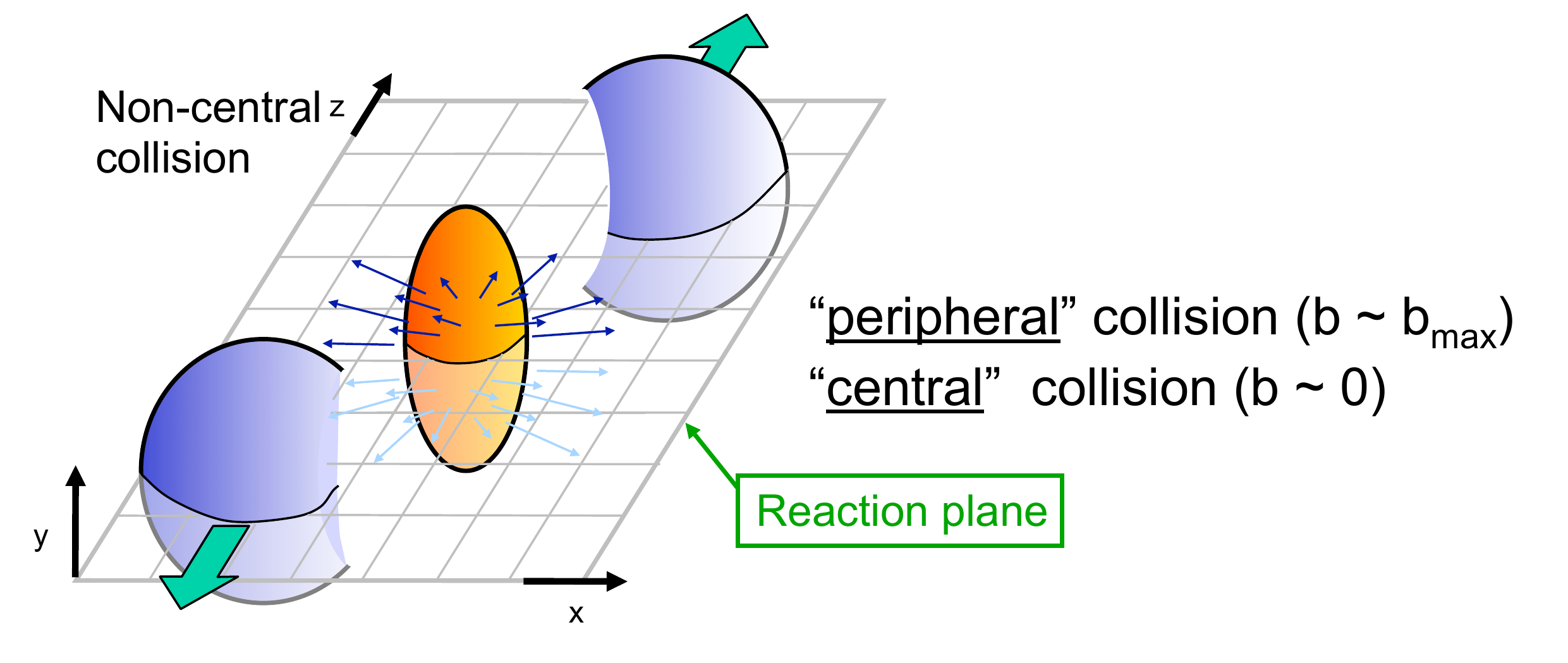}}
 \caption{\sl A sketch of the geometry of a heavy ion collision. The
 collision (or longitudinal) axis is denoted as $z$, while $x$ and $y$
 are the transverse coordinates. The interaction region
  (the almond--shape
 region where the two nuclei overlap which each other) is singled out.
 $N_{\rm part}$ is the number of nucleons in this region.
 The interaction region is horizontally cut by the reaction plane $(x,z)$.
  \label{fig:geometry}}
\end{center}\vspace*{-.8cm}
\end{figure}

\subsection{The Glasma}
\label{sec:glasma}

By the uncertainty principle, it takes a time $\Delta t\sim 1/Q_s$ to
liberate a particle with transverse momentum $p_\perp\sim Q_s$ at
midrapidities ($\eta\simeq 0$). So, by that time, the small--$x$ gluons
which were originally confined within the wavefunctions of the incoming
nuclei, are released by the collision. What is the subsequent evolution
of these gluons ? To answer this question one needs a better
understanding of their configuration at the time of emission. This can be
inferred from the solution $A^\mu_a$ to the Yang--Mills equation
\eqref{YMAA}, or, more precisely of the associated chromo--electric and
magnetic fields, ${\bm E}_a$ and  ${\bm B}_a$.

\begin{figure}[t]
\begin{center}\centerline{
\includegraphics[width=.7\textwidth]{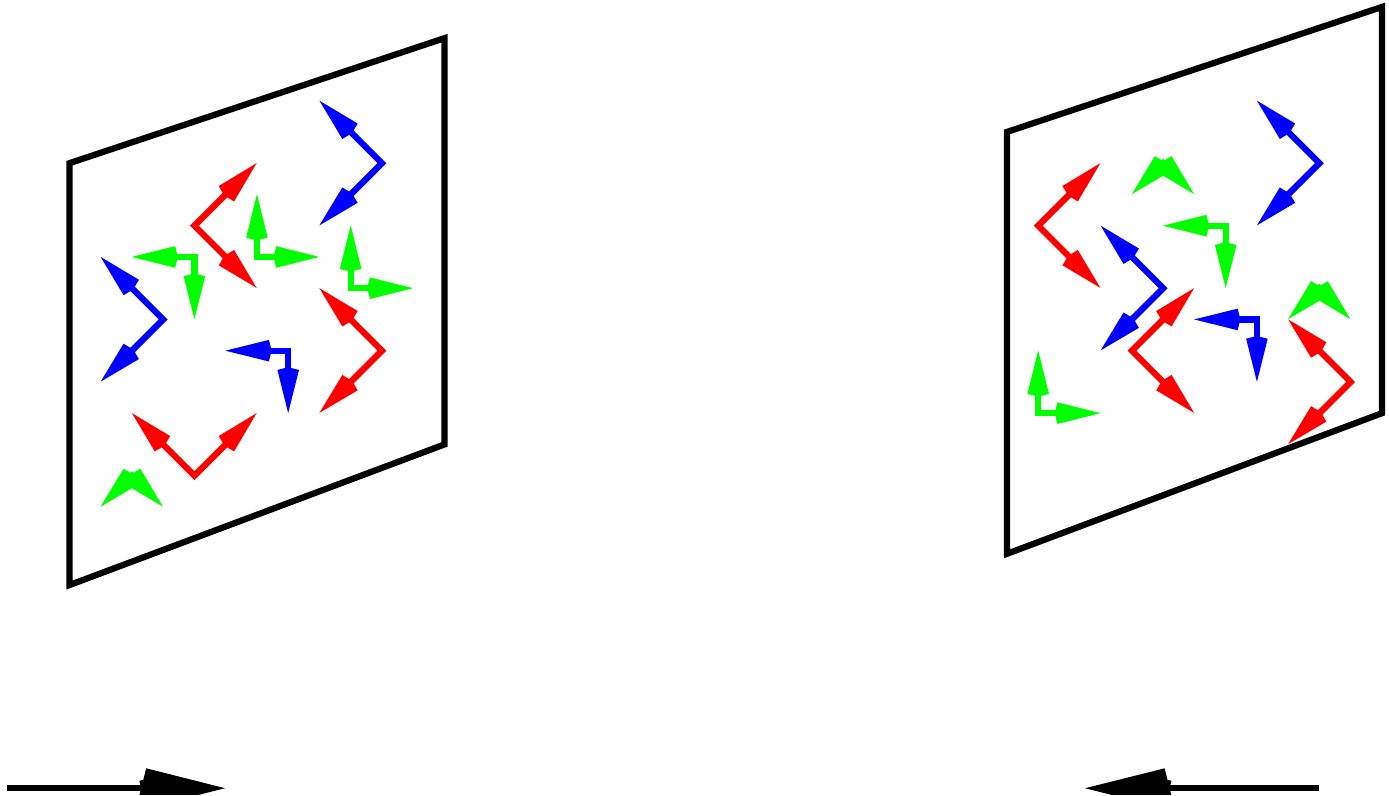}
}  \caption{\sl The chromo-electric and chromo-magnetic field
configurations in the two nuclei prior to the collision.
 \label{fig:sheet}} \end{center}   \vspace*{-.5cm}\end{figure}

Prior to the collision ($t <0$), these fields describe the two `colour
glass condensates' of the incoming nuclei, which by causality are
independent of each other. Due to the high--energy kinematics, the CGC
fields turn out to be quite simple (see \Fref{fig:sheet}) : for each
nucleus, the respective vectors ${\bm E}_a$ and ${\bm B}_a$ have only
transverse components, $E^i_a$ and $B^i_a$ with $i=1,2$, meaning that
they are orthogonal to the collision axis $x^3$. Besides, they are also
orthogonal to each other, ${\bm E}_a\cdot {\bm B}_a=0$ (for each of the
$N_c^2-1=8$ values of the colour index $a$), and they have equal
magnitudes:
 \beq
 {\bm E}_a\,\perp\,{\bm B}_a\,\perp\,x^3,\quad |{\bm E}_a|\,=\,
 |{\bm B}_a|\qquad\mbox{(prior to the collision)}\,.\eeq
These initial electric and magnetic fields are localized near $x^-=0$ for
the left--moving nucleus and, respectively, $x^+=0$ for the right--moving
one, like the respective colour charges. In a given event, their values
and orientations can randomly vary from one point to the other in the
transverse plane. But on the average, the fields at different points are
correlated due to `memory' effects in the high energy evolution, in
particular, due to saturation. The correlations, which are encoded in the
respective CGC weight functions, are typically restricted to a {\em
saturation disk}, i.e. to transverse areas with radius $\sim 1/Q_s$~:
domains separated by transverse distances $\Delta x_\perp \gg 1/Q_s$
evolve independently from each other, since saturation prohibits the
emission of gluons with momenta $k_\perp\ll Q_s$. Within a saturation
disk, gluons arrange themselves in such a way to shield their colour
charges and thus minimize their mutual repulsion; accordingly, a
saturation disk has zero overall colour charge (see the right figure in
\Fref{fig:hadron}). Also, gluons can be correlated with each other {\em
in rapidity}, due to the fact that they have common ancestors, i.e. they
belong to the same parton cascade. Such correlations extend over a
rapidity interval $\Delta Y\sim 1/\alpha_s$, since this is the typical
value of $Y$ which is required to build parton cascades according to the
BFKL evolution. These correlations, which were built in the initial
wavefunctions via the high--energy evolution, get transmitted to the
gluons liberated by the collision and thus have consequences for the
distribution of the particles in the final state.

In view of the above, a A+A collision can be viewed as the scattering
between two sheets of coloured glass, as illustrated in \Fref{fig:sheet}.
Incidentally, a similar structure for the incoming fields --- electric
and magnetic fields which are orthogonal to the beam axis and to each
other, and which are localized near the respective light--cone and frozen
by Lorentz time dilation --- would also hold if the nuclei were made with
{\em electric} (rather than {\em colour}) charges. In both QED and QCD,
such field configurations --- known as the Weizs\"{a}cker--Williams
fields --- represent the boosted version of the Coulomb fields created by
the ensemble of (electric or colour) charges in their rest frame.
However, the non--Abelian structure of QCD is essential for having a {\em
collision}: the Abelian version of \Eref{YMAA} would be linear and it
would not describe a scattering process. (The total electromagnetic field
at $t>0$ would be simply the sum of the individual fields of the two
nuclei.) The non--linear effects encoded in the Yang--Mills equation
\eqref{YMAA} describe the scattering between the small--$x$ gluons in the
two CGC's. The corresponding solution at $t>0$ (more precisely, in the
forward light cone at $x^+>0$ and $x^->0$, which is the space--time
region causally connected to the collision) represents the gluonic matter
produced by the collision.

\begin{figure}[t]
\begin{center}\centerline{
\includegraphics[width=0.5\textwidth]{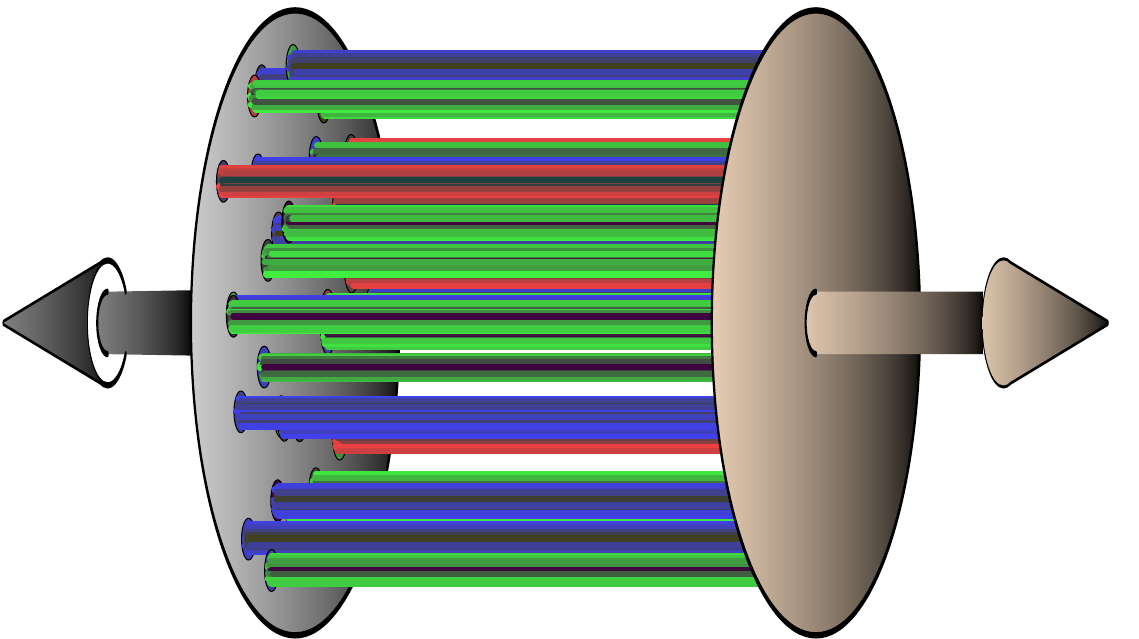}\quad
 \includegraphics[width=0.5\textwidth]{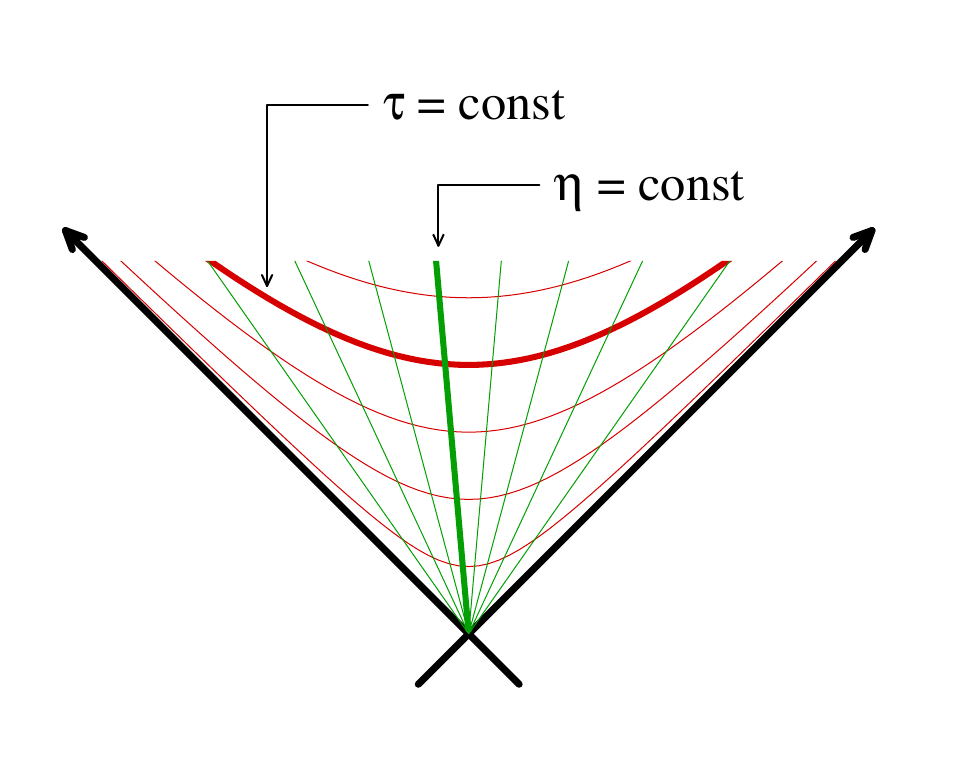}
 }
 \caption{\sl Left: the longitudinal colour flux tubes which develop
 in between the remnants of the two nuclei in the early
 stages ($\tau\lesssim 1/Q_s$) of a A+A collision.
 Right: various choices of coordinates in the future light cone.
 \label{fig:flux}}
\end{center}\vspace*{-.8cm}
\end{figure}

This solution exhibits a very interesting structure: in addition to the
{\em transverse} fields on the two sheets, which after the collision are
separating from each other, there are also {\em longitudinal}, electric
and magnetic fields, $E^3_a$ and $B^3_a$, which extend along the
collision axis. The latter give rise to {\em colour flux tubes} (or
`strings') with the endpoints on the two sheets and a typical transverse
radius $1/Q_s$ (see \Fref{fig:flux} left). Right after the collision
($t=0_+$), these fields are quite strong, $E^3\sim B^3\sim 1/g$, since
they carry most of the energy of the original CGC fields. At such early
times, the gluonic matter is still in a high--density, coherent state,
for which a description in terms of classical fields is better suited
than one in terms of particles. But with increasing time, the system
expands, its density decreases, and so does the strength of the fields.
After a time $t\sim 1/Q_s$, the magnitudes of all the fields (transverse
and longitudinal) becomes of order one, meaning that, from now on, these
fields can be also interpreted as incoherent superpositions of {\em
particles}. The spectrum of these particles (mostly gluons) is obtained
from the Fourier modes of the colour fields at time $t\gtrsim 1/Q_s$.
These particles can interact with each other, as their density is still
quite high (albeit decreasing with time, due to expansion). As we shall
see, these interactions are expected to lead to a phase of local thermal
equilibrium --- the quark--gluon plasma (QGP).

The intermediate, non--equilibrium, form of matter, which interpolates
between the CGC in the initial wavefunctions and the QGP at later stages
is known as the {\em glasma} (a name coined as a combination of `glass'
and `plasma') \cite{Lappi:2006fp}. The main, qualitative, feature of the
glasma is the presence of longitudinal colour flux tubes with transverse
area $\sim 1/Q_s^2$, which are {\em boost invariant}~: the colour fields
depend upon the {\em proper time} $\tau$ but not also upon the {\em
space--time rapidity} $\eta_s$. The variables $\tau$ and $\eta_s$, which
can be used instead of $t$ and $z\equiv x^3$ in the forward LC, are
defined as
 \beq\label{propert}
 \tau\,=\,\sqrt{t^2-z^2}\,,\qquad
 \eta_s\,=\,\frac{1}{2}\,\ln\frac{t+z}{t-z}\,\Longrightarrow\,
 \tanh\eta_s\,=\,\frac{z}{t}\,.\eeq
Under a boost along the $z$ axis, $\tau$ is invariant while $\eta_s$ is
shifted by a constant. Lines of constant $\tau$ and of constant $\eta_s$
are shown in \Fref{fig:flux} (right). The fact that the glasma fields
depend upon $\tau$ but not upon $\eta_s$ is a consequence of the
symmetries of the collision, as encoded in the classical field equations
\eqref{YMAA}. This is also consistent with the hypothesis of {\em uniform
longitudinal expansion}, as originally formulated by Bjorken.
Specifically, Bjorken has assumed that \texttt{(i)} after being produced
at $t\simeq z\simeq 0$, the particles undergo free longitudinal
streaming, meaning that they keep a constant velocity along the $z$ axis;
accordingly, the particles that can be found at some later time $t$ at
point $z$ are those with a longitudinal velocity $v_z=z/t$; \texttt{(ii)}
the distribution of the produced particles is uniform in $v_z$. Together,
\texttt{(i+ii)} imply that the distribution at time $t$ is independent of
$z/t$, hence of $\eta_s$. Note that this argument identifies the {\em
momentum} rapidity $y$ (cf. \Eref{ydef}) of the produced particles with
their {\em space--time} rapidity $\eta_s$ : $ \tanh y \equiv v_z = z/t
\equiv \tanh\eta_s$. Hence, the boost invariance of the glasma fields
implies that the distribution of the particles produced by the decay of
these fields is independent of $y$. This is a generic feature of the
particle production at the {\em classical} level, that is, on an
event-by-event basis: the associated spectra are boost invariant. But the
{\em physical} spectra, as obtained after averaging the classical results
with the CGC weight functions of the incoming nuclei, cf. \Eref{CGCfact},
{\em are} rapidity--dependent, because of the respective dependencies of
the weight functions, as introduced by the quantum evolution with $Y$.

\begin{figure}[t]
\begin{center}\centerline{
\includegraphics[width=0.4\textwidth]{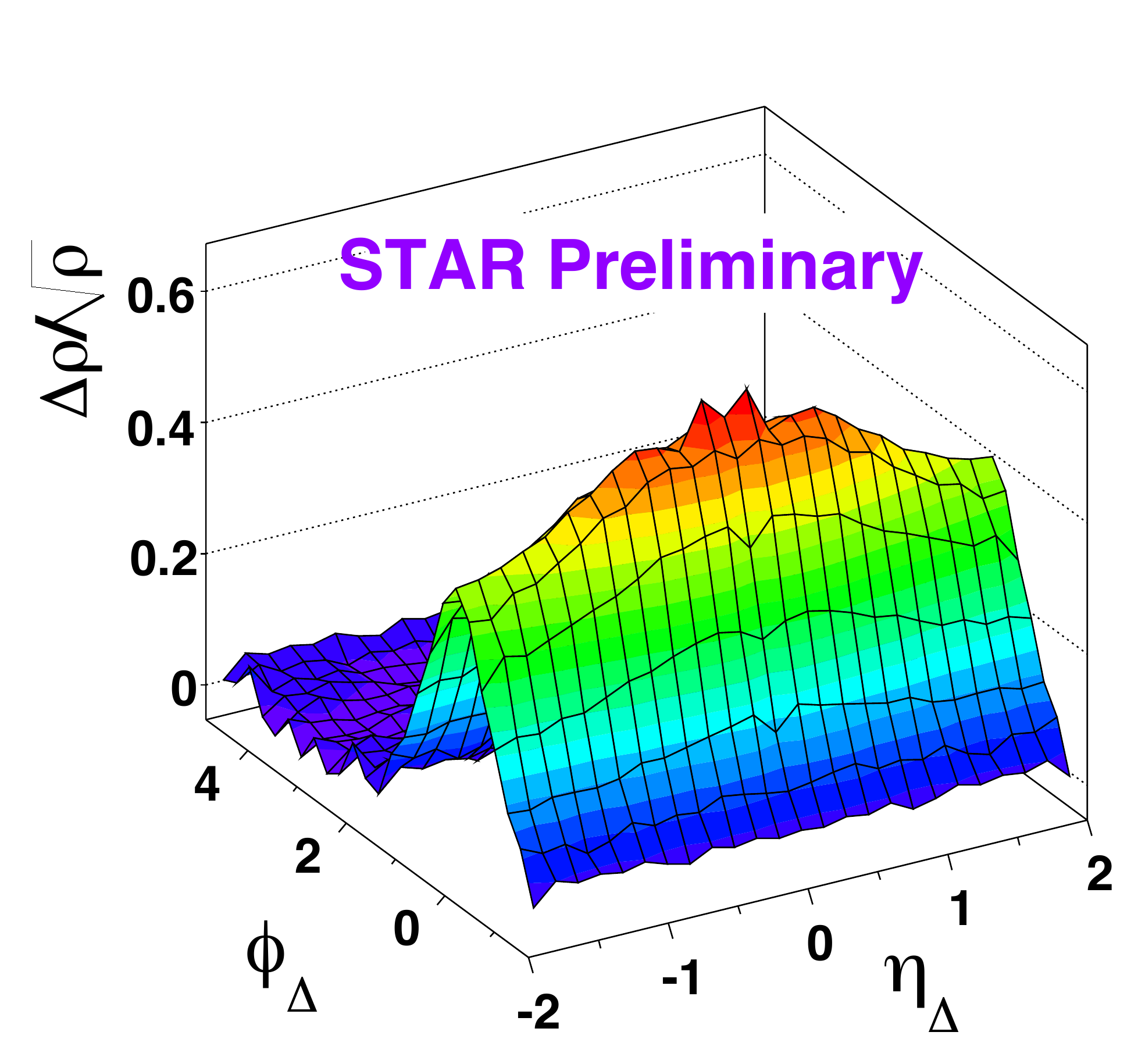}\quad
 \includegraphics[width=0.65\textwidth]{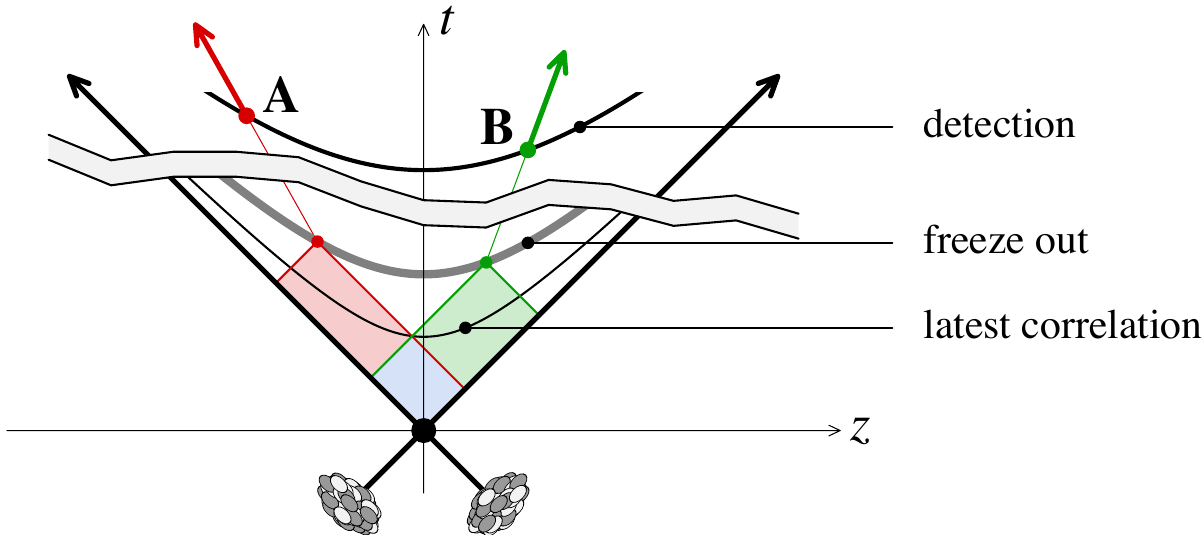}
 }
 \caption{\sl Left: the `ridge' in the di--hadron correlations as measured by
 RHIC (STAR).
 Right: the causal structure of the di--hadron
 rapidity correlations; the 2 hadrons $A$ and $B$ propagating with different
 rapidities were causally connected in the space--time region where the
 respective backward light--cones with vertices on the freeze--out surface
 were overlapping with each other.
 \label{fig:ridge}}
\end{center}\vspace*{-.8cm}
\end{figure}

An interesting consequence of the above considerations, which might be
related to a remarkable phenomenon seen in the RHIC
\cite{Adams:2005ph,Adare:2008cqb,Abelev:2009af,Alver:2009id} and the LHC
data \cite{ATLAS:2011hfa,CMS:2011mp,ALICE:2011kv} and known as the {\em
ridge}, refers to the rapidity dependence of the two particle
correlation. The latter is defined as
 \beq\label{2partC}
 C_2(\eta_a, \bmp_{a\perp};\eta_b, \bmp_{a\perp})\,=
 \left\langle\frac{\rmd N_2}{\rmd \eta_a\rmd^2\bmp_{a\perp}
 \,\rmd \eta_b\rmd^2\bmp_{b\perp}}\right\rangle
- \left\langle\frac{\rmd N}{\rmd \eta_a\rmd^2\bmp_{a\perp}}\right\rangle
\left\langle\frac{\rmd N}{\rmd
 \eta_b\rmd^2\bmp_{b\perp}}\right\rangle\,,\eeq
and what is generally plotted is the ratio $\mathcal{R}$ between this
correlation and is disconnected part (below, $N_a$ denotes the number of
particles of type $a$ in a given bin in pseudo--rapidity and azimuthal
angle),
 \beq
 \mathcal{R}\,\equiv\,\frac{\langle N_a\,N_b\rangle
- \langle N_a\rangle\,\langle N_b\rangle} {\langle N_a\rangle\,\langle
 N_b\rangle}\,,\eeq
as a function of the rapidity and the azimuthal separations between the
two particles, $\Delta\eta=\eta_a-\eta_b$ and $\Delta\phi=\phi_a-\phi_b$.
Remarkably, the data for A+A collisions at both RHIC and the LHC show the
existence of correlations which extend over a large rapidity interval
$\Delta\eta\simeq 4\div 8$, but restricted to small azimuthal separations
$\Delta\phi\simeq 0$ (see \Fref{fig:ridge} left). This means that
particles which propagate along very different directions with respect to
the collision axis preserve nevertheless a common direction of motion in
the transverse plane. By causality, such a correlation must have been
produced at early times, when these particles --- which rapidly separate
from each other --- were still causally connected (see the right panel of
Fig.~\ref{fig:ridge}).  A simple estimate gives
\begin{equation}
\tau_{\rm max}=\tau_{\rm freeze-out}\; \rme^{-\frac{|\Delta\eta|}{2}}\; ,
\end{equation}
for the latest time at which these particles could have been correlated.
For a freeze--out time $\tau_{\rm freeze-out}\approx 10~$fm/c, and
rapidity separations $\Delta\eta\geq 4$, one sees that these correlations
must have been generated before $1~$fm/c.

Long--range rapidity correlations are natural in the glasma picture,
where all the spectra (in particular, the 2--particle ones) are
independent of $y$ --- at least, at the classical level. After averaging
with the CGC weight functions, as in \Eref{CGCfact}, the 2--particle
spectrum acquires a dependence upon the {\em average} rapidity of the two
particles $(\eta_a+\eta_b)/2$, but not upon their difference
$\Delta\eta$. This last argument remains correct so long as one can
neglect quantum corrections due to soft gluon emissions within the
rapidity interval $\Delta\eta$, which in turn requires
$\Delta\eta\lesssim 1/\alpha_s$. Concerning the azimuthal collimation of
the ridge, this is not a consequence of the glasma --- the particles
produced via the decay of the classical fields are emitted isotropically
in the transverse plane in the rest frame of the medium ---, but can be
generated via {\em radial flow} : the local fluid element has some
transverse velocity $\bm{v}_\perp$, which introduces a bias in the
azimuthal distribution of the particles produced by the decay of a same
flux tube (the final correlation is peaked around the direction of
$\bm{v}_\perp$). Flow phenomena will be discussed in more detail in the
next section.

For sufficiently high energy, the long--range rapidity correlations
invoked for A+A collisions should be also present in the p+p collision.
In that case, one expects no flow, as there are fewer produced particles
and the freeze--out time is shorter. Yet, a small ridge has been measured
in p+p collisions by the CMS collaboration at the
LHC~\cite{Khachatryan:2010gv}, but only in the high--multiplicity events
(which are believed to be more central) and within a limited range in
$p_\perp$ (from 1 to 3~GeV), which is in the ballpark of the proton
saturation momentum at the LHC. In that case, the azimuthal collimation
could be explained by an intrinsic angular correlation between the
emission of two particles from the glasma field \cite{Dumitru:2010iy}.

Let us also note another possible consequence of the glasma flux tubes:
the presence of longitudinal electric and magnetic fields implies the
existence of topologically non--trivial configurations, characterized by
a large density of {\em Chern--Simons topological charge}. Such
configurations are interesting in that they break the charge--parity (CP)
symmetry: via the chiral anomaly, they generate a difference between the
number of quarks with right--handed and respectively left--handed
helicity. In the context of HIC, they may generate a new phenomenon,
known as the {\em chiral magnetic effect} \cite{Kharzeev:2007jp}~: quarks
with opposite helicities can be separated by the ultra strong magnetic
fields ($B\sim 10^{18}$~Gauss) created in the peripheral
ultrarelativistic A+A collisions, thus leading to a charge asymmetry
between the two sides of the reaction plane. Since the direction of the
magnetic field varies from one collision to another, this effect leads to
fluctuations in the distribution of the electric charge of the final
hadrons. These theoretical expectations appear to be supported by
measurements at RHIC (STAR) \cite{Abelev:2009txa}.

\section{The Quark Gluon Plasma}
\label{sec:QGP}

The main topic of this chapter is the {\em quark--gluon plasma} (QGP)
--- the partonic form of QCD matter in thermal equilibrium which exists
for sufficiently large temperatures, as demonstrated by numerical
calculations in lattice QCD. This form of matter is expected to be
created during the intermediate stages of a ultrarelativistic HIC, albeit
only for a short lapse of time and in a state of only {\em local} thermal
equilibrium. We shall first review the main experimental evidence in
favour of QGP in a HIC, namely the observation of {\em flow} in the
particle production and its successful description in terms of {\em
hydrodynamics}. Then we shall discuss the QGP {\em thermodynamics} from
the viewpoints of lattice theory and perturbative QCD. We shall also
mention the difficulty of perturbation theory to describe the dynamics
{\em out-of-equilibrium} (in particular, the transport coefficients and
the process of thermalization). Then we shall consider the phenomenon of
{\em jet quenching} (the energy loss by an energetic parton via
interactions in the plasma), which is an important tool for exploring the
deconfined matter produced in HIC's. At several places in what follows,
we shall encounter situations where perturbation theory appears to be
insufficient and which may signal a regime of {\em strong coupling}. To
address such situations from the opposite limit --- that of a coupling
which is {\em arbitrarily} strong ---, one can rely on techniques
borrowed from string theory, via the {\em AdS/CFT correspondence}. This
will be briefly discussed (in relation with the physics of HIC's) in the
last section of these lectures.
\begin{figure}[t]
\begin{center}\centerline{
\includegraphics[width=0.8\textwidth]{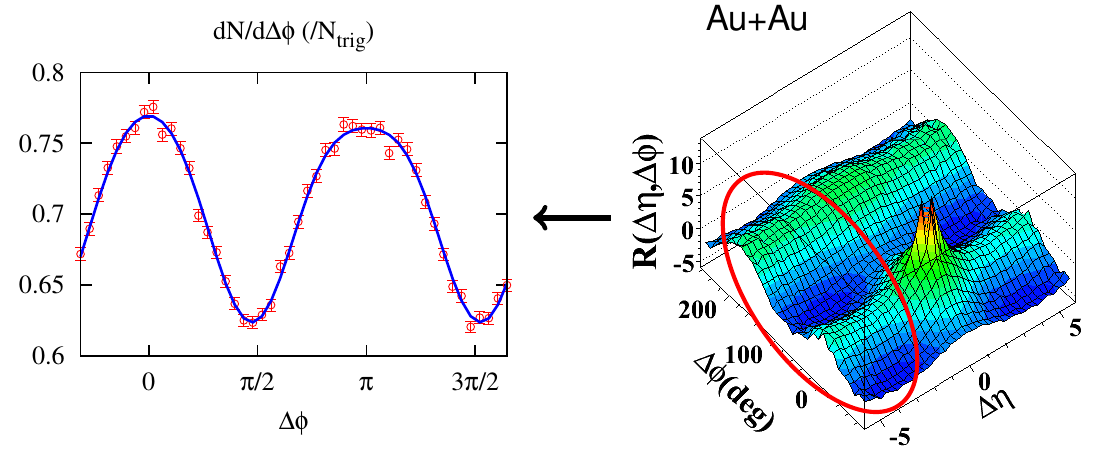}\quad
 }
 \caption{\sl The double--peak structure characteristic of elliptic flow:
 two wide peaks at
 $\Delta\phi=0$ and $\Delta\phi=\pi$ which extend over a wide
 interval $\Delta\eta$ in rapidity.
The strong correlation peak visible at $(\Delta\phi,\Delta\eta)\simeq (0,0)$
is associated with the fragmentation of the trigger jet.
 \label{fig:ridge-flow}}
\end{center}\vspace*{-.8cm}
\end{figure}

\subsection{Correlations and flow in HIC}
\label{sec:flow}

In our previous discussion of the `ridge', in \Sref{sec:glasma}, we
focused on the long--range rapidity correlations which are rather
strongly peaked in $\Delta\phi=\phi_a-\phi_b$ near $\Delta\phi=0$,
leading to the ressemblance with a mountain ridge. But as a matter of
fact, the di--hadron correlations measured in A+A collisions at RHIC and
the LHC show an even more pronounced double--peak structure, visible in
\Fref{fig:ridge-flow}, with large but wider peaks at $\Delta\phi=0$ and
$\Delta\phi=\pi$. (In the analysis leading to \Fref{fig:ridge}, this
structure has been subtracted away to render the ridge effect visible.)
This is known as {\em elliptic} (or `transverse') {\em flow}
\cite{Ollitrault:1992bk}. As also visible in \Fref{fig:ridge-flow}, this
double peak structure can be well parameterized as
 \beq\label{v2}
 \left\langle \frac{\rmd N_{pairs}}{\rmd
   \Delta\phi}\right\rangle\,\propto\,v_2^2\,\cos(2\Delta\phi)\,,
   \eeq
with $v_2$ the `magnitude of the elliptic flow'.

\begin{figure}[t]
\begin{center}\centerline{
\includegraphics[width=0.35\textwidth]{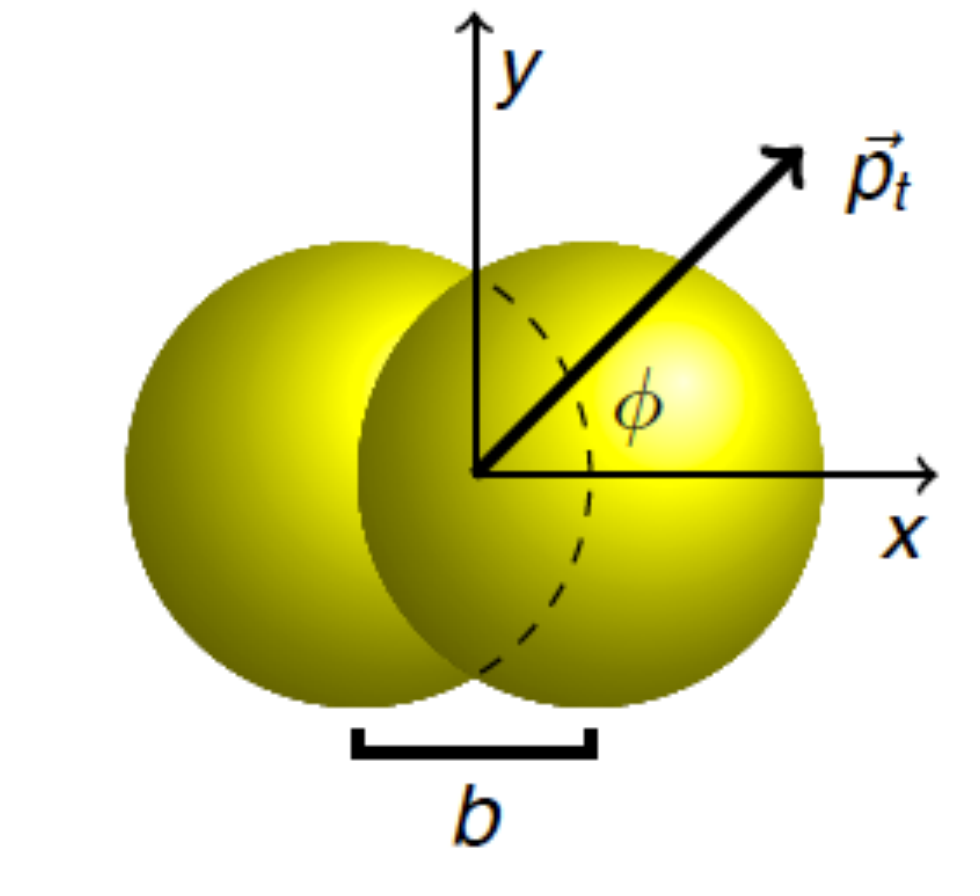}
\includegraphics[width=0.44\textwidth]{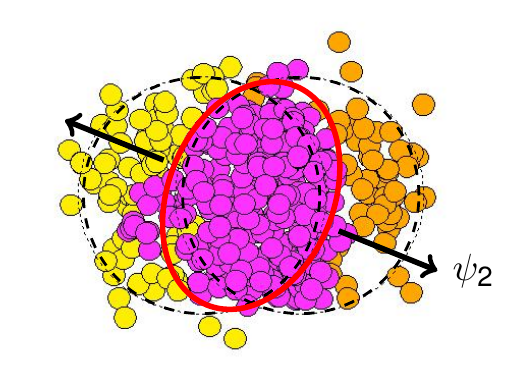}
 \includegraphics[width=0.4\textwidth]{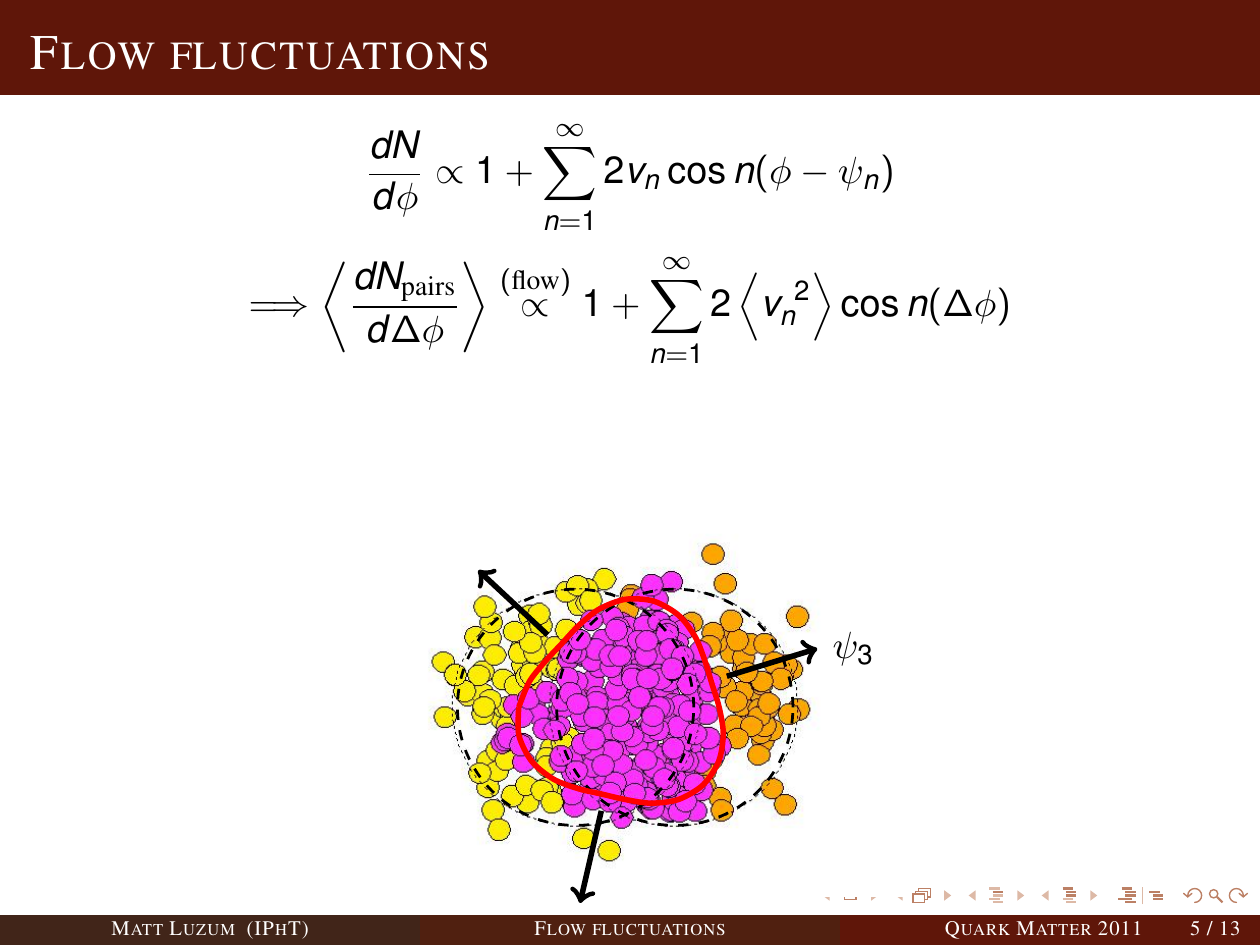}}
 \caption{\sl Left: a frontal view of a HIC, illustrating
 the impact parameter $b_\perp$, the almond--shape interaction region,
 and the azimuthal angle $\phi$.
 Central and right: the distribution of nucleons within the interaction region
 in a given event can include elliptic ($v_2$), triangular
 ($v_3$), or even higher harmonic modes. The reference angles $\psi_2$, $\psi_3$
 show the tilt of the interaction region with respect to the
 geometrical `reaction plane' (cf. \Fref{fig:geometry}).
 \label{fig:ellipse}}
\end{center}\vspace*{-.8cm}
\end{figure}

The explanation of this phenomenon turns out to be quite simple: it
reflects the anisotropy of the interaction region --- the almond--shape
region where the two nuclei overlap with each other; see
\Fref{fig:geometry} --- for non--central collisions. This anisotropy
entails a pressure gradient in the initial conditions: the pressure is
larger along the minor axis of the ellipse (the $x$ axis in the left
panel of \Fref{fig:ellipse}) rather than along the major one;
accordingly, more particles will be emitted in the direction of the
largest gradient. This ultimately generates an anisotropy in the
azimuthal distribution of the produced particles, which for symmetry
reasons is of the form shown in \Eref{v2}:
 \beq\label{flowv2}
 \frac{\rmd N}{\rmd
   \phi}\,\propto\, 1 \,+\,2v_2\cos 2\phi\,.\eeq
This argument looks simple, as anticipated, but there is something deep
about it: the role played by {\em collective phenomena} like pressure
gradients or flow. Such phenomena are natural for many--body systems
which relatively {\em strong interactions}
--- sufficiently strong to be able to transmit the asymmetry of the
initial geometry into properties of the final state. This is an important
point to which we shall return.

The qualitative arguments above suggests that the coefficient $v_2$
characterizing the strength of the anisotropy should increase with
centrality. This trend is indeed seen in the data (at least, for not too
peripheral collisions, for which the interaction region becomes tiny and
dilute). Also, particles which experience a stronger pressure gradient
are expected to have a larger transverse momentum, as they inherit the
velocity of the fluid; so, $v_2$ should rise with $p_\perp$. This
expectation, too, is confirmed by the data, at least for not too large
$p_\perp\lesssim 5$~GeV: very hard particles cannot be driven by the
medium, so for them $v_2$ is naturally small. The measurements of $v_2$
(say, via 2--hadron correlations) yield very similar results at RHIC and
the LHC (see the left panel of \Fref{fig:vn}), showing that $v_2$ is
roughly independent of the COM energy. Note also that the typical values
of $v_2$ for semi--hard momenta are relatively large, $v_2\sim 0.2$,
meaning that the collective phenomena alluded to above are indeed quite
strong.

\begin{figure}[t]
\begin{center}\centerline{
\includegraphics[width=0.9\textwidth]{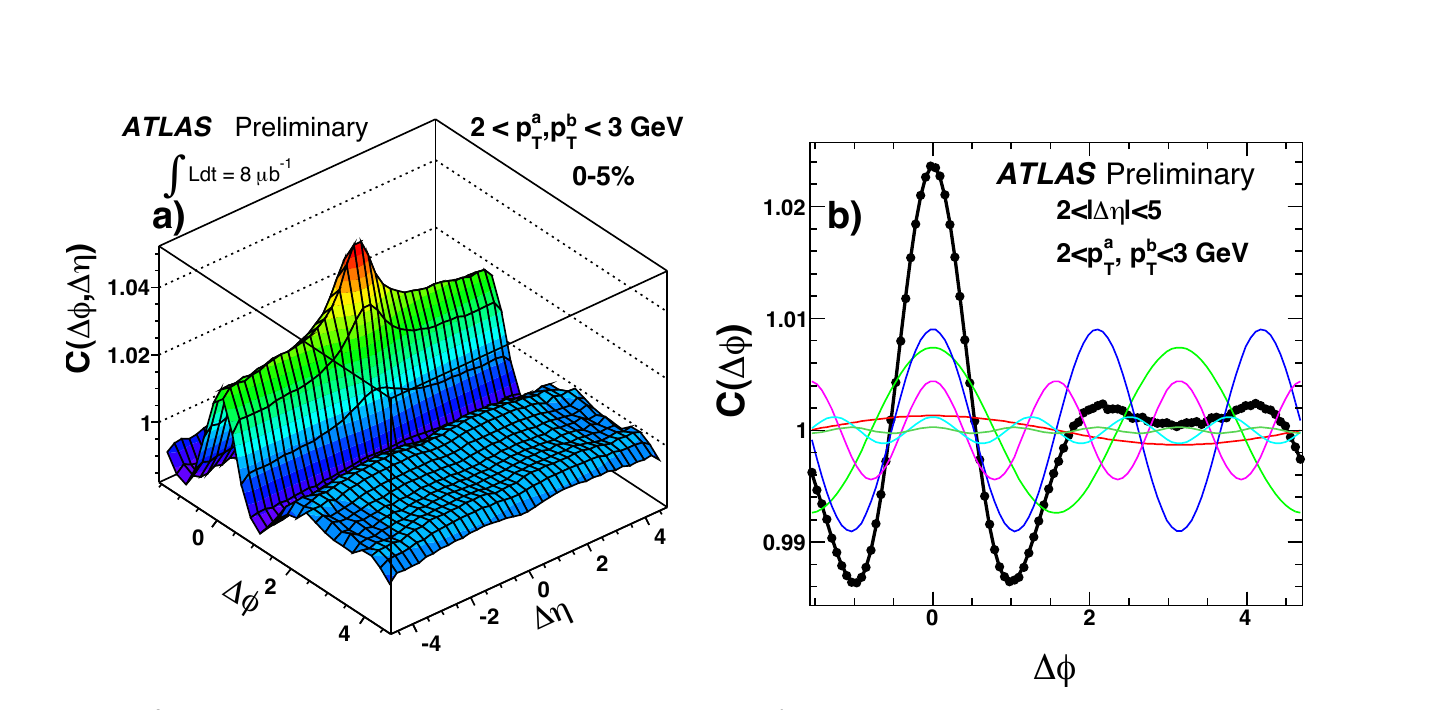}}
 \caption{\sl The `ridge' at $\Delta\phi\simeq 0$ and the `double--hump' at
$|\Delta\phi-\pi|\simeq 1.1$ as visible in the LHC data for di--hadron
correlations in the 5\% most central collisions (left panel), together
with the harmonic decomposition of these correlations, cf. \Eref{vn}
(right panel) \cite{collaboration:2011hfa}.
\label{fig:hump}}
\end{center}\vspace*{-.8cm}
\end{figure}

\begin{figure}[thb]
\begin{center}\centerline{
  \includegraphics[width=0.55\textwidth]{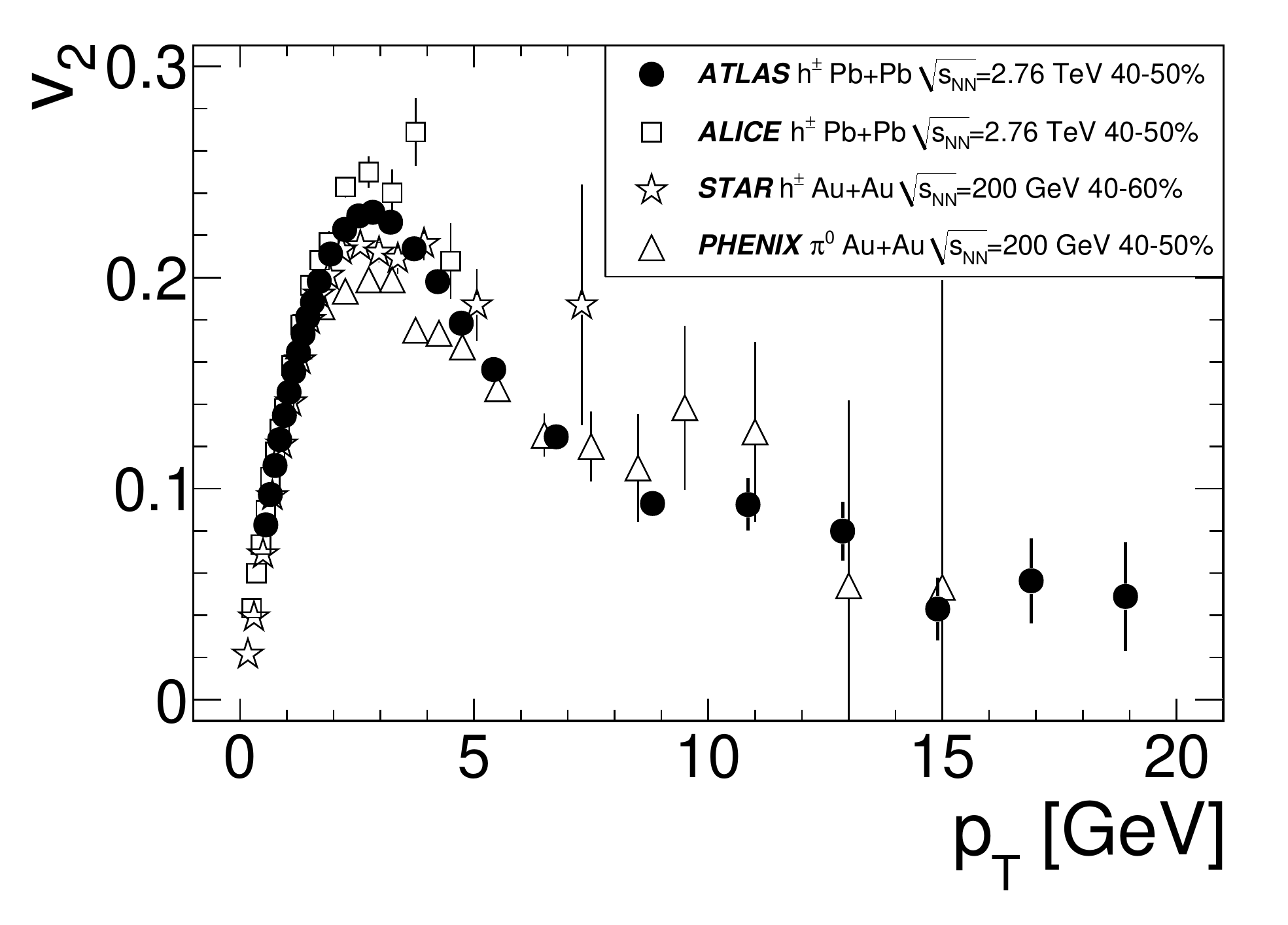}\quad
 \includegraphics[width=0.45\textwidth]{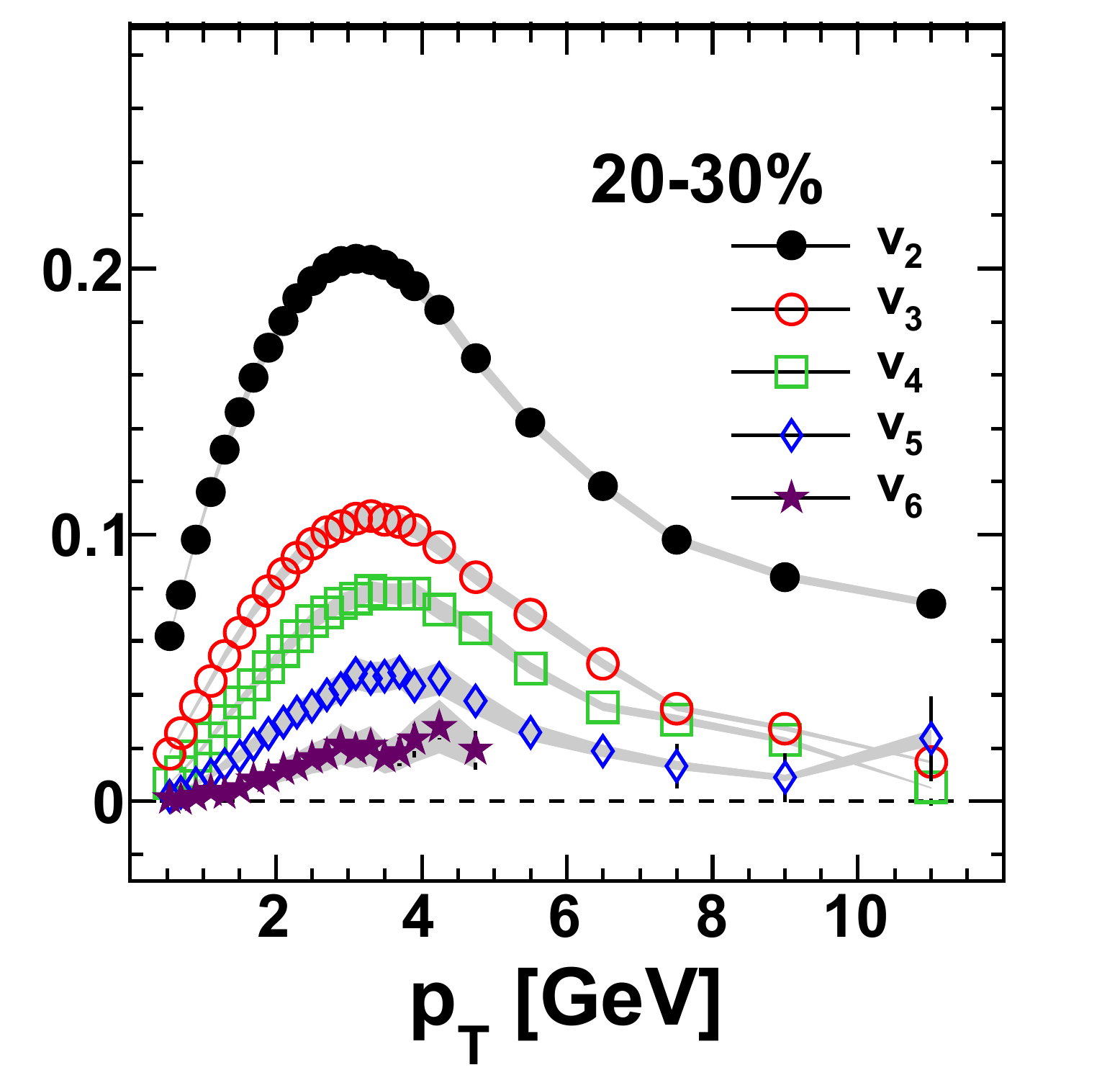}
 }
 \caption{\sl Left: the RHIC and LHC results for $v_2$ as a
 function of $p_\perp$, for collisions within
 the centrality bin $40\%-50\%$; from Ref.~\cite{ATLAS:2011ah}.
 Right: the LHC results (ATLAS)
 for $v_n(p_\perp)$, for $n=2,3,4,5$ and 6, and the
 centrality bin $20\%-30\%$ \cite{collaboration:2011hfa}.
 \label{fig:vn}}
\end{center}\vspace*{-.8cm}
\end{figure}

But the elliptic flow and the ridge are not the only collective phenomena
hidden in the di--hadron correlations illustrated in
\Fref{fig:ridge-flow}. By looking at the most central collisions where
$v_2$ is relatively small,  one sees not only the narrow `ridge' at
$\Delta\phi\simeq 0$, but also a `double--hump' on the away side, at
$|\Delta\phi-\pi|\simeq 1.1$, which extends too over a large interval
$\Delta\eta$ (see \Fref{fig:hump} left). The harmonic decomposition of
this signal reveals higher Fourier modes with significant strengths, as
illustrated in the right panel of \Fref{fig:hump}. This leads to the
following generalization of \Eref{flowv2} :
 \beq\label{vn}
 \left\langle \frac{\rmd N_{pairs}}{\rmd
   \Delta\phi}\right\rangle\,\propto\, 1 \,+\,2\sum_{n=1}^{\infty}
   \big\langle  v_n^2 \big \rangle\cos (n\Delta\phi)
   \eeq
where the various coefficients $v_n$ up to $v_6$ have been extracted from
the LHC data and they are compared to $v_2$ in the right panel of
\Fref{fig:vn}. (All these coefficients are roughly independent of $\eta$,
meaning that they describe correlations over a wide interval
$\Delta\eta$.)

What is the physics of such higher harmonics ? It is generally believed
that they are the consequence of {\em fluctuations} in the distribution
of nucleons within the interaction region, as illustrated in
\Fref{fig:ellipse} (right panel) \cite{Alver:2010gr}. Namely, even though
the overlapping region between the two nuclei has an elliptic shape, the
nuclear matter inside it is neither homogeneous, nor strictly
ellipsoidal, because of fluctuations in the particle distribution. The
azimuthal distribution in a given event can be decomposed into harmonics,
with coefficients $v_n$ and reference angles $\psi_n$ :
 \beq
  \frac{\rmd N}{\rmd
   \phi}\,\propto\, 1 \,+\,\sum_{n=1}^{\infty}
   2v_n\cos n(\phi-\psi_n)\,.\eeq
The reference angles $\psi_n$ are generally different from the
conventional reaction plane ($\psi_{RP}=0$) and are difficult to measure,
but they drop out in the 2--particle correlations, as manifest in
\Eref{vn}.

The fact that the initial geometry of the interaction region can have
complicated fluctuations is not necessarily a surprise, given the
granularity of the nucleons. What is remarkable though is the ability of
the system to transmit these fluctuations into the distribution of the
produced particles, via {\em transverse flow}. The effective theory for
flow is {\em hydrodynamics} and will be succinctly discussed in the next
section.

\subsection{Hydrodynamics and kinetic theory}
\label{sec:hydro}

{\em Hydrodynamics} is the theory which describes the flow of a fluid
independently of its detailed microscopic structure. More precisely, the
equations of hydrodynamics have an universal form (at least, for a given
underlying fundamental theory, like QCD), but they involve a few
`parameters' which depend upon the nature of the fluid and can in
principle be computed via microscopic calculations. The scope of
hydrodynamics can be most easily explained with reference to
thermodynamics. The latter describes a many--body system in {\em global
thermal equilibrium}, in which the intensive quantities, like
temperature, pressure and the chemical potentials associated with the
various conserved charges (electric charge, baryonic charge, etc) are
{\em time--independent} and {\em uniform} throughout the volume $V$ of
the system. Hydrodynamics can be viewed as a generalization of this
picture towards a state of {\em local} equilibrium: the intensive
quantities alluded to above can vary in space and time, but they do that
so slowly that one can still assume thermal equilibrium to hold locally,
in the vicinity of any point. Gradients of pressure and thermodynamics
naturally lead to {\em flow}, with a local fluid velocity ${\bm v}$ which
is itself slowly varying in space and time.

The equations of hydrodynamics are simply the ensemble of the relevant
{\em conservation laws} --- for the energy, momentum and the other
conserved charges:
 \beq\label{conservation}
 \partial_\mu\,T^{\mu\nu}\,=\,0\,,\qquad
 \partial_\mu J^\mu_B\,=\,0\,,\quad \cdots
 \eeq
where $T^{\mu\nu}$ is the energy--momentum tensor, $J^\mu_B$ is the
density of the baryonic current (the volume integral of $J^0_B$ is the
difference between the number of baryons and the number of antibaryons),
and the dots stand for other conserved charges. These densities depend
upon the intensive (local) quantities describing the state of the fluid:
the energy density $\varepsilon=E/V$, the pressure $P$, the 4--velocity
$u^\mu=\gamma(1,{\bm v})$ (with $\gamma = 1/\sqrt{1-v^2}$), and a set of
`friction coefficients' known as {\em viscosities}, which characterize
the dissipative properties of the medium.

The relations between the densities of the conserved charges
($T^{\mu\nu}$, $J^\mu_B$, ...) and the intensive quantities are obtained
via a {\em gradient expansion} with respect to the slow space--time
variations of the latter. More precisely, this amounts to an expansion in
powers of $\ell/R$, where $R$ is a characteristic size of the system (in
a HIC, $R$ is the transverse size of the interaction region) and $\ell$
is the {\em mean free path} of the particles composing the fluid (the
typical distance between two successive collisions). This quantity will
play an important role in what follows, so let us open here a parenthesis
and discuss it in more detail.

At least for sufficiently weak coupling, the mean free path $\ell$ can be
estimated using {\em kinetic theory}. This is an effective theory too,
but it applies at shorter, microscopic, scales: in that context, the mean
free path is typically the {\em largest} scale in the problem  (it is
much larger than the Compton wavelength $\lambda\sim 1/k$ of a particle
or the typical duration of a scattering processes). Kinetic theory allows
one to follow the evolution of the particle distributions in {\em
phase--space} --- i.e. in space--time and in momentum space. To that aim,
this theory involves more information about the microscopic dynamics,
like cross--sections for the particles interactions described via the
`collision term' in the {\em Boltzmann equation} --- the central equation
of kinetic theory. But even without solving that equation, one can deduce
an estimate for $\ell$ via simple considerations: the {\em collision
rate} (the inverse of the typical time $\tau_{\rm coll}$ between two
successive collisions) scales like $\tau^{-1}_{\rm coll}\sim nv_{\rm
rel}\sigma$, where $n$ is the particle density, $v_{\rm rel}$ is their
average relative velocity, and $\sigma$ is the cross--section for their
mutual interactions. The mean free path is then obtained as
 \beq \label{mfp}
 \ell\,\sim\,v\tau_{\rm coll}\,\sim\,
 \frac{v}{v_{\rm rel} n\,\sigma}\,\sim\,\frac{1}{n\,\sigma}
 \,,\eeq
where $v$ is the average velocity of the particles, so $v/v_{\rm rel}$ is
a number of order one. Since $\sigma$ is naturally proportional to some
power of the coupling constant, \Eref{mfp} shows that the mean free path
becomes smaller --- meaning that {\em the hydrodynamical description
works better} --- when the coupling is {\em strong}.

\begin{figure}[t]
\begin{center}\centerline{
\includegraphics[width=0.9\textwidth]{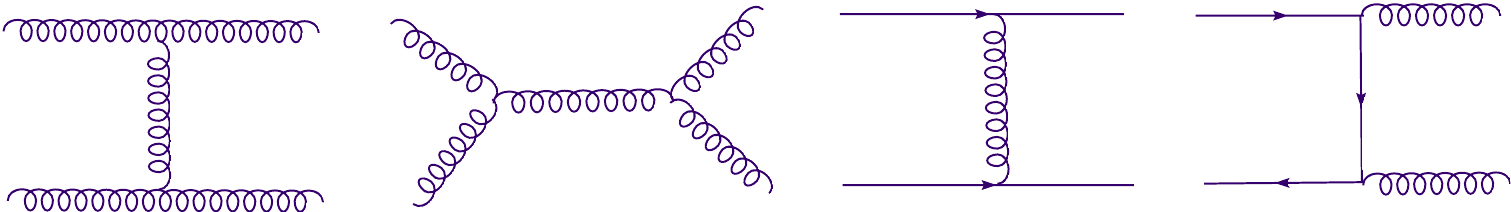}}
 \caption{\sl $2 \to 2$ partonic processes dominating the cross--section for
 (elastic) scattering at weak coupling.
\label{fig:elastic}}
\end{center}\vspace*{-.8cm}
\end{figure}

To be more specific, consider a system that will play an important role
in what follows: a {\em weakly coupled quark--gluon plasma} with (local)
temperature $T$. This is a nearly ideal gas of ultrarelativistic
particles, so the particle densities scale like $n\sim T^3$ separately
for quarks and gluons. To leading order in $\alpha_s$, scattering is
controlled by the $2\to 2$ elastic collisions shown in
\Fref{fig:elastic}, where the external lines represent thermal particles
with typical energies and momenta of order $T$. These processes yield
$\sigma\propto \alpha_s^2$. However, for the processes involving the
exchange of a gluon in the $t$ channel, there is a logarithmic
enhancement associated with the singularity of the Coulomb scattering at
small angles: the Rutherford formula reads $\rmd \sigma/\rmd
\Omega\propto \alpha_s^2/(T^2\sin^4\theta)$, with $\theta$ the scattering
angle, and it is strongly divergent when $\theta\to 0$. The
cross--section $\sigma$ which is relevant for computing the mean free
path \eqref{mfp} is not the {\em total} cross--section $\sigma_{\rm
tot}=\int\rmd\Omega \big(\rmd \sigma/\rmd \Omega\big)$, but rather the
{\em transport cross--section}~:
 \beq\label{sigma}
 \sigma\,=\,\int\rmd\Omega \big(1-\cos\theta\big)\,\frac{\rmd
\sigma}{\rmd \Omega}\,\propto\,\int\rmd\theta\,\sin\theta
\big(1-\cos\theta\big)\,\frac{\alpha_s^2}{T^2\sin^4\theta} \,\sim\,
 \frac{\alpha_s^2}{T^2}\int_g \frac{\rmd\theta}{\theta}\,\sim\,
 \frac{\alpha_s^2}{T^2}\,\ln\frac{1}{\alpha_s}\,,\eeq
which more properly characterizes the efficiency of the interactions in
redistributing energy and momentum. The factor $1-\cos\theta$, which
vanishes as $\theta^2/2$ at small angles, accounts for the fact that the
small--angle scattering is inefficient in that sense, as intuitive from
the fact that one cannot equilibrate an anisotropic energy--momentum
distribution via collinear scattering. Due to this factor, the integral
in \Eref{sigma} is only {\em logarithmically} divergent as $\theta\to 0$
(unlike $\sigma_{\rm tot}$, which would be {\em quadratically}
divergent). In reality, this divergence is screened by plasma effects
which occur at the momentum scale $gT$ (see the discussion in
\Sref{sec:thermo}). This implies that the minimal collision angles are
$\theta\sim gT/T\sim g$, corresponding to transferred momenta of order
$gT$. Hence, to leading logarithmic accuracy, the integral can be
estimated as shown in the r.h.s. of \Eref{sigma}. In turn, this implies
the following estimate for the mean free path in a weakly--coupled QGP
(cf. \Eref{mfp})
 \beq \label{mfpQGP}
 \ell\,\sim\,\frac{1}{T}\,\frac{1}{\alpha_s^2\ln(1/\alpha_s)}\,,
 \eeq
which as long as $g\ll 1$ is indeed much larger than both the Compton
wavelength $\lambda\sim 1/T$ of the thermal particles and the typical
duration $\sim 1/gT$ of a scattering process.

We now close the parenthesis dedicated to the mean free path and return
to the discussion of hydrodynamics. As already mentioned, this is a
legitimate effective theory for flow when $\ell\ll R$. The {\em
constitutive relation} allows one to relate the energy--momentum tensor
to the velocity field $u^\mu$ and its gradients, via an expansion in
powers of $\ell/R$. The powers of $\ell$ are associated with dissipative
phenomena, while those of $1/R$ with the gradients in the fluid. To
zeroth order in this gradient expansion one obtains the {\em ideal
hydrodynamics}. This is `ideal' in the sense that there is no
dissipation. The corresponding structure of $T^{\mu\nu}$ follows entirely
from the assumption of local thermal equilibrium. Namely, in the {\em
local rest frame} of a fluid element ($u^\mu_{RF}=(1, 0,0,0)$), the
energy--momentum tensor has the diagonal structure familiar from the
thermodynamics: $T^{\mu\nu}_{RF}=\mbox{diag}(\varepsilon,P,P,P)$.
Boosting to the laboratory frame, where the fluid 4--velocity is $u^\mu$,
this yields
 \beq\label{Tmunu}
 T^{\mu\nu}\,=\,
  (\varepsilon +P)u^\mu u^\nu-P g^{\mu\nu}\,,\eeq
where $g^{\mu\nu}=\mbox{diag}(1,-1,-1,-1)$ is the Minkowski metric
tensor. The r.h.s. of \Eref{Tmunu} involves 5 independent quantities:
$\varepsilon$, $P$, and the 3 components $v^i$ of the (local) velocity.
Their space--time evolution is determined by the respective conservation
law in \Eref{conservation}, which yields 4 equations, together with the
assumed {\em equation of state}, which specifies the functional relation
$\varepsilon(P)$ between the energy density and the pressure (e.g.,
$\varepsilon=3P$ for an ideal gas of massless particles).

Since ideal hydrodynamics ignores dissipation, one might think that it
corresponds to a situation where the coupling is weak, but that would be
wrong: it rather corresponds to {\em strong coupling}. This may seem
counterintuitive but it can be understood as follows: the dissipative
phenomena are proportional to the ability of the system to transfer
momentum in a direction {\em perpendicular} to the fluid velocity (since
such a transfer results in slowing down the flow). Within kinetic theory
at least, this transfer is realized by particles moving throughout the
fluid in between successive collisions. Hence, the rate for transfer is
proportional to the mean free path \eqref{mfp} and thus to the inverse of
the coupling. As an example, consider the {\em shear viscosity}: this
characterizes the friction force between two neighboring layers of fluid
which propagate, say, along the $x$ axis, but at slightly different
velocities (so there is a non--zero gradient $\partial u_x/\partial y$).
There is friction because some longitudinal momentum $p_x$ gets
transferred from the faster layer to the slower one, at a rate
proportional to the velocity gradient:
 \beq\label{shear}
    \frac{1}{A}\, \frac{\rmd p_x}{\rmd t} \,=\,-\eta\,\frac{\partial u_x}
    {\partial y}\,,\eeq
where $A$ is the contact area between the two layers and $\eta$ is the
shear viscosity. Within kinetic theory, $\eta \simeq \ell\rho v \sim
(\rho/n)(v/\sigma)$  where $\rho$ is the {\em mass} density in the fluid
and the second estimate follows after using \Eref{mfp}. For a
non--relativistic fluid $\rho/n=m$, so $\eta\sim mv/\sigma$, whereas for
the weakly--coupled QGP, $\rho=\varepsilon \simeq 3nT$ and $v=1$ and
therefore
 \beq\label{shearQGP}
 \eta\,\sim \,\frac{T}{\sigma}\,\sim
 \,\frac{T^3}{\alpha_s^2\ln(1/\alpha_s)}\,.
 \eeq
In both cases, the fluid density has canceled in the ratio $\rho/n$, so
the viscosity is {\em independent of the density}, or, equivalently, of
the {\em pressure}. (Recall that $P=nk_BT$, with $k_B$ the Boltzmann
constant.) This remarkable conclusion has been first derived by Maxwell
in 1860 via kinetic theory, and then confirmed by him experimentally.

The l.h.s. of \Eref{shear} represents a contribution to the component
$T_{xy}$ of the energy--momentum (or `shear') tensor : the flux of the
$x$ component of the momentum vector across a surface with constant $y$.
So, \Eref{shear} displays a dissipative correction to $T_{\mu\nu}$~; as
expected, this is of linear order in the gradient expansion and it scales
like $\ell/R$ (since $\eta\sim \ell$ and $\partial_y\sim 1/R$). While
first--order gradient corrections (leading to the Navier--Stokes
equation) are sufficient to describe dissipation for a non--relativistic
fluid, this is not true anymore for a {\em relativistic} one: to be
consistent with causality and Lorentz invariance, one must use a {\em
second--order formalism}, which also includes {\em quadratic} terms in
the gradient expansion.

To summarize, for the problem of hydrodynamics to be well defined, one
needs to specify \texttt{(i)} the equation of state $\varepsilon(P)$
(this is generally taken from lattice QCD calculations; see below),
\texttt{(ii)} the time $\tau_0$ at which hydrodynamical evolution can be
turned on (meaning that local thermal equilibrium has been reached),
\texttt{(iii)} the initial conditions at $\tau_0$ for the energy density
$\varepsilon({\bm x})$ and the velocity ${\bm v}({\bm x})$ fields, and
\texttt{(iv)} the various viscosities like $\eta$ which characterize the
dissipative properties of the medium. Note that, in this context, the
`initial time' $\tau_0$ is not the same as the time $\tau_s\sim 1/Q_s$ at
which the CGC formalism provides the `initial conditions' (in the sense
of the discussion in \Sref{sec:glasma}), but it is the {\em a priori}
larger {\em equilibration time} $\tau_{\rm eq}$. Within the hydro
simulations, this is a free parameter, like the viscosities or the
parameters which enter the equation of state. These parameters are fixed
{\em a posteriori}, by matching the results of the hydro evolution at the
time of freeze--out onto some of the experimental results for particle
production, like the centrality dependence of the particle multiplicities
and of their average transverse momentum.

For quite some time, roughly until 2007, it seemed that the RHIC data can
be well accounted for (within the error bars) by ideal
hydrodynamics~\cite{Kolb:2003dz}. This led to the conjecture that the
deconfined matter produced at the intermediate stages of a HIC might be
{\em strongly interacting} (`strongly coupled quark--gluon plasma' or
sQGP). In order to test this conjecture, and also to describe the more
accurate, recent data, it became necessary to include dissipative
effects, within the second--order formalism. Full calculations in that
sense, including comparison with RHIC data, became available only
recently \cite{Luzum:2008cw,Ollitrault:2010tn,Luzum:2011mm} (and refs.
therein). They are all consistent with a non--zero, albeit small, {\em
relative} value of the viscosity, as measured by the ratio $\eta/s$.
Here, $s$ is the entropy density, and the ratio $\eta/s$ is dimensionless
in natural units (in general, it has the dimension of $\hbar$). This
ratio is a natural measure of the deviations from ideal hydro, as we
explain now. The entropy density $s$ is proportional to the particle
density; e.g., $s=4n$ for an ideal gas of massless particles. Thus,
$\eta/s\sim \ell v (\rho/n) \sim \ell/\lambda$, where $\lambda$ is the
Compton wavelength of a particle in the fluid: $\lambda=1/(mv)$ in the
non--relativistic case and $\lambda\sim 1/T$ for a weakly coupled QGP. By
the uncertainty principle, the ratio $\ell/\lambda$ cannot be smaller
than $\hbar$ times a number of $\order{1}$. So, the ratio $\eta/s$ cannot
become arbitrary small, even when increasing the coupling. In that sense,
a physical fluid can never be ideal.

\begin{figure}[thb]
\begin{center}\centerline{
  \includegraphics[width=0.5\textwidth]{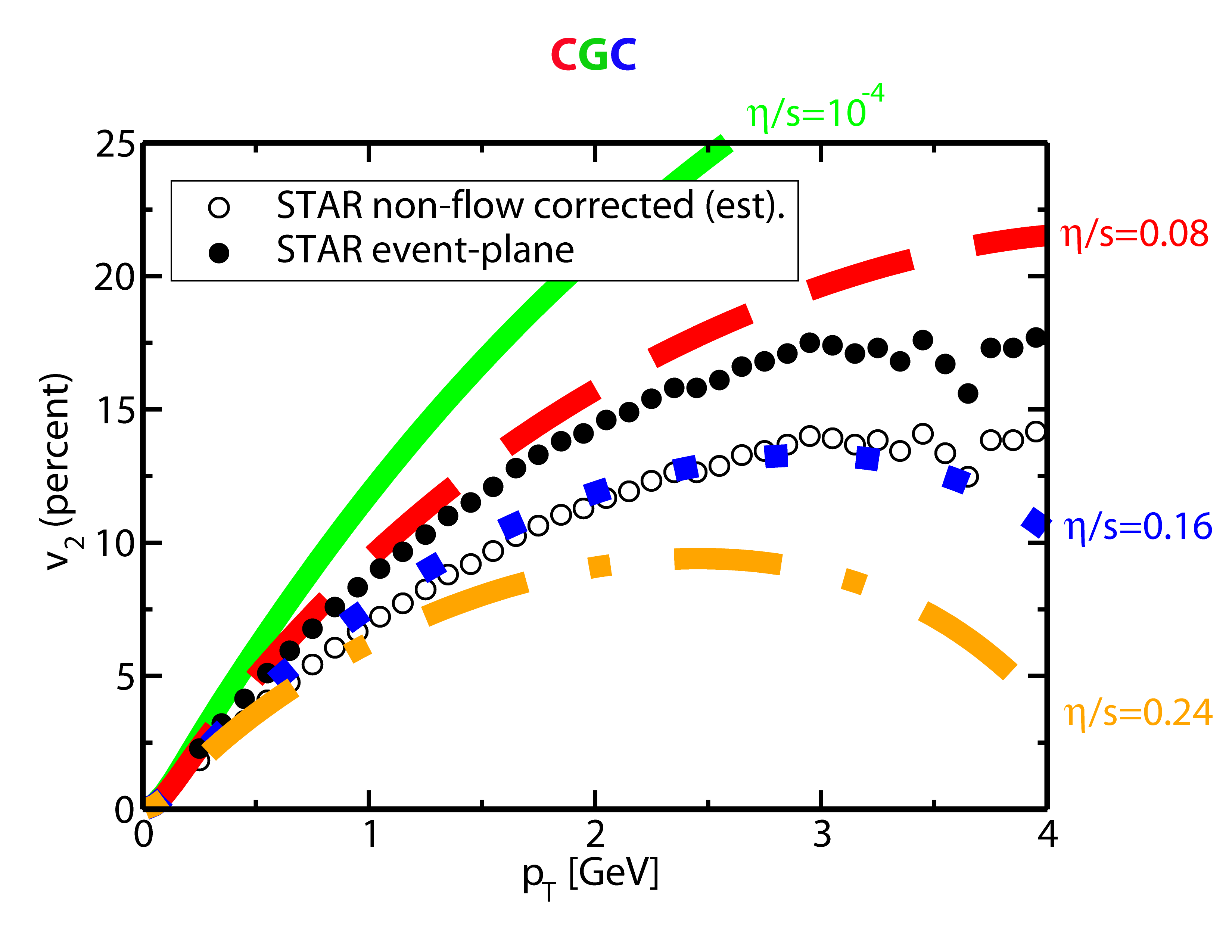}\quad
 \includegraphics[width=0.5\textwidth]{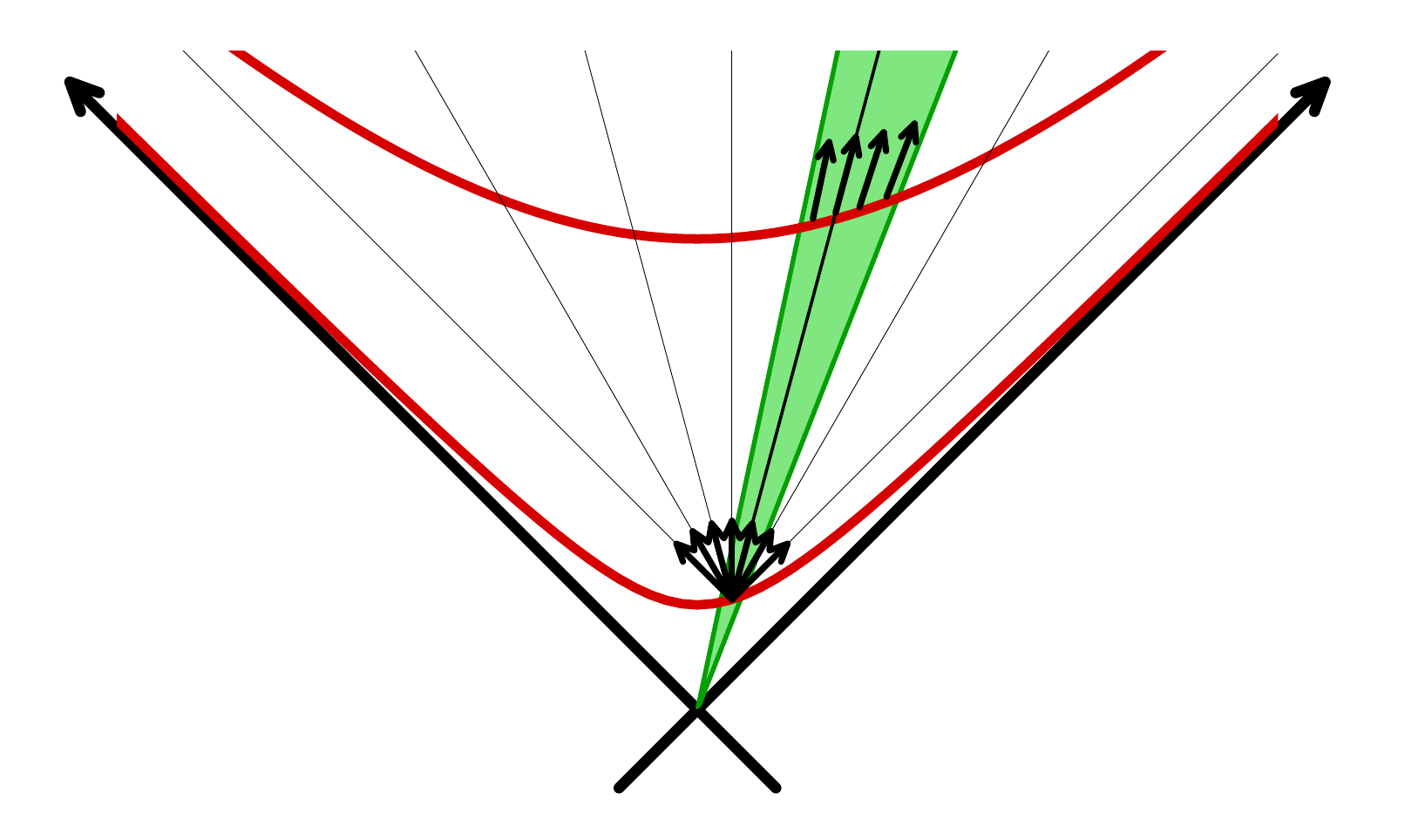}
 }
 \caption{\sl Left panel: the $v_2$ results of hydro calculations using
 the second order formalism with CGC initial conditions and various
 values of $\eta/s$ \cite{Luzum:2008cw}.
 The comparaison with the RHIC data favors
 $\eta/s\simeq 0.16$, which is about twice the lower limit \eqref{Soneta}
 predicted by AdS/CFT at infinitely strong coupling. Right label: the
 longitudinal expansion has the tendency to collimate nearby particles,
 thus opposing to the evolution towards isotropy.
 \label{fig:hydro}}
\end{center}\vspace*{-.8cm}
\end{figure}

An additional argument in that sense comes from the study of a
strongly--coupled theory via the AdS/CFT correspondence. At least for the
more symmetric, conformal, field theories to which it applies, this
formalism predicts a lower bound on the ratio $\eta/s$, namely
\cite{Policastro:2001yc,Kovtun:2004de} (the subscript `CFT' refers to a
conformal field theory; see \Sref{sec:AdS} for details)
 \beq\label{Soneta}
 \frac{\eta}{s}\bigg|_{\rm CFT}\,\ge\,\frac{\hbar}{4\pi}\,,\eeq
with the lower bound being reached in the limit of an infinitely strong
coupling (in a sense to be characterized in \Sref{sec:AdS}). One
remarkable thing about the heavy--ion data at RHIC and the LHC is that
they seem to require a value $\eta/s$ almost as small as this absolute
lower bound: $\eta/s\simeq 0.08\div 0.20$ depending upon the details of
the `initial conditions' at time $\tau_0$. (For instance, the analysis in
Ref. \cite{Luzum:2008cw} favors a value $\eta/s\simeq 0.16$ for initial
conditions of the CGC type, as shown in the left panel of
\Fref{fig:hydro}.) The other remarkable thing is that, in order to be
successful, the hydro descriptions of the data must assume a very small
equilibration time $\tau_0\lesssim 1$~GeV/c. These are both hallmarks of
a system with {\em strong interactions}. Indeed, the particles thermalize
by exchanging energy and momentum (and other quantum numbers) with each
other, via their mutual collisions. So, we expect the thermalization time
to be shorter for strongly interacting systems. This expectation is
supported by kinetic theory, which yields a thermalization time
$\tau_{\rm eq}\simeq \ell/v\propto [T\alpha_s^2 \ln(1/\alpha_s)]^{-1}$ to
leading--order at weak coupling. For realistic values of $\alpha_s$, this
perturbative estimate is too large to be consistent with the data (even
when corrected for inelastic processes like $2\to 3$, which turn out to
be important \cite{Baier:2000sb}).

In the context of heavy ion collisions, the evolution towards (local)
thermal equilibrium is furthermore hindered by the extreme anisotropy of
the initial conditions and also by the anisotropy of the early--time
expansion, which is predominantly longitudinal. Recall the glasma picture
of the initial conditions (at times $\tau_s\sim 1/Q_s$), which is that of
colour flux tubes extending along the collision $x$, cf. \Fref{fig:flux}.
Flux tubes have an internal tension opposing to their longitudinal
extension, like a string. Accordingly, the longitudinal component of the
energy--momentum tensor associated with the glasma is {\em negative}~;
one finds $T^{\mu\nu}_{\rm
glasma}=\mbox{diag}(\varepsilon,\varepsilon,\varepsilon,-\varepsilon)$ at
such early times, which is extremely anisotropic, as anticipated.
Subsequent interactions among particles are supposed to restore isotropy,
but this is rendered difficult by the {\em longitudinal expansion}, as
illustrated in the right panel of \Fref{fig:hydro}. Namely, even if the
particle distribution turns out to be locally isotropic at a given
position in space and time, the subsequent anisotropic expansion rapidly
separates the particles from each other according to the directions of
their velocities: only those particles remain close to each other which
had nearly parallel velocities. In other terms, by itself, the
longitudinal expansion would naturally build a particle distribution in
which nearby particles move along quasi--parallel directions, thus
opposing isotropy. To beat this tendency and ensure isotropy, one needs
strong interactions which continuously randomize the directions of motion
of the particles. This would be natural at strong coupling, as alluded to
above. But there are also other scenarios which are currently explored,
including weak--coupling ones. One promising mechanism in that sense
refers to {\em plasma (Weibel) instabilities}~: due to the anisotropy of
the expanding parton distribution, the soft colour fields radiated by
these partons can develop unstable modes, that is, modes whose amplitudes
grow exponentially with time (at least, during a limited time interval).
So far, it is not clear whether this mechanism can lead to rapid
isotropisation in the presence of longitudinal expansion, but its studies
are under way (see
Refs.~\cite{Ipp:2010uy,Epelbaum:2011pc,Dusling:2011rz,Kurkela:2011ti,Kurkela:2011ub,Berges:2012iw}
for recent work and related references). Recent developments include a
calculation (similar to previous work in inflationary dynamics) of the
spectrum of initial quantum fluctuations in the
glasma~\cite{Dusling:2011rz}, a parametric analysis of the interplay
between plasma instabilities and Bjorken expansion in the weak--coupling
limit \cite{Kurkela:2011ti,Kurkela:2011ub}, and an interesting scenario
(still at weak coupling) in which the elastic scattering between the
highly occupied glasma fields leads to the formation of a transient
Bose--Einstein condensate \cite{Blaizot:2011xf}.

\subsection{QGP: Thermodynamics and collective excitations}
\label{sec:thermo}

In this section, we shall deviate from the experimental situation in
HIC's, where the partonic medium is rapidly expanding, and focus on a
quark--gluon plasma at rest, in thermal and chemical equilibrium. The
existence of such a deconfined phase in QCD at finite temperature has
been unambiguously demonstrated via numerical calculations on a lattice,
which have also given a lot of information about the thermodynamics of
this system. Some of this information has been corroborated via analytic
calculations at weak coupling, which turned out to be very non--trivial.
The analytic methods become essential when one is interested in {\em
real--time} phenomena, like the response of the system to time--dependent
external perturbations, as characterized by {\em transport coefficients}.
Indeed, real--time phenomena cannot be (easily) studied via lattice
calculations\footnote{There is some recent progress in computing
transport coefficients on the lattice (see e.g. the review paper
\cite{Meyer:2011gj} and Refs. therein), but although promising, this
method is still inaccurate and very fastidious.}, which are {\em a
priori} formulated in a space--time with Euclidean signature (`imaginary
time'). It should be also stressed that, even though weak--coupling
techniques appear to be quite successful in reproducing the lattice
results for the QGP thermodynamics, the hypothesis that the coupling be
{\em strong} is not yet totally excluded (within the temperature range
relevant for the phenomenology at RHIC and the LHC): indeed,
weak--coupling calculations seem unable to explain the small $\eta/s$
ratio supported by the data (cf. \Sref{sec:hydro}). In what follows, all
that will be explained in some detail.

\begin{figure}[t]
\begin{center}\centerline{
\includegraphics[width=0.75\textwidth]{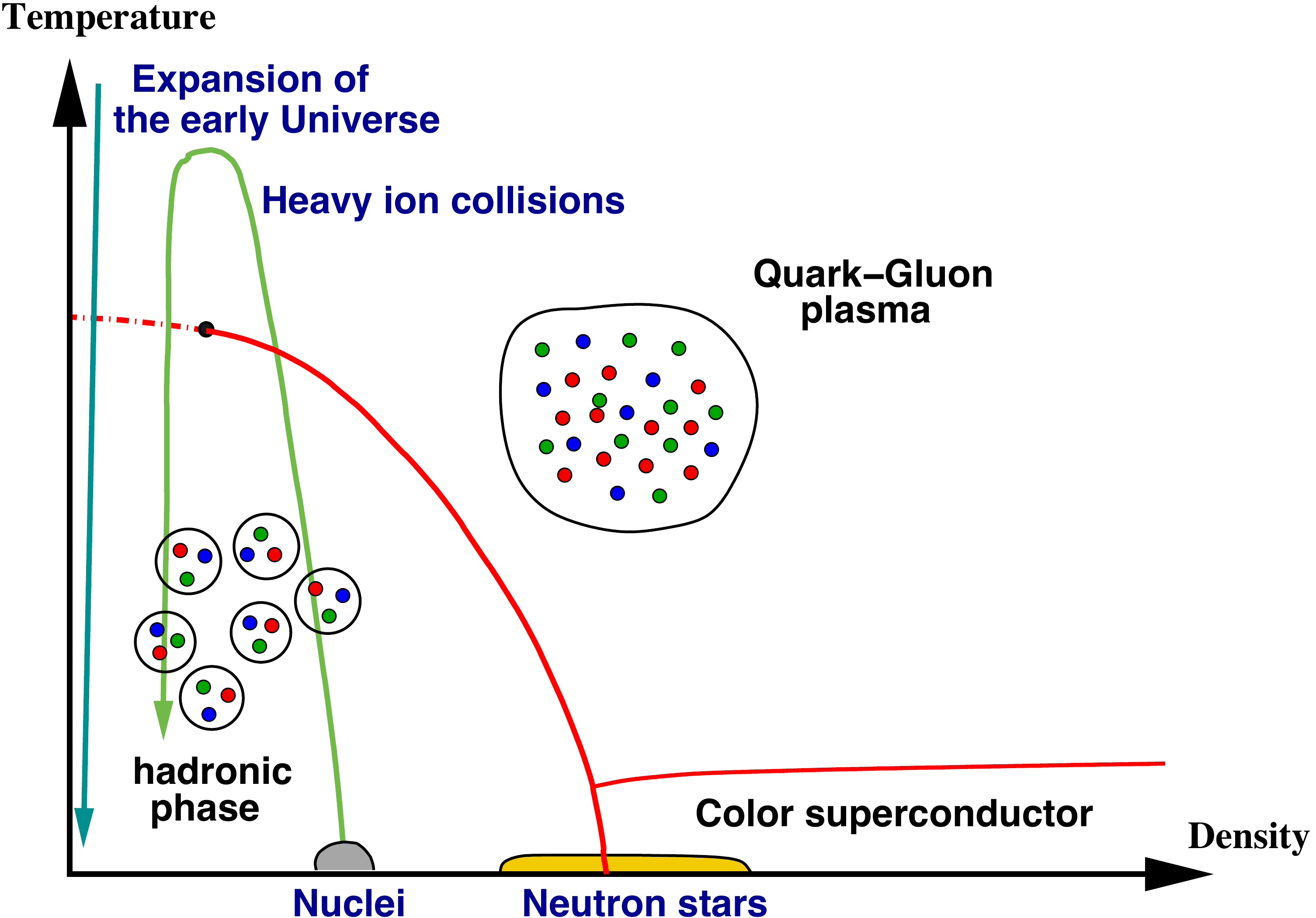}}
 \caption{\sl Schematic representation of the phase--diagram in QCD
 at finite temperature and non--zero quark density,
 as emerging from lattice calculations at zero (or small) quark density and
 from various theoretical considerations (like pQCD) in the other domains.
\label{fig:phase}}
\end{center}\vspace*{-.8cm}
\end{figure}

\Fref{fig:phase} shows a cartoon of the phase--diagram expected in QCD
when varying the temperature $T$ and the net quark density (or the quark
chemical potentials $\mu_f$), by which one means the difference between
the density of quarks and that of the antiquarks. (For simplicity,
\Fref{fig:phase} treats the three light quark flavors
--- the only ones to be relevant for the phase diagram --- on the same
footing.) This diagram has been actually {\em demonstrated} only in
special corners, like the {\em deconfinement phase transition} with
increasing $T$ at zero (or small) fermionic density, that has been
established on the lattice, and the islands denoted as `nuclei' or
`neutron stars', which are rather well understood within nuclear theory.
The `colour superconductivity' phase at high quark density, which is
predicted by pQCD (at least for $\mu\gg \Lam$), will not be discussed
here, as it is not expected to play any role in the ultrarelativistic
HIC's. (See the review papers \cite{Rajagopal:2000wf,Alford:2007xm} and
Refs. therein for detailed discussions of this phase.)

As also illustrated in \Fref{fig:phase}, the deconfinement phase
transition has been first explored during the expansion of the Early
Universe: the high temperature `soup' of matter created right after the
Big Bang was originally in the deconfined, QGP, phase; due to its rapid
expansion, this matter has cooled down and thus crossed into the
confined, hadronic, phase, at a very short time $\sim 10^{-5}$ seconds
after the Big Bang. In the context of HIC's, this transition is being
probed the other way around: to start with, the partons are confined
within the nucleons composing the two nuclei; the collision liberates
these partons and, if their energy density is high enough, they can
thermalize at a temperature superior to the critical temperature for
deconfinement. If so, they form a transient QGP phase which cools down
via expansion and eventually `evaporates' into hadrons. In both
scenarios, the net quark density is small and plays no role for the
transition. In the Early Universe, the excess in the number of quarks
over antiquarks was negligible (if any !) \cite{Rubakov}. In HIC's, there
is of course a net baryon number, due to the $2A\simeq 400$ nucleons
within the incoming nuclei; however, this excess is small compared to the
number of hadrons (a few thousand) produced in the final state. This
implies that most of the partons which exist in the intermediate stages
of the collision are actually gluons or `sea' quark--antiquark pairs.

\begin{figure}[t]
\begin{center}\centerline{
\includegraphics[width=0.38\textwidth]{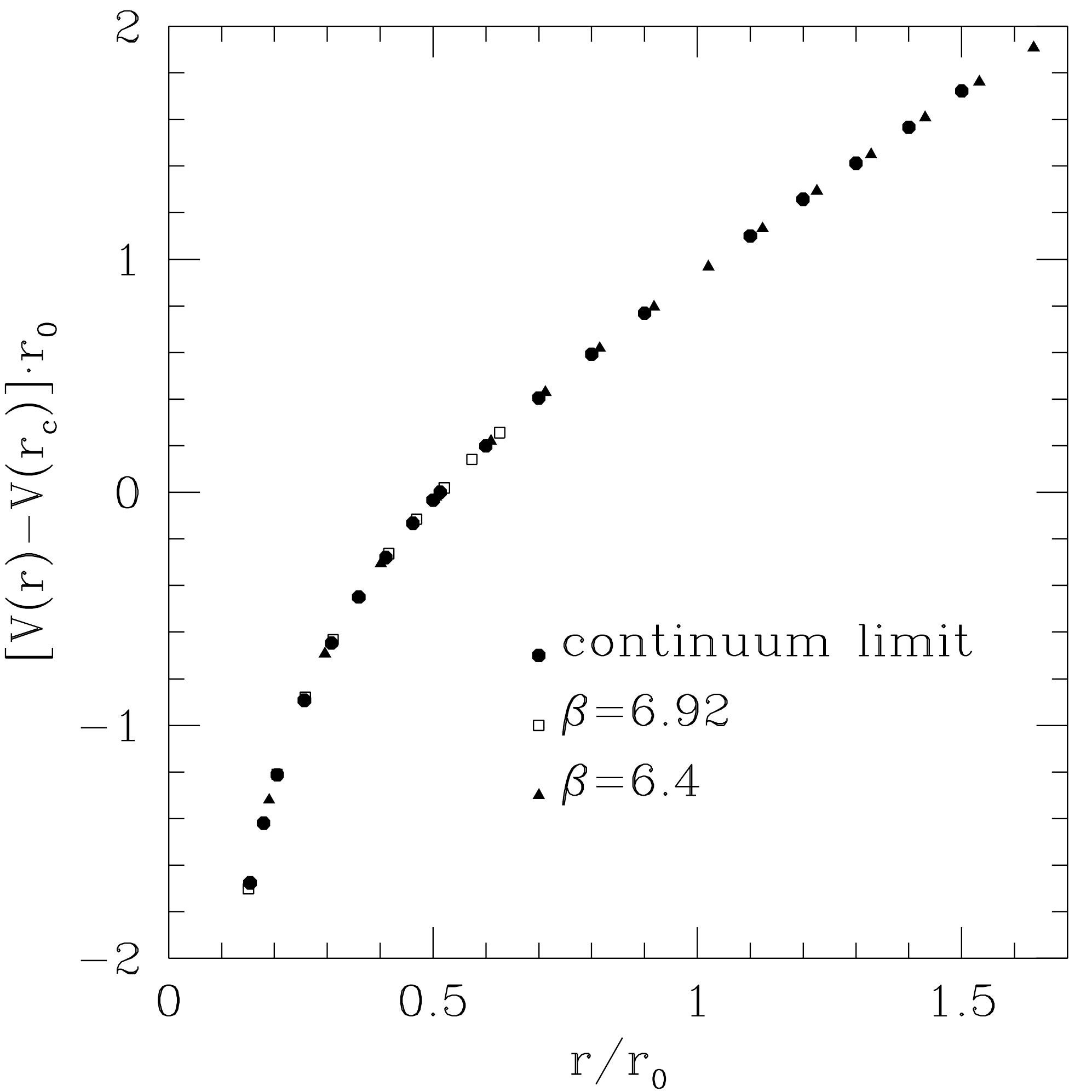}
\quad\includegraphics[width=0.52\textwidth]{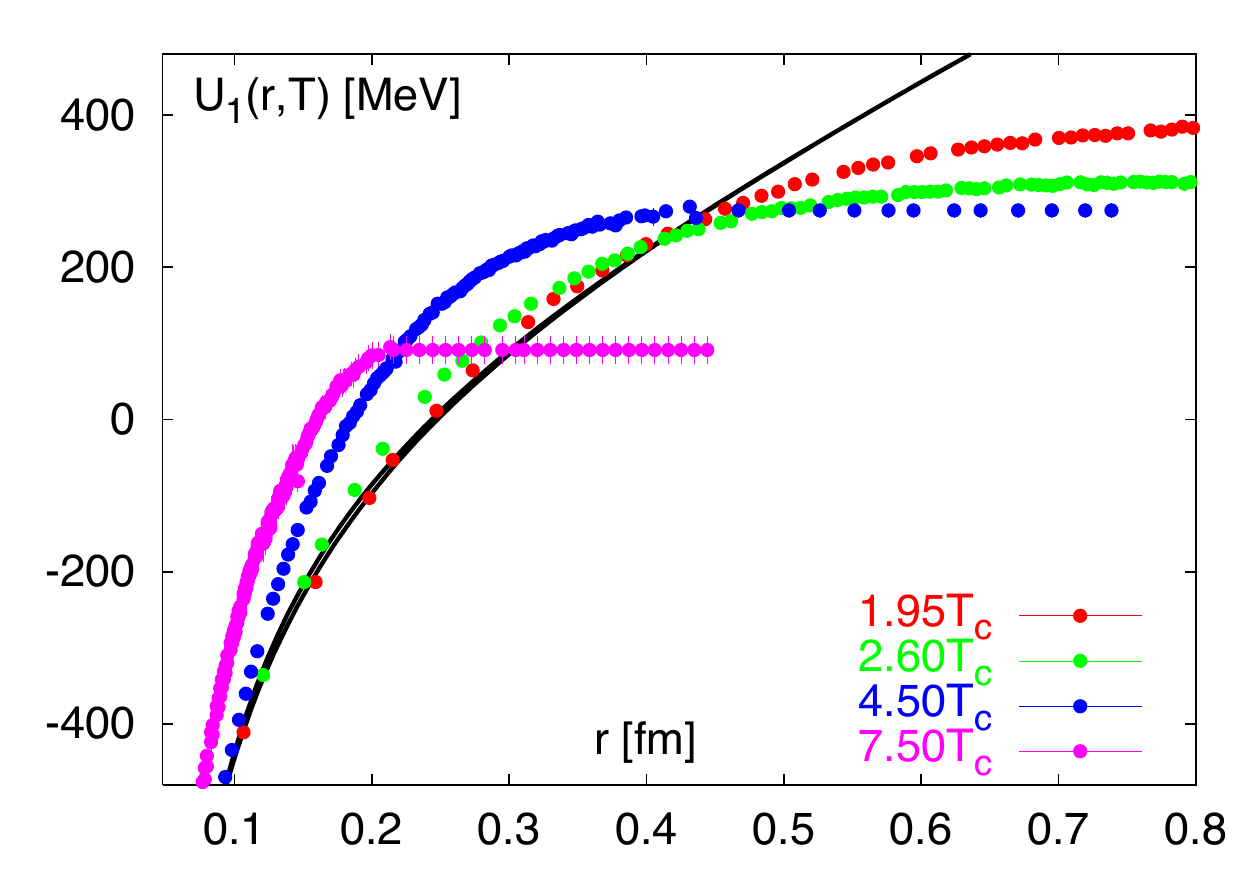}}
 \caption{\sl The quark--antiquark potential between two heavy quarks,
 as computed in QCD on the lattice. Left panel: $T=0$.
 Right panel: various temperatures, which are all larger than the
 critical temperature for deconfinement $T_c$ (for comparison, the $T=0$
 potential is also shown, as the continuous line which keeps rising)
 \cite{Karsch:2004ik}.
\label{fig:confinement}}
\end{center}\vspace*{-.8cm}
\end{figure}

The fundamental property of the QGP is, of course, {\em deconfinement} :
quarks and gluons can move (more or less freely, depending upon their
mutual interactions) throughout the whole volume of the plasma, without
being confined within hadrons with radia $\sim 1/\Lam$. How is this
possible ? A quark and an antiquark in isolation (i.e. at zero
temperature) attract each other via a force which becomes roughly
constant --- corresponding to a linear potential; see
\Fref{fig:confinement} left --- at distances $r\gtrsim 1/\Lam$. Due to
this force, the $q\bar q$ pair is tightly bound (`confined') into a
meson. This meson can be broken, say, via a hard scattering, but only at
the expense of producing additional gluons and $q\bar q$ pairs which
`glue' to the original quark and antiquark, in such a way to form colour
singlet states (new hadrons). This is the situation in the `usual'
hadronic processes, including p+p collisions at the LHC, where the parton
density right after the collision is not very high --- so, these partons
can evolve and eventually hadronize independently of each other. But in
HIC's, the density of the liberated partons is such that the typical
interparticle separation is much shorter than $1/\Lam$. At such short
distances, the attraction force between these partons is smaller than
their kinetic energy, so the partons can move around each other and
arrange themselves in such a way to minimize their mutual repulsion. The
net result is that the colour charge gets {\em screened} over relatively
short distances $r\ll 1/\Lam$, thus preventing the development of
confining forces at larger distances.

\begin{figure}[t]
\begin{center}\centerline{
\includegraphics[width=0.5\textwidth]{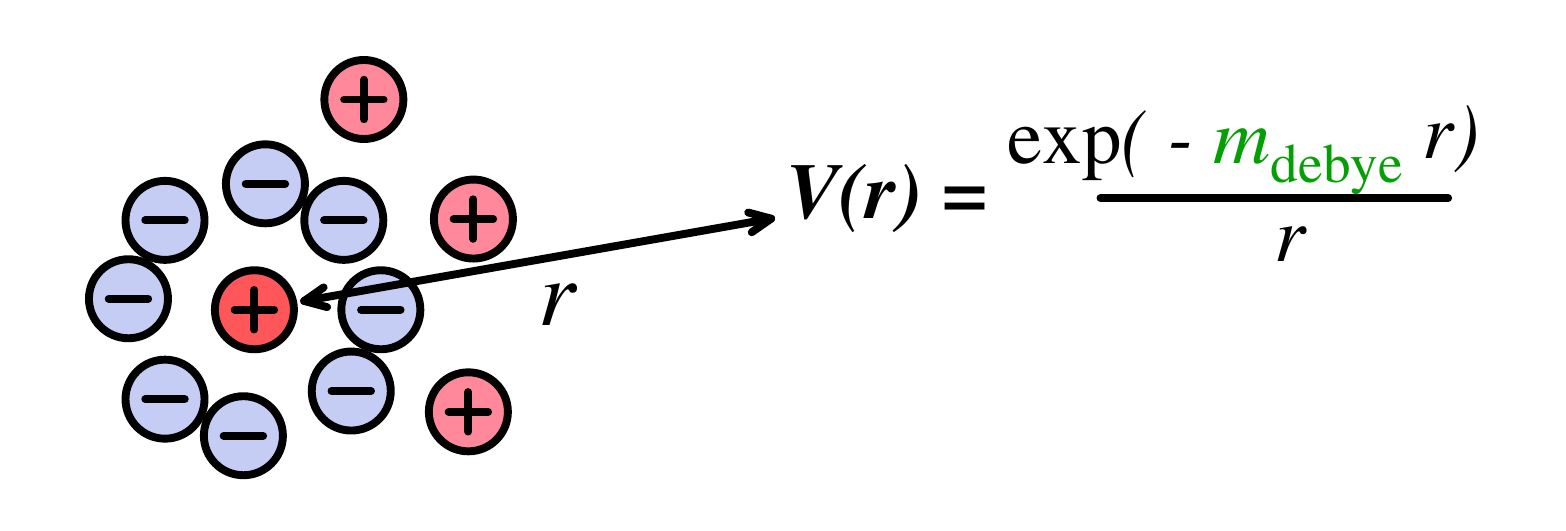}
\quad\includegraphics[width=0.5\textwidth]{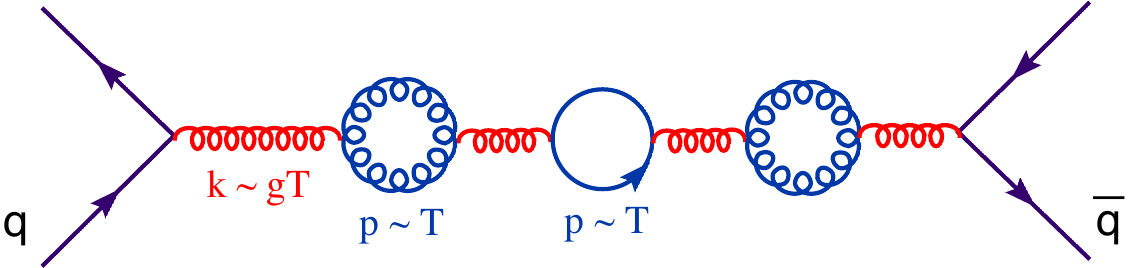}}
 \caption{\sl Left panel: illustration of the Debye screening
 in a QED plasma; the screening length is the inverse of the
 Debye `mass': $r_D=1/m_D$.
  A similar mechanism is active in the QGP. Right panel:
 the Debye screening in QCD is the result of one--loop corrections due
 to the `hard' ($k\sim T$) thermal particles (quark and
 gluons) which are resummed in the
 propagator of the `soft' ($k\sim gT$) electric gluon exchanged
 between the quark and the antiquark.
\label{fig:debye}}
\end{center}\vspace*{-.8cm}
\end{figure}

This colour screening in the QCD plasma is very similar to electric
(Debye) screening in ordinary, electromagnetic, plasmas, or in
electrolytes. Ions with positive electric charge attract ions or
electrons with negative charge, in such a way to form clouds of particles
which look electrically neutral when seen from far away: the net charge
decreases exponentially with the distance from the central charge (see
\Fref{fig:debye} left). In the context of QCD, the `positive and negative
electric charges' are replaced by the $N_c^2-1=8$ `colour' charges
carried by quarks and gluons, but the exponential screening of the
chromo--electric charges works in a similar way. As a consequence of
that, the quark--antiquark potential flattens out (meaning that there is
no attraction force) at large distances $r\gtrsim r_D$ with $r_D$ the
Debye radius. This flattening is clearly seen in the lattice calculations
at finite temperature (see e.g. the discussion in \cite{Karsch:2004ik}),
as illustrated in the right panel of \Fref{fig:confinement}. This also
suggests that in a finite temperature plasma one cannot have quarkonia
(bound states made with a heavy quark ($Q$) and a heavy antiquark ($\bar
Q$), like $J/\psi$, with size $r_{Q\bar Q}>r_D$. This observation
\cite{Matsui:1986dk} led to the fertile idea of {\em quarkonia melting}
in a quark--gluon plasma, a very active field of research for both
theoretical (including lattice) and experimental studies of HIC's. (See
\cite{Mocsy:2009ca} for a recent overview of the theory and more
references.)

\begin{figure}[t]
\begin{center}\centerline{
\includegraphics[width=1.\textwidth]{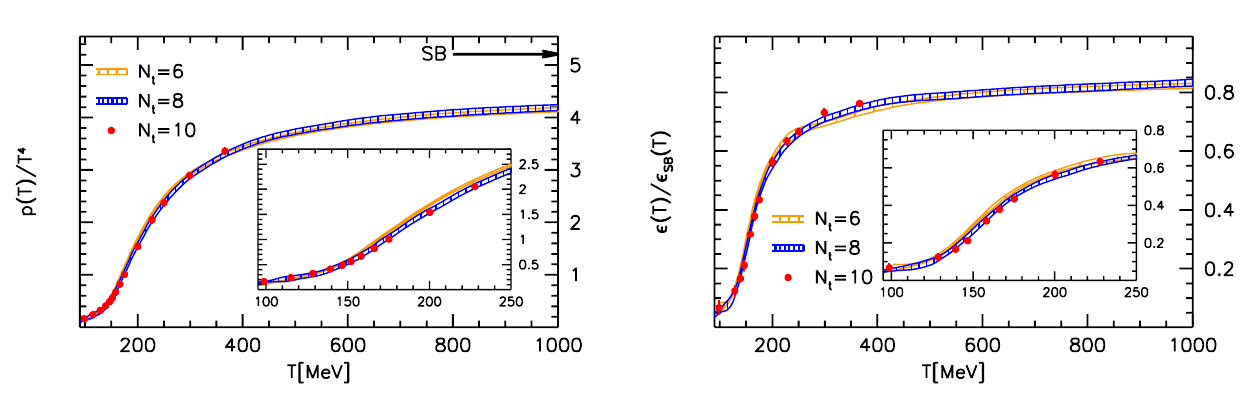}}
 \caption{\sl The lattice results (as obtained by the Budapest--Wuppertal
 collaboration \cite{Fodor:2009ax}) for the QCD pressure and energy density
 as a function of the temperature.
\label{fig:pressure}}
\end{center}\vspace*{-.8cm}
\end{figure}

The lattice calculations also allow one to study the deconfinement phase
transition with increasing temperature. The respective results for the
pressure and energy density are illustrated in \Fref{fig:pressure}. They
show a sudden increase around a {\em critical temperature} $T_c\simeq
160\div 180$~MeV, interpreted as the result of the rise in the number of
degrees of freedom (d.o.f.), due to the liberation of quarks and gluons.
For $T\le T_c$, the only `thermodynamically active', hadronic d.o.f.
(those whose masses are not much higher than $T$) are the 3 pions:
$\pi^0$ and $\pi^{\pm}$. For $T> T_c$, this number jumps from 3 to 52:
the gluons, which appear in 8 colours and 2 transverse polarizations
($8\times 2=16$ d.o.f.), and the 3 light quarks and antiquarks, each of
them having 2 spin states and 3 possible colours ($3\times 3\times
2\times 2=36$ d.o.f.). Lattice calculations become more tedious for light
quark masses and the extrapolation to physical quark masses has become
possible only recently \cite{Aoki:2009sc,Cheng:2009zi}. This is important
since both the actual value of the critical temperature and the nature of
the phase transition are strongly influenced by the values of these
masses. A phase transition is said to be of $n$th order if it involves a
discontinuity in the derivative of order $(n-1)$ of the pressure. For
instance, if the QCD phase transition was of {\em first--order}, then it
would proceed via a mixed phase where hadronic bubbles coexist with
regions of QGP. But this is not what happens in QCD: recent lattice
calculations \cite{Aoki:2006we} show that, for physical quark masses, the
deconfinement phase transition is truly a {\em cross--over}, that is, a
relatively smooth process during which the pressure and all its
derivatives remain continuous across the transition.

\begin{figure}[t]
\begin{center}\centerline{
\includegraphics[width=0.45\textwidth]{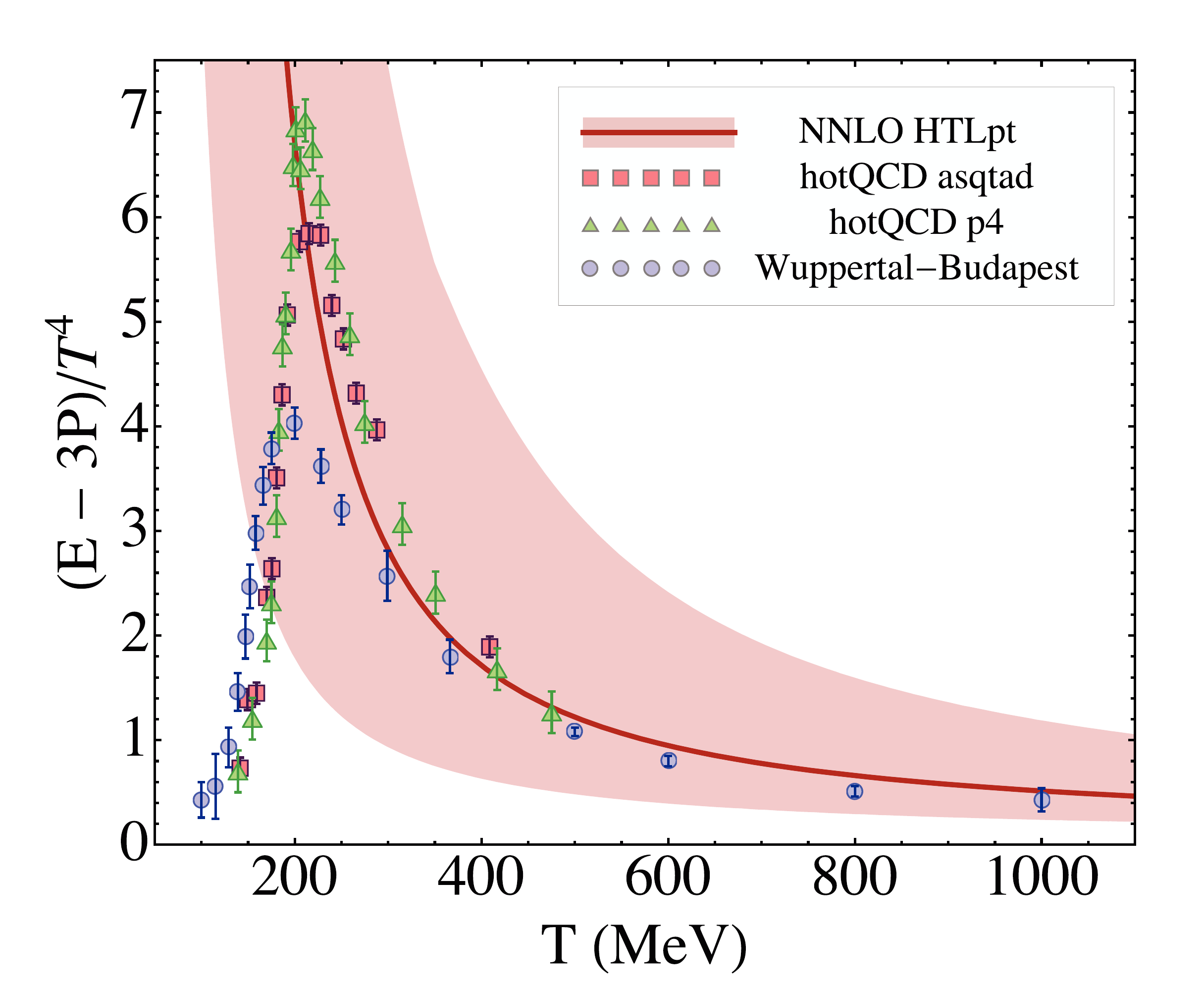}
\quad\includegraphics[width=0.55\textwidth]{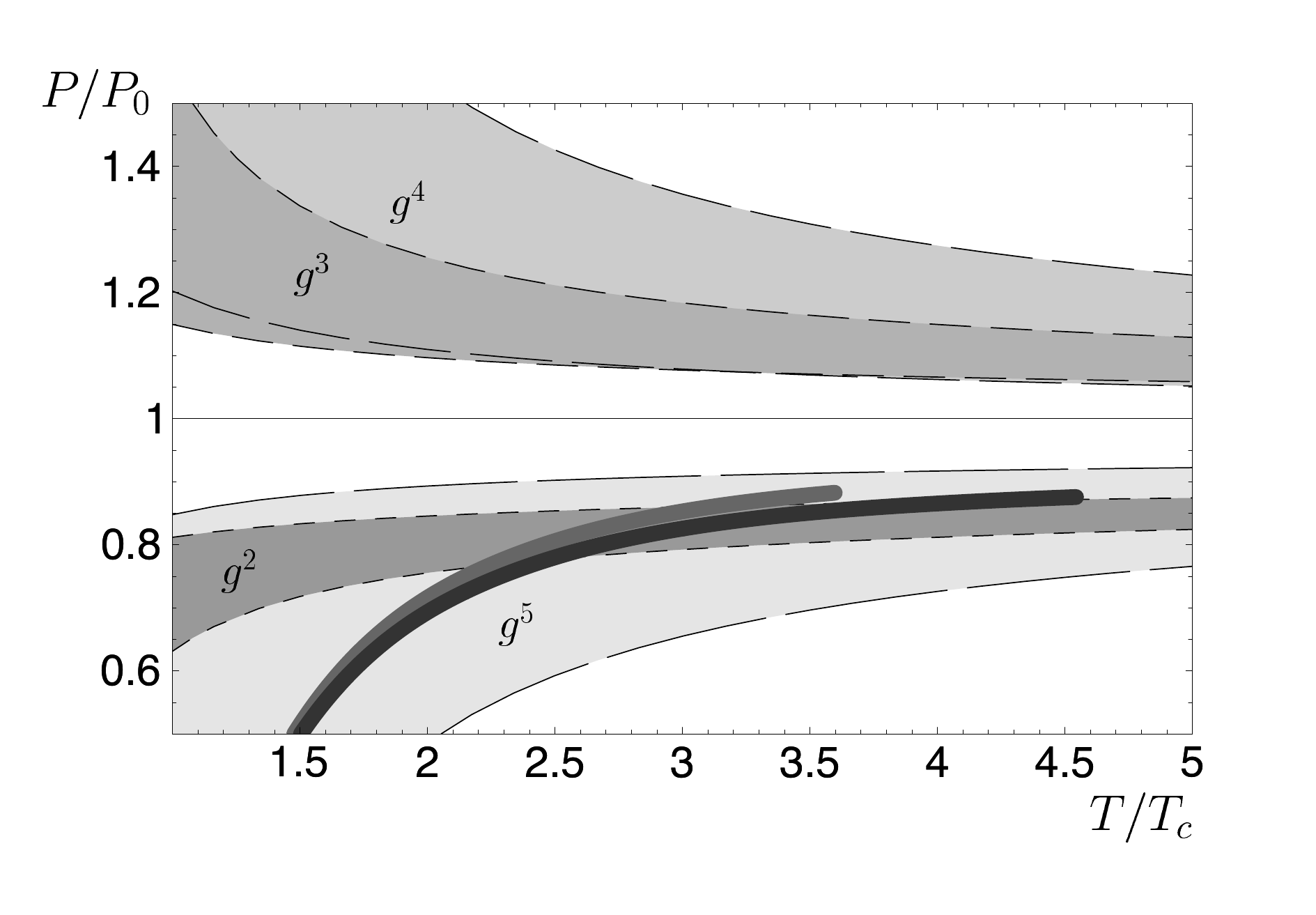}}
 \caption{\sl Left panel: the trace anomaly in QCD, as numerically
 computed on the lattice
 \cite{Aoki:2009sc,Cheng:2009zi}, together with the respective predictions
 of a `HTL--resummed' perturbation theory \cite{Andersen:2011sf}
 (see the text for details), shown
 as a band (due to the uncertainty in the choice of
 the renormalization scale $\mu_{\rm ren}$). Right panel: the predictions
 of (strict) perturbation theory for the pressure, as computed
 up to order $g^2$, $g^3$, $g^4$, and $g^5$, respectively. Whereas the
 $\mathcal{O}(g^2)$--result appears to match the
 lattice results (the dark grey band) rather closely, this agreement
 is spoilt after including higher order corrections; the ensuing series
 in powers of $g$ shows no sign of convergence.
\label{fig:trace}}
\end{center}\vspace*{-.8cm}
\end{figure}

But albeit smooth, the phase transition represents a genuinely
non--perturbative phenomenon, which cannot be described within
perturbative QCD. This can be appreciated e.g. by inspection of the
lattice results for the {\em trace anomaly} $(\varepsilon-3P)/T^4$, which
exhibit a sharp peak around $T_c$, as visible in the left panel of
\Fref{fig:trace}. As mentioned in \Sref{sec:hydro}, $\varepsilon=3P$ for
an ideal gas of massless particles, meaning that the corresponding
energy--momentum tensor $T^{\mu\nu}=\mbox{diag}(\varepsilon,P,P,P)$ is
traceless: $T^{\mu}_{\ \mu}=0$. This property is in fact a general
consequence of {\em conformal symmetry}~: it holds for any theory which
involves no intrinsic mass parameter and hence is invariant under
dilations. This is in particular the case for QCD with massless quarks at
the {\em classical} level. However, at the {\em quantum} level, conformal
symmetry in QCD is broken by the radiative corrections responsible for
the running of the coupling, which introduce the mass scale $\Lam$. So,
not surprisingly, the corresponding `trace anomaly' (the deviation of
$T^{\mu}_{\ \mu}$ from zero) is proportional to the $\beta$--function,
which measures the running of the coupling :
 \beq\label{trace}
 T^{\mu}_{\ \mu}\,=\,\varepsilon-3P\,\,=\,\beta(g)\,\frac{\partial P}
 {\partial
 g}\,,\qquad \beta(g)\,\equiv\,\frac{\partial g}{\partial\ln \mu_{\rm
 ren}}\,. \eeq
Here $\mu_{\rm ren}$ is the renormalization scale, as introduced by the
subtraction of the ultraviolet divergences. In perturbation theory at
finite temperature, it is convenient to choose $\mu_{\rm ren} = 2\pi T$
as the central value and study the dependence of the results upon
variations in $\mu_{\rm ren}$ (typically by a factor of 2) around this
central value. These variations measure the stability of the calculations
against higher order corrections and thus are indicative of the
theoretical uncertainties. They are shown as `error bands' for the
theoretical results in Figs.~\ref{fig:trace} and \ref{fig:resum}.

The  peak in the l.h.s. of \Fref{fig:trace} is a hallmark of the phase
transition and is clearly non--perturbative. But for temperatures above
$T_c$, the `trace anomaly' is rapidly decreasing (its relative strength
becomes of order 10\% for $T\gtrsim 3T_c$), thus suggesting that a
perturbative approach may become viable. This is furthermore indicated by
the fact that, for $T\gtrsim 3T_c$, the pressure and the energy density
reach about 80\% of the respective values for an ideal gas of quarks and
gluons, denoted as `SB' (from `Stefan--Boltzmann') in
\Fref{fig:pressure}. A deviation of 20\% may seem sufficiently small to
be easily accommodated in perturbation theory, but this turns out not to
be the case. There are two main reasons for that. First, unlike what
happens at $T=0$, where perturbation theory in QCD is an expansion in
powers of $\alpha_s\equiv g^2/4\pi$, at finite temperature this is rather
an expansion in powers of $g$, for reasons to be shortly explained.
Second, the relevant values of the QCD running coupling are not that
small: for $T\simeq 3T_c$ and hence $2\pi T\simeq 2$~GeV, one has
$\alpha_s\simeq 0.25$ and hence $g=1.5\div 2$. For such large valus of
$g$, there is no reason why an expansion in powers of $g$ should
converge, and indeed it does not: as visible in the right panel of
\Fref{fig:trace}, the successive corrections of $\mathcal{O}(g^2)$,
$\mathcal{O}(g^3)$, $\mathcal{O}(g^4)$, and $\mathcal{O}(g^5)$ jump up
and down, without any sign of convergence. (The weak--coupling expansion
of the pressure in QCD is known to $\mathcal{O}(g^6\ln (1/g))$
\cite{Kajantie:2002wa}, which is the highest order that can be computed
in perturbation theory: the corrections of $\mathcal{O}(g^6)$ and higher
are afflicted with severe infrared divergences due to magnetic gluons;
see below.)

The reason why, at finite temperature, perturbation theory is an
expansion in powers of $g$ rather than $\alpha_s$ is because the quantum
corrections associated with {\em soft gluons} --- those with momenta $k$
much smaller than $T$ --- are amplified by the Bose--Einstein thermal
distribution function:
 \beq
 n_B(k)\,=\,\frac{1}{\rme^{\beta E_k}-1}\,\simeq\,
  \frac{T}{E_k}\,\gg\,1\quad\mbox{when}\quad E_k=|\bmk|\,\ll\,T\,.\eeq
This property is generic: it holds for any field theory which involves
massless bosons (e.g., it holds for photons in a QED plasma). When $k\to
0$, the thermal factors $n_B(k)\simeq T/k$ lead to infrared divergences
in the calculation of Feynman graphs, which are regulated by plasma
effects, like Debye screening. These effects typically enter at the
`soft' scale\footnote{This scale $gT$ is truly `soft' only so long as
$g\ll 1$~; as we shall shortly argue, results obtained under the
assumption that the coupling is weak can be extrapolated towards $g\sim
1$ provided the plasma effects are properly taken into account.} $gT$.
For instance, the {\em Debye mass} $m_D\equiv 1/r_D$ which characterizes
the exponential screening of the electric colour charge by the plasma
constituents, is generated by the one--loop diagrams illustrated in the
right panel of \Fref{fig:debye}, which yield $m_D\sim gT$. The
resummation of these diagrams within the propagator of the exchanged
gluon, as also illustrated in \Fref{fig:debye}, renders the (electric)
gluons effectively massive\footnote{There is strictly speaking a
difference between the {\em Debye mass} $m_D$, which governs the infrared
($k\to 0$) limit of the {\em static} ($k_0=0$) propagator for the
electric gluons, and the {\em thermal mass} $m_g$, which enters the
dispersion relation for the {\em on--shell} gluons; but these quantities
are proportional with each other and are both of order $gT$; see e.g.
\cite{Blaizot:2001nr,Blaizot:2003tw,Kraemmer:2003gd}.}
 : $E_k=\sqrt{k^2+m_D^2}$. Hence, when $k\to 0$, the
Bose--Einstein occupation number remains finite, but it is parametrically
large: $n_B(k)\simeq T/m_D \sim 1/g$. This inverse power of $g$ changes
the perturbative order of the 2--loop correction to the pressure with one
`hard' loop ($k\sim T$) and one `soft' ($k\lesssim gT$) from
$\alpha_s^2\sim g^4$ to $g^4 n_B(k)\sim g^3$. This is the origin of the
odd powers of $g$ in the perturbative expansion.

One should also mention here that Debye screening, as illustrated in the
left panel of \Fref{fig:debye}, is operational for the {\em electric}
gluons (i.e. for the Coulomb interactions), but not also for the {\em
magnetic} ones --- those having transverse polarizations. For
non--relativistic plasmas, magnetic interactions are suppressed by powers
of the velocities, but for (ultra)relativistic plasmas, like the QGP,
they are as important as the electric ones. One expects magnetic
interactions in the QGP to be screened at the `ultrasoft' scale $m_{\rm
mag}\sim g^2T$, but the associated physics --- in particular, the
contribution of the `ultrasoft' magnetic gluons to thermodynamics, which
starts at $\mathcal{O}(g^6)$ --- cannot be computed in perturbation
theory. Indeed, each additional `ultrasoft' loop is accompanied by a
factor $\sim g^2n_B(k)\sim g^2(T/m_{\rm mag})\sim 1$, meaning that
diagrams with arbitrarily many such loops contribute at the same order in
$g$. This explains why the corrections of $\mathcal{O}(g^6)$ are
non--perturbative, as alluded to above.

\begin{figure}[t]
\begin{center}\centerline{
\includegraphics[width=0.9\textwidth]{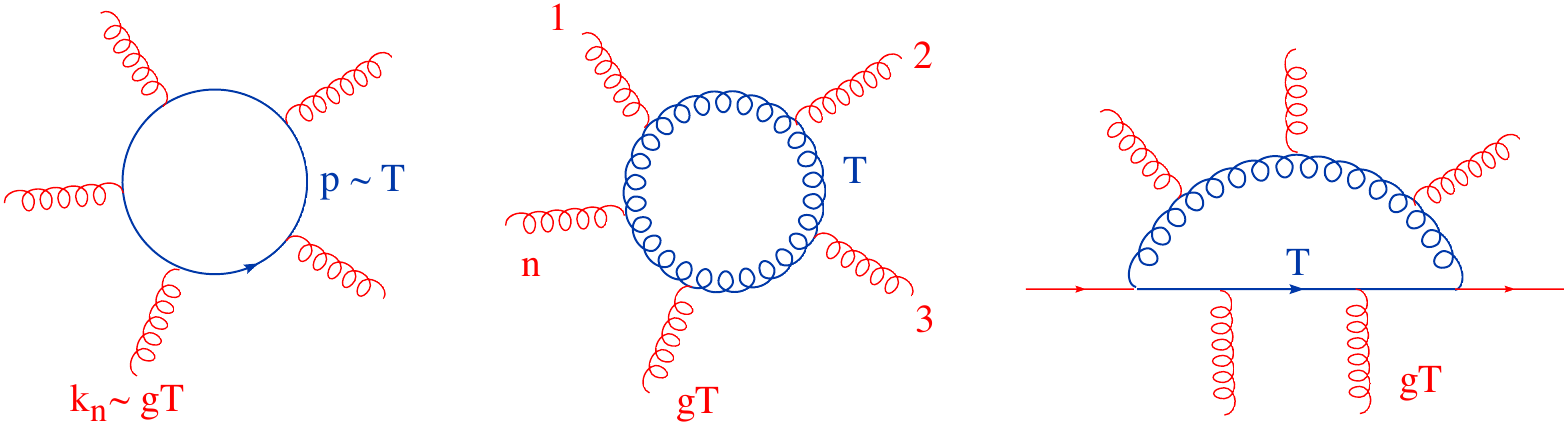}}
 \caption{\sl A few examples of hard thermal loops (HTL's). All the external
 momenta are soft ($k_i\sim gT$) and the loop integrations are dominated
 by relatively hard modes with $p\sim T$.
\label{fig:HTL}}
\end{center}\vspace*{-.8cm}
\end{figure}

\begin{figure}[t]
\begin{center}\centerline{
\includegraphics[width=0.45\textwidth]{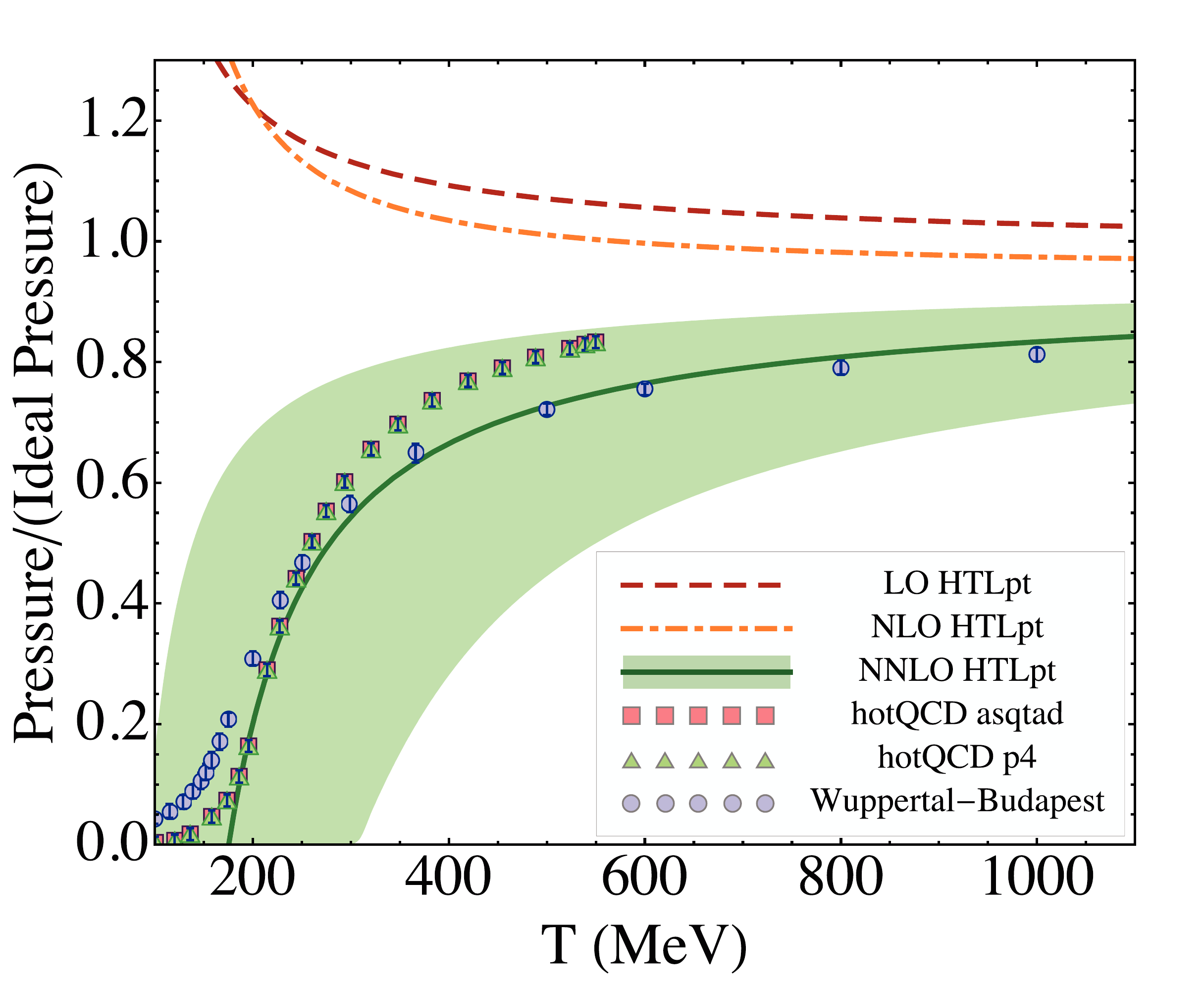}
\quad\includegraphics[width=0.55\textwidth]{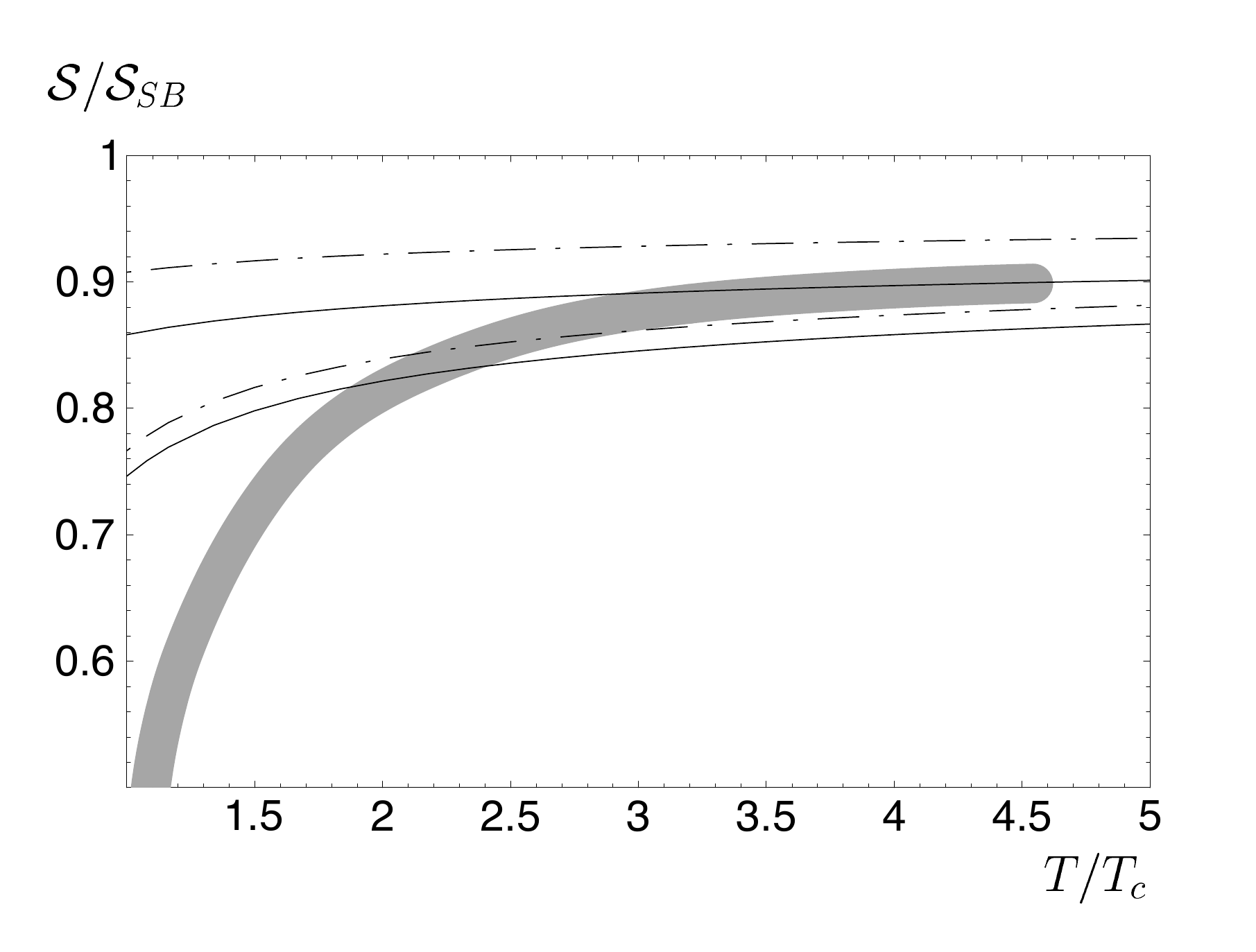}}
 \caption{\sl Comparison between the predictions of two versions of
 HTL--resummed perturbation theory  for
 thermodynamics and the respective lattice results.
 Left panel: the predictions of HTLpt \cite{Andersen:1999fw} for the pressure
 at one--loop (LO), 2--loop (NLO) and, respectively, 3--loop
 (NNLO) order \cite{Andersen:2011sf} vs.
 the lattice results (small cercles,
 triangles or squares) from two different collaborations.
 Right panel: the 2--loop result of the
 `2--particle irreducible resummation' of the entropy
 \cite{Blaizot:1999ip} --- solid and dotted lines correspond to two successive
 approximations for the thermal masses  ---
 vs. the lattice results shown as the grey band.
 In both cases, a good agreement with the lattice
results is observed for temperatures $T\gtrsim 2.5T_c$.
The theoretical `error bands' follows from varying the renormalization scale
 in the range $\pi T\le \mu_{\rm ren} \le 4\pi T$.
\label{fig:resum}}
\end{center}\vspace*{-.8cm}
\end{figure}

In QCD, the Debye mass is only one example of a class of one--loop
`corrections' which are non--perturbative at the `soft' scale $gT$ and
should be viewed as a part of the {\em leading--order theory} at that
scale, and not as corrections \cite{Braaten:1989mz,Blaizot:2001nr}. These
diagrams are known as {\em hard thermal loops}, since the typical momenta
within the loop are of order $T$ (the value preferred by the statistical,
Bose--Einstein and Fermi--Dirac factors) and thus are hard compared to
the soft ($k\sim gT$) momenta flowing along the external legs. There are
HTL's with any number $n$ of external gluons lines and with either zero,
or two, quark external lines (see \Fref{fig:HTL} for some examples). They
are generally {\em non--local}, that is, they depend upon the external,
soft, momenta. In particular, the HTL's for the 2--point functions (the
quark and gluon self--energies) encode phenomena like Debye screening for
the electric gluons, dynamical screening (or Landau damping) for the
magnetic gluons, and the dispersion relations for on--shell quanta with
momenta of order $gT$. Such quanta have wavelengths $\lambda\sim 1/gT$
which are parametrically larger than the typical separation $\sim 1/T$
between the typical plasma constituents --- quarks and gluons with
momenta of order $T$. Accordingly, the soft modes are truly {\em
collective excitations}  (or `plasma waves'), with either quark or gluon
quantum numbers.

As already mentioned, the HTL's are of the same order as the respective
tree--level amplitudes with `soft' external legs, so they cannot be
expanded out in perturbation theory. Rather, one needs to perform a {\em
reorganization} of the perturbation theory in which the HTL's are viewed
as a part of the leading--order theory for the soft modes. Roughly
speaking, this amounts to expanding around a gas of {\em dressed
quasi--particles} whose zeroth--order properties (propagators and
interaction vertices) are encoded in the HTL's. In practice, there are
various ways to perform such reorganizations and it is quite reassuring
that all the methods that have been proposed so far
\cite{Blaizot:1999ip,Blaizot:2003tw,Kraemmer:2003gd,Andersen:1999fw,Andersen:2011sf,Kajantie:2002wa}
appear to be successful in describing the lattice data (although with
considerably different amounts of efforts).


One of these methods, known as `HTL perturbation theory' (HTLpt)
\cite{Andersen:1999fw}, consists in including the HTL's in the
`tree--level' effective theory, by adding and subtracting
$\mathcal{L}_{\rm HTL}$ (the sum of the HTL amplitudes) to/from the
original Lagrangian:
 \beq\label{screened}
 \mathcal{L}_{\rm QCD} = \mathcal{L}_0+\mathcal{L}_{\rm int}
 = \big(\mathcal{L}_0+\mathcal{L}_{\rm HTL}\big)
 +\big(-\mathcal{L}_{\rm HTL} +\mathcal{L}_{\rm int}\big)
 =\mathcal{L}^\prime_0+\mathcal{L}_{\rm int}^\prime\,.\eeq
In this equation, $\mathcal{L}_0$ is the free ($g=0$) piece of the QCD
Lagrangian, $\mathcal{L}_{\rm int}$ is the respective interaction piece,
$\mathcal{L}^\prime_0\equiv \mathcal{L}_0+\mathcal{L}_{\rm HTL}$
represents the new `tree--level Lagrangian' which defines the Feynman
rules (HTL--resummed propagators and vertices) of HTLpt, and, finally,
$\mathcal{L}_{\rm int}^\prime\equiv  \mathcal{L}_{\rm int}
-\mathcal{L}_{\rm HTL}$ is the new `interaction Lagrangian'. The
subtracted piece $-\mathcal{L}_{\rm HTL}$ within $\mathcal{L}_{\rm
int}^\prime$ acts as a `counterterm' to prevent double counting (the
HTL's have been already included in the zeroth--order theory and they
should not be regenerated via loop corrections in HTLpt) and also to
correct for the fact that, within $\mathcal{L}^\prime_0$, the HTL's are
used for {\em all} the modes, including the hard modes to which they do
not really apply (this introduces spurious contributions at lower orders
which are compensated by the `counterterm' only in higher orders). In
order for such compensations to efficiently work, one needs to go up to
relatively high orders in HTLpt, which involve very tedious calculations
(due to the non--local nature of the HTL's). It was only recently, after
pushing such calculations up to three loop order \cite{Andersen:2011sf},
that one has finally reached a good agreement with the respective lattice
results for $T\gtrsim 2.5T_c$. This agreement is visible in the left
panel of \Fref{fig:trace} for the trace anomaly and in the left panel of
\Fref{fig:resum} for the pressure.

A more economical approach, in which a similarly good agreement with the
lattice results (see the right panel of \Fref{fig:resum}) has been
obtained via a simpler, 2--loop, calculation, is the `2--particle
irreducible (2PI) resummation' of the entropy
\cite{Blaizot:1999ip,Blaizot:2003tw}. In that approach, most of the
difference between the 2PI result for $S$ and the corresponding result
$S_{SB}$ for the ideal gas comes from the thermal masses $m_q,\,m_g\sim
gT$ acquired by the {\em hard} ($k\sim T$) quarks and gluons via
interactions in the plasma. Albeit formally small ($m_{q,g}\sim gT\ll
k\sim T$), these masses cannot be expanded in perturbation theory, since
such an expansion would generate powers of $m^2/k^2$ leading to infrared
divergences in the integral over $k$. The success of the 2PI description
supports the physical picture of the QGP in terms of {\em
quasi--particles} --- quarks and gluons with typical momenta $k\sim T$,
which are dressed by the medium (in particular, in the sense of acquiring
thermal masses), but whose residual interactions are relatively small.
Moreover, a substantial part of these residual interactions can be
associated with collective excitations and screening effects at the
`soft' scale $gT$, as encoded in the HTL--resummed propagators.

It is furthermore interesting to notice that the HTL resummation is based
on the separation of scales $gT\ll T$ which is {\em a priori} valid at
weak coupling ($g\ll 1$), yet this turns out to rather successfully
describe the lattice data in a temperature range where $g=1.5\div 2$.
This confirms that the failure of ordinary perturbation theory (cf.
\Fref{fig:trace}, right panel) is not imputable to the fact that the
coupling is relatively strong, but rather it is a consequence of
expanding out the medium effects in powers of $g$ in a kinematical domain
where they are truly non--perturbative.

Yet, the issue of the strength of the coupling remains open in so far as
the study of {\em dynamical phenomena} is concerned. These phenomena
refer to non--trivial evolutions in time, say off--equilibrium deviations
in response to external perturbations. As long as the perturbations are
small, their effects can be computed within the {\em linear response
theory}, via the Kubo formula: the response of the plasma is linear in
the strength of the perturbation, with a proportionality, or `transport',
coefficient which represents a correlation function in thermal
equilibrium. For instance, a constant electric field ${\bf E}$ acting on
the quark constituents of the plasma (which carry electric charge)
induces an electromagnetic current with density $\langle j^i_{\rm
em}\rangle =\sigma E^i$, where $\sigma$ is the {\em electrical
conductivity}. The respective Kubo formula relates $\sigma$ to the
long--wavelength ($k^i\to 0$) and zero--frequency ($\omega\to 0$) limit
of the current--current correlator in thermal equilibrium. Other
transport coefficients include the (quark) flavor diffusion coefficients
and the shear viscosity $\eta$ introduced in \Eref{shear}. The use of
Kubo formul\ae{} for perturbative calculations at weak coupling turns out
to be quite tedious, because of the need to perform sophisticated
resummations \cite{Jeon:1994if}. However, these formul\ae{} are very
useful for non--perturbative calculations of the transport coefficients,
either on the lattice \cite{Meyer:2011gj}, or via the AdS/CFT
correspondence at strong coupling \cite{CasalderreySolana:2011us}.

For a weakly coupled QGP and to leading order in the coupling, the
transport coefficient can be alternatively, and more efficiently,
computed from the Boltzmann equation (linearized with respect to the
off--equilibrium perturbation). This amounts to solving a linear integral
equation which effectively resums an infinite number of diagrams of the
ordinary perturbation theory in thermal equilibrium. These diagrams
describe multiple scattering via soft gluon exchanges and can be
generated by iterating the $2\to 2$ elastic processes shown in
\Fref{fig:elastic} arbitrarily many times. The ensuing transport
coefficients are of the parametric form anticipated (on the example of
the shear viscosity) in \Eref{shearQGP}, but the use of the Boltzmann
equation allows one to obtain more precise results, which are complete to
leading order in $\alpha_s$ \cite{Arnold:2000dr,Arnold:2003zc}. Yet,
these results are deceiving with respect to the heavy--ion phenomenology:
as already mentioned in \Sref{sec:hydro}, the leading order estimate for
$\eta/s$ is too large to be consistent with the hydrodynamical
description of the data.

The last observation raises the question of the next--to--leading order
corrections. Their calculation is extremely complicated and so far this
has been accomplished for just one quantity: the diffusion coefficient
$D$ for a heavy quark with mass $M\gg T$. In the context of HIC, this
quantity controls the collisional energy loss and the thermalization of
heavy quarks like the charm or the bottom. Once again, the LO
perturbative estimate for $D$ \cite{Moore:2004tg} appears to be too large
to be consistent with the data. The NLO correction to $D$ is of relative
order $g$ and has been computed in Ref.~\cite{CaronHuot:2008uh}. This
appears to go in the right direction (it diminishes the value of $D$),
but the effect is extremely large for realistic values of $g$ --- the NLO
`correction' is almost an order of magnitude larger than the respective
LO result ! ---, thus rising doubts about the reliability of the whole
scheme. It looks like the perturbative series suffers from a
lack--of--convergence problem similar to that noticed for the pressure.
It might be that this problem too will be cured by all--orders
resummations of the HTL's; but this issue is still open since such
resummations have not yet been performed for {\em dynamical} quantities.
Alternatively, there is the possibility that the transport phenomena,
which involve long--range dynamics, be sensitive to rather large values
of the QCD running coupling, which exclude weak--coupling techniques. If
so, one could search for physical guidance in the corresponding results
at strong coupling, as obtained via the AdS/CFT correspondence (see
\Sref{sec:AdS} below). Finally, let us notice that the first lattice
results for the transport coefficients have started to emerge, although
the current errors bars are still quite large. These calculations are
very difficult as they require to numerically perform an analytic
continuation (from imaginary time to real time), which in turns requires
very precise numerical data. In view of that, it is quite encouraging
that the recent lattice results for the heavy quark diffusion coefficient
\cite{Francis:2011gc,Banerjee:2011ra} appear to be consistent with the
heavy--ion phenomenology, within the (lattice and experimental) errors
bars.

\subsection{Jet quenching}
\label{sec:jet}

In \Sref{sec:pA} we have mentioned two interesting phenomena occurring in
`dense--dilute' (p+A or d+A) collisions --- the suppression of particle
production and that of azimuthal di--hadron correlations {\em at forward
rapidities} ---, which in that context have been interpreted as
consequences of gluon saturation in the wavefunction of the nuclear
target: the larger the rapidity, the smaller the values of the
longitudinal momentum fraction that are probed in the nucleus, and hence
the stronger the saturation effects. On the other hand, the RHIC data for
d+Au collisions at {\em central} rapidities ($\eta\le 1$) show no similar
suppression (see the 2 left-most plots in \Fref{fig:dAu} and also the
corresponding data in the right plot in \Fref{fig:azimuthal} below),
which implies that nuclear saturation effects are not important in the
central--rapidity kinematics at RHIC. (But this is likely to change at
the LHC; see e.g. \cite{Gelis:2010nm,Albacete:2010bs,ALbacete:2010ad}.)
The situation is however different for the `dense--dense' A+A collisions:
the respective data at RHIC and the LHC show a strong suppression of
particle production and of the azimuthal di--hadron correlations {\em
already for central rapidities} (and for relatively hard transverse
momenta). These phenomena cannot be related to `initial--state',
saturation, effects in the incoming nuclei (at least not fully), since
they do not show up in the mid--rapidity d+Au data at RHIC. Rather, they
must correspond to interactions in the {\em final state}, that is,
interactions with the dense partonic medium (the glasma and the
quark--gluon plasma) which exists at intermediate stages. The change in
the properties of a hard particle or of the associated jet induced by its
interactions in the medium is generally referred to as {\em jet
quenching} (see e.g.
\cite{CasalderreySolana:2007zz,d'Enterria:2009am,Wiedemann:2009sh,Majumder:2010qh}
for recent reviews).

\begin{figure}[t]
\begin{center}\centerline{
\includegraphics[width=.55\textwidth]{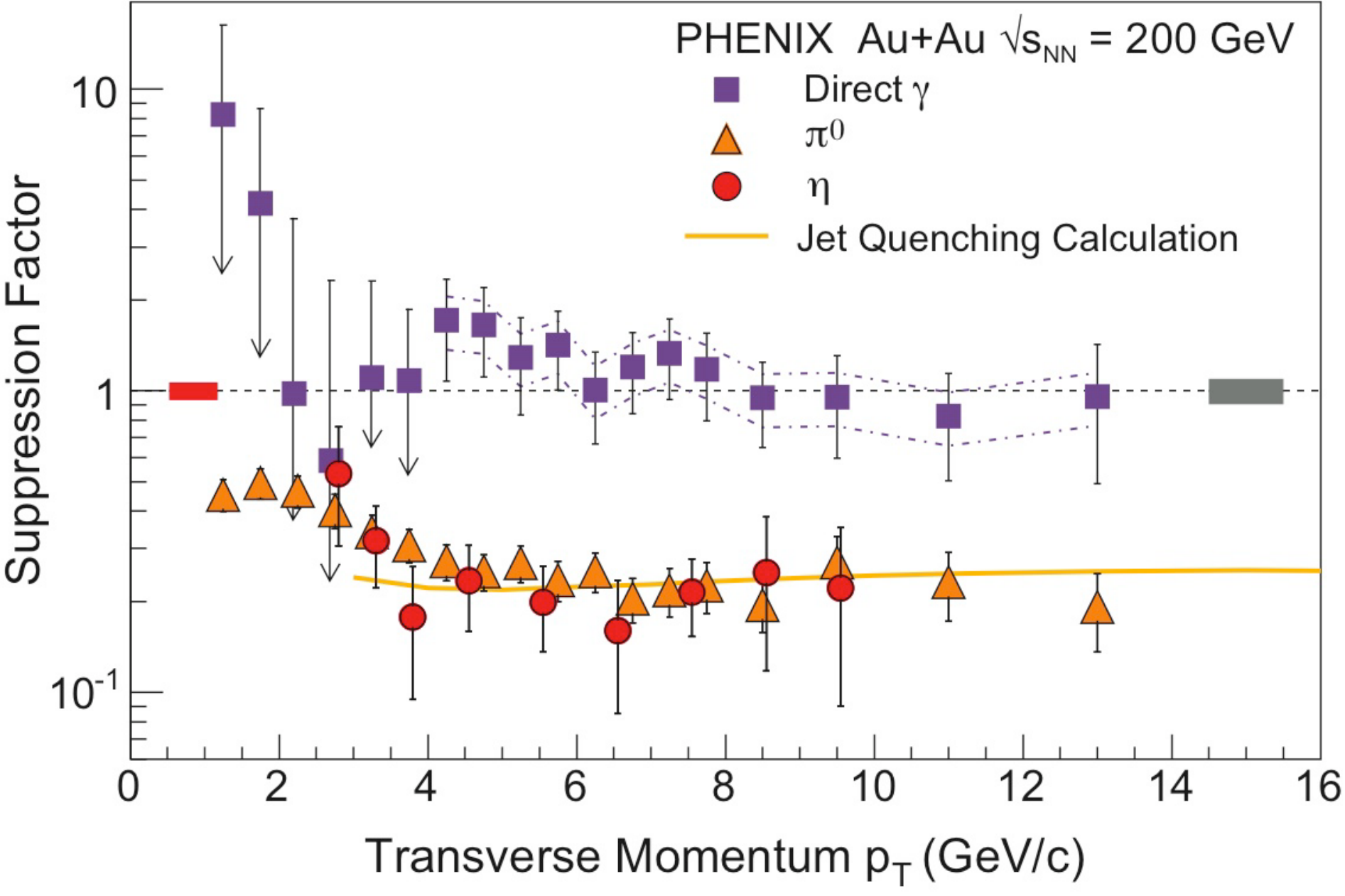}\qquad
\includegraphics[width=.45\textwidth]{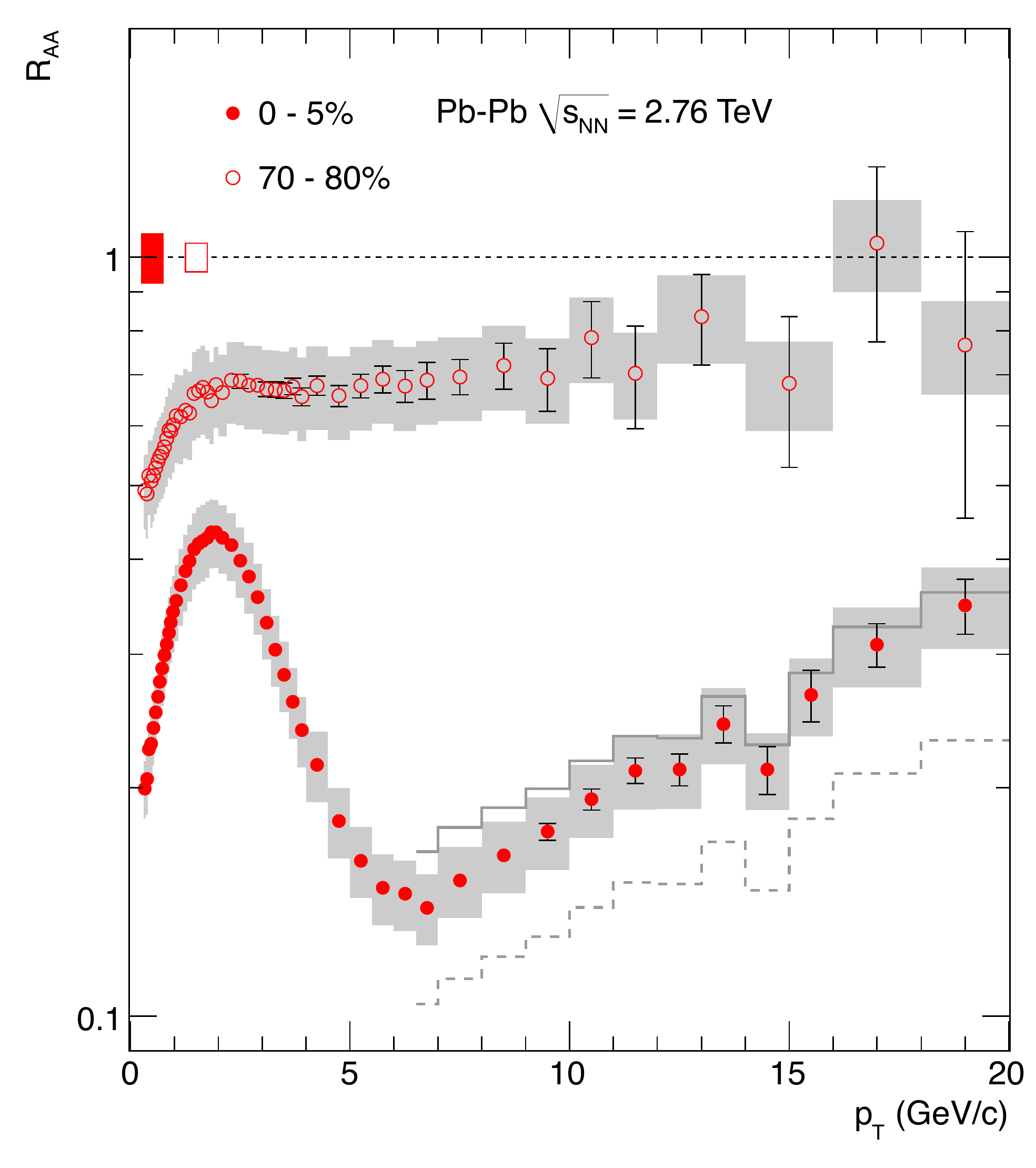}
}
\caption{\sl The nuclear modification factor $R_{AA}$, as measured in
central Au+Au
collisions at RHIC (PHENIX, left panel) and in Pb+Pb collisions
at the LHC (CMS, right panel). For comparaison, one also shows the
data for the production of direct photons, which show no nuclear effect
as expected (left panel) and for hadron production in peripheral collisions,
for which the nuclear effects are quite small (right panel).} \label{fig:RAA}
\end{center}\vspace*{-.8cm}
\end{figure}

The suppression of particle production in A+A collisions is best
characterized by the ratio $R_{AA}$ (the `nuclear modification factor'),
defined by analogy with \Eref{RpA}, that is
\beq\label{RAA}
 R_{AA}(\eta,p_\perp) \,\equiv\,\frac{1}{N_{coll}}\,
 \frac{
\frac{\rmd N_h}{\rmd^2p_\perp
   \rmd \eta}\Big |_{AA}}
   {\frac{\rmd N_h}{\rmd^2p_\perp \rmd
 \eta}\Big |_{pp}}\,,\eeq
where the number $N_{coll}$ of binary collisions at a given impact
parameter scales like $A^{4/3}$ (for relatively central collisions):
indeed, there is a factor $A^{1/3}$ associated with the longitudinal
width of each of the two nuclei and an additional factor of $R_A^2\propto
A^{2/3}$ coming from the integral over all the impact parameters. The
experimental results for $R_{AA}$ (at mid--rapidities) in Au+Au
collisions at RHIC ($\sqrt{s_{NN}}=200$~GeV) and in Pb+Pb collisions at
the LHC ($\sqrt{s_{NN}}=2.76$~TeV) are shown in the left and right panels
of \Fref{fig:RAA}, respectively. As anticipated, they show a substantial
suppression of the hadron production as compared to p+p collisions, which
persists up to $p_\perp\simeq 20$~GeV, at least. The interpretation of
this suppression as a dense--medium effect is furthermore supported by
the fact that \texttt{(i)} the direct photons (which do not interact with
the hadronic matter) show indeed no suppression (cf. the left figure),
and \texttt{(ii)} even for hadrons, the suppression is considerably
smaller in the peripheral collisions (cf. the right figure), in agreement
with the fact that the density and the size of the produced medium are
much smaller in that setup. Note also that the suppression at
intermediate values $p_\perp\simeq 6\div 7$~GeV is stronger at the LHC
than at RHIC, indicating that the medium produced there is denser, as
expected.

\begin{figure}[t]
\begin{center}\centerline{
\includegraphics[width=.35\textwidth]{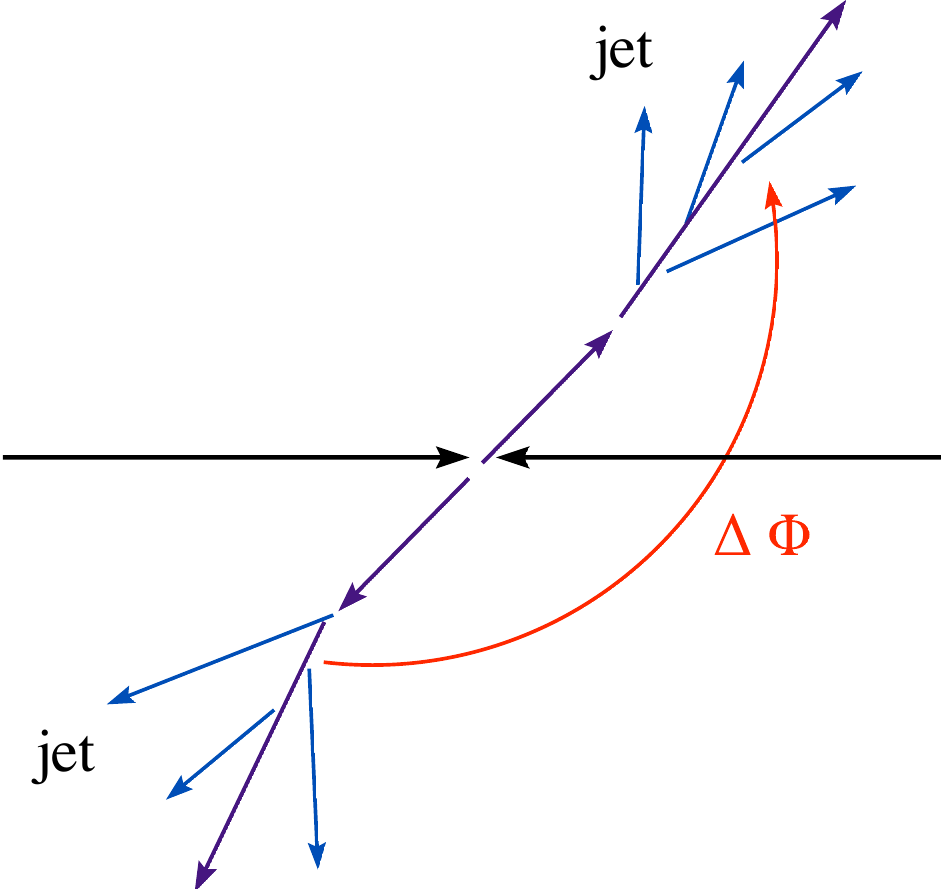}\qquad
\includegraphics[width=.6\textwidth]{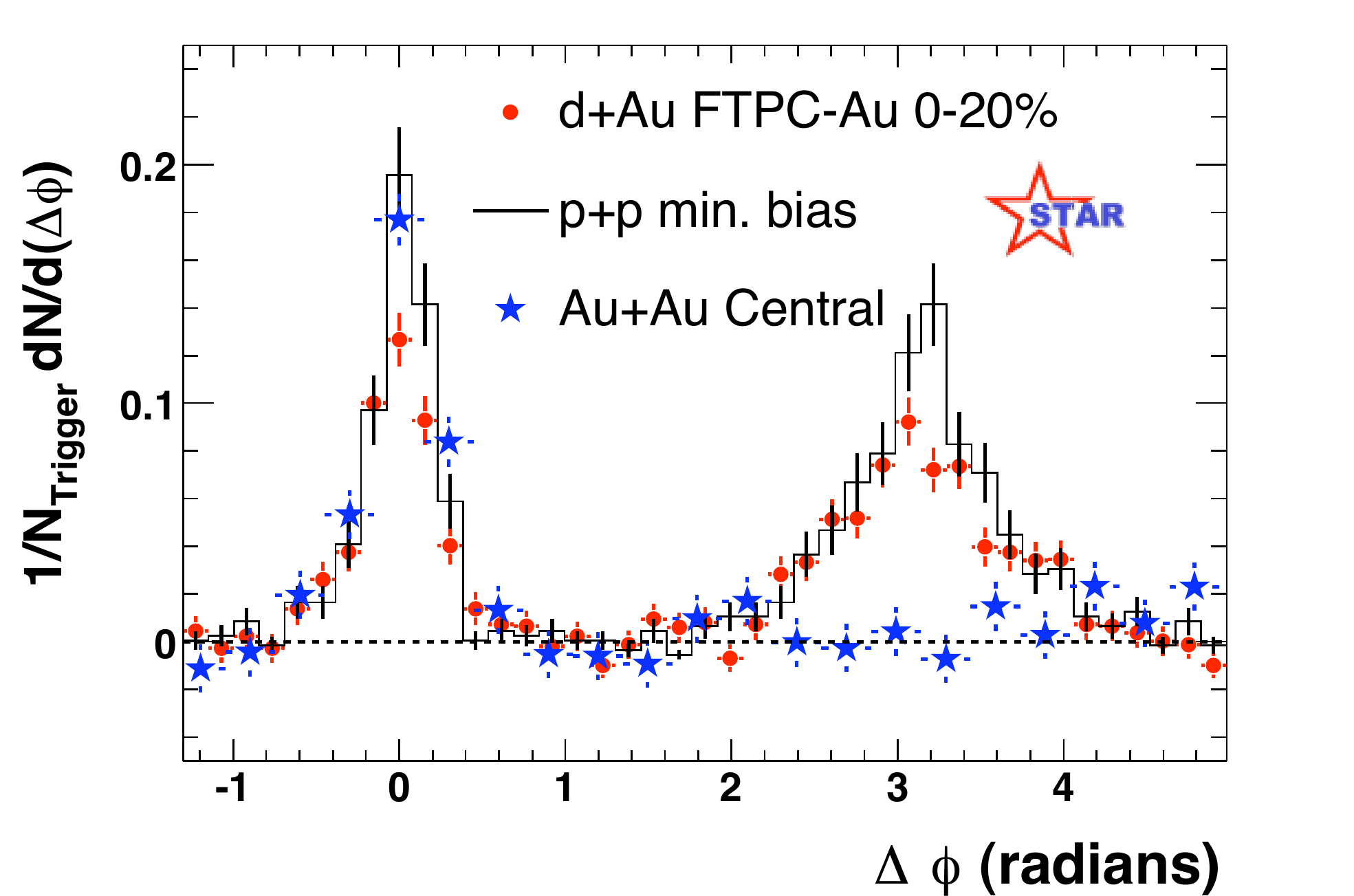}
}
\caption{\sl Left: cartoon of a typical di--jet event as produced
by a hard scattering. Right: azimuthal distribution of hadrons
with $p_\perp\ge 2$~GeV relative to a trigger hadron with $p_\perp\ge 4$~GeV,
as measured at RHIC (STAR), in p+p, d+Au and central Au+Au collisions
\cite{Adams:2005dq}.} \label{fig:azimuthal}
\end{center}\vspace*{-.8cm}
\end{figure}

Concerning the suppression of azimuthal correlations in the di--hadron
production, this is clearly visible in the right plot in
\Fref{fig:azimuthal}, which shows data taken at RHIC for hadrons with
$p_\perp\gtrsim 2$~GeV~: unlike for p+p and d+Au collisions, where one
can see a peak at $\Delta\Phi=\pi$, as expected for a pair of hadrons
which are produced back--to--back, there is no such a peak in the central
Au+Au collisions. This is interpreted as the consequence of the
interactions suffered by the `away' particle (the one that would have
normally emerged at $\Delta\Phi=\pi$) while propagating through the
medium. Via such interactions, the particle transverse momentum has been
degraded and/or the particle has been deviated towards different
directions, so it will not show up around $\Delta\Phi=\pi$, nor in the
original bin in $p_\perp$.

\begin{figure}[ht]
\begin{center}\centerline{
\includegraphics[width=0.65\textwidth]{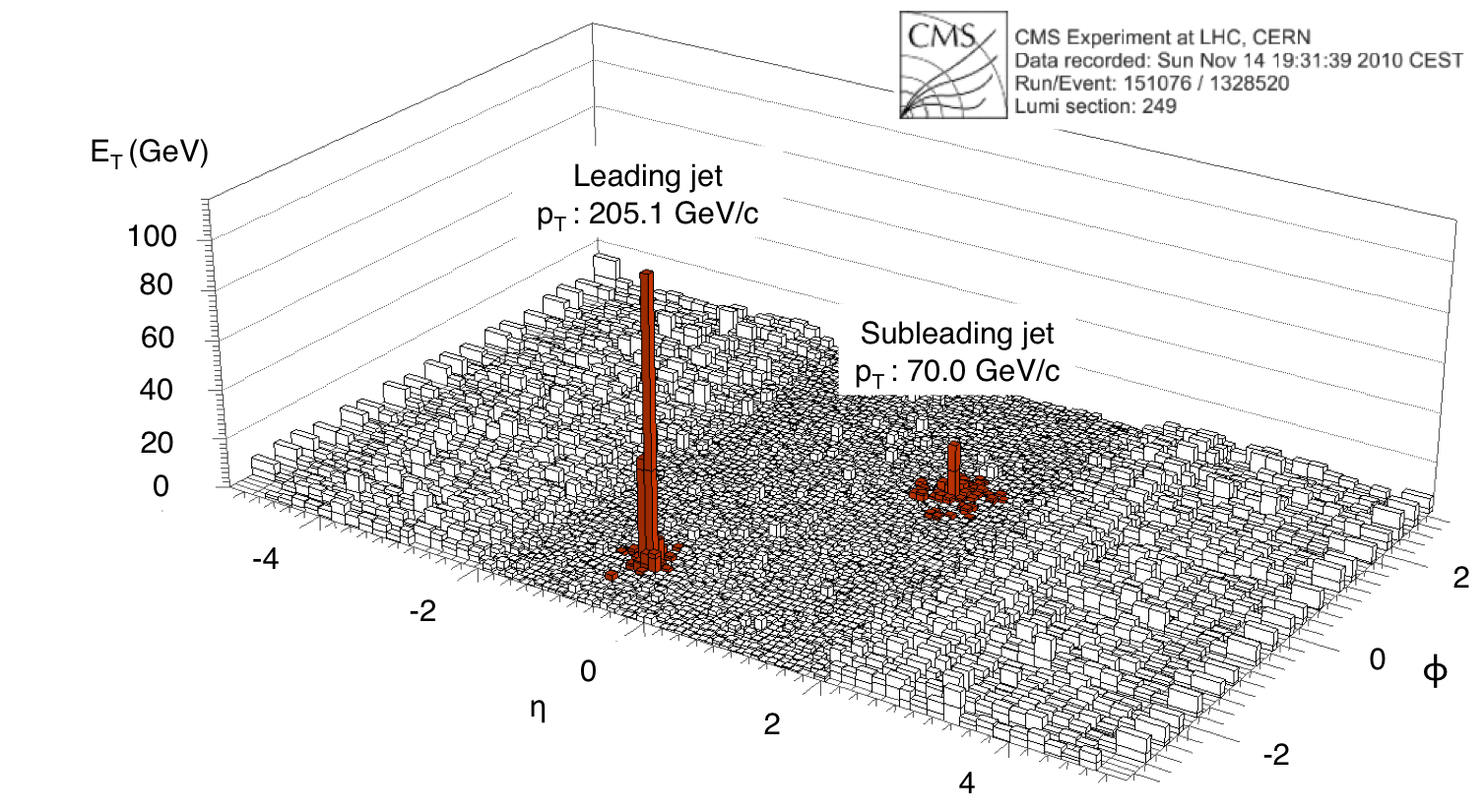}\quad
\includegraphics[width=.35\textwidth]{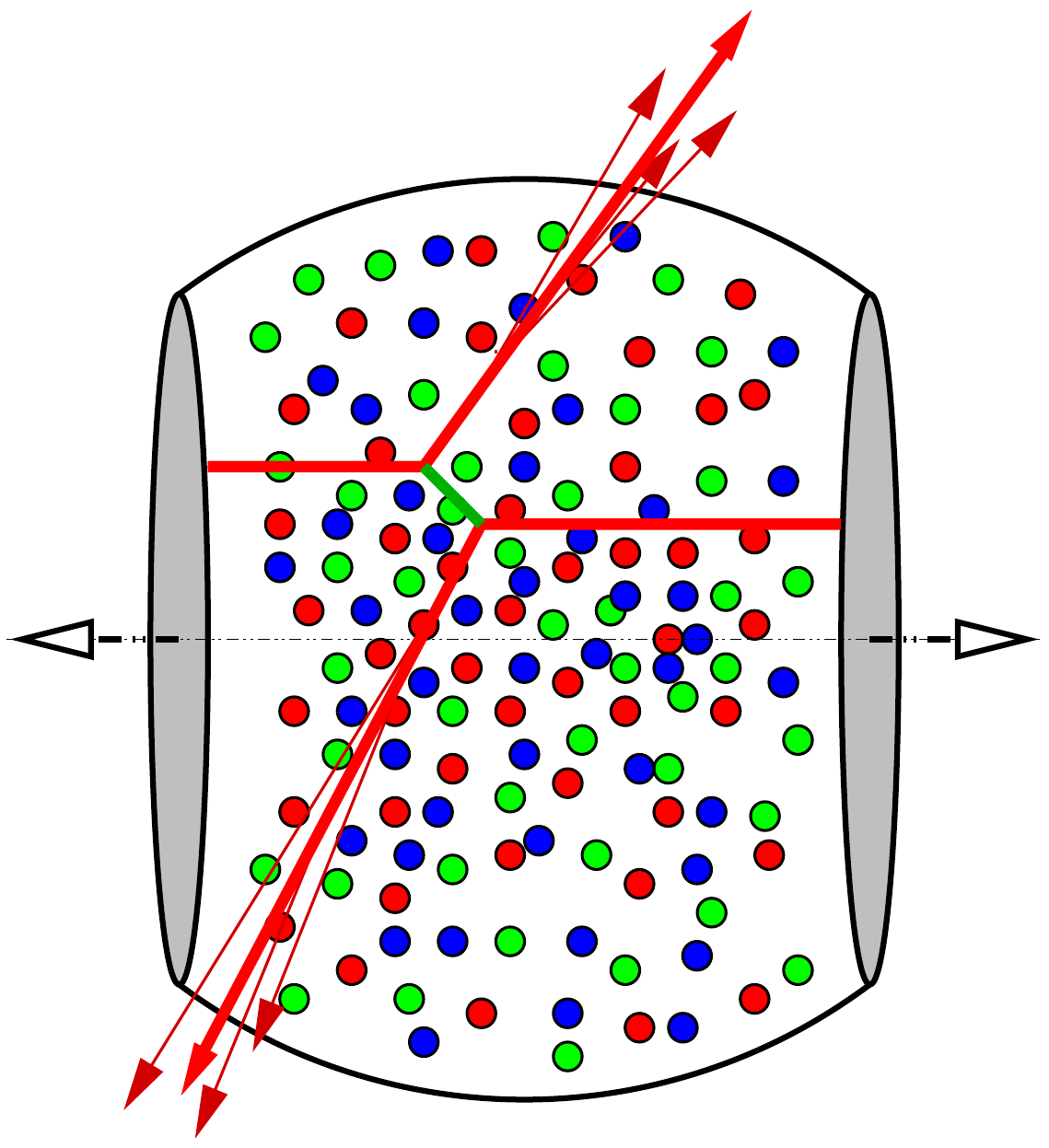}
}
\caption{\sl Left panel: a highly asymmetric di--jet event in
Pb+Pb collisions at the LHC as measured by CMS. Right panel: a cartoon
of an asymmetric di--jet event in A+A collisions. The hard scattering
producing the jets occurs near the edge of the fireball. One
of the jets (the `trigger jet') leaves the medium soon after its
formation and thus escapes unscattered, while the other one
(the `away jet') crosses the medium and
is strongly modified by the latter.} \label{fig:dijets}
\end{center}\vspace*{-.8cm}
\end{figure}

\begin{figure}[t]
\begin{center}\centerline{
\includegraphics[width=0.5\textwidth]{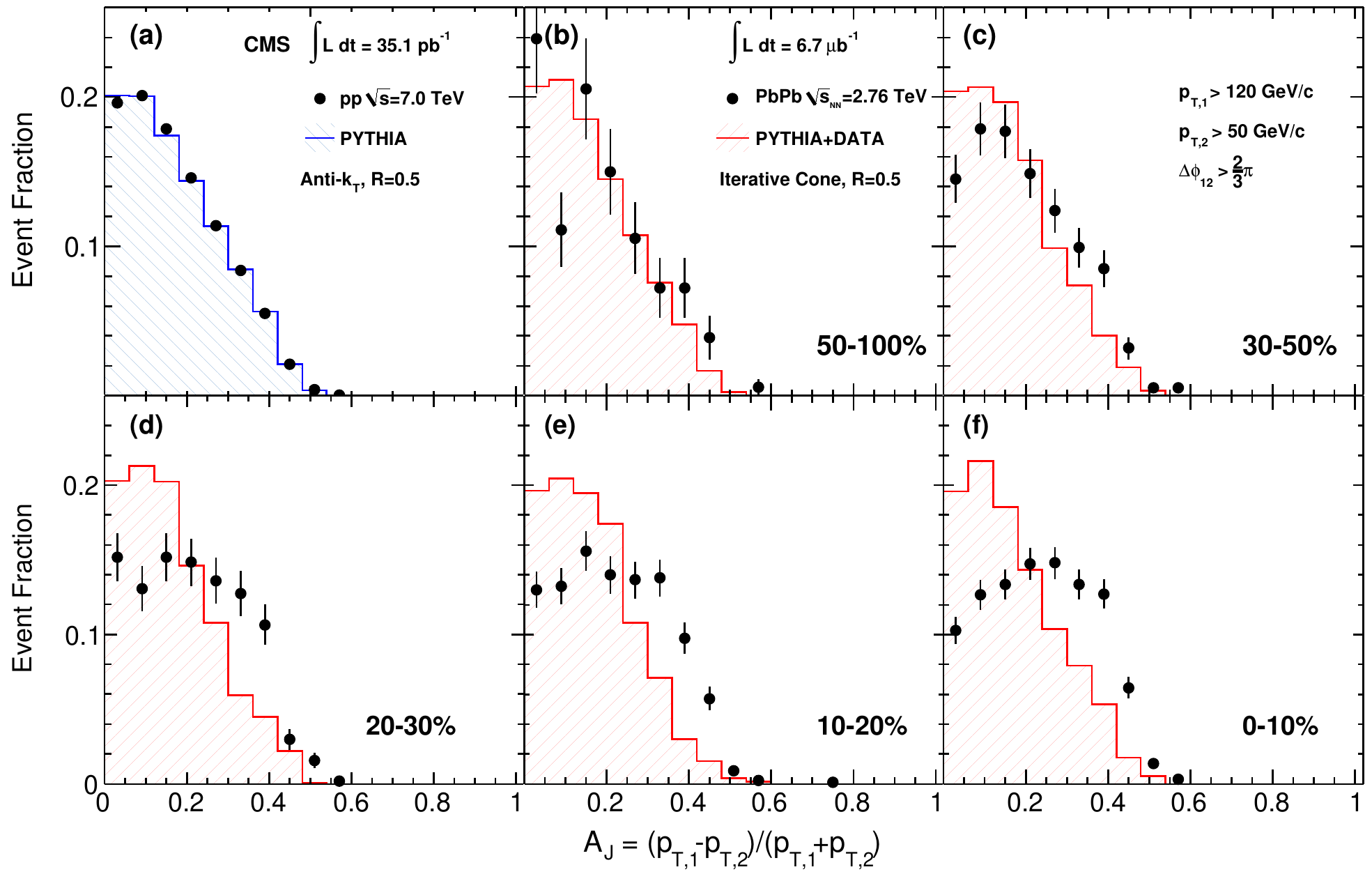}\quad
\includegraphics[width=.5\textwidth]{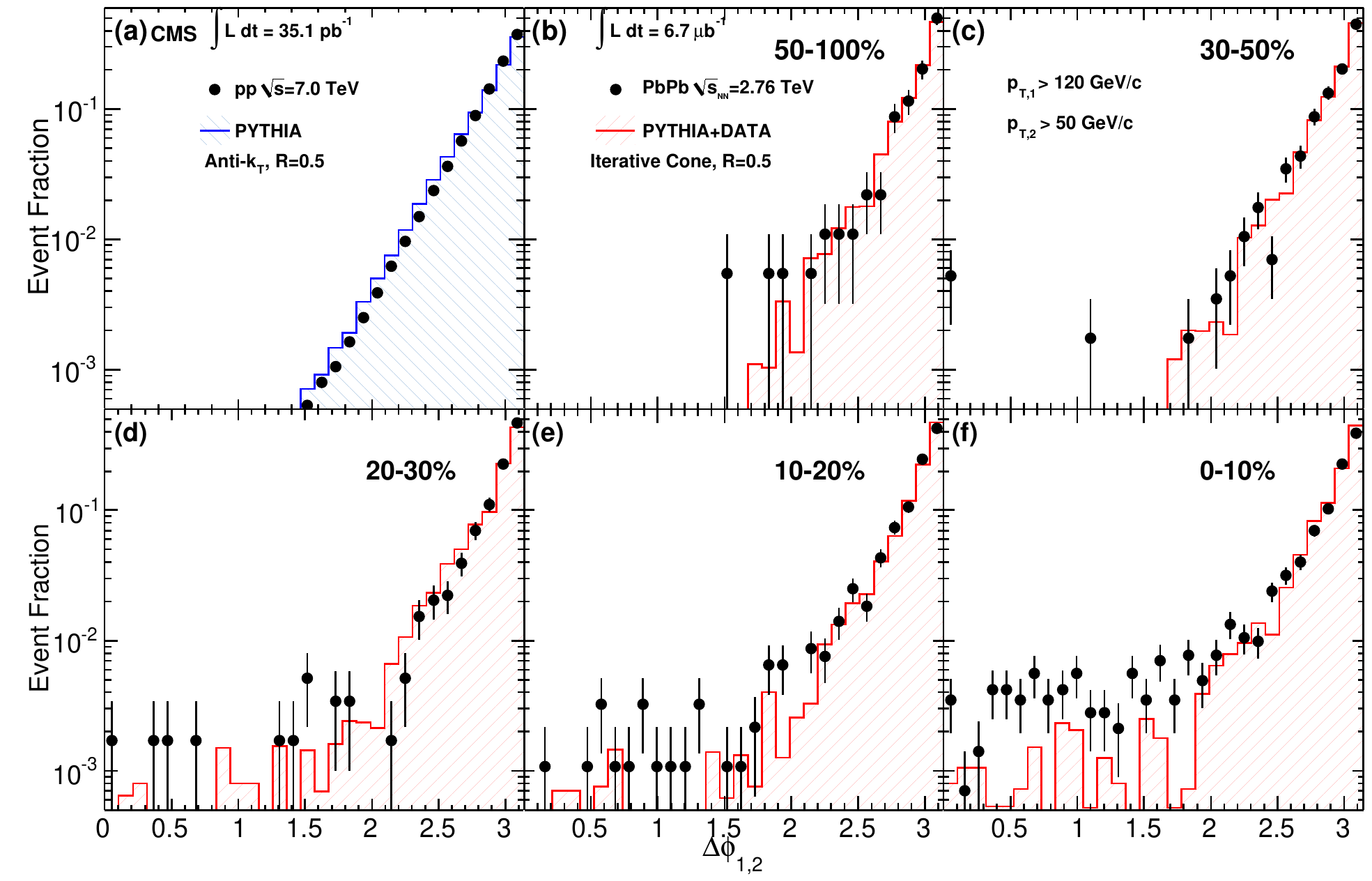}
}
\caption{\sl The distribution of the CMS data for p+p collisions
($\sqrt{s}=7$~GeV) and Pb+Pb collisions ($\sqrt{s_{NN}}=2.76$~TeV)
as a function of $A_{\rm J}$ (left panel) and respectively of the
azimuthal separation $\Delta\Phi_{12}$ w.r.t. the leading--jet axis
(right panel) in different bins of centrality: (a) refers to p+p, (f)
to the 10\% most central Pb+Pb, (e) to the centrality bin 10-20\%, etc.
The histograms show the respective predictions of PYTHIA, which agree well
with the data for p+p and for peripheral Pb+Pb, but not also for central
Pb+Pb. From Ref.~\cite{Chatrchyan:2011sx}.} \label{fig:AJ}
\end{center}\vspace*{-.8cm}
\end{figure}

Besides confirming and sharpening the discoveries at RHIC, the first
heavy ion data at the LHC revealed a new phenomenon, whose observation
was possible because of the unprecedented ability of the detectors there
(notably the calorimeters at ATLAS and CMS, with a wide coverage in
rapidity) to reconstruct jets: the {\em di--jet asymmetry}
\cite{Aad:2010bu,Chatrchyan:2011sx}. Namely, a significant fraction of
the di--jet events in Pb+Pb collisions at $\sqrt{s_{NN}}=2.76$~TeV shows
a large transverse energy imbalance between the trigger (or `leading')
jet and the away (or `subleading') jet. (The left panel in
\Fref{fig:dijets} and the right panel in \Fref{fig:events} display such
asymmetric di--jet events, as measured by CMS.) One should stress that
the criterion used to define a `jet' --- the value of the product
$R=\Delta\Phi\times \Delta\eta$ between the spreadings of the hadron
yield in azimuthal angle and in pseudo--rapidity --- is the same for the
leading and subleading jets. Moreover, a substantial asymmetry between
the two jets exists already in p+p collisions, because of the bias
introduced by the trigger process (see the histograms in \Fref{fig:AJ}).
But the heavy ion collisions show a significant increase in this
asymmetry, which becomes more pronounced with increasing centrality. A
quantitative way to characterize this asymmetry is via the {\em
transverse energy imbalance},
 \beq\label{AJ}
 A_{\rm J}=\frac{p_{\perp 1}-p_{\perp 2}}
 {p_{\perp 1}+p_{\perp 2}}\,,\eeq
where $p_{\perp 1}$ ($p_{\perp 2}$) is the transverse momentum of the
leading (subleading) jet. The normalization in \Eref{AJ} is useful for
removing uncertainties in the overall jet energy scale. \Fref{fig:AJ}
(left panel) shows the distribution of the Pb+Pb events as a function of
$A_{\rm J}$, in different bins of centrality: for the most peripheral
collisions, this is quite similar to the respective distribution for p+p
collisions, as shown in figure (a). But for the more central collisions,
there is an increase in the fraction of events with relatively large
$A_{\rm J}=0.3\div 0.4$, which significantly exceeds the respective
prediction of the PYTHIA Monte--Carlo event generator (which neglects the
medium effects). This demonstrates that there is additional energy loss
by the jet, estimated as 20 to 30 GeV, due to its interactions in the
medium. But this energy loss does not lead to significant angular
decorrelations: as visible in the right panel of \Fref{fig:AJ}, the
distribution of the subleading jet is still peaked at
$\Delta\Phi_{12}=\pi$, like in p+p collisions (and in good agreement with
the PYTHIA simulations). This implies that the additional energy loss is
not to be attributed to rare, hard, emissions (which would typically lead
to 3--jet events). A careful analysis of the background around the away
side jet allowed one to establish that the missing energy is in fact
associated with relatively {\em soft particles} ($p_\perp\le 2$~GeV)
emitted {\em at large angles} with respect to the jet axis
\cite{Chatrchyan:2011sx}. This rises the question about the physical
mechanism which is responsible for such in--medium emissions at large
angles.

\begin{figure}[t]
\begin{center}\centerline{
\includegraphics[width=.58\textwidth]{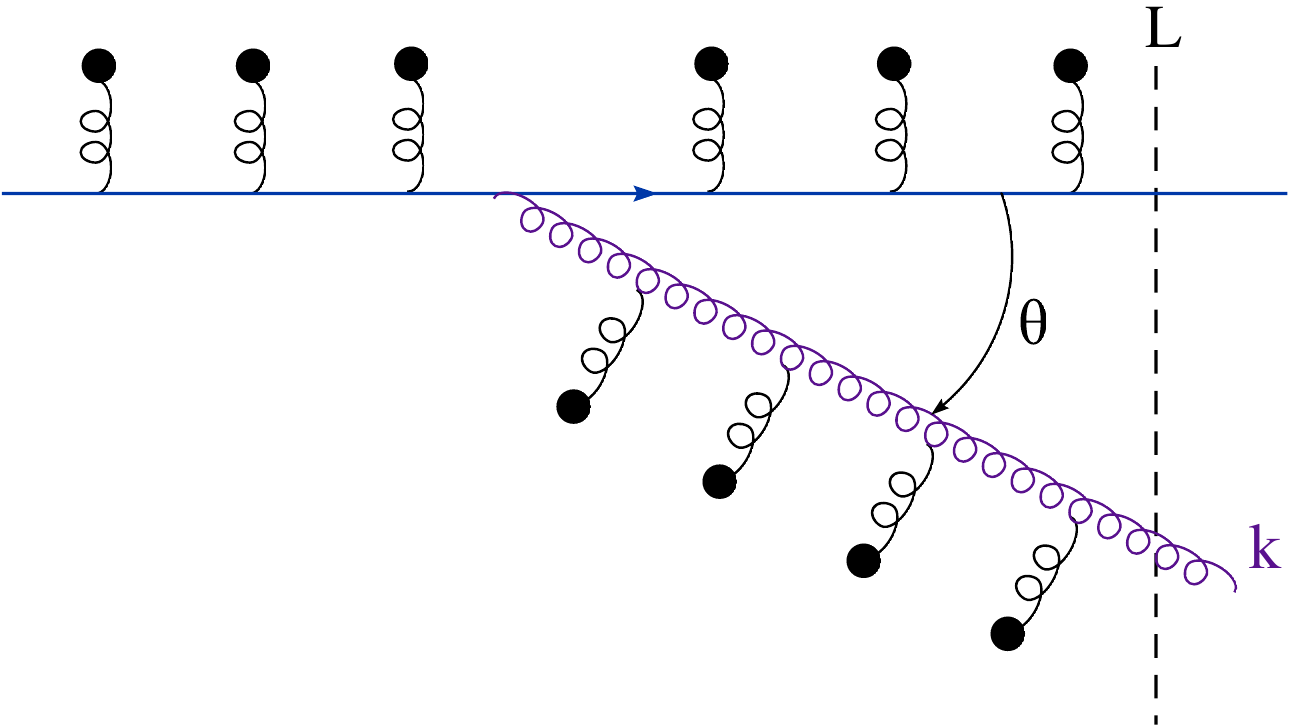}\qquad
 \includegraphics[width=.35\textwidth]{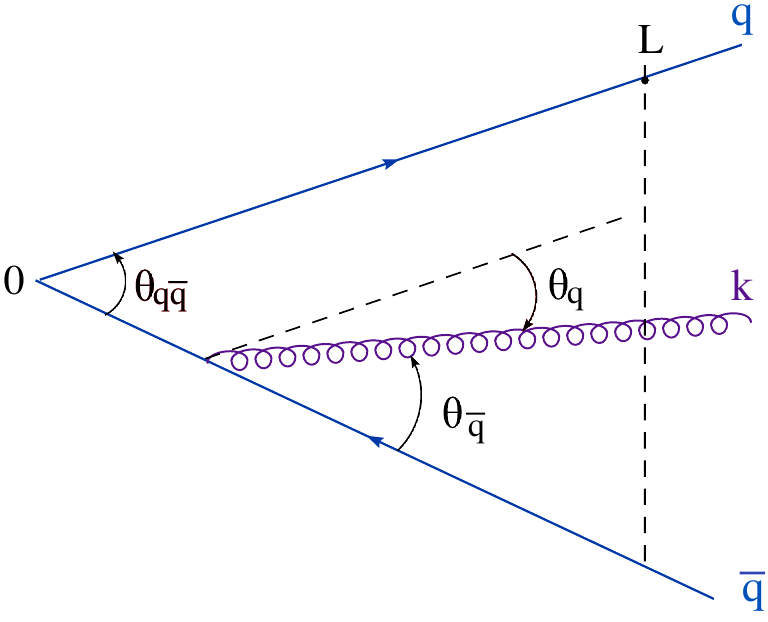}
 }
\caption{\sl Right: A cartoon illustrating medium--induced gluon radiation by
an energetic quark propagating through a dense QCD medium. Both the quark and the emitted gluon undergo multiple scattering
off the medium constituents (represented by the black blobs). Left:
A quark-antiquark antenna emitting a gluon (here, from the antiquark leg)
within the medium with size $L$. The interactions with the medium are not
explicitly shown.}
\label{fig:BDMPS}
\end{center}\vspace*{-.8cm}
\end{figure}

As we shall now explain, a natural mechanism in that sense exists in QCD
at weak coupling: this is {\em medium--induced gluon radiation}, that is,
the emission of gluons stimulated by the interactions between the partons
composing the jet and the constituents of the medium. The most
interesting situation --- originally studied by Baier, Dokshitzer,
Mueller, Peign\'e, and Schiff \cite{Baier:1996kr,Baier:1996sk} and
independently by Zakharov \cite{Zakharov:1996fv,Zakharov:1997uu},
following pioneering work by Gyulassy and Wang \cite{Gyulassy:1993hr}
(see also
\cite{Wiedemann:2000za,Gyulassy:2000er,Wang:2001ifa,Arnold:2002ja,Salgado:2003rv,Arnold:2008iy,CasalderreySolana:2011rz})
---, is when the medium is so dense that the gluon {\em formation time}
is much larger than the {\em mean free path} of a parton propagating
through the medium. In that case, there are many collisions which
coherently contribute to the emission of a single gluon (see
\Fref{fig:BDMPS}), leading to a suppression of the radiation spectrum as
compared to the (Bethe--Heitler) spectrum that would be produced via
independent multiple scattering. This suppression is known as the
Landau--Pomeranchuk--Migdal (LPM) effect.

The `gluon formation time' is the typical time that it takes in order to
emit a gluon with a given kinematics. This concept is quite similar to
the `fluctuation lifetime' introduced in \Eref{lifetime}, but it is
instructive to present here an alternative derivation for it, which is
adapted to the problem at hand. To be specific, consider the emission of
a gluon with momentum $\bmk$ and energy $\omega=|\bmk|$ by an energetic
quark propagating along the $z$ axis. Even though the $q\bar q g$ vertex
in QCD is local, the emission process is truly non--local, as it takes
some time for the emitted gluon to lose coherence w.r.t. its parent
quark. Namely, when the gluon starts being emitted, its wavefunction is
still overlapping with that of the quark, so the two quanta cannot be
distinguished from each other. But with increasing time, the gluon
separates from the quark and their quantum coherence gets progressively
lost. When the quark is very energetic, the gluon is typically emitted at
a very small angle $\theta\simeq k_\perp/\omega\ll 1$ and the coherence
between the two quanta is measured by their overlap in the transverse
space. The gluon is considered as being `formed' (or `fully emitted')
when its transverse separation $b_\perp\simeq \theta \Delta t$ from the
quark becomes larger than its transverse Compton wavelength
$\lambda_\perp=1/k_\perp \simeq 1/(\omega\theta)$. This condition is
satisfied after a time
 \beq\label{tau}
 \Delta t_{\rm form}\,\simeq\,\frac{2\omega}{k_\perp^2}
 \,\simeq\,\frac{2}{\omega\theta^2}\,.
 \eeq
(The factor of 2 in the numerator is conventional.) The above argument is
completely general: it holds for gluon emissions in the medium or in the
vacuum. What is different, however, in the two cases is the mechanism
causing the radiation and the associated gluon spectrum.

To better appreciate this difference, remember first that an on--shell
quark cannot radiate: it can produce virtual fluctuations and thus
develop a partonic substructure, as discussed in \Sref{sec:CGC}, but
energy--momentum conservation prevent these quanta to become on--shell,
and hence to separate from the parent quark. For the radiation to be
possible, the quark and/or the emitted gluon must suffer additional
interactions, which provide the energy deficit.

Consider first the situation in the {\em vacuum}~: the quark is produced
in an off--shell state via a hard scattering and then evacuates its
virtuality via bremsstrahlung. Namely, it emits a gluon with energy
$\omega$ and transverse momentum $k_\perp$ within a time interval $\Delta
t\sim {2\omega}/k_\perp^2$ after the original scattering. These values
$\omega$ and $k_\perp$ are arbitrary (subjected to energy--momentum
conservation) and independent of each other. However the bremsstrahlung
spectrum, \Eref{brem}, favors the emission of soft ($x\ll 1$, or
relatively small $\omega$) and nearly collinear ($\theta\to 0$, or small
$k_\perp$) gluons, for which the formation time is long.

The situation is very different for a quark propagating through a {\em
dense medium}~: the quark undergoes collisions with the medium
constituents, with a typical distance $\ell$ (the mean free path) between
two successive collisions. Any such a collision provides a small
acceleration, thus allowing the quark to radiate. Accordingly, the
initial virtuality of the quark is not essential anymore: the quark can
now radiate {\em anywhere} within the medium, and not only within a
distance $\sim \Delta t_{\rm form}$ after the original hard scattering.
This implies that the phase--space for in--medium emissions is {\em
enhanced} by a factor $L/\Delta t_{\rm form}$ with respect to emissions
in the vacuum. Here $L$ is the longitudinal extent of the medium as
crossed by the quark and is typically much larger than $\Delta t_{\rm
form}$, as we shall see. Moreover, the emission mechanism and the
associated formation time are influenced by the {\em gluon} interactions
in the medium, which destroy the coherence between the gluon and the
parent quark and thus facilitate the radiation.

To estimate $\Delta t_{\rm form}$ for medium--induced emissions, we also
need the rate at which the (virtual) gluon accumulates transverse
momentum via rescattering in the medium. We shall later check that the
successive collisions proceed independently from each other and thus
provide transverse momenta which are randomly oriented and add in
quadrature. This implies that the {\em average} transverse momentum {\em
squared} grows linearly with time: $\langle k_\perp^2 \rangle \simeq \hat
q\Delta t$, where $\Delta t$ is the lifetime of the virtual gluon, as
measured from the emission vertex, and $\hat q$ is a medium--dependent
transport coefficient known as the {\em jet quenching parameter}, to be
specified later. Hence, during its formation, the gluon acquires a
typical transverse momentum squared $k_f^2 \simeq \hat q \Delta t_{\rm
form}$ via scattering within the medium. On the other hand, the condition
for quantum decoherence requires the relation \eqref{tau} between $\Delta
t_{\rm form}$ and $k_f^2$. Together, these two conditions determine {\em
both} the formation time and the typical transverse momentum of the gluon
at the time of emission:
 \beq
  \Delta t_{\rm form}\,\simeq\,\frac{2\omega}{k_f^2}\quad \& \quad
k_f^2 \simeq \hat q \Delta t_{\rm form} \quad \Longrightarrow \quad
 \Delta t_{\rm
form}\,\simeq \,\sqrt{\frac{2\omega}{\hat q}} \quad \& \quad
 k_{f}^2 \,\simeq\,
 (2\omega\hat q)^{1/2}\,.\eeq
We thus see that, for medium--induced radiation, $k_\perp$ and $\omega$
are {\em not} independent kinematical variables anymore: the transverse
momentum of the emitted gluon is acquired via interactions in the medium
during the formation time which grows with the energy like $\Delta t_{\rm
form}\propto \omega^{1/2}$.

This mechanism for gluon production is operational provided the formation
time is much larger than the mean free path $\ell$, but smaller than the
size $L$ of the medium which is available for the emission process (the
distance traveled through the plasma by the parent quark):
 \beq\label{omegac}
 \ell\,\ll\,\Delta t_{\rm form}\,\le\,L \quad \Longrightarrow \quad
 \omega_{\rm min}\,\equiv\,\frac{1}{2}\, \hat q\, \ell^2 \,\ll\,\omega\,\le\,
 \omega_c\, \equiv\, \frac{1}{2}\, \hat q L^2\,.\eeq
These arguments imply that the typical emission angle at the time of
formation, $\theta_f\simeq {k_{f}}/{\omega}$, cannot be arbitrarily
small:
 \beq\label{thetaf} \theta_f\,\simeq\,\frac{k_{f}}{\omega}\,\simeq\,
 \left(\frac{2 \hat q}{\omega^3}\right)^{1/4}
 \quad \Longrightarrow \quad \theta_c\, \equiv\,
 \frac{2}{\sqrt{\hat q L^3}}\,\le\,\theta_f\,\ll\,\theta_{\rm max}
 \, \equiv\,\frac{2}{\sqrt{\hat q \ell^3}}\,.\eeq
Unlike bremsstrahlung, the in--medium radiation does {\em not} favour
collinear radiation. In fact, the smaller is the gluon energy $\omega$,
the larger is the emission angle $\theta_f$ and the shorter the emission
time $\Delta t_{\rm form}$. The final spectrum favors indeed the emission
of {\em relatively soft} gluons with $\omega \ll \omega_c$, for which the
formation time is much smaller than the size of the medium, $\Delta
t_{\rm form}\ll L$, and the emission angle is quite large:
$\theta_f\gg\theta_c$. Moreover, after being emitted, the gluons keep
interacting with the medium and thus get deflected at even larger angles:
their average transverse momentum can rise up to a final value $\langle
k_\perp^2 \rangle \simeq \hat q(L-t_0) \sim \hat q L$, where $t_0$ is the
time at the emission vertex. This phenomenon is known as {\em transverse
momentum broadening}.

The above considerations show that the medium--induced radiation is very
efficient in broadening the jet energy in the transverse plane, via the
emission of soft ($\omega \ll \omega_c$) gluons, in qualitative agreement
with the LHC data for di--jet asymmetry. On the other hand, the $R_{AA}$
data for `high--$p_\perp$ suppression' are probably more sensitive to the
emission of harder gluons, with $\omega\sim\omega_c$, which dominate the
energy loss by the `leading particle' (the parton which has initiated the
jet). Accordingly, the total energy loss $\Delta E\sim \omega_c\sim \hat
q L^2$ scales like the {\em square} of the medium length $L$, and not
like $L$ (as one would expect for a mechanism where the energy is lost
{\em locally}, say via elastic collisions in the medium). The reason for
this scaling with $L^2$ is, of course, the fact that the actual mechanism
at work is {\em non--local}~: it takes a time $\Delta t_{\rm
form}(\omega)$ to emit a gluon and for $\omega\sim\omega_c$ this time is
of the order of $L$. Since moreover the emission can be initiated at any
point within $L$, the overall energy loss scales like $L^2$. Reversing
the argument, one concludes that the {\em stopping length} for a particle
which loses all its energy inside the medium scales like $L_{\rm
stop}\sim E^{1/2}$, where $E$ is the initial energy of that particle.

For more quantitative studies and applications to phenomenology, one
still needs an estimate for the jet quenching parameter $\hat q$. Let us
assume, for definiteness, that the medium is a quark--gluon plasma with
(local) temperature $T$ and weak coupling. The typical collisions which
matter for the problem at hand are {\em soft}, in the sense that the
typical transferred momentum is of the order of the Debye mass $m_D\sim
gT$ introduced in \Sref{sec:QGP}. This is so since the soft collisions
occur much more frequently than the hard ones: the corresponding
cross--section is not the {\em transport} cross--section evaluated in
\Eref{sigma} (that was the cross--section for scattering at {\em large
angles}), but rather the {\em total} cross--section
 \beq\label{sigmatot}
 \sigma_{\rm tot}\,=\,\int\rmd\Omega\,\frac{\rmd
\sigma}{\rmd \Omega}\,\propto\,\int\rmd\theta\,\sin\theta
\,\frac{\alpha_s^2}{T^2\sin^4\theta} \,\sim\,
 \frac{\alpha_s^2}{T^2}\int_g \frac{\rmd\theta}{\theta^3}\,\sim\,
 \frac{\alpha_s}{T^2}\,\ln\frac{1}{\alpha_s}\,.\eeq
This is dominated by small--angle scattering --- the integral over
$\theta$ is cut off at $\theta\sim g$ by the plasma effects --- and is
larger by a factor $1/\alpha_s$ than the transport cross--section
\eqref{sigma}. The relevant mean free path is obtained by inserting the
total cross--section in \Eref{mfp} : $\ell\sim 1/(n \sigma_{\rm tot})
\sim [T \alpha_s\ln({1}/{\alpha_s})]^{-1}$. As anticipated, this is
parametrically larger than the interaction range $1/m_D\sim 1/gT$,
meaning that successive collisions can be treated as independent. Since
on the average there is a transfer $\Delta k_\perp^2 \sim m_D^2$ of
transverse momentum squared per collision, we finally conclude that
 \beq\label{dkdt}
 \hat q\,\equiv\,\frac{\rmd \langle k_\perp^2 \rangle}{\rmd t}\,\simeq\,
 \frac{m_D^2}{\ell}\,\sim\, \alpha_s^2 T^3 \ln({1}/{\alpha_s})
 \,,
 \eeq
for a weakly--coupled QGP. This is merely a parametric estimate, valid to
leading logarithmic accuracy, and as such it suffers from the same
lack--of--accuracy drawback as discussed for other transport coefficients
towards the end of \Sref{sec:thermo}: it cannot be trusted for
phenomenological applications. In fact, the whole set--up above described
is a bit too idealized to correspond to the actual experimental
situation. To cope with that, more sophisticated, phenomenological models
have been proposed which treat the geometry of the collision in a more
realistic way (finite volume, longitudinal expansion, time and point
dependent jet quenching parameter) and include also free parameters. Such
models are quite successful in describing the data, for both the $R_{AA}$
ratio \eqref{RAA} and the di--jet asymmetry \eqref{AJ} (see
\cite{Renk:2012cx} for a recent discussion and more references), but at
the expense of using a rather large (average) value for $\hat q$ ---
considerably larger than the corresponding perturbative estimate to
leading logarithmic accuracy. This situation is sometimes viewed as an
argument in favor of the strong coupling scenario, but it might simply
reflect the inaccuracy of the current perturbative results.

Note finally that the theory discussed above has addressed the
(medium--induced) emission of a single gluon, whereas in reality one
expects the in--medium evolution of a jet to involve several successive
emissions --- both by the hard parton which has initiated the jet and by
its descendants. (Multiple emissions become important when the quantity
$\alpha_s(L/\Delta t_{\rm form})$ --- which is roughly the probability
for one gluon emission --- becomes of order one.) Phenomenological models
generally assume that successive emissions proceed independently from
each other, but this is still to be demonstrated: {\em a priori}, there
could be interference effects between emissions by different sources (the
various partons forming the jet). For jet evolution in the {\em vacuum},
one knows that such interference effects are indeed important: they lead
to {\em angular ordering} of the subsequent emissions --- the successive
emission angles are smaller and smaller
\cite{Dokshitzer:1991wu,Ellis:1991qj}. For the case of {\em in--medium}
radiation, there is so far no explicit calculation of two (or more)
successive emissions, but there are studies of interference effects in
the emission by two sources: a quark and an antiquark forming a `colour
antenna' (see the right panel of \Fref{fig:BDMPS})
\cite{MehtarTani:2010ma,MehtarTani:2011tz,CasalderreySolana:2011rz}. In
particular, the analysis in \cite{CasalderreySolana:2011rz} shows that
the interference effects are negligible so long as the antenna opening
angle (the angle $\theta_{q\bar q}$ in \Fref{fig:BDMPS}) is much larger
than the minimal angle $\theta_c$ introduced in \eqref{thetaf}. As
previously explained, the typical emission angles obey this condition
already at the time of formation and they become even larger at later
times, due to the momentum broadening by the medium. This suggests that
successive medium--induced gluon emissions can be effectively treated as
{\em independent}, thus justifying a {\em probabilistic} approach to
in--medium jet evolution, like in
Refs.~\cite{Zapp:2008af,Schenke:2009gb,Zapp:2011ya}.

\subsection{The AdS/CFT correspondence: insights at strong coupling}
\label{sec:AdS}

At several points in the previous presentation, we pointed out
observables whose values as extracted from the heavy--ion data seem
difficult to understand if the coupling is weak, but would be more
naturally accommodated at strong coupling. These observables include the
viscosity-over-entropy ratio $\eta/s$, the thermalization time $\tau_{\rm
eq}$, the jet quenching parameter $\hat q$, and the heavy quark diffusion
coefficient $D$. In all these cases, the hypothesis of a strong coupling
must be subjected to caution. First, these quantities are measured only
{\em indirectly}, that is, they are extracted from fits to the data based
on complex analyses which involve theoretical prejudices (notably, on the
overall physics scenario), various assumptions which are difficult to
check, and a considerable amount of model--building (concerning e.g. the
initial conditions for hydrodynamics, the geometry of the collision, the
theoretical description of multi--particle interactions). So, it is fair
to say that the systematic uncertainties on these observables are still
quite large (even though, within a {\em given} scenario, they might be
strongly constrained by the data). Second, the weak--coupling results
which serve as benchmarks for comparison are generally leading--order
results in the perturbative expansion. But, as emphasized in
\Sref{sec:thermo}, the standard perturbation theory (i.e. the strict
expansion in powers of the coupling) is not reliable for the description
of transport phenomena, even if the coupling is weak. This is so because
of the need to resum finite--density effects like the Hard Thermal Loops
to all orders, and this has not been done so far for dynamical
quantities. The situation becomes even more complicated for the
far--from--equilibrium situations, as relevant for the phenomenology,
where the medium effects are not well understood.

This being said, the hypothesis of a strong coupling is both interesting
and intriguing, and not easy to refute on the basis of asymptotic freedom
or of the current lattice data. Indeed, as already noted in
\Sref{sec:thermo}, the QCD coupling $g$ is quite large when evaluated for
temperatures a few times  $T_c$  (the critical temperature for
deconfinement): $g=1.5\div 2$. As also mentioned there, the perturbative
series at finite temperature is truly an expansion in powers of $g$ (and
not of $\alpha_s=g^2/4\pi$), so for that purpose the coupling is
moderately strong. Reorganizing the perturbation theory via appropriate
resummations of medium effects is one of the possible strategies to cope
with this problem. But performing fully non--perturbative calculations,
whenever possible, is clearly interesting. For thermodynamics, lattice
QCD is the obvious and pertinent non--perturbative tool. As discussed in
\Sref{sec:thermo}, its results are roughly consistent with those of HTL
resummations at weak coupling. Yet, as we shall later argue, the lattice
QCD results for the pressure do not totally exclude a strong coupling
scenario. For real--time quantities and the non--equilibrium evolution,
lattice methods become unapplicable (or, at least, inefficient), so it
has become common practice to rely on the {\em AdS/CFT correspondence}
for guidance as to general properties of strongly coupled field theories
at finite temperature. (See \cite{Aharony:1999ti} for a general review on
AdS/CFT and
Refs.~\cite{Son:2007vk,Iancu:2008sp,Gubser:2009md,CasalderreySolana:2011us}
for recent reviews of its applications to a finite--temperature plasma.)

The AdS/CFT  correspondence (or `gauge/string duality') does not apply to
QCD, but to a `cousin' of it, the ${\mathcal N}=4$ supersymmetric
Yang--Mills (SYM) theory, which has a non--Abelian {\em gauge} symmetry
with the `colour' group SU$(N_c)$, like QCD, but also additional {\em
global} symmetries (notably, supersymmetry), which strongly constrain the
dynamics. These additional symmetries ensure that the {\em conformal
invariance} of the classical Lagrangian is preserved after including
quantum corrections --- meaning that, unlike in QCD, the coupling is
fixed and there is no confinement. Accordingly, this theory has probably
little to say about the zero--temperature, hadronic, phase of QCD, where
the non--perturbative aspects of QCD are controlled by confinement.
Moreover, this is probably not a good model for the QCD dynamics in the
vicinity of the deconfinement phase transition, where the running
coupling effects are known to be important, as shown by the lattice
results for the `trace anomaly' in \Fref{fig:trace} (left). However, as
also manifest in that figure, the relative `anomaly' $(\varepsilon
-3P)/\varepsilon$ decreases very fast with increasing $T$ above $T_c$ and
becomes unimportant (smaller than 10\%) for $T\gtrsim 2 T_c\simeq
400$~MeV. Hence, there is a hope that, within the intermediate range of
temperatures at $2T_c\lesssim T \lesssim 5T_c$, which is the relevant
range for heavy ion collisions at RHIC and the LHC, the dynamics in QCD
may be at least qualitatively understood by analogy with the ${\mathcal
N}=4$ SYM theory at strong coupling.

Specifically, the AdS/CFT correspondence is a {\em duality}, that is, an
equivalence between two theories which {\em a priori} look very different
from each other: \texttt{(i)} the ${\mathcal N}=4$ SYM gauge theory
mentioned above (the `conformal field theory', or CFT) and \texttt{(ii)}
a special, `type II B', string theory, leaving in a curved 10 dimensional
space--time with Anti-de-Sitter (AdS) geometry\footnote{More precisely,
this 10 dimensional space--time is the direct product AdS$_5\times S^5$,
where AdS$_5$ is the 5--dimensional Anti-de-Sitter space--time, with
constant negative curvature, and $S^5$ is the 5--dimensional sphere, with
constant positive curvature; see e.g. \cite{Aharony:1999ti} for
details.}. This duality is interesting in that it maps the strong
coupling sector of ${\mathcal N}=4$ SYM onto the weak coupling sector of
the string theory. Accordingly, it allows one to compute observables in
the CFT at strong coupling via perturbative calculations in the string
theory. More precisely, the `strong coupling limit' to which refers the
duality is the special limit ($g$ denotes the gauge coupling in
${\mathcal N}=4$ SYM)
 \beq \lambda\,\equiv\,g^2 N_c\,\to\,\infty\quad\mbox{with}\quad
 g^2\,\ll\,1\,,\eeq
that is, the limit of a large number of colours ($N_c\to\infty$) taken
for a fixed, and relatively small, value of the gauge coupling $g$. This
defines indeed a regime of strong coupling (despite $g$ being small)
because when $N_c\gg 1$ the effective coupling in the gauge theory is the
't Hooft coupling $\lambda=g^2 N_c$. (This is also true for QCD with
colour group SU$(N_c)$.) For instance, the perturbation theory in the
multi--colour limit is dominated by planar Feynman graphs which are such
that each additional loop brings a factor of $\lambda$. As long as
$\lambda\ll 1$, the large--$N_c$ limit of the theory can be studied in a
perturbative expansion in powers of $\lambda$. In the opposite limit
$\lambda\to\infty$ (but with $g\ll 1$), one can rely on the AdS/CFT
correspondence. In that limit, the dual string theory reduces to
`supergravity' (or SUGRA) --- a classical field theory in a curved
space--time with 10 dimensions. From the solutions to the classical
equations of motion (e.g., Einstein equations), one can unambiguously
construct, via the AdS/CFT dictionary, the correlations in the ${\mathcal
N}=4$ SYM theory at infinitely strong coupling.

The ${\mathcal N}=4$ SYM theory at finite temperature and $\lambda\gg 1$
provides a model for the strongly--coupled quark--gluon plasma (sQGP).
The corresponding string--theory dual is obtained by adding a {\em black
hole} into the AdS space--time. This is somewhat natural, since, as we
know from Hawking, a black hole has entropy and generates black--body
radiation, so in that sense it behaves indeed like a thermal system. The
entropy of a black hole is proportional to the area of its event horizon.
(No information can escape from the volume inside the horizon, so this
volume cannot contribute to the entropy.) The corresponding,
Bekenstein--Hawking, formula can be adapted to supergravity, to yield
 \beq\label{sbh}
 S_{\rm
    BH}\,=\,\frac{\mbox{Horizon area}}{4G_{10}} \Longrightarrow
    s\,\equiv\,
\frac{S_{\rm BH}}{V_3}\,=\,\frac{\pi^2}{2}\,N_c^2
 T^3\,=\,\frac{3}{4}\,s_0 \,,\eeq
where $G_{10}$ is Newton constant in 10 dimensions and $V_3$ is the
volume of the physical 3--dimensional space (as usual, we set
$\hbar=c=k_B=1$). The last equality in \Eref{sbh} shows that the entropy
density $s$ of the ${\mathcal N}=4$ SYM plasma at {\em infinitely strong}
coupling is 3/4 of the corresponding quantity $s_0$ at {\em zero}
coupling ! Hence, in spite of the interactions being so strong, the
entropy does not deviate strongly from that of an ideal gas. The first
correction to this result at strong coupling, of order $1/\lambda^{3/2}$,
is also known \cite{Aharony:1999ti} and it is positive
--- meaning that the NLO result for $s$ is even closer to the respective
Stefan--Boltzmann limit. Even though such results cannot be directly
applied to QCD, they nevertheless suggest that the relatively small
deviations --- about 20\% in the temperature range relevant for HIC's,
cf. \Fref{fig:pressure} --- between the lattice results for the pressure
in QCD and the respective ideal gas limit are not necessarily in
contradiction with a strong coupling scenario.

We have previously mentioned that one important prediction of AdS/CFT is
the limiting value \eqref{Soneta} for the ratio $\eta/s$, which has been
conjectured to be a lower bound of nature \cite{Kovtun:2004de} (as it
holds at infinitely strong coupling for all the gauge theories having a
gravity dual). Let us sketch here the derivation of this result
\cite{Policastro:2001yc,Kovtun:2004de}. As explained around \Eref{shear},
the shear viscosity $\eta$ describes the response of the plasma to `shear
forces' (its ability to transfer momentum $p_x$ along the $y$ direction).
This is made precise by the Kubo formula which expresses $\eta$ as a
2--point function of the shear tensor $T_{xy}$ in thermal equilibrium:
 \beq\label{Kubo}
 \eta\,=\,\lim_{\omega\to 0}\,\frac{1}{2\omega}
\int \rmd t\rmd^3{\bm x}\,\rme^{-i\omega t}\,
 \langle \, [T_{xy}(t,{\bm x}), T_{xy}(0,{\bm 0})]
 \,\rangle_T\,.
 \eeq
This representation is exact, so in particular it holds at strong
coupling. In that case, the string--theory dual of the 2--point function
in the r.h.s. is the cross--section for the absorption of a  soft {\em
AdS graviton} (the supergravity field dual to the energy--momentum tensor
in the CFT) by the black hole. This cross--section is known from general
relativity: it is proportional to the area of the event horizon, like the
entropy. Thus, in this context, $\eta$ and $s$ are naturally proportional
with each other. The proportionality coefficient can be explicitly
computed, with the result that $\eta/s=1/(4\pi)$. It is remarkable that
the heavy ion data seem to favour a value which is close to this
conjectured lower bound.

From \Sref{sec:flow} we recall that the shear viscosity enters the
equations of hydrodynamics at {\em linear} order in the gradient
expansion. This corresponds to the fact that, in the respective Kubo
formula \eqref{Kubo}, $\eta$ is extracted from the term linear in
$\omega$ in the small frequency expansion of the 2--point function of the
shear tensor. By going up to the second order in this expansion, one can
similarly extract the transport coefficients for the {\em second--order}
formalism (which, we recall, are essential to provide a consistent
formulation of relativistic hydrodynamics). Interestingly, the
calculation of these coefficients turns out to be simpler at {\em strong
coupling}, where one can rely on AdS/CFT for that purpose, than at {\em
weak coupling}, where the respective calculations (say, using kinetic
theory) require the resummation of infinitely many Feynman graphs. And as
a matter of fact, the general structure of the second--order terms in the
equations of hydrodynamics has been clarified only recently, via AdS/CFT
calculations at strong coupling \cite{Baier:2007ix,Hubeny:2011hd}. Thus,
thanks to AdS/CFT, one can study the emergence of hydrodynamics from the
underlying fundamental field theory in a controlled way (at least on the
example of ${\mathcal N}=4$ SYM).

Since the string theory dual of a finite--$T$ plasma is a black hole in
AdS$_5$, it is natural that the phenomenon of thermalization at strong
coupling can be studied as the emergence of an event horizon in the
solution to the Einstein equations. We more precisely mean the equations
describing the deviation in the AdS$_5$ metric generated by some matter
distribution, which is initially out of equilibrium. For instance, the
collision between two heavy ions can be modeled as the scattering between
two gravitational shock waves --- two Lorentz contracted shells of matter
which propagate against each other and scatter via gravitational
interactions. This problem has been addressed within the supergravity
context, via analytic approximations
\cite{Albacete:2009ji,Gubser:2009sx,Lin:2009pn} and via exact, numerical,
calculations \cite{Chesler:2010bi}, with the conclusion that the system
evolves rather fast towards a (locally) isotropic distribution.
Interestingly, it appears that the only acceptable solution to the
Einstein equations which is boost invariant is the one describing the
emergence of perfect hydrodynamics (in the sense of \Eref{Tmunu}) at
asymptotically large times \cite{Janik:2005zt,Beuf:2009cx}.

Another phenomenon which is interesting to study at strong coupling is
jet quenching --- the energy loss by a `hard probe' (energetic parton)
propagating through a strongly coupled plasma. The corresponding AdS/CFT
calculations have been performed both for a very heavy quark (which loses
only a tiny fraction of its total energy) and for a light parton (quark,
gluon, or virtual photon), which can be totally stopped in the medium
\cite{Iancu:2008sp,Gubser:2009md,CasalderreySolana:2011us} (and Refs.
therein). Here we shall focus on the second case --- that of a light, but
very energetic, parton with original energy $E\gg T$. The corresponding
`dual' object on the supergravity side can be a semi--classical string
falling into AdS$_5$ (in the case of a light quark), a pair of such
strings (to describe a gluon), or a falling wavepacket carrying the
photon quantum numbers (for a virtual photon). The respective AdS/CFT
calculations
\cite{Hatta:2008tx,Gubser:2008as,Chesler:2008uy,Arnold:2010ir} revealed
that, in all such cases, the stopping distance over which the light
parton loses most of its energy through interactions in the medium scales
like
 \beq\label{stop}
  L_{\rm stop}\,\sim\,\frac{1}{T}\left(\frac{E}{T}\right)^{1/3}.
 \eeq
Note the difference w.r.t. the corresponding result at weak coupling,
which in \Sref{sec:jet} has been found to scale like $E^{1/2}$. This
reflects the difference between the respective mechanisms for energy
loss, that we shall now explain
\cite{Iancu:2008sp,Hatta:2008tx,Dominguez:2008vd}.

From \Sref{sec:jet}, we recall that the mechanism at work at weak
coupling is {\em medium--induced radiation} ---  the emission of gluons
stimulated by the interactions between the radiating system (the `hard
probe' and its partonic descendants) and the individual constituents of
the medium. In general, several such interactions can contribute to the
emission of a single gluon (the LPM effect), but the role of the
individual interactions is nevertheless well identified: they provide
transverse momentum kicks at a rate measured by the jet quenching
parameter, \Eref{dkdt}. This is in agreement with the fact that, at weak
coupling, the plasma is a collection of elementary constituents, or
`quasi--particles' (cf. \Sref{sec:thermo}), which are pointlike and
quasifree. But at strong coupling, we do not expect such a
quasi--particle picture to hold anymore --- rather, the plasma should
look homogeneous, without any microscopic substructure. And indeed, the
AdS/CFT results like \Eref{stop} can be understood by assuming that the
plasma acts on the external probe with a {\em uniform force} $F_T\sim
T^2$. This is like a gravitational force in the sense that it is fully
determined by the local energy density $\sim T^4$ in the plasma,
irrespective of its microscopic nature. The effect of this force on a
virtual parton (the `hard probe') is to stimulate gluon emission, via
{\em medium--induced parton branching}
\cite{Iancu:2008sp,Hatta:2008tx,Dominguez:2008vd}.

Specifically, a partonic fluctuation with energy $\omega$ and virtuality
$Q$ can decay under the action of the plasma force $F_T$ provided the
mechanical work $W=L\,F_T$ furnished by this force over a distance $L$ of
the order of the lifetime of the fluctuation ($L\sim \omega/Q^2$) is
large enough to compensate the parton virtuality. This condition implies
 \beq\label{QsAdS}
 \frac{\omega}{Q^2}\ T^2 \,\sim\ Q\ \Longrightarrow\ \ Q= Q_s(\omega)\,
 \sim\big(
 \omega
 T^2\big)^{1/3}\quad \& \quad \ L\,\simeq \,\frac{\omega}{Q^2_s(\omega)}
 \,\sim\,
 \frac{1}{T}\left(\frac{\omega}{T}\right)^{1/3},
 \eeq
in agreement with \Eref{stop}. More precisely, the above argument
provides the typical distance $L$ for the occurrence of one branching,
but this is of the same order of magnitude as the overall stopping
distance; indeed, the subsequent branchings involve gluons which are
softer and softer, and thus proceed faster and faster. For a given energy
$\omega\gg T$, any parton with initial virtuality $Q_0\le Q_s$ can decay
in this way, including the {\em space--like} photon exchanged in DIS (cf.
\Sref{sec:DIS}). Accordingly, the quantity $Q_s(\omega)$ plays also the
role of the {\em saturation momentum} for the finite--$T$ plasma at
strong coupling. However, unlike at weak coupling, where the phenomenon
of saturation requires large gluon occupation numbers $n\sim 1/\alpha_s$,
cf. \Eref{Qsat}, at strong coupling one can argue that it occurs for
occupation numbers of order one \cite{Hatta:2007cs,Iancu:2008sp}.


In the previous discussion we have implicitly assumed the plasma to be
infinite (or, at least, much larger than the stopping distance
\eqref{stop}). This means that, on the supergravity side, one has studied
the propagation of the `dual' objects in a metric describing a black hole
into AdS$_5$. The corresponding calculations for a {\em finite--size}
medium are more difficult, in particular because the corresponding metric
is more complicated. But one can at least heuristically revert the logic
leading to \Eref{stop} and conclude that, if an energetic parton
propagates through the medium over a finite distance $L$ without being
stopped, then the amount of energy lost by the particle scales like
$\Delta E \sim L^3$ \cite{Dominguez:2008vd}. Interestingly, this scaling
appears to be supported by some of the RHIC data \cite{Marquet:2009eq}.
This result at strong coupling should be contrasted with the
corresponding scaling--law at weak coupling, namely $\Delta E \sim \hat q
L^2$ (cf. \Sref{sec:jet}). This difference reflects the fact that the
medium--induced parton branching is not a {\em local} phenomenon (unlike
transverse momentum broadening at weak coupling), but is delocalized over
a distance of the order of the lifetime $E/Q^2_s$ of the decaying parton,
which in turn is commensurable with its stopping distance.

The above picture of medium--induced parton branching can also explain
the AdS/CFT results for the energy loss and the transverse momentum
broadening of a {\em heavy quark} propagating through a strongly coupled
plasma \cite{Gubser:2009md,CasalderreySolana:2011us} (and Refs. therein).
In that case, the variables $\omega$ and $Q$ which appear in \Eref{QsAdS}
refer to any of the quanta emitted by the heavy quark: among all the
virtual fluctuations of the latter, the only ones which can decay (and
thus take away energy and momentum) are those which, for a given energy
$\omega$, have a relatively small virtuality $Q\lesssim Q_s(\omega)$.
(Quanta with $Q\gg Q_s(\omega)$ cannot significantly interact with the
plasma and hence they are reabsorbed by the heavy quark.) The energy loss
is dominated by the most energetic among the emitted quanta
--- those having a boost factor $\omega/Q$ comparable to that
(denoted as $\gamma$) of the heavy quark. These two conditions,
$\omega/Q\simeq\gamma$ and $Q\lesssim Q_s(\omega)$, imply the following
upper limits on the energy and transverse momentum that can be taken away
by one emitted parton: $\omega\le \omega_{max}$ with $\omega_{max}\sim
\gamma Q_s(\omega_{max}) \sim \gamma^{3/2} T$ and, respectively, $\Delta
k_{\perp}\lesssim Q_{s}(\omega_{max})\sim \gamma^{1/2} T$. These maximal
values control the energy loss and the transverse momentum broadening of
the heavy quark. By also taking into account the typical duration
$\omega/Q^2$ of an emission, one finally deduces the following
expressions
\begin{eqnarray}\label{dEdt}
 -\,\frac{{\rm d} E}{{\rm d} t}\,\simeq\,\sqrt{\lambda}\,
 \frac{ \omega}{(\omega/Q_s^2)}\bigg |_{\omega_{max}}
 \,\simeq\,\sqrt{\lambda}\,Q_s^2 \,\sim\,\sqrt{\lambda}\,\gamma\,T^2
 \,.\end{eqnarray}
 \begin{eqnarray}\label{dpTdt}
 \frac{{\rm d} \langle k_\perp^2\rangle}{{\rm d} t}\,\sim\,
\frac{\sqrt{\lambda}\,Q_s^2}{(\omega/Q_s^2)} \,\sim\,
\sqrt{\lambda}\,\frac{Q_s^4}{\gamma Q_s}\,\sim\,
 \sqrt{\lambda}\,\sqrt{\gamma}\,T^3\,,\end{eqnarray}
for the respective rates. The factor $\sqrt{\lambda}$ in the r.h.s.'s of
these equations appears because the heavy quark is a semi--classical
object which acts as a colour source with a strength of order
$\sqrt{\lambda}$ at strong coupling --- meaning that it emits a number of
quanta (with given $\omega$ and $Q$) of order $\sqrt{\lambda}$ during the
formation time $\omega/Q^2$ of one such a quanta. As anticipated,
Eqs.~\eqref{dEdt}--\eqref{dpTdt} agree at parametric accuracy with the
respective results of the AdS/CFT calculations
\cite{Gubser:2009md,CasalderreySolana:2011us}.

\section*{Acknowledgments}
I would like to thank the organizers of the 2011 European School of
High--Energy Physics, notably Nick Evans, Christophe Grojean and Ionut
Ursu, for inviting me to present this series of lectures. During the
preparation of the lecture notes, I have benefited from many exchanges
and related discussions. I am particularly grateful to my colleagues who
patiently read these pages and helped me improved my presentation and
correct the misprints: Fabio Dominguez, Hanna Gr\"onqvist, Jean-Yves
Ollitrault, Dionysis Triantafyllopoulos, and Raju Venugopalan. Also, many
of the figures which appear in these notes have been graciously provided
by my colleagues in Saclay, notably Fran{\c c}ois Gelis (Figs.~1, 6 left,
15, 17 left, 25, 26 right, 32 right, and 42 right) and Matt Luzum (Fig.
28).


\providecommand{\href}[2]{#2}\begingroup\raggedright\endgroup

\end{document}